\documentclass[11pt]{article}

\usepackage[margin=1in]{geometry}
\usepackage{titlesec}
\titlespacing*{\paragraph}{0pt}{0.5ex}{1em} %controls the whitespace before paragraph
\usepackage{microtype,needspace}
\usepackage{times}

\usepackage{amsmath,amssymb,amsfonts,latexsym,amsthm,amsxtra}
\usepackage{mathtools,mathrsfs,bm,bbm,nicefrac}

\usepackage{enumerate,enumitem}
\usepackage{array,tabularx,longtable,booktabs,caption}
\usepackage{mdframed,pifont}

\usepackage{graphicx,xcolor,tikz}
\usetikzlibrary{arrows.meta,decorations.pathreplacing,patterns}
\usepackage{float}

\usepackage{subfiles}

\titleformat{\part}[hang]
  {\normalfont\huge\bfseries}
  {\partname~\thepart:}
  {0.5em}
  {}
\titlespacing*{\part}{0pt}{4ex}{3ex}
\usepackage[
  backend=biber,
  style=alphabetic,
  sorting=nyt,
  maxnames=7,
  maxalphanames=7,
  backref=true,
  doi=false,
  isbn=false,
  url=false
]{biblatex}
\usepackage[colorlinks=true]{hyperref}
\hypersetup{
  linkcolor={red!50!black},
  citecolor={blue!50!black},
  urlcolor={blue!80!black}
}
\usepackage[nameinlink,noabbrev,capitalize]{cleveref}

\newtheorem{globtheorem}{Theorem}

\newtheorem{counter}{Counter}[section]
\newtheorem{theorem}[counter]{Theorem}
\newtheorem{lemma}[counter]{Lemma}

\newtheorem{claim}[counter]{Claim}

\newtheorem{corollary}[counter]{Corollary}
\newtheorem{definition}[counter]{Definition}

\newtheorem{remark}[counter]{Remark}

\newtheorem{assumption}[counter]{Assumption}
\newtheorem{construction}[counter]{Construction}

\crefname{construction}{Construction}{Constructions}
\Crefname{construction}{Construction}{Constructions}
\crefname{assumption}{Assumption}{Assumptions}
\Crefname{assumption}{Assumption}{Assumptions}

\definecolor{figblue}{RGB}{76,114,176}
\definecolor{figorange}{RGB}{221,132,82}
\definecolor{figgreen}{RGB}{85,168,104}
\definecolor{figred}{RGB}{196,78,82}
\definecolor{figpurple}{RGB}{129,114,179}
\definecolor{figteal}{RGB}{100,181,205}
\definecolor{figyellow}{RGB}{204,185,116}

\newcommand{\N}{\mathbb{N}}

\newcommand{\F}{\mathbb{F}}

\newcommand{\zo}{\ensuremath{\{0,1\}}}
\newcommand{\Bi}{\{0,1\}} %Ronen's macro, used in intro

\newcommand{\cC}{\mathcal{C}}

\newcommand{\from}{\leftarrow}
\newcommand{\abs}[1]{\ensuremath{\left\vert#1\right\vert}}

\newcommand{\concat}{\mathbin\Vert}

\newcommand{\poly}{\operatorname{poly}}
\newcommand{\polylog}{\operatorname{polylog}}

\DeclareMathOperator*{\E}{\mathbb{E}}   %expectation
        
\DeclareMathOperator{\supp}{Supp}
\newcommand{\ind}{\mathbf 1}
\newcommand{\minH}{\mathbf{H}_{\infty}} %min-entropy
\newcommand{\Hi}{\mathbf{H}_{\infty}}   %min-entropy, used in intro

\newcommand{\TV}{\mathsf{SD}}           %statistical distance

\newcommand{\A}{\mathbf{A}}
\newcommand{\B}{\mathbf{B}}
\newcommand{\X}{\mathbf{X}}
\newcommand{\Y}{\mathbf{Y}}
\newcommand{\ZZ}{\mathbf{Z}}
\newcommand{\W}{\mathbf{W}}
\newcommand{\V}{\mathbf{V}}
\newcommand{\U}{\mathbf{U}}

\newcommand{\bL}{\mathbf{L}}
\newcommand{\bP}{\mathbf{P}}
\newcommand{\bR}{\mathbf{R}}
\newcommand{\bS}{\mathbf{S}}

\newcommand{\bv}{\mathbf{v}}

\newcommand{\bu}{\mathbf{u}}
\newcommand{\bb}{\mathbf{b}}

\newcommand{\sfE}{\mathsf{E}}
\newcommand{\EXP}{\mathsf{EXP}}
\newcommand{\NP}{\mathsf{NP}}
\newcommand{\Pclass}{\mathsf{P}}
\newcommand{\BPP}{\mathsf{BPP}}
\newcommand{\AM}{\mathsf{AM}}
\newcommand{\SAT}{\mathsf{SAT}}

\DeclareMathOperator{\DTime}{\mathsf{DTIME}}
\DeclareMathOperator{\DTIME}{\mathsf{DTIME}}    %used in intro
\newcommand{\NDH}{\mathsf H_{\mathrm{NP}}}      %H_NP
\newcommand{\SH}{\mathsf H_{\Sigma_2}}          %H_sigma2

\DeclareMathOperator{\Samp}{\mathsf{Samp}}
\DeclareMathOperator{\Post}{\mathsf{Post}}

\newcommand{\Src}{\operatorname{Src}}
\newcommand{\NPSources}{\mathcal C^{\mathrm{NP}}}

\newcommand{\Ext}{\operatorname{\mathsf{Ext}}}
\newcommand{\SExt}{\operatorname{\mathsf{SExt}}}    %seeded extractor

\newcommand{\TExt}{\operatorname{\mathsf{TExt}}}    %two-source
\newcommand{\IExt}{\operatorname{\mathsf{IExt}}}

\newcommand{\NMExt}{\operatorname{\mathsf{2NMExt}}}

\newcommand{\Cond}{\operatorname{\mathsf{Cond}}}
\newcommand{\rbCond}{\operatorname{\mathsf{rbCond}}}
\newcommand{\TwoCond}{\operatorname{\mathsf{2Cond}}}

\newcommand{\CB}{\mathsf{CB}}

\newcommand{\IP}{\mathsf{IP}}
\newcommand{\Maj}{\mathsf{Maj}}

\newcommand{\CBlab}{\CB_{\mathrm{lab}}}
\newcommand{\CBext}{\CB_{\mathrm{ext}}}
\newcommand{\CBmult}{\CB_{\mathrm{mult}}}

\newcommand{\RandLine}{\mathsf{RandLine}}
\newcommand{\Thr}{\operatorname{Thr}}
\newcommand{\Test}{\operatorname{Test}}
\newcommand{\ACCEPT}{\operatorname{ACCEPT}}

\newcommand{\hatf}{\widehat{f}}

\newcommand{\eps}{\varepsilon}

\newcommand{\good}{\text{good}}

\newcommand{\rest}{\text{rest}}

\newcommand{\out}{\text{out}}
\newcommand{\inn}{\text{in}}

\newcommand{\Low}{\text{Low}}
\newcommand{\size}{\mathrm{size}}

\newcommand{\dobib}{
    \printbibliography
}
\newcommand\numberthis{\addtocounter{equation}{1}\tag{\theequation}}

\title{Multiplicative and Additive Extractors for Low Entropy Samplable Sources Under Reduced Assumptions}

\author{%
\makebox[0pt][c]{
   \begin{tabular}{ccc}
    Marshall Ball\thanks{Supported by NSF Awards CCF-2443735 and CCF-2552209.} &
    Eshan Chattopadhyay\thanks{Supported by  NSF Award CCF-2514586.} &
    Mohit Gurumukhani\footnotemark[2] \\
     New York University  & Cornell University & Cornell University \\
   \small\texttt{marshall.ball@cs.nyu.edu} & \small\texttt{eshan@cs.cornell.edu}  & \small\texttt{mgurumuk@cs.cornell.edu}
  \end{tabular}
}
  \\[2em]
\makebox[0pt][c]{
  \begin{tabular}{cc}
    Ronen Shaltiel\thanks{Supported by ISF grant 1006/23 and also co-funded by the European Union (ERC, NFITSC, 101097959). Views and opinions expressed are however those of the author(s) only and do not necessarily reflect those of the European Union or the European Research Council. Neither the European Union nor the granting authority can be held responsible for them.}  &
    Yunya Zhao\footnotemark[2]  \\
    University of Haifa &  Cornell University \\
    \small\texttt{ronen@cs.haifa.ac.il} & \small\texttt{yunya@cs.cornell.edu} 
  \end{tabular}
}
}

\date{}

\begin{document}
% Suppress bibliographies in included sections; print one bibliography below.
\renewcommand{\dobib}{}

\pagenumbering{Roman}

\maketitle

\vspace{-0.2in}
\begin{abstract}
Trevisan and Vadhan [TV, FOCS'00] constructed (seedless) extractors for distributions samplable by poly-size circuits that have very high min-entropy, under the very strong complexity theoretic hardness assumption that $\mathsf{E}=\mathsf{DTIME}(2^{O(n)})$ requires size $2^{\Omega(n)}$ circuits that have oracle access to $\Sigma_6^{\text{P}}$. 
In this paper we give three new explicit constructions of extractors for samplable distributions.

First, we construct an extractor that extracts one bit, with constant error, for min-entropy $k=O(\log n)$, which matches the non-constructive probabilistic argument. This result uses hardness for nondeterministic circuits, and achieves the same parameters as the recent extractor of Oh and Shaltiel [OS, STOC'26] under an incomparable hardness assumption. In contrast to the extractor of [OS26], our extractor (and, similarly, the constructions below) also works for the more general class of distributions samplable with postselection (a model introduced by [BGDM23] that generalizes samplable and recognizable sources).
    
Second, we construct an extractor that extracts almost all the entropy, with polynomially small error, for min-entropy $k=n^{\alpha}$, where $\alpha>0$ is an arbitrary constant. This result also uses hardness for nondeterministic circuits, and significantly improves upon the recent extractor of [Shaltiel, FOCS'25] which required a much stronger assumption. This extractor is also \emph{multiplicative}: a stronger notion suggested in [AASY, Comp. Complexity'16; Shaltiel, CCC'25] to make extractors for samplable distributions suitable for selecting keys for cryptographic protocols.
    
Third, we construct an extractor that extracts almost all the entropy, with polynomially small error, for min-entropy $k=\operatorname{polylog}(n)$. This result uses hardness for circuits that have oracle access to $\Sigma_2^{\text{P}}$. 

Prior to this work, the lowest min-entropy threshold $k$ that was achieved using hardness for nondeterministic circuits (or even circuits with oracle access to $\Sigma_2^{\text{P}}$) was larger than $0.99n$ [BGDM, FOCS'23]. Our results significantly improve both the min-entropy threshold and the hardness assumptions compared to all past work that extracts more than one bit.

Our constructions build upon earlier work [TV, FOCS'00; BGDM, FOCS'23; BSS, STOC'25; Shaltiel, FOCS'25] while introducing several new ideas: (1) We simplify and build upon the ideas of [BGDM, FOCS'23] to construct our extractors under nondeterministic hardness. (2) We introduce a notion of robust errorless condensers and construct and use them to obtain our multiplicative extractors. (3) We adapt the two-source extractor recipe from [CZ, Ann. of Mathematics'19] to construct our additive extractors. We do this by constructing and using a correlation breaker against small circuits. Our correlation breaker extends recent constructions of non-malleable extractors against small circuits [BDL, CRYPTO'22; BD, CRYPTO'25]. 
\end{abstract}

% Insert any abstract here!

\clearpage
\setcounter{tocdepth}{3}
\tableofcontents
\newpage
\pagenumbering{arabic}

\section{Introduction}
\label{sec:introduction}

\subsection{Background}

Randomized algorithms and cryptographic protocols typically assume access to uniform random bits. Available sources of randomness, however, may be biased and exhibit dependencies. Randomness extractors aim to convert such ``weak sources of randomness'' into distributions that are close to uniform, assuming that the source contains sufficient randomness (measured by min-entropy). Deterministic extraction is impossible for general sources with high min-entropy. We therefore restrict the class of sources, seeking a single extractor that works for every source in the class that has sufficient min-entropy.

\begin{definition}[Seedless extractor]
\label{dfn:intro:seedless extractor}
A function $\Ext:\Bi^n \to \Bi^m$ is a \emph{$(k,\eps)$-extractor} for a class $\mathcal{D}$ of distributions, if for every distribution $X \in \mathcal{D}$ over $\Bi^n$ with $\Hi(X) \ge k$, $\Ext(X)$ is $\eps$-close to~$U_m$.\footnote{See \cref{sec:standard-definitions}
 for the standard definitions of min-entropy and statistical distance.}
\end{definition}

An influential paper by Trevisan and Vadhan \cite{TV00} introduced the notion of (seedless) extractors for samplable distributions.
The goal of Trevisan and Vadhan was to identify a class of distributions that contains sources of randomness that are ``available to computers'', and allows seedless extractors that run in poly-time.

\begin{definition}[Samplable distributions]
\label{dfn:intro:samplable}
We say that a distribution $X$ over $\Bi^n$ is \emph{sampled} by a function $\Samp:\Bi^r \to \Bi^n$ if $X=\Samp(U_r)$. $X$ is \emph{samplable} by a class $\mathcal{C}$ of functions, if there exists $\Samp \in \mathcal{C}$ that samples~$X$.
\end{definition}

Trevisan and Vadhan made the philosophical assumption that ``nature is efficient'' and that therefore, weak sources of randomness that are ``available to computers'' are efficiently samplable. They considered extractors for distributions that are samplable by poly-size circuits, namely distributions samplable by circuits of size $n^c$ for some constant parameter $c$. They showed that such extractors cannot run in time smaller than $n^c$, and therefore considered extractors that run in time $n^d$ for a constant $d>c$.
They showed that such extractors imply circuit lower bounds, and so, motivated by the hardness vs. randomness paradigm, they gave a conditional construction based on hardness assumptions.

\paragraph{Hardness assumptions against various types of nondeterministic circuits.}
We say that ``$\sfE$ is hard for exponential size circuits of some type'', if there exists a problem $L \in \sfE=\DTIME(2^{O(n)})$ and a constant $\beta>0$, such that for every sufficiently large $n$, circuits of size $2^{\beta \cdot n}$ (of the specified type) fail to compute the characteristic function of $L$ on inputs of length $n$. (See \cref{sec:hardness} for a more formal definition).

The assumption that $\sfE$ is hard for exponential size (deterministic) circuits was used by the celebrated paper of Impagliazzo and Wigderson \cite{impagliazzo1997p} to imply that $\BPP=\Pclass$.
The stronger assumption that $\sfE$ is hard for exponential size nondeterministic circuits\footnote{A precise definition of nondeterministic circuits appears in \cref{sec:hardness}.}, originated in works on hardness versus randomness for $\AM$, and is used in many results \cite{AK97,KvM,MV99,SU05,BarakOngVadhan,GW02,GST03,SU06,
SU07,Drucker,
AASY16,BitanskyV17,AIKS16,HubacekNY17,DMOZ,BDL22,ChenTellAM22,BDGM23,BallSS24,SS24,Shaltiel25multiplicative}. It can be viewed as a scaled, nonuniform version of the widely believed assumption that $\EXP \ne \NP$.\footnote{We remark that the recent refutation of the ETH assumption, does not refute any of these assumptions.}

In their seminal paper on extractors for samplable distributions, Trevisan and Vadhan \cite{TV00} introduced a version of the assumption for a stronger circuit class. A $\Sigma_i$-circuit, is a circuit that in addition to the standard gates, is also allowed to use a special gate (with large fan-in) that solves the canonical complete language for the class $\Sigma_i^{\text{P}}$ (the $i$'th level of the polynomial time hierarchy).\footnote{A $\Sigma_i$-circuit is a nonuniform analogue of the class $\Pclass^{\Sigma_i^{\text{P}}}$ that contains $\Sigma_i^{\text{P}}$, and recall that $\Pclass=\Sigma_0^{\text{P}}$ and $\NP=\Sigma_1^{\text{P}}$. See \cref{def: sigma i circuit} for a formal definition.}
The extractor of Trevisan and Vadhan \cite{TV00} relies on the strong assumption that
$\sfE$ is hard for exponential size $\Sigma_6$-circuits (which can be viewed as a scaled, nonuniform version of the widely believed assumption that $\EXP \ne \Sigma_6^{\text{P}}$). 

\paragraph{Previous work on extractors for samplable distributions.}

The main result of Trevisan and Vadhan \cite{TV00} is that under a hardness assumption for $\Sigma_6$-circuits, there is an extractor for distributions samplable by poly-size circuits with $k=(1-\eta) \cdot n$, for some small constant $\eta>0$. Below is a precise statement.

\begin{theorem}[\cite{TV00}]
\label{thm:TV}
If $\sfE$ is hard for exponential size $\Sigma_6$-circuits then for every constant $\gamma>0$, and every constant $c>1$, there are constants $\eta>0$ and $d>c$ such that for every sufficiently large $n$, and every $(1-\eta)n \le k\le n$ there is a function $\Ext:\Bi^n \to \Bi^{m}$ that is a $(k,\eps)$-extractor for distributions samplable by circuits of size $n^c$, where $\eps=n^{-c}$ and $m=(1-\gamma)k$. Furthermore, $\Ext$ is computable in time $n^d$.
\end{theorem}

\noindent
Recent work on extractors for samplable distributions improved \cref{thm:TV} in several respects.
\begin{description}
\item[Weaker hardness assumptions:] Ball, Goldin, Dachman-Soled and Mutreja \cite{BDGM23} (see also Shaltiel \cite{Shaltiel25multiplicative}) achieved the conclusion of \cref{thm:TV} under the weaker, and more standard assumption that $\sfE$ is hard for exponential size nondeterministic circuits. Prior to this work, this was the lowest min-entropy threshold achieved under this hardness assumption. The objective of this paper is to construct extractors for sources with substantially lower min-entropy under the same assumption.
\item[Extractors for samplable distributions with low min-entropy:] Ball, Shaltiel and Silbak \cite{BSS25} improved the min-entropy threshold of Trevisan and Vadhan to $k=n^{1-\gamma}$ under a hardness assumption for $\Sigma_5$-circuits. Shaltiel \cite{shaltiel25extract} improved the min-entropy threshold to $k=n^{1/i}$ for an arbitrary constant $i$, at the cost of a hardness assumption for $\Sigma_{i+3}$-circuits. 
\item[Different hardness assumptions:] Oh and Shaltiel \cite{OS2026}  gave an extractor construction under a different and incomparable hardness assumption; the extractor of \cite{OS2026} has the advantage that it achieves very low min-entropy of $k=O(\log n)$, but it only extracts one bit, with constant error.
\item[Samplable distributions with postselection:] Ball, Goldin, Dachman-Soled and Mutreja \cite{BDGM23} considered a generalization of samplable distributions which allows ``postselection''. In this model, there are two functions: a sampler $\Samp:\Bi^r \to \Bi^n$, and a post-selector, $\Post:\Bi^r \to \Bi$, and we say that $X$ is sampled by $\Samp$ with postselection by $\Post$, if $X$ is the distribution $(\Samp(R)|\Post(R)=1)$, for $R \from U_r$. Note that this is indeed a generalization of samplable distributions, as we can take the circuit $\Post(R)$ that outputs one on every input. One motivation for such sources, suggested in \cite{BDGM23}, is that this experiment captures a scenario where information about the sampled value $\Samp(R)$ is leaked to an adversary in the form of an efficiently computable leakage function $L(R)$. Postselection captures conditioning on any specific value of $L(R)$. This means that extraction remains possible even given the leakage, provided sufficient min-entropy remains.

With the exception of the aforementioned extractor of Oh and Shaltiel \cite{OS2026}, all constructions in the literature (including the constructions of this paper) apply also to samplable distributions with postselection. In fact, this holds even if the circuit $\Post$ is allowed to be a size $n^c$ nondeterministic circuit. 
Shaltiel \cite{Shaltiel25multiplicative} shows that extractors for this richer class of distributions imply lower bounds against nondeterministic circuits.\footnote{It is not known whether explicit extractors for distributions samplable by deterministic circuits with post selection by nondeterministic circuits imply that $\sfE$ is hard for exponential size nondeterministic circuits. What is shown in \cite{Shaltiel25multiplicative} is that such extractors cannot be computed (even on the average) by nondeterministic circuits of size slightly smaller than $n^c$.}
\end{description}
A detailed comparison of past work and our new results appears in \cref{tab:samplable-extractors}.

\subsection{Our Results}
\label{sec:intro:our results}

In this paper we give constructions of both standard extractors, and of multiplicative extractors (that are defined in \cref{sec:intro:results mult}).

\subsubsection{Standard (Additive) Extractors}

We construct extractors for samplable distributions with min-entropy $k=O(\log n)$. This min-entropy threshold $k$ matches the one achieved by the nonconstructive probabilistic argument (up to constants). Our first result assumes hardness for nondeterministic circuits, but only extracts one bit with constant error.

\begin{globtheorem}[extracting one bit for $k=O(\log n)$ under nondeterministic hardness]
\label{thm:intro:main:one bit}
If $\sfE$ is hard for exponential size nondeterministic circuits then for every constant $\eps>0$, and every constant $c>1$, there are constants $d,a$ such that for every sufficiently large $n$, and every $a\log n \le k\le n$ there is a function $\Ext:\Bi^n \to \Bi$ that is a $(k,\eps)$-extractor for distributions samplable by circuits of size $n^c$. Furthermore, $\Ext$ is computable in time $n^d$.
\end{globtheorem}

The parameters of the extractor of \cref{thm:intro:main:one bit}  match the parameters of the extractor of \cite{OS2026}. The hardness assumptions in these two results are incomparable. 

The extractor of \cref{thm:intro:main:one bit} works for distributions that are samplable by deterministic circuits with postselection by nondeterministic circuits (see \cref{thm:log-extractor} for a precise statement) 
whereas that of \cite{OS2026} is not known to allow samplable distributions with postselection, even if postselection is by deterministic circuits.

Our next result gives an extractor that outputs almost all the randomness with polynomially small error. This is achieved under hardness for nondeterministic circuits, but applies only for $k=n^{\alpha}$ for an arbitrary constant $\alpha>0$.

\begin{globtheorem}[extracting many bits for $k=n^{\alpha}$ under nondeterministic hardness]
\label{thm:intro:main:middle}
If $\sfE$ is hard for exponential size nondeterministic circuits then for all constants $\alpha,\gamma>0$, and every constant $c>1$, there is a constant $d$ such that for every sufficiently large $n$, and every $n^{\alpha} \le k \le n$ there is a function $\Ext:\Bi^n \to \Bi^m$ that is a $(k,\eps)$-extractor for distributions samplable by circuits of size $n^c$, with $\eps=n^{-c}$ and $m=(1-\gamma)k$. Furthermore, $\Ext$ is computable in time $n^d$.
\end{globtheorem}

This extractor has exactly the same parameters as the extractor construction of Shaltiel \cite{shaltiel25extract}. However, whereas Shaltiel's extractor assumes hardness for $\Sigma_i$-circuits with $i=3+1/\alpha$, \cref{thm:intro:main:middle} assumes only hardness for nondeterministic circuits.
See \cref{thm:additive from multiplicative} for the general statement.

We are also able to construct extractors that extract almost all of the randomness from sources with min-entropy $k=\polylog(n)$ with error $\eps=n^{-c}$. However, for this result, we use a hardness assumption against $\Sigma_2$-circuits.

\begin{globtheorem}[extracting many bits for $k=\polylog(n)$ under $\Sigma_2$-hardness]
\label{thm:intro:main:many bits}
If $\sfE$ is hard for exponential size $\Sigma_2$-circuits then there exists a constant $a$ such that for every constant $\gamma>0$, and every constant $c>1$, there is a constant $d$ such that for every sufficiently large $n$, and every $\log^a n \le k \le n$ there is a function $\Ext:\Bi^n \to \Bi^m$ that is a $(k,\eps)$-extractor for distributions samplable by circuits of size $n^c$, with $\eps=n^{-c}$ and $m=(1-\gamma)k$. Furthermore, $\Ext$ is computable in time $n^d$.
\end{globtheorem}

Compared with previous constructions that extract many bits \cite{BSS25,shaltiel25extract}, \cref{thm:intro:main:many bits} achieves both a lower min-entropy threshold and a weaker hardness assumption. See \cref{poly:thm:main} for the general statement.

We remark that the hardness assumption in \cref{thm:intro:main:many bits} can be weakened to hardness for nondeterministic circuits, at the cost of allowing the extractor to run in time $2^{\log^2 n}$. See a more general statement in \cref{many:thm:almost-all}.

\begin{table}[t]
  \centering
  \small
  \def\bcell#1{{\bfseries\boldmath #1}}
  \captionsetup{font=small,position=top,skip=6pt}
  \setlength{\tabcolsep}{3pt}
  \renewcommand{\arraystretch}{1.2}

  \caption{Constructions of extractors for samplable distributions.\\
    Below, $\gamma>0$ is an arbitrarily small constant,
    $\eta>0$ is a sufficiently small constant,
    $\alpha>0$ and $c\ge 1$ are arbitrary constants, and
    $i>1$ is an arbitrary integer.}
  \label{tab:samplable-extractors}

  \begin{tabularx}{\textwidth}{
    |>{\raggedright\arraybackslash}p{0.13\textwidth}
    |>{\raggedright\arraybackslash}X
    |c|c|c|c|
  }
    \hline
    Reference
    & Hardness assumption
    & \shortstack{Threshold\\$k$}
    & \shortstack{Output\\$m$}
    & \shortstack{Error\\$\eps$}
    & \shortstack{Post-\\selection?}
    \\
    \hline

    \cite{TV00}
    & $\sfE$ is hard for exponential-size $\Sigma_6$-circuits
    & $(1-\eta)n$
    & $(1-\gamma)k$
    & $n^{-c}$
    & Yes
    \\
    \hline

    \cite{BDGM23}
    & $\sfE$ is hard for exponential-size nondeterministic circuits
    & $(1-\eta)n$
    & $(1-\gamma)k$
    & $n^{-c}$
    & Yes
    \\
    \hline

    \cite{BSS25}
    & $\sfE$ is hard for exponential-size $\Sigma_5$-circuits
    & $n^{1-\gamma}$
    & $(1-\gamma)k$
    & $n^{-c}$
    & Yes
    \\
    \hline

    \cite{shaltiel25extract}
    & $\sfE$ is hard for exponential-size $\Sigma_{i+3}$-circuits
    & $n^{1/i}$
    & $(1-\gamma)k$
    & $n^{-c}$
    & Yes
    \\
    \hline

    \cite{OS2026}
    & An incomparable hardness assumption
    & $O(\log n)$
    & $1$
    & $\Theta(1)$
    & No
    \\
    \hline\hline

    \bcell{This work\newline
      Thm~\ref{thm:intro:main:one bit}}
    & \bcell{$\sfE$ is hard for exponential-size nondeterministic circuits}
    & \bcell{$O(\log n)$}
    & \bcell{$1$}
    & \bcell{$\Theta(1)$}
    & \bcell{Yes}
    \\
    \hline

    \bcell{This work\newline
      Thm~\ref{thm:intro:main:middle}}
    & \bcell{$\sfE$ is hard for exponential-size nondeterministic circuits}
    & \bcell{$n^\alpha$}
    & \bcell{$(1-\gamma)k$}
    & \bcell{$n^{-c}$}
    & \bcell{Yes}
    \\
    \hline

    \bcell{This work\newline
      Thm~\ref{thm:intro:main:many bits}}
    & \bcell{$\sfE$ is hard for exponential-size $\Sigma_2$-circuits}
    & \bcell{$\polylog(n)$}
    & \bcell{$(1-\gamma)k$}
    & \bcell{$n^{-c}$}
    & \bcell{Yes}
    \\
    \hline
  \end{tabularx}
\end{table}

\subsubsection{Multiplicative Extractors}
\label{sec:intro:results mult}

A signature application of seedless extractors is choosing keys for cryptographic protocols by extracting randomness from weak random sources. (Indeed, this was the original motivation of Trevisan and Vadhan \cite{TV00} for introducing extractors for samplable distributions). More specifically, consider a cryptographic protocol which is known to be secure when the key of an honest party is chosen according to $U_m$. That is, the probability that an adversary can steal the honest party's money is smaller than some ``negligible'' $\mu>0$.

If the key is chosen using the output of an extractor (which is only $\eps$-close to uniform) then we are only guaranteed that the adversary's probability to cheat is smaller than $\mu + \eps$, which may be unacceptable if $\eps$ is ``large'' compared to $\mu$.

Applebaum, Artemenko, Shaltiel and Yang \cite{AASY16} showed that ``current techniques'' cannot yield extractors for samplable distributions with error $\eps=n^{-\omega(1)}$, under a hardness assumption against $\Sigma_i$-circuits, and this holds for every constant $i$. (See \cite{AASY16} for a precise formulation). This means that with current extractors for samplable distributions (which only achieve $\eps=n^{-c}$) we cannot expect cryptographic protocols to maintain negligible probability of cheating, when selecting the key with an extractor.

In order to address this problem, Applebaum et al. \cite{AASY16}, and later Shaltiel \cite{Shaltiel25multiplicative}, introduced a notion of ``multiplicative extractors'' which are designed to maintain negligible probability of cheating in cryptographic protocols. Below, we use the definition of \cite{Shaltiel25multiplicative} which is inspired by choices made in definitions in differential privacy \cite{DworkMNS06}.

\begin{definition}[Multiplicative seedless extractor \cite{Shaltiel25multiplicative}]
\label{dfn:intro:mult extractor}
A function $\Ext:\Bi^n \to \Bi^m$ is a \emph{$(k,\eps)$-multiplicative extractor} for a class $\mathcal{D}$ of distributions, if for every distribution $X \in \mathcal{D}$ over $\Bi^n$ with $\Hi(X) \ge k$, and every event $A \subseteq \Bi^m$
\[ \Pr[\Ext(X)\in A] \le e^{\eps} \cdot \Pr[U_m \in A]. \]
\end{definition}

It is easy to see that every multiplicative extractor is also a standard extractor.\footnote{Indeed, it is easy to see that a $(k, \eps)$ multiplicative extractor is a $(k,O(\eps))$ additive extractor, see \cite{Shaltiel25multiplicative}.}
However, unlike regular extractors, multiplicative extractors apply for the application of selecting keys for cryptographic protocols, even if $\eps>0$ is a small constant. This is because such extractors give that the adversary's probability to cheat is at most $e^{\eps} \cdot \mu \le (1+O(\eps)) \cdot \mu$.\footnote{This argument applies to ``unpredictability security games'' where one bounds the probability that the adversary can cheat, but not necessarily to ``indistinguishability security games'' where one bounds the probability that the adversary can distinguish between two distributions. See e.g. \cite{DY13} for a discussion.}

Multiplicative extractors for samplable distributions are harder to construct. Before this paper, the only construction for min-entropy $k<n/2$ was given by Shaltiel \cite{shaltiel25extract}, who achieved $k=n^\alpha$ for every constant $\alpha>0$.

In this paper, we give a multiplicative extractor with the same parameters.   Shaltiel’s extractor assumes hardness for $\Sigma_i$-circuits with $i=1/\alpha+O(1)$; in contrast, our extractor assumes only hardness for nondeterministic circuits.

\begin{table}[t]
  \centering
  \small
  \def\bcell#1{{\bfseries\boldmath #1}}
  \captionsetup{font=small,position=top,skip=6pt}
  \setlength{\tabcolsep}{3pt}
  \renewcommand{\arraystretch}{1.2}

  \caption{Constructions of multiplicative extractors for samplable distributions.\\
    Below, $\eta>0$ is a sufficiently small constant,
    $\alpha>0$ and $c\ge 1$ are arbitrary constants, and
    $i>1$ is an arbitrary integer.}
  \label{tab:multiplicative-samplable-extractors}

  \begin{tabularx}{\textwidth}{
    |>{\raggedright\arraybackslash}p{0.13\textwidth}
    |>{\raggedright\arraybackslash}X
    |c|c|c|c|
  }
    \hline
    Reference
    & Hardness assumption
    & \shortstack{Threshold\\$k$}
    & \shortstack{Output\\$m$}
    & \shortstack{Error\\$\eps$}
    & \shortstack{Post-\\selection?}
    \\
    \hline

    \cite{AASY16}
    & $\sfE$ is hard for exponential-size $\Sigma_4$-circuits
    & $(1-\eta)n$
    & $\eta n$
    & $n^{-c}$
    & Yes
    \\
    \hline

    \cite{Shaltiel25multiplicative}
    & $\sfE$ is hard for exponential-size nondeterministic circuits
    & $(1-\eta)n$
    & $\eta n$
    & $n^{-c}$
    & Yes
    \\
    \hline

    \cite{shaltiel25extract}
    & $\sfE$ is hard for exponential-size $\Sigma_{i+O(1)}$-circuits
    & $n^{1/i}$
    & $k^{\Omega(1)}$
    & $n^{-c}$
    & Yes
    \\
    \hline\hline

    \bcell{This work\newline
      Thm~\ref{thm:intro:main:mult}}
    & \bcell{$\sfE$ is hard for exponential-size nondeterministic circuits}
    & \bcell{$n^\alpha$}
    & \bcell{$k^{\Omega(1)}$}
    & \bcell{$n^{-c}$}
    & \bcell{Yes}
    \\
    \hline
  \end{tabularx}
\end{table}

\begin{globtheorem}[extracting many bits, under nondeterministic hardness]
\label{thm:intro:main:mult}
If $\sfE$ is hard for exponential size nondeterministic circuits then for every constant $\alpha>0$, and every constant $c>1$, there is a constant $d$ such that for every sufficiently large $n$, and every $n^{\alpha} \le k \le n$ there is a function $\Ext:\Bi^n \to \Bi^m$ that is a $(k,\eps)$-multiplicative extractor for distributions samplable by circuits of size $n^c$, with $\eps=n^{-c}$ and $m=k^{\Omega(1)}$. Furthermore, $\Ext$ is computable in time $n^d$.
\end{globtheorem}

As in our previous theorems, this extractor can handle distributions that are samplable by deterministic circuits with postselection by nondeterministic circuits (see \cref{thm:multiplicative}).

We remark that in the intended application of selecting keys for cryptographic protocols, we typically need $m=\sigma^{\Theta(1)}$ random bits, where $\sigma$ is the security parameter. As we are only allowed to run in time $\poly(\sigma)$, the source length $n$ must be polynomially bounded in $\sigma$. We therefore need an extractor with output length $m=n^{\Omega(1)}$, which requires min-entropy $k=n^{\Omega(1)}$. \cref{thm:intro:main:mult} matches these parameters.

\dobib

\section{Overview of the Techniques}
\label{sec: proof overview}

In this section, we give a detailed informal overview of the main ideas that we use to construct our multiplicative extractors (\cref{subsec: mult proof overview}) and additive extractors (\cref{subsec: additive proof overview}).
The later technical sections contain full definitions, statements and proofs and do not build on the informal explanation of this section.
The readers can skip to the technical section if they wish.

\subsection{Proof Overview for Multiplicative Extractors}
\label{subsec: mult proof overview}

We provide a sketch of our multiplicative extractor construction, as guaranteed by \cref{thm:intro:main:mult}, for $n^{0.01}$ entropy under nondeterministic hardness.
The same construction directly yields additive extractors with the same parameters as well (see \cref{thm:additive from multiplicative}).
Our construction synthesizes and builds on several ideas across prior works \cite{BDGM23, BSS25, shaltiel25extract}.

We follow the high level strategy from \cite{BSS25,shaltiel25extract}, of constructing errorless condensers and using them to construct multiplicative extractors. 

\begin{definition}[Errorless Condenser, Informal version of \cref{def:errorless-condenser}]
We say $\Cond: \zo^n\to \zo^m$ is a $(k_{\inn}, k_{\out})$-errorless condenser for a class $\cC$ of sources if for all $\X\in \cC$ with $\minH(\X) \ge k_{\inn}$, we have that $\minH(\Cond(\X)) \ge k_{\out}$.
\end{definition}

Unlike traditional goals of condensers where one wants to increase the entropy rate, our goal with an errorless condenser is to make the output length as small as possible, while retaining some non-trivial entropy in the output distribution.
For our applications, we require the output length of $\Cond$ to be smaller than the min-entropy of the input distribution $\X$.
We will construct the following errorless condensers.
\begin{theorem}[Informal version of \cref{thm:errorless-condenser}]
\label{thm:overview:errorless-condenser}
Assume $\sfE$ is hard for exponential-size nondeterministic circuits. Then, there exists $\delta > 0$ such that for all $\gamma > 0$, we can construct an explicit $(n^{\gamma}, n^{\gamma\cdot \delta})$-errorless condenser $\Cond: \zo^n \to \zo^{n^{\gamma/2}}$ for samplable sources.
\end{theorem}

Equipped with this errorless condenser, we will build on ideas from \cite{BDGM23} to construct our multiplicative extractor under nondeterministic hardness. 
To construct such an errorless condenser, as well as to construct our extractor from errorless condenser, we introduce a new pseudorandom object which we call a \emph{robust errorless block condenser}. This is an errorless condenser for block sources which are subsources of a joint distribution.\footnote{$(\A, \B)$ form a block source if $\A$ has entropy, $\B$ has entropy conditioned on every fixing of $\A$. See \cref{def:block-source} for a formal definition.} Formally,

\begin{definition}[Robust errorless block condenser, Informal version of \cref{def:block-robust}]
We say $\rbCond\colon\zo^{n_y+n_x}\longrightarrow\zo^m$ is a $(\lambda,k_y,k_y^{\ge},k_x,k_{\out})$-robust errorless block condenser for class of sources $(\mathcal{Y, X})$ if for every $(\Y,\X)\in(\mathcal{Y, X})$ and every decomposition
\[
 (\Y,\X)=\lambda(\Y_{\good},\X_{\good}) +(1-\lambda)(\Y_{\rest},\X_{\rest})
\]
satisfying:
(1) $\minH(\Y_{\good})\ge k_y$;
(2) $\Pr[\Y_{\good}=y]\ge 2^{-k_y^{\ge}}$ for every $y\in\supp(\Y_{\good})$;
(3) $\minH(\X)\ge k_x$,
% \begin{itemize}
%  \item $\minH(\Y_{\good})\ge k_y$;
%  \item $\Pr[\Y_{\good}=y]\ge 2^{-k_y^{\ge}}$ for every
%        $y\in\supp(\Y_{\good})$;
%  \item $\minH(\X)\ge k_x$;
% \end{itemize}
we have
\[
 \minH\bigl(\rbCond(\Y_{\mathrm{good}},\X_{\mathrm{good}})\bigr)
 \ge k_{\mathrm{out}}.
\]
\end{definition}

We will ensure that $k_y^{\ge} < k_x$ so that $(\Y_\good, \X_\good)$ form a block source.\footnote{Loosely speaking, conditioning on fixing of $\Y_\good$, $\X_\good$ will output an element with probability at most $\lambda^{-1}\cdot 2^{-k_x + k_y^{\ge}}$, bringing its conditional entropy to $k_x - k_y^{\ge} - \log(1/\lambda)$.} We were inspired by a similar notion of robust extractors introduced by \cite{BKMO26sample} to define these objects.

We will construct the following robust errorless block condensers.
\begin{theorem}[Informal version of \cref{thm:block-robust}]
\label{thm:overview:block-robust}
Assume $\sfE$ is hard for exponential-size nondeterministic circuits. Then, for all $\gamma > 0$ and $k_x>k_y> n^\gamma$
there is an explicit polynomial-time computable function
\[
 \rbCond\colon\zo^n\times\zo^n\longrightarrow\zo^m
\]
that is a
\[
 \bigl(\lambda=2^{-m/4},\ k_y,\ k_y^{\ge}=k_x/2,\ k_x,\
       k_{\mathrm{out}}=m/4\bigr)
\]
robust errorless block condenser for samplable sources for any $(\log(n))^2\le m \le k_y^{0.01}$.
\end{theorem}

To construct such a robust errorless block condenser, we will utilize and build on ideas from \cite{BSS25, BDGM23}. Using this, we will use an iterative win-win strategy, that has featured in several works in extractor theory such as \cite{RSW06extractor, shaltiel25extract}, to construct the errorless condenser of \cref{thm:overview:errorless-condenser}.

To summarize, our construction of multiplicative extractors has three technical steps:
\begin{enumerate}
\item 
Construct robust errorless block condensers under nondeterministic hardness.

\item
Use robust errorless block condensers, combined with win-win arguments to construct errorless condensers.

\item
Construct multiplicative extractors using errorless condensers and robust errorless block condensers under nondeterministic hardness.
\end{enumerate}

The first and the third steps share many ideas and we will provide a combined technical overview of them in \cref{subsubsec:overview:BGDM++}. We will separately provide a technical overview of the second step in \cref{subsubsec:overview:errorless condenser from robust block errorless condenser}.

\subsubsection{Constructing robust block errorless condenser, and multiplicative extractor from errorless condenser}
\label{subsubsec:overview:BGDM++}

This part is quite technical and builds on a long line of work \cite{STV01PRG, TV00, BDGM23, BSS25, shaltiel25extract}, introducing several new technical insights. To save space, we only give a high level overview of them; for a more detailed exposition, see \cref{sec:proof-overview-BGDM}.

We sketch how we construct our multiplicative extractor for samplable source $\X$ with sufficient min-entropy, given the output of an errorless condenser $\W = \Cond(\X)$, such that $|\W| \ll \minH(\X)$.

\paragraph{Warm-up: Assume $\W$ has high entropy}
We first assume that $\W$ has entropy rate $0.999$, and sketch how we construct multiplicative extractors in this case.
We will follow the same recipe as laid out by Trevisan and Vadhan \cite{TV00}.
For this, we require access to a low-error two-source extractor $\TExt: \zo^n \times \zo^n \to \zo^m$ which has the property that for any two independent sources $\A, \B$ with sufficient entropy, $\TV(\TExt(\A, \B), \U_m) \le \eps$ where $\eps > 0$ is very small.
Also, from our assumption that $\sfE$ is hard for exponential size nondeterministic circuits, we have that there exists a  constant $d > 0$ and a function $f: \zo^{d\log(n)} \to \zo$ such that $f$ is computable in $\poly(n)$ time, and it is hard to compute for size $n^{\gamma d}$ circuits for some constant $\gamma > 0$. 
With this, we define a size $2^{O(n)}$ finite field $\F_q$ and obtain a $d$-variate low degree extension of $f$, $\hatf: \F_q^d\to \F_q$.\footnote{This low degree extension has the property that to efficiently compute $f(x)$, it suffices to efficiently compute $\hatf(\widehat x)$ for a related $\widehat x$.} We can ensure that $\hatf$ has degree at most $O(n)$.

Equipped with these, we construct our extractor by computing:
\[
    \Ext(\X) = \TExt(\hatf(\W), \X).
\]
To argue about correctness, we proceed by contradiction and assume there exists a heavy outcome $\sigma\in \zo^m$ such that $\Pr[\Ext(\X) = \sigma] > (1 + \eps)\cdot 2^{-m}$.
We will show how using this, we can construct a small $\poly(n)$ size nondeterministic circuit $C$ such that $C(x) = \hatf(x)$ for all $x\in \F_q^d$, contradicting the hardness of $f$.
In fact, by standard arguments, it suffices to find $C$ such that $C(x) = \hatf(x)$ for $15/16$ fraction of the inputs $x\in \F_q^d$.

We now sketch how to construct such $C$.
On input $x$, the circuit $C$ will choose a random point $z\in \F_q^d$ and consider the line $L: \F_q\to \F_q^d$ between $x$ and $z$. The line will be such that $L(0) = x, L(1) = z$. Then, $\hatf\circ L: \F_q\to \F_q$ will be a univariate degree $O(n)$ polynomial, and $\hatf\circ L(0) = \hatf(x)$, our desired answer.
We show how $C$ will reconstruct this polynomial $\hatf\circ L$.
To do so, $C$ will nondeterministically guess a degree $O(n)$ univariate polynomial $p: \F_q\to \F_q$, and devise a test such that the following holds:
1) Completeness: If $p$ equals $\hatf\circ L$, then the test accepts.
2) Soundness: If $p$ does not equal $\hatf\circ L$, then the test rejects.
In fact, it suffices to devise such a test with relaxed soundness requirement: $C$ accepts some small ($\ll q$) number of polynomials $p$ which do not equal $\hatf\circ L$.

The original test of \cite{TV00} reconstructed small size $\Sigma_5$ circuits (which required initial hardness assumption against the same). Later, \cite{BDGM23} constructed such a test so that it results in a small nondeterministic circuit.
As a first step, we revisit and simplify the test of \cite{BDGM23}.
This simplification is key for obtaining a generalized version of this test, that we will be able to apply to the case that the samplable source has low min-entropy; and this is a key core contribution of our work.
For our test, we introduce a quantity defined for $w\in \F_q^d, a\in \F_q$ as:
\[
    J(w, a) = \Pr[\W = w, \TExt(a, \X) = \sigma].
\]
We easily see that $J(w, \hatf(w)) = \Pr[\W = w, \Ext(\X) = \sigma]$.
From this, we obtain that
$$
\E_{w\leftarrow\F_q^d}[J(w,\widehat f(w))] =\frac{1}{q^d}\sum_{w\in\F_q^d}\Pr[\W=w,\ \Ext(\X)=\sigma] =\frac{1}{q^d}\Pr[\Ext(\X)=\sigma].\footnote{We use $w\leftarrow\F_q^d$ to denote sampling $w$ uniformly from $\F_q^d$.}
$$

With this, we define our test to be that it accepts polynomial $p$, as a guess for $\hatf\circ L$ if
\[
\E_{t\leftarrow \F_q}[J(L(t), p(t))] \ge \frac{1}{q^d}(1+\eps/2)2^{-m}.
\]
We can implement this test since $\X$ is samplable by an efficient deterministic circuit, and we can use the Goldwasser-Sipser set lower bound protocol \cite{GoldwasserSipser} to execute this test by a small nondeterministic circuit (instead of cycling over all the seeds used to sample $\X$, which will blow up circuit size).
Completeness easily follows since a random line is a good sampler; to argue about soundness, we will use properties of a random line, as well as combinatorial list-recovery bounds on Reed Solomon codes \cite{Sudan97}. For soundness, we will also use the fact that in a two-source extractor with error at most $\eps\cdot 2^{-m}$, the set $B_w = \{a: \Pr[\TExt(a, \X) = \sigma \mid (\W = w)] > (1 + \eps)\cdot 2^{-m}\}$ is such that $|B_w| < 2^{k_1}$, where $k_1$ is the entropy requirement of the first source of $\TExt$.

\paragraph{Handling the case when $\W$ has low entropy}

It is important to note that the parameters of the low degree extension (a.k.a. Reed-Muller code) are not suitable to handle the case where $\W$ has min-entropy smaller than $|\W|/2$. As our aforementioned condenser does not achieve such output min-entropy, we will have to modify the construction and test to handle the case where $\W$ has low min-entropy.

Recall that our test was over the uniform distribution over sampling $t$. This is the reason why we required $\W$ to have very high entropy (entropy rate $0.99$). 
Following an approach from \cite{BSS25}, we modify our construction as follows to handle low entropy case.
We use a strong seeded extractor  $\SExt$ (see \cref{def:seeded-extractor}) with $\W$ as a source to obtain a list of $D = \poly(n)$ many sources $\SExt(\W, 1) = \W_1, \dots, \SExt(\W, D) = \W_D$ such that there exists some $i^*$ (dependent on $\W$) for which $\W_{i^*}$ is statistically close to the uniform distribution.
With this, our construction is:
\[
    \Ext(\X) = \TExt(\hatf(\W_1)\concat \dots \concat\hatf(\W_D), \X).
\]
Observe that the random variable $(\W_1, \dots, \W_D)$ itself may not have high entropy, and all we need to use is that $\W_{i^*}$ is statistically close to the uniform distribution.\footnote{We also crucially rely on the fact that conditioned on all fixings of $(\W_1, \dots, \W_d)$, $\X$ still has entropy.
This is because $(\W_1, \dots, \W_D)$ are a deterministic function of $\W$, and $\W$ has length smaller than the entropy in $\X$.}
When attempting to apply all the pseudorandom properties from above to $\hatf(\W_{i^*})$, we run into a big problem: we also need to recover values of $\hatf(\W_i)$ for all $i\ne i^*$, which are not accessible to us.
To overcome these, we modify our test by defining a generalized version $J(w,a)$ for $w\in \F_q^d, a\in \F_q$:
\[
 J(w,a)=\sum_{y:\SExt(y,i^*)=w}
 \max_{v\in\F_q^{[D]\setminus i^*}}
 \Pr[\W=y,\ \TExt((v,a),\X)=\sigma].
\]
Crucially in this definition, $v$ can depend on $y$, but not on a sample from $\X$.
Using this, we similarly devise an appropriately modified test on $J(L(t), p(t))$ where we nondeterministically guess the values for $v$; here we do use the fact that the Goldwasser-Sipser protocol can tolerate when the membership function is nondeterministic. We also carefully modify other parts of the test to ensure our reconstruction can indeed succeed.\footnote{For this step, we require low error two source extractors with very low entropy requirements. While we have constructions of two-source extractors which require low entropy, they do not attain low enough error of less than $2^{-m}$, for our applications as multiplicative extractors. To overcome this, we use low error two-source condensers \cite{BCDT19condense}, and low error two source extractors with structural requirements \cite{Li15threesource}.} A more detailed overview of the test of \cite{BDGM23}, our simplifications and generalizations, and an overview of the analysis appears in \cref{sec:proof-overview-BGDM}.

\paragraph{Constructing robust errorless block condenser}
To extend this to constructing robust errorless block condensers, we will use a similar construction and analogously modify the analysis.
The primary challenge here is that $\rbCond$ needs to provide guarantees for when the input distribution is the block source $(\Y_{\good}, \X_{\good})$, a distribution that may not even be samplable since all we are guaranteed is that it is a subsource of the samplable source $(\Y, \X)$. We overcome this with careful conditioning arguments.

\subsubsection{Constructing errorless condenser from robust block errorless condenser}
\label{subsubsec:overview:errorless condenser from robust block errorless condenser}
We here sketch how, using a robust block errorless condenser $\rbCond$ from \cref{thm:overview:block-robust}, we can construct errorless condensers as specified in \cref{thm:overview:errorless-condenser}, with output length smaller than the input entropy.

\paragraph{Warm-up: Errorless condenser for entropy $n^{0.51}$}
We will build up towards our construction of errorless condensers for entropy $n^{0.01}$ samplable sources by first constructing an errorless condenser for samplable sources with entropy $n^{0.51}$. Given such a source $\X\sim \zo^n$, we will chop $\X$ into $\ell = n^{0.5}$ parts of size $n / \ell = n^{0.5}$ each, to obtain $\X_1, \dots, \X_{\ell}$. With this, we define
\[
\Cond(\X) = \rbCond(\X_1, \X)\concat \rbCond(\X_{[1, 2]}, \X)\concat\dots\concat \rbCond(\X_{[1, \ell]}, \X).
\]
We will set the output length of $\rbCond$ such that the total output length of our errorless condenser is $\ell\cdot n^{0.001} = n^{0.501}$, which is much smaller than the input entropy of $n^{0.51}$, as desired.
The remaining task is for us to show that $\minH(\Cond(\X)) \ge k_{\out} = n^{\alpha}$ for some constant $\alpha > 0$.

To do that, we use a classical technique, by \cite{SSZ98disperser, tashma1996extractor}, which proved properties of the source obtained by dividing $\X$ into these $\ell$ blocks. Informally, it states that $\X$ can be expressed as a convex combination of sources $\X = \sum_{i=0}^{\ell} \lambda_i \X^{(i)}$ with the following properties, where $\eps = 2^{-m/16}$:
\begin{itemize}
\item 
The sets $\supp(\X^{(0)}), \dots, \supp(\X^{(\ell)})$ are disjoint and form a partition of the support of $\X$.

\item
For $i > 0$, either $\lambda_i > \eps / 2n$, or $\lambda_i = 0$. Also, $\lambda_0 \le \eps$.

\item
For all $i > 0$, $\minH\left(\X^{(i)}_{[1, i]}\right) \ge k/5$.

\item
For all $i > 0$, and $x_i \in \supp\left(\X^{(i)}_{[1, i]}\right)$, we have that $\Pr\left[\X^{(i)}_{[1, i]} = x_i\right] \ge 2^{-k/2}$.

\item
For all $i\ge 1$, it holds that $\X^{(i)}_{[1, i]} \coloneqq \X_{[1, i]} \mid (\X\in \supp(\X^{(i)}))$.
\end{itemize}
In words, $\lambda_0$ is the ``error'' part of the decomposition, and $\X^{(i)}$ is such that the prefix block $\X^{(i)}_{[1, i]}$ has min-entropy, and there is a lower bound on the output probabilities.
See \cref{lem: chop source convex entropy bounds} for a formal statement, and a self-contained proof of this decomposition statement.

Using this decomposition lemma, we argue about the output entropy of $\Cond(\X)$ as follows. We will show that for all $i > 0$, $\minH(\Cond(\X) \mid (\X\in \supp(\X^{(i)}))) \ge k_{\out}+1$.
This suffices to prove the desired claim since we can ``absorb'' the error $\lambda_0$ by using the fact that  $\lambda_0 \le \eps = 2^{-m/16} \le 2^{-k_{\out}-1}$ (as $k_{\out}\le m/32$), bringing our output min-entropy to $k_{\out}$.

To prove that $\minH(\Cond(\X) \mid (\X\in \supp(\X^{(i)}))) \ge k_{\out}+1$, we will show that the $i^{\text{th}}$ output block has the desired entropy guarantees, i.e., that $\minH\left(\rbCond\left(\X_{[1, i]}, \X\right) | \left(\X\in \supp\left(\X^{(i)}\right)\right)\right) \ge k_{\out}+1$.
By the last property above, we have that $\X^{(i)}_{[1, i]} \coloneqq \X_{[1, i]} \mid (\X\in \supp(\X^{(i)}))$. Moreover, we have lower bounds on min-entropy, lower bounds on output probability on each outcome in the support of $\X^{(i)}_{[1, i]}$, and lower bound that $\lambda_i \ge \eps/2n \ge 2^{-m/4}$. Therefore, we satisfy all of the requirements to apply $\rbCond$ and can indeed conclude that $\minH(\Cond(\X) \mid (\X\in \supp(\X^{(i)}))) \ge k_{\out}+1$, as desired.

\paragraph{Generalizing to obtain mild robust condenser}

For our low-entropy condenser construction, we will need to generalize the condenser from above.
First, we make the condenser itself robust, i.e., for any samplable source $\A = p\cdot \A_{\good} + (1-p)\cdot \A_{\rest}$ with $\minH(\A_{\good}) > n^{0.51}$, we guarantee that $\minH(\Cond(\A_{\good})) \ge n^{\alpha}$ for some constant $\alpha > 0$ and where $p = 2^{-m/20}$.\footnote{Of course, $\A_{\good}$ need not be samplable at all.}
Second, we generalize the parameters in the construction from above, to obtain a mildly shrinking robust errorless condenser. 
Concretely, for any constant $\gamma_{\inn} > 0$ we set $\ell = n^{1 - 3\gamma_{\inn}/4}$, and output length of the robust errorless block condenser to $n^{\gamma_{\inn}/4}$ in the construction above, to obtain a $(n^{\gamma_{\inn}}, n^{0.01\gamma_{\inn}})$ robust errorless condenser $\Cond: \zo^n \to \zo^{n^{1 - \gamma_{\inn}/2}}$. See \cref{lem: mild block errorless condenser by chopping} for the general statement.

\paragraph{Constructing $n^{0.01}$ entropy errorless condenser}

We finally sketch how to construct errorless condensers for entropy $n^{0.01}$ sources, with output length much smaller than $n^{0.01}$.
We will require a mild robust condenser $\Cond$ from above, as well as a robust block errorless condenser $\rbCond$. We construct our desired errorless condenser, which has output length smaller than $n^{0.01}$ as follows.

Our high level strategy will be to use an iterative construction and a win-win argument, that has featured in several works in extractor theory including \cite{NT1999extracting, RSW06extractor, shaltiel25extract}.
Given a source $\X$ with $\minH(\X) = n^{0.01}$, we first set parameters of $\Cond$ appropriately to obtain $\Y_1 = \Cond(\X)$ with $\minH(\Y_1) \ge n^{\alpha}$ for some constant $\alpha > 0$ with $|\Y_1| = n^{0.995}$.

Now, consider the source $(\Y_1, \X)$. Ideally, we would like to show that this is a block source. We know that $\Y_1$ has entropy, but it is not clear whether $\X$ has entropy conditioned on $\Y_1$. This is because the length of $\Y_1$ is much larger than the entropy guarantees on $\X$, and hence it may ``steal'' all the entropy from $\X$.
To overcome this, we will therefore split $(\Y_1, \X)$ into a convex combination of sources, one of which will form a block source and the other, we will directly be able to condense from (note that the components need not be efficiently samplable).
To obtain this convex combination, we partition $\supp(\Y_1)$ into medium and light weight sets $M_1, L_1 \subset \supp(\Y_1)$ and show that on the medium part, the resultant sources will form a block source where we will apply $\rbCond$, while the light part will have non-trivial entropy guarantees, and we will be able to recursively condense by applying $\Cond$.

Formally, we let $M_1 = \{y_1\in \supp(\Y_1): \Pr[\Y_1 = y_1] \ge 2^{-n^{0.005}}\}$, and let $L_1 = \supp(\Y_1)\setminus M_1$.
Without loss of generality, we assume that $\Pr[\Y_1\in M_1]$ is large enough to satisfy the requirements of $\rbCond$ (if $\Y_1$ outputs an element from $M_1$ with very low probability, then we will ``pay'' an additive factor to the final output probability, which will have next to no effect on the output min-entropy).
With this, we carefully check that we satisfy all the requirements to apply our robust block errorless condenser $\rbCond$ to $(\Y_1, \X)$ to conclude that $\ZZ_1 = \rbCond(\Y_1, \X) \mid (\Y_1 \in M_1)$ is such that $\minH(\ZZ_1) > n^{\beta}$ for some constant $\beta > 0$.
By definition of $L_1$, neglecting the minor factor that depends on $\Pr[\Y_1\in L_1]$, we have that $\minH(\Y_1 \mid (\Y_1\in L_1)) \ge n^{0.005}$, and we recurse on this source.

We then repeat this process: We let $\Y_2 = \Cond(\Y_1)$ with $|\Y_2|= n^{0.99}$; we  split the support of $\Y_2 \mid (\Y_1\in L_1)$ into $M_2, L_2$; we apply appropriate $\rbCond$ to obtain $\ZZ_2 = \rbCond(\Y_2, \X) \mid (\Y_1 \in L_1, \Y_2\in M_2)$ with $\minH(\ZZ_2) \ge n^{\beta}$; and have that $\minH(\Y_2 \mid (\Y_1\in L_1, \Y_2\in L_2)) \ge n^{0.005}$.
We continue repeating this until we obtain a source $\Y_t$ with $|\Y_t| < n^{0.005}$, and then we directly apply robust block errorless condenser to the resultant source.
We obtain that each of $\ZZ_1, \dots, \ZZ_t$ has small length, and entropy at least $n^{\beta}$. Moreover, our overall output distribution can be written as a convex combination of $t$ many sources, where the source $j$ will receive its entropy from the block $\ZZ_j$. Note that $t = O(1)$, and so any of the intermediate sources are still samplable by $\poly(n)$ size circuits. See \cref{subsec: final errorless condenser} for complete details.

\dobib

\subsection{Proof Overview for Additive Extractors}
\label{subsec: additive proof overview}

 We provide a proof overview of the additive extractors guaranteed by \cref{thm:intro:main:one bit}. We will also give the proof ideas for \cref{thm:intro:main:many bits}, in particular, we will introduce the high level ideas to extract many bits from polylogarithmic entropy, under nondeterministic hardness, with the caveat that the extractor runs in quasipolynomial time. Detailed discussion on how to get efficient extractors under a stronger hardness assumption is given in \cref{sec:poly-construction}. 

We state the informal versions of both results below.
\begin{theorem}[One-bit, informal]\label{thm:ov-one-bit-informal}
    If $\sfE$ is hard for exponential size nondeterministic circuits then for every constant $\eps>0$, and entropy $k=O(\log n)$, there is an efficiently computable function $\Ext_{\mathrm{one}}:\zo^n \to \zo$ that is a $(k,\eps)$-extractor for distributions samplable by polynomial-size circuits.
\end{theorem}

\begin{theorem}[Many-bit, informal]\label{thm:ov-many-bit-informal}
    If $\sfE$ is hard for exponential-size nondeterministic circuits, then for all constants $b>0$, there is a constant $a$, such that for $k=\polylog n$, there is a $(k,n^{-b})$-extractor $\Ext_{\mathrm{many}}:\zo^n\to\zo^{k^a}$ for distributions samplable by polynomial-size circuits.
    Moreover, $\Ext_{\mathrm{many}}$ is computable in time $2^{O(\log^2 n/\log k)}$.
\end{theorem}

Our strategy follows a familiar paradigm from independent-source extraction: first reduce to a \emph{non-oblivious bit-fixing source}, in which the good coordinates are approximately $t$-wise independent while the remaining coordinates may depend arbitrarily on them, and then extract using a suitable resilient function. In the two-source construction of Chattopadhyay and Zuckerman \cite{CZ19}, a seeded non-malleable extractor produces a table in which a large set of rows has approximately $t$-wise independent entries; the second source is then used to sample a smaller table, from which a resilient function extracts a nearly uniform bit. Ben-Aroya, Doron, and Ta-Shma~\cite{BDT17} refine the sampling step so that, after combining small groups of outputs, Majority suffices. 

We pursue an analogous strategy using only a single samplable source $\X$. Exploiting samplability and our hardness assumption, we first transform $\X$ into many correlated rows, most of which have almost full min-entropy. Crucially, this guarantee is \emph{errorless}: each \emph{good} row itself has high min-entropy, rather than merely being statistically close to a high-entropy source. We turn this entropy into approximate independence using a \emph{local correlation breaker}, constructed under the same hardness assumption by adapting a template for \emph{seedless non-malleable extractors for polynomial-size tampering}~\cite{BDL22,BD25}. The correlation breaker is used to process each indexed row separately, ensuring that any $t$ good output rows are jointly close to independent uniform strings. Thus, both the high-entropy rows and the randomness used to break their correlations come from $\X$ itself.

For one-bit extraction, we arrange that sufficiently few output bits are bad and apply Majority. For many-bit extraction, inspired by the techniques in \cite{Li13FOCS,Li13STOC,Li15threesource}, we use the approximately independent outputs to select a small group containing a good row via Feige's lightest bin protocol \cite{Fei99}, and then apply correlation breaking again to extract a longer string. Here the source itself drives the selection, so we must preserve enough entropy and account for the correlations that selection introduces. 

We first introduce errorless seeded condensing and local correlation breaking, then sketch both constructions, and finally explain how to construct the local correlation breaker.

\subsubsection{Errorless seeded condensing to obtain many good rows.}\label{sec:ov-hash-rows}
In this subsection, we state the common starting point of the one-bit and the many-bit constructions: converting the source into polynomially many rows, \textit{most} of which have almost full entropy. 
\begin{lemma}\label{overview:hash-to-rows}
Assume $\sfE$ is hard for exponential-size nondeterministic circuits. For $k\ge C\log n$, we construct a fixed family of $R=\poly(n)$ hashes $h_i:\zo^n\to\zo^m$, where $m=k-\log k+O(1)$. For every samplable source $\X$ over $\zo^n$ with $\minH(\X)\geq k$, let $\X_i = h_i(\X)$, then at least $1-R^{-\eta}$ fraction of rows $\X_i$ satisfy $\minH(\X_i)\ge m-O(1)$, for a constant $\eta>0$. \end{lemma}
Note that the rows remain jointly samplable, and the entropy guarantee holds for each good row itself, without error.

\begin{proof}[Proof sketch]
We obtain our list of condensers in three steps.

\emph{Random hashing gives errorless condensation.}
Fix a source $\X$ with $\minH(\X)\ge k$. Choose a random $k$-wise independent hash $h:\zo^n\to\zo^m$, where $m=k-\log k+\Delta$ for a sufficiently large constant $\Delta$. A collision bound shows that, except with probability at most $n^{-2}$ over $h$, we have $\minH(h(\X))\ge k-\log k=m-\Delta$. This guarantee is errorless: it bounds the largest probability in $h(\X)$ itself. Thus we have a large family of efficiently computable candidates, most of which are good for each fixed source.

\emph{Samplability lets us test the candidates.}
Write $\X=\Samp(\U_r)$ for a sampling circuit $\Samp$ of the prescribed size. If $\minH(h(\X))<k-\log k$, then there must be some very heavy output, forcing an unusually large probability that $k$ distinct samples all hash to the same value. This collision event can be checked efficiently on independent seeds of $\Samp$. The Goldwasser--Sipser counting protocol~\cite{GoldwasserSipser}, followed by amplification and nonuniform fixing of randomness, therefore gives a polynomial-size nondeterministic test on hash descriptions. It accepts every $h$ violating the entropy bound. The collision estimate also shows that it accepts at most an $n^{-2}$ fraction of random hashes.

\emph{A PRG reduces the family to polynomial size.}
Under our hardness assumption, the PRG of~\cite{SU05} fools all these tests with error $n^{-2}$ using $O(\log n)$ seed bits. Enumerating its seeds gives $R=\poly(n)$ hashes, at most $2n^{-2}R\le R^{1-\eta}$ of which violate the entropy bound for any fixed source, for a sufficiently small constant $\eta>0$. The same list works for every source because the generator fools every test of the prescribed size.
\end{proof}

\subsubsection{Local correlation breaking}\label{sec:ov-cb-statement}
In this subsection, we state the guarantees of our local correlation breaker and explain its role in our extractor constructions. We will sketch its construction and analysis in \cref{sec:ov-construction-cb} and give full details and proofs in \cref{sec:cb}.

After obtaining the rows from \cref{sec:ov-hash-rows}, we want to extract from a small collection containing at least one good row. A natural approach is to apply a local map to each row and XOR the outputs. This works if one output is almost uniform even given the others. Our \emph{local correlation breaker} provides precisely this guarantee for sufficiently small collections of distinctly indexed rows, despite their correlations.

\begin{lemma}[Local correlation breaker, informal]\label{lem:overview:correlation breaker}
Assume $\sfE$ is hard for exponential-size nondeterministic circuits. For appropriate parameters, there exists an efficiently computable function $\CB:\zo^L\to\zo^q$ with the following guarantee. For every jointly samplable tuple $(\V_1,\ldots,\V_D)$ whose inputs contain distinct fixed indices, and every $g\in[D]$ such that $\V_g$ has almost full min-entropy,
\[
\bigl(\CB(\V_g),\CB(\V_{-g})\bigr)
\approx_{\varepsilon}
\bigl(U_q,\CB(\V_{-g})\bigr).
\]
Here $\CB(\V_{-g})$ denotes the tuple of all other outputs, and $U_q$ is uniform and independent of that tuple. The inputs may be arbitrarily correlated within the sampling model, and only $\V_g$ is required to have high entropy.
\end{lemma}
In our constructions, we will instantiate the local correlation breaker with different parameters including the output length $q$, the number $D$ of rows it can handle, the entropy deficiency of an ``almost full entropy" row, etc.
We now describe our extractor constructions using the above common tools.

\subsubsection{The one-bit construction}
To prove \cref{thm:ov-one-bit-informal}, we turn the rows from \cref{overview:hash-to-rows} into an NOBF source with almost $t$-wise independent good coordinates: outside a small fixed set of bad coordinates, every $t$ bits are jointly close to uniform. Here $t$ is a sufficiently large constant depending on the desired error; the majority function will extract the final bit.

\paragraph{Grouping.}
The seeded errorless condenser gives $R$ rows $\X_1 ,\dots, \X_R$, at most $R^{1-\eta}$ of which are bad. Following~\cite{BDT17}, we use Zuckerman's disperser~\cite{Zuckerman07} to form $M=\Theta(R^8)$ groups of constant size $d_0$. By the guarantee of the disperser, any $M^{1/4}$ groups collectively touch more than $R^{1-\eta}$ original rows, so fewer than $M^{1/4}$ groups can contain only bad rows. We call a group good if it contains at least one good row. The grouping is fixed independently of the source. Since each row $\X_i$ might appear in multiple groups, we give each row copy a distinct index specifying its group $u$ and position in the group $a$, and denote it by $\V_{u,a}$.
\paragraph{Local correlation breaking.}
Instantiate $\CB$ for $D=d_0t$ indexed rows and $q=1$ output bit, with sufficiently small inverse-polynomial error, and for each group $u$ define
\[
\Y_u=\bigoplus_{a=1}^{d_0}\CB(\V_{u,a}).
\]
We claim that $\Y=(\Y_1,\ldots,\Y_M)$ has the required NOBF guarantee that except for fewer than $M^{1/4}$ bad coordinates, all the good coordinates are almost $t$-wise independent. To see this, fix any $t$ good groups and choose one good row from each. Since each group contains $d_0$ rows, they supply $d_0t$ indexed inputs; note that copies of the same original row are distinct because we gave them distinct indices. Correlation breaking makes each chosen output nearly uniform given all the others in this collection. XORing within each group and applying a hybrid argument across groups shows that the $t$ group bits are jointly almost uniform. 

\paragraph{Extracting from the NOBF source.}
Finally, output
\[
\Ext_{\mathrm{one}}(\X)=\Maj(\Y).
\]
By Viola's majority bound~\cite{viola2014extractors}, in the form stated in~\cite[Lemma~4.28]{CGL21}, majority achieves any desired constant error while tolerating a sufficiently small constant multiple of $\sqrt M$ bad coordinates, for constant $t$ and sufficiently small almost-independence error. Our bound of $M^{1/4}$ bad coordinates therefore suffices.

\subsubsection{The many-bit construction}
We give the construction of $\Ext_{\mathrm{many}}$ in \cref{thm:ov-many-bit-informal}. For simplicity, let's fix a sufficiently small constant $a>0$ and given samplable source $\X$ over $\zo^n$ of min-entropy $k$, we would like to extract $k^a$ bits from it. 

\paragraph{Extracting from a small collection.}
Suppose we have a correlation breaker $\CB_{\mathrm{ext}}:\zo^L\to\zo^q$, with $L=\Theta(k)$, $q=k^a$, and inverse-polynomial error $\zeta$, that handles $D=O(k^a)$ rows. Then we can extract from these rows:
\begin{claim}
    Let $\V_1, \dots, \V_D$ be jointly samplable, distinctly indexed rows with nondeterministic postselection. If there exists a row $\V_g$ such that $\V_g$ has almost full entropy, then XOR of the correlation breaker outputs on each row extracts $q=k^a$ almost-uniform bits:
        \[
        \bigoplus_{j=1}^{D}\CB_{\mathrm{ext}}(\V_j)\approx_\zeta\U_q.
        \]
\end{claim}
Indeed, the guarantee of the correlation breaker ensures that the good output is $\zeta$-close to uniform even given all the other outputs so the XOR of them is also close to uniform. Such a map $\CBext$ can be obtained by instantiating \cref{lem:overview:correlation breaker}:
\begin{lemma}
Assume $\sfE$ is hard for exponential-size nondeterministic circuits. For $k = \poly(\log n)$, $L=\Theta(k)$, there is a map
$
\CBext:\zo^L\to\zo^{k^a}
$
with any desired inverse-polynomial error $\zeta$. For every collection of $D=O(k^a)$ distinctly indexed rows $\V_1,\ldots,\V_D$, jointly samplable with nondeterministic postselection, and every good row $\V_g$ with $\minH(\V_g)\ge L-O(\log n)$, we have
\[
\bigl(\CBext(\V_g),(\CBext(\V_i))_{i\ne g}\bigr)
\approx_\zeta
\bigl(\U_q,(\CBext(\V_i))_{i\ne g}\bigr).
\]
Consequently, $\bigoplus_{i=1}^D\CBext(\V_i)\approx_\zeta\U_q$.
\end{lemma}

 \cref{thm:ov-many-bit-informal} will follow if we have the appropriate $\V_1, \dots, \V_D$ with the following two guarantees (i) $D=O(k^a)$; (ii) at least one row has high entropy.
    
\paragraph{Selecting a small collection.}
The errorless seeded condenser gives $R=\poly(n)$ rows $\X_1,\ldots,\X_R$ of length $\Theta(k)$, all but $R^{1-\eta}$ of which have constant entropy deficiency, for a constant $\eta>0$. Following Feige's lightest-bin protocol~\cite{Fei99}, assign each row a label in $[T]$, where $T\approx R/k^a$, and select the least populated bin. By averaging, its size is always at most $D=R/T=O(k^a)$, which satisfies guarantee (i). It remains to ensure guarantee (ii)---the lightest bin contains at least one good row. To show this, it suffices to bound the probability that every bin contains at least one good row. In an ideal world, independent uniform labels on the good rows would provide this guarantee: each bin receives $\Theta(k^a)$ good rows in expectation, and the probability that there exists a bin that receives no good row is small. However, we don't have independent uniform bin labels---we instead use another correlation breaker to implement enough of the random labeling on the good rows. 

\begin{lemma}[]
Assume $\sfE$ is hard for exponential-size nondeterministic circuits. 
Let $T={R}/{k^a}, d=O({\log n}/{\log k})$. There is a map $\CB_{\mathrm{lab}}:\zo^L\to[T]$
computable in time $2^{O(\log^2 n/\log k)}$, such that for any $d$ distinctly indexed rows $\V_1,\ldots,\V_d$, jointly samplable with nondeterministic postselection, if every row has entropy deficiency $O(\log n)$, then with small (at most inverse polynomial) error $\delta$
\[
\bigl(\CB_{\mathrm{lab}}(\V_1),\ldots,\CB_{\mathrm{lab}}(\V_d)\bigr)
\approx_\delta \U_{[T]^d}.
\]
\end{lemma}
Under assignment $\CBlab$, any labeling of $d$ good rows acts like independent uniform labeling.
In reality, we instantiate $\CB_{\mathrm{lab}}$ with sufficiently small error, and prove using a standard moment bound from \cite{BR94} that every bin contains a good row, except with inverse-polynomial probability.

So far, the lightest-bin rule guarantees we select a small collection of size $O(k^a)$ and the independence property of $\CB_{\mathrm{lab}}$ guarantees a good row in that collection.

\paragraph{Construction: two applications of correlation breaking} To recap, let $\X_1,\ldots,\X_R$ be the condensed rows from \cref{overview:hash-to-rows}. Our plan is to use $\CBlab$ and the lightest-bin rule to select at most $D=O(k^a)$ rows, including a good row with high probability, and then apply $\CBext$ and XOR its outputs to extract $k^a$ bits.

To implement the two applications, split each row as $\X_i=\mathbf A_i\concat\mathbf B_i$ into nearly equal halves, with $|\mathbf B_i|=m_1$. Assign row $i$ the bin label $\Y_i=\CBlab(i\concat\mathbf A_i)\in[T]$, and let $i_1<\cdots<i_s$ be the indices selected by the lightest bin. Retain their second halves and index them by their positions in the bin: set $\W_j=j\concat\mathbf B_{i_j}$ for $j\le s$, and pad with $\W_j=j\concat0^{m_1}$ for $s<j\le D$. For example, suppose $D=6$ and the lightest bin contains exactly rows $\X_4,\X_9,\X_{17},\X_{23}$. Then $s=4$ and $(i_1,i_2,i_3,i_4)=(4,9,17,23)$, so we construct
\[
\begin{aligned}
\W_1&=1\concat\mathbf B_4,  & \W_2&=2\concat\mathbf B_9,\\
\W_3&=3\concat\mathbf B_{17}, & \W_4&=4\concat\mathbf B_{23},\\
\W_5&=5\concat0^{m_1},       & \W_6&=6\concat0^{m_1}.
\end{aligned}
\]
Thus, $\W_3$ contains the second half of original row $17$, prefixed by its selected position $3$. The last two rows are padding, ensuring that the tuple always contains exactly $D$ rows. We also make sure all position indices are encoded using the same number of bits. Finally, output
\[
\Ext_{\mathrm{many}}(\X)
=\bigoplus_{j=1}^{D}\CBext(\W_j).
\]

\paragraph{Preserving entropy after selection.}
The construction as stated above has one major problem: the half of the input used for bin selection and the other half used for extraction are not independent! We split the rows into two halves, and while both halves of a good row have constant deficiency---their indexed versions have $O(\log n)$ deficiency---they remain highly correlated. Bin selection using the first halves can therefore bias the second halves. We must show that enough entropy remains after the first application of the correlation breaker. Moreover, the position of a good selected row may depend on the input. We handle both issues by conditioning on the first good row in the selected bin.
For each good original row $\X_i$ and position $j$ in the selected bin, let $E_{i,j}$ be the event that $\X_i$ is the first good row in the ordered selected list and occupies position $j$. On this event, we have $\W_j=j\concat\mathbf B_i$, while the other selected rows may vary. We emphasize that these events are for the purpose of analysis only, the extractor need \textit{not} identify the good row. In the indexing example above, if rows $4$ and $9$ are bad and row $17$ is good, then $E_{17,3}$ occurs: the first good row selected is $\X_{17}$, and it is on position $3$ in the bin. 

These sub-events partition the event that the lightest bin contains a good row. There are at most $RD$ such events. Fix an inverse polynomial $\gamma>0$, and discard those events with probability below $\gamma/(RD)$; their total probability is at most $\gamma$. 

Let $h=\log D$ be the length of the position index, so each indexed row has length $L=m_1+h$. Every remaining event satisfies $\Pr[E_{i,j}]\ge\gamma/(RD)$, and hence, for every $b$,
\[
\Pr[\mathbf B_i=b\mid E_{i,j}]
\le \frac{\Pr[\mathbf B_i=b]}{\Pr[E_{i,j}]}
\le \frac{RD}{\gamma}\Pr[\mathbf B_i=b].
\]
Thus conditioning loses at most $\log(RD/\gamma)$ bits of min-entropy. On $E_{i,j}$, we have $\W_j=j\concat\mathbf B_i$. Appending the fixed index $j$ leaves min-entropy unchanged but increases the row length by $h$. Since $\mathbf B_i$ initially has constant deficiency,
\[
\begin{aligned}
\minH(\W_j\mid E_{i,j})
&=\minH(\mathbf B_i\mid E_{i,j})\\
&\ge m_1-O(1)-\log(RD/\gamma)\\
&=L-\bigl(h+O(1)+\log(RD/\gamma)\bigr)\\
&=L-O(\log n),
\end{aligned}
\]
where the last equality uses that $R,D$ are polynomial in $n$ and $\gamma$ is inverse-polynomial. Therefore, the indexed row has deficiency $O(\log n)$ even after conditioning. This is the deficiency that $\CBext$ is instantiated to tolerate, so we are allowed to apply the correlation breaker after bin selection and get close to uniform bits. This concludes the proof for \cref{thm:ov-many-bit-informal}. The final error of the extractor will accumulate $\gamma$ above, the error of $\CBlab$, and the error of $\CBext$.
\paragraph{Running time} The total running time is $2^{O(\log^2 n/\log k)}$, which is polynomial when $k=n^{\Omega(1)}$, and quasipolynomial when $k=\poly( \log n)$. In \cref{sec:poly-construction}, we obtain polynomial running time for polylogarithmic entropy under the stronger hardness assumption that $\sfE$ is hard for exponential-size $\Sigma_2$ circuits. There, $\Sigma_2$-hardness is used to get a two-sided multiplicative (or relative-error following~\cite{AASY16}) guarantee on the correlation breaker which can then be exploited via refined analysis of the lightest bin protocol via a (truncated) inclusion--exclusion argument.

\subsubsection{Construction of the correlation breaker}\label{sec:ov-construction-cb}
In the rest of this section, we build the local correlation breaker used in our constructions. Recall that a local correlation breaker makes a good row's output nearly uniform and independent of the other outputs: for a tuple $\V_1,\ldots,\V_D$ and any good row $\V_g$,
\[
   \bigl(\CB(\V_g),(\CB(\V_i))_{i\ne g}\bigr) \approx
   \bigl(\U,(\CB(\V_i))_{i\ne g}\bigr)
\]
where $\U$ is uniform and independent of the companion outputs. For \emph{additive decorrelation}, we obtain a statistical-distance guarantee between LHS and RHS above. The additive guarantee suffices for our one-bit extractor and quasipolynomial-time many-bit extractors under the assumption that $\sfE$ is hard for exponential-size nondeterministic circuits. Here, we give the construction with additive guarantee, following the template of Ball, Dachman-Soled, and Loss~\cite{BDL22} and Ball and Dachman-Soled~\cite{BD25}. In \cref{sec:decorrelation}, we additionally provide more details for \emph{multiplicative decorrelation}.

\paragraph{The construction.}
At entropy threshold $k$, the errorless seeded condenser of \cref{overview:hash-to-rows} gives $R=\poly(n)$ rows. After indexing and/or splitting, the row length remains $L=\Theta(k)$. We construct $\CB:\zo^L\to\zo^q$ with output length $q$. Its guarantee concerns a tuple of $D$ indexed rows $\V_1,\ldots,\V_D\in\zo^L$ from this collection, jointly samplable with nondeterministic postselection by circuits of total size at most $S$. The indices are fixed and distinct. At least one row has nearly full min-entropy; the others may have arbitrarily low entropy. We use three ingredients:
\begin{itemize}[leftmargin=1.5em,itemsep=4pt,topsep=4pt,parsep=0pt]
\item A \emph{preliminary extractor} $\Ext_{\mathrm{pre}}:\zo^L\to\zo^\ell$ for postselecting sources extracts a short nearly uniform seed of length $\ell=O(\log S)$ from a good row. Short seeds limit the entropy lost when we fix all $D$ seeds.
\item A \emph{seed-extending PRG} $G:\zo^\ell\to\zo^L$, with $\ell=O(\log S)$ and $G(\sigma)=\sigma\concat G'(\sigma)$. We get $G$ by composing the seed-extending PRG of \cite{Shaltiel25multiplicative} with the long-stretch PRG of \cite{DS26}, with inverse-polynomial error against nondeterministic circuits of size $\poly(S)$.

\item A \emph{strong two-source non-malleable extractor} $\NMExt$ whose output on independent high-entropy inputs stays nearly uniform 
given its first input, namely the expanded seed, as well as the outputs on $t$ separately tampered pairs. We use different $\NMExt$ for our two constructions. For the one-bit extractor at logarithmic entropy, we use an extractor based on Li's construction~\cite{li2023two}, supporting constant $t$ and one output bit. For the many-bit extractor at polylogarithmic entropy, we use the construction of Chattopadhyay, Goyal, and Li~\cite{CGL20}, allowing both $t$ and the output length $q$ to grow as sufficiently small fixed powers of $L$.
\end{itemize}
Combining the above, for a complete indexed row $v$, define
\[
 \CB(v)=\NMExt\bigl(G(\Ext_{\mathrm{pre}}(v)),v\bigr).
\]

Building on~\cite{BDL22}, Ball and Dachman-Soled~\cite{BD25} establish non-malleability under nondeterministic hardness against bounded-complexity tamperings of sources with efficiently computable postselection predicates. Our application requires local decorrelation for distinctly indexed rows that are jointly samplable with \emph{nondeterministic} postselection. Moreover, a companion row need not be efficiently computable from the good row. The local decorrelation viewpoint is already implicit in~\cite{BD25}. We instantiate their framework with stronger ingredients and extend the analysis to handle nondeterministic postselection under nondeterministic hardness. We outline the analysis below following the template of \cite{BD25}.

\paragraph{The analysis: failed $\CB$ implies distinguisher for $G$}
The central obstacle in the analysis is that the first input to $\NMExt$ is a deterministic function of the second. Following~\cite{BDL22,BD25}, we overcome this by reducing a failure of correlation breaking to a violation of $G$'s pseudorandomness against nondeterministic circuits:
\[
\exists g, (\CB(\V_g),(\CB(\V_i))_{i\ne g})\not\approx(\U_q,(\CB(\V_i))_{i\ne g}) \implies  \text{nondeterministic distinguisher } T \text{ for } G
\]
Naively, a distinguisher $T$ might try to determine whether $w=\sigma\concat y$ is generated or uniform by simulating the correlation breaker. It samples rows $v_i$ and computes their seeds $p_i=\Ext_{\mathrm{pre}}(v_i)$. If $p_g\ne\sigma$, it rejects. Otherwise, it treats $w$ as the generator output for the good row, computing $z_g=\NMExt(w,v_g)$, while computing $z_i=\NMExt(G(p_i),v_i)$ for each companion. However, computing the generator will exceed $T$'s size budget.

\paragraph{Interactive protocol} To avoid this expensive simulation, \cite{BDL22,BD25} followed \cite{AASY16}'s observation that a constant-round private-coin protocol suffices to yield a (nondeterministic) distinguisher: an unbounded prover, Merlin, supplies the evaluations and performs the statistical test for a randomized verifier, Arthur. Since constant-round interactive proofs can be simulated by nondeterministic circuits, i.e. $\mathsf{IP}[O(1)]\subseteq\mathsf{NP}/\mathsf{poly}$~\cite{GoldwasserSipser,GL20,BM88,Adleman1978}, we convert this protocol into a polynomial-size nondeterministic circuit $T$ that guesses Merlin's reply. This circuit $T$ is our distinguisher for $G$.

The protocol is as follows,
\begin{figure}[H]
\centering
\setlength{\fboxsep}{8pt}
\fbox{%
\begin{minipage}{\dimexpr\linewidth-2\fboxsep-2\fboxrule\relax}
\textbf{Fixed data:} The joint sampler and a good index $g$ witnessing the assumed failure of $\CB$.

\smallskip
\textbf{Common input:} $w=\sigma\concat y$.

\begin{enumerate}[label=(\roman*),leftmargin=*,itemsep=0.5em,topsep=0.5em]
    \item \textbf{Sample and match the seed.}
    Arthur privately samples random coins $r$, computes candidate rows $v_i$ and seeds $p_i=\Ext_{\mathrm{pre}}(v_i)$, and rejects unless $p_g=\sigma$. He sends the companion seeds $(p_i)_{i\ne g}$ to Merlin.

    \item \textbf{Supply generator evaluations.}
    Merlin returns claimed generator suffixes $(a_i)_{i\ne g}$. Arthur computes
    $z_i=\NMExt(p_i\concat a_i,v_i), \text{ for each }i\ne g.$

    \item \textbf{Guess the hidden bit.}
    Arthur chooses a uniform private bit $c$. If $c=1$, he sets $z=\NMExt(w,v_g)$; otherwise, he samples $z\leftarrow\U_q$ independently. He sends $(z,(z_i)_{i\ne g})$ to Merlin, who returns a guess $\widehat c$.

    \item \textbf{Certify postselection.}
    Only after receiving $\widehat c$ does Arthur reveal $r$ and request a witness certifying successful postselection. He accepts exactly when the witness is valid and $\widehat c=c$.
\end{enumerate}
\end{minipage}%
}
\caption{The private-coin protocol used in the reduction.}
\label{cbov:fig:interaction}
\end{figure}
We show that if correlation breaking fails at the good row $g$, then honest generator evaluations give Merlin a noticeable guessing advantage on a noticeable fraction of generated inputs $w=G(\sigma)$. On uniform inputs $w\leftarrow\U_L$, strong non-malleability bounds every prover's average guessing advantage, even when Merlin supplies arbitrary claimed evaluations. In particular, conditioning on the ``seed'' $\sigma$ and Arthur's message to Merlin (which is short) induces independent high min-entropy distributions and a split-state tampering strategy: Merlin ``tampers'' the LHS, $(\sigma,y)$, and Arthur ``tampers'' the RHS, $v_g$, by sampling the $v_i$'s conditioned on $v_g$. This latter sampling is not efficiently computable, but we observe that by the strong multiplicative guarantee of Goldwasser--Sipser~\cite{GoldwasserSipser,GL20}, it suffices to construct a protocol where Arthur simply gives up whenever the conditional sampling fails. After suitable amplification, this completeness--soundness gap yields a nondeterministic circuit $T$ that distinguishes $G(\U_\ell)$ from $\U_L$, contradicting the security of $G$.

\subsection{Organization}
% The rest of the paper is organized as follows. We provide an overview of our techniques for multiplicative extractors and additive extractors in \cref{sec: proof overview}.
% Our AI Usage statement is at \cref{subsec: AI}
We provide a statement on AI use in \cref{sec:AI}. We will provide preliminaries for the technical sections in \cref{sec:preliminaries}.
The technical sections are divided into a multiplicative extractor part (\cref{part: mult}) and an additive extractor part (\cref{part: additive}).\footnote{Our additive extractor for $n^{\alpha}$ entropy samplable sources, for any $\alpha > 0$, will follow directly from our multiplicative extractor, and is presented there.}
Each of these will contain several sections that will present the full details of the constructions of these extractors.

\section{AI Use Statement}
\label{sec:AI}

The work on this project started in 2025, and some (weaker versions) of the current results were already obtained by the authors without any use of AI.

Once AI became more useful, it was used to assist with proofreading, developing proofs of technical lemmas, and in helping to write them. 

The use of AI was also specifically helpful in obtaining some improvements over the preliminary results that we had. More specifically:
\begin{itemize}
    \item A long conversation with AI (in which we discussed our improved \cite{BDGM23} test, that we already had at that point) helped in obtaining an analysis of the improved test that applies to multiplicative extractors with large output length. This is part of the proof of \cref{thm:intro:main:mult} and specifically affected the proofs in \cref{sec:robust-proof} and \cref{sec:mult-proof}. Before this improvement, we could only prove \cref{thm:intro:main:mult} under a hardness assumption for $\Sigma_3$ circuits, with a larger min-entropy threshold.
    \item While the main ideas behind the additive extractors came from the authors, AI (particularly OpenAI's GPT~5.6) proved immensely helpful in rapidly exploring the delicate parameter space and working out the technical details. This assistance was especially useful in developing the inclusion--exclusion analysis of Feige's protocol so that it was compatible with the parameters achievable by our relative-error correlation breaker.
\end{itemize}

\section{Preliminaries}
\label{sec:preliminaries}\label{sec:setup}

We fix the notation and source models, then state the tools used in
the constructions.

\paragraph{Notation.}
Logarithms are base~$2$, and $e$ is Euler's number.
Write $[D]=\{1,\ldots,D\}$.
Random variables are boldface, and lowercase letters denote their
values; $x\leftarrow\X$ denotes sampling. For a finite set $S$,
$\U_S$ is uniform on $S$, and $\U_m\coloneqq\U_{\zo^m}$.
Uniform random variables appearing in a product distribution are
independent. We identify a random variable with its distribution
when no confusion can arise.
Min-entropy thresholds are denoted by $k$ and errors by $\eps$, with
subscripts naming the object.
A sampler has $r$ input bits; $\bR\leftarrow\U_r$ is its random input
and $\rho\in\zo^r$ a fixed one.
For $x\in(\zo^b)^\ell$, $x_{[1,i]}=(x_1,\ldots,x_i)$ denotes its first
$i$ blocks. Tuples are written $(u_1,\ldots,u_D)$.
We write $\concat$ for concatenation and reserve $\circ$ for
composition.
The size of a circuit, deterministic or nondeterministic, includes its
input, output, and witness lengths.

\subsection{Basic Probability}

\subsubsection{Standard definitions}
\label{sec:standard-definitions}

\begin{definition}[Statistical distance]
\label{def:statistical-distance}
For distributions $\X,\Y$ on a common finite set $\Omega$, their
statistical distance is
\[
 \TV(\X,\Y)=\frac12\sum_{a\in\Omega}
       \bigl|\Pr[\X=a]-\Pr[\Y=a]\bigr|
 =\max_{A\subseteq\Omega}\bigl|\Pr[\X\in A]-\Pr[\Y\in A]\bigr|.
\]
They are $\eps$-close if $\TV(\X,\Y)\le\eps$.
\end{definition}

\begin{definition}[Min-entropy]
\label{def:min-entropy}
The min-entropy of a finite random variable $\X$ is
\[
 \minH(\X)=-\log\max_x\Pr[\X=x].
\]
An $(n,k)$-source is a distribution on $\zo^n$ with min-entropy at
least $k$. Its entropy deficiency is $n-\minH(\X)$, and its support is
$\supp(\X)=\{x:\Pr[\X=x]>0\}$.
\end{definition}

\subsubsection{Useful probability lemmas}
\label{sec:conditioning-density}

\begin{definition}[A subsource of density $\lambda$]
\label{def:subsource}
Let $\ZZ$ and $\ZZ'$ be distributions on the same finite set and let
$0<\lambda\le1$. We say that $\ZZ'$ is a subsource of $\ZZ$ of density
$\lambda$ if there is a distribution $\ZZ''$ such that
\[
 \ZZ=\lambda\ZZ'+(1-\lambda)\ZZ''.
\]
The same definition applies to joint distributions. In particular,
for every event $A$ on their common domain,
$\Pr[\ZZ\in A]\ge\lambda\cdot\Pr[\ZZ'\in A]$.
\end{definition}

We use the following simple bounds when conditioning on the seeds
extracted from the rows.

\begin{lemma}[Conditioning on an event]
\label{lem:conditioning}
If $\minH(\Y)\ge k$ and $E$ is an event jointly defined with $\Y$ of
probability $p>0$, then
\[
 \minH(\Y\mid E)\ge k-\log(1/p).
\]
\end{lemma}
\begin{proof}
For every $y$, $\Pr[\Y=y\mid E]\le\Pr[\Y=y]/p\le2^{-k}/p$.
\end{proof}

\begin{lemma}[Density bounds from statistical distance]
\label{lem:density}
Let $\bP$ be a distribution on a finite set $\Omega$ of size $M$
with $\TV(\bP,\U_\Omega)\le\eps$.
Write $p(w)=\Pr[\bP=w]$. For every $\lambda>0$,
\[
 \Pr_{w\leftarrow\U_\Omega}
 \left[p(w)<\frac{1-\lambda}{M}\right]\le\frac\eps\lambda,
 \qquad
 \Pr_{w\leftarrow\U_\Omega}
 \left[p(w)>\frac{1+\lambda}{M}\right]\le\frac\eps\lambda.
\]
\end{lemma}
\begin{proof}
The sum of the positive deficits $1/M-p(w)$ is at most $\eps$, as is
the sum of the positive excesses $p(w)-1/M$. Each point counted in
either event contributes more than $\lambda/M$ to its respective sum.
\end{proof}

We record a simple fact that follows from the hybrid argument.
\begin{lemma}\label{pre:lem:joint-uniformity}
Let $\mathcal A$ be a finite alphabet and
$\X=(\X_1,\ldots,\X_t)\in\mathcal A^t$.
Suppose that, for every $i\in[t]$,
\[
 \TV\Bigl(
   \bigl(\X_i,(\X_j)_{j\ne i}\bigr),
   \bigl(\U_{\mathcal A},(\X_j)_{j\ne i}\bigr)
 \Bigr)\le\eps,
\]
where $\U_{\mathcal A}$ is uniform on $\mathcal A$ and
independent of $\X$. Then
$\TV(\X,\U_{\mathcal A^t})\le t\eps$.
\end{lemma}

\subsection{Circuits and Hardness assumptions}
\label{sec:hardness}

\subsubsection{Circuit definitions}

\begin{definition}[Nondeterministic circuits]
For $n, w\in \N$, a nondeterministic circuit $C$ is a Boolean circuit that takes in two inputs $(x, \mathrm{wit})\in \zo^{n + w}$ and computes a Boolean function $f: \zo^n \to \zo$ such that $f(x) = 1$ if and only if there exists $\mathrm{wit}\in \zo^w$ such that $C(x, \mathrm{wit}) = 1$.
\end{definition}

\begin{definition}[$\Sigma_i$ circuits]
\label{def: sigma i circuit}
For $n\in \N$, a $\Sigma_i$ circuit $C(x)$ is a Boolean circuit equipped with special oracle gates which can query a canonical $\Sigma_i$ complete problem.
\end{definition}

We will additionally find it useful to define single valued nondeterministic circuits

\begin{definition}[Nondeterministic outputs, PSV, and SV circuits]
\label{def:sv-circuits}
Let $\mathcal O$ be an output set with a fixed binary encoding.
A multi-output nondeterministic circuit is a deterministic circuit $C(x,\mathrm{wit})$ that,
on input $x$ and witness $\mathrm{wit}$, either rejects or accepts and outputs an
element of $\mathcal O$. Define the set of accepted outputs by
\[
    \ACCEPT_C(x)\coloneqq\{a\in\mathcal O:\exists\mathrm{wit}\textup{ such that }C(x,\mathrm{wit})
    \textup{ accepts and outputs }a\}.
\]
\begin{enumerate}[label=(\arabic*)]
    \item $C$ is \emph{partial single-valued} (PSV) if
    $|\ACCEPT_C(x)|\le1$ for every input $x$.
    \item $C$ is \emph{single-valued} (SV) if
    $|\ACCEPT_C(x)|=1$ for every input $x$.
\end{enumerate}
For a function $g\colon\mathcal X\to\mathcal O$, we say that $C$ is
\emph{correct and single-valued at $x$} if
$\ACCEPT_C(x)=\{g(x)\}$. If this holds for every $x\in\mathcal X$,
then $C$ is an SV circuit computing $g$.
Thus PSV permits inputs with no successful witness, whereas SV requires
at least one successful witness on every input.
\end{definition}

\subsubsection{Hardness assumptions}

Write $\sfE=\bigcup_{c>0}\DTime(2^{c n})$.

\begin{assumption}[Nondeterministic hardness, $\NDH$]
\label{def:nd-hardness}\label{ass:ndh}
There are a Boolean function $f\in\sfE$ and a constant
$\beta_{\mathrm{hard}}>0$ such that, for every sufficiently large
input length $\ell$, no nondeterministic circuit of size at most
$2^{\beta_{\mathrm{hard}}\ell}$ computes $f$ on all inputs in
$\zo^\ell$.
\end{assumption}

When $\NDH$ holds, we also say that $\sfE$ is hard for exponential-size
nondeterministic circuits.

\begin{assumption}[$\Sigma_i$ hardness, $\mathsf{H}_{\Sigma_i}$]
\label{ass:sigma}
There is a Boolean function $f\in\sfE$ and a constant $\beta_{\mathrm{hard}}>0$ such that, for every sufficiently large
input length $\ell$, no $\Sigma_i$ circuit of size at most $2^{\beta_{\mathrm{hard}}\ell}$ computes $f$ on all inputs in
$\zo^\ell$.
\end{assumption}

Observe for instance that the assumption $\SH$ implies $\NDH$.

\begin{remark}
Note that $\NDH$ hardness assumption is against nondeterministic circuits while $\SH$ hardness assumption is against Boolean circuits equipped with $\Sigma_2$ oracle gates. Moreover, to prove our later theorems stated under $\SH$, it suffices to establish them under the weaker assumption that $\sfE$ is hard for exponential-size nondeterministic circuits with $\SAT$ oracles, since this assumption follows from $\SH$.
\end{remark}

\subsection{Extractors, Condensers, and Models of Randomness}

\subsubsection{Extractors and condensers for a class of distributions}
\label{sec:extractors-classes}

\begin{definition}[Extractors for a class of distributions]
\label{def:extractors-class}\label{def:class-extractor}\label{def:samplable-extractor}
Let $\mathcal C$ be a class of distributions on $\zo^n$.
A function $\Ext\colon\zo^n\to\zo^m$ is a $(k,\eps)$-extractor for
$\mathcal C$ if, for every $\X\in\mathcal C$ with $\minH(\X)\ge k$,
\[
 \TV(\Ext(\X),\U_m)\le\eps.
\]
We also call this the \emph{additive} guarantee.
\end{definition}

\begin{definition}[Multiplicative extractors for a class of distributions]
\label{def:multiplicative}
Let $\mathcal C$ be a class of distributions on $\zo^n$.
A function $\Ext\colon\zo^n\to\zo^m$ is a $(k,\eps)$-multiplicative
extractor for $\mathcal C$ if, for every $\X\in\mathcal C$ with
$\minH(\X)\ge k$ and every $\sigma\in\zo^m$,
\[
 \Pr[\Ext(\X)=\sigma]\le e^\eps\cdot2^{-m}.
\]
Equivalently, for every such $\X$ and every $A\subseteq\zo^m$,
\[
 \Pr[\Ext(\X)\in A]\le e^\eps\cdot\Pr[\U_m\in A].
\]
\end{definition}

Observe that $e^{\eps} \le 1 + 2\eps$ for $\eps \le 1$.

\begin{definition}[Errorless condensers for a class of distributions]
\label{def:errorless-condenser}
Let $\mathcal C$ be a class of distributions on $\zo^n$.
A function $\Cond\colon\zo^n\to\zo^{m_{\Cond}}$ is an errorless
$(k,k_{\Cond})$-condenser for $\mathcal C$ if, for every $\X\in\mathcal C$,
\[
 \minH(\X)\ge k\quad\Longrightarrow\quad
 \minH(\Cond(\X))\ge k_{\Cond}.
\]
The min-entropy guarantee holds for the output distribution itself.
\end{definition}

\begin{definition}[Errorless strong seeded condenser]
\label{def:errorless-strong-seeded-condenser}
Let $\mathcal C$ be a class of distributions on $\zo^n$ and
$0\le\delta<1$. A function
$\Cond\colon\zo^n\times[R]\to\zo^m$ is an errorless strong
seeded $(k,k',\delta)$-condenser for $\mathcal C$ if, for every
$\X\in\mathcal C$ with $\minH(\X)\ge k$, at least a
$1-\delta$ fraction of seeds $i\in[R]$ satisfy
$\minH(\Cond(\X,i))\ge k'$.
\end{definition}

\subsubsection{Seeded extractors}
\label{sec:seeded-prelim}

\begin{definition}[Seeded extractors]
\label{def:seeded-extractor}\label{def:strong-seeded-extractor}
A function $\SExt\colon\zo^n\times\zo^s\to\zo^m$ is a seeded
$(k,\eps)$-extractor if, for every $(n,k)$-source $\X$ and
an independent uniform seed $\bS\leftarrow\U_s$,
\[
 \TV(\SExt(\X,\bS),\U_m)\le\eps.
\]
We say that $\SExt$ is \emph{strong} if
\[
 \TV\bigl((\bS,\SExt(\X,\bS)),(\bS,\U_m)\bigr)\le\eps.
\]
\end{definition}

We use the following strong seeded extractor from \cite{GUV09extract}.

\begin{theorem} 
\label{thm:guv}
For every constant $\delta\in(0,1)$, there is a constant $C$
such that, for input length $n$, error $0<\eta<1/2$, and
$r \ge C(\log n+\log(1/\eta))$, there is an explicit
strong $(r,\eta)$-extractor $\SExt:\zo^n \times \zo^d \to \zo^m$ with seed length
$d=O(\log n+\log(1/\eta))$ and output length
$m = (1-\delta)r$.
The extractor is computable in time polynomial in $n$ and $1/\eta$.
\end{theorem}

\subsubsection{Block-wise sources}

\begin{definition}[Two-block source]
\label{def:block-source}
A pair $(\A,\B)$ is a $(r_1,r_2)$-block-wise source if
$\minH(\A)\ge r_1$ and, for every $a\in\supp(\A)$,
$\minH(\B\mid\A=a)\ge r_2$.
A distribution is $\eta$-close to such a source if its statistical
distance from some such joint distribution is at most $\eta$.
\end{definition}

\subsubsection{Two-source extractors and condensers}
\label{sec:two-source-prelim}

\begin{definition}[Two-source extractors]
\label{def:two-source-extractor}
A function $\TExt\colon\zo^{n_1}\times\zo^{n_2}\to\zo^m$ is a
$(k_1,k_2,\eps)$-two-source extractor if, for every pair of
independent sources $\X_1,\X_2$ with $\minH(\X_i)\ge k_i$,
\[
 \TV(\TExt(\X_1,\X_2),\U_m)\le\eps.
\]
\end{definition}

\begin{definition}[Two-source condensers]
\label{def:two-source-condenser}
A function $\TwoCond\colon\zo^{n_1}\times\zo^{n_2}\to\zo^m$ is a
$(k_1,k_2,k_{\mathrm{out}},\eta)$-two-source condenser if, for every
pair of independent sources $\X_1,\X_2$ with $\minH(\X_i)\ge k_i$,
the distribution $\TwoCond(\X_1,\X_2)$ is $\eta$-close to a distribution
of min-entropy at least $k_{\mathrm{out}}$. It is \emph{errorless}
if $\eta=0$; in this case, we omit $\eta$ from the parameters.
\end{definition}

\begin{definition}[An extractor for one source and one block-wise source]
\label{def:independent-block-extractor}
A function
\[
 \IExt\colon\zo^{n_1}\times(\zo^{n_2}\times\zo^{n_3})\to\zo^m
\]
is a $(k_1,k_2,k_3,\eps)$-extractor for one source and one
block-wise source if, whenever $\U$ is independent of $(\A,\B)$,
$\minH(\U)\ge k_1$, and $(\A,\B)$ is a $(k_2,k_3)$-block-wise source,
\[
 \TV\bigl(\IExt(\U,(\A,\B)),\U_m\bigr)\le\eps.
\]
We also write $\IExt(\U,\A,\B)$ for $\IExt(\U,(\A,\B))$.
\end{definition}

We use a two-source condenser that guarantees high output min-entropy
with no statistical error. The following is a convenient consequence
of Ben-Aroya, Cohen, Doron, and Ta-Shma~\cite{BCDT19condense}.

\begin{theorem}[Errorless two-source condensation
{\cite[Theorem~27]{BCDT19condense}}]
\label{thm:two-source-condenser}
There is a universal constant $C_{\TwoCond}\ge1$ such that, for every
sufficiently large $n$ and all $r,m$ satisfying
\[
 \log n\le m,\qquad (m\log n)^{C_{\TwoCond}}\le r\le n,
\]
there is an explicit polynomial-time function
\[
 \TwoCond_0\colon\zo^n\times\zo^n\longrightarrow\zo^m
\]
such that, for every pair of independent sources $\U,\V$ on $\zo^n$
with $\minH(\U),\minH(\V)\ge r$,
\[
 \minH\bigl(\TwoCond_0(\U,\V)\bigr)\ge3m/4.
\]
\end{theorem}

To obtain this consequence, apply the cited theorem with error
$2^{-m}$. For a universal constant $C_0$, its min-entropy requirement
is $r\ge[m\log(n2^m)]^{C_0}$. This follows from the displayed
hypothesis after choosing the universal constant $C_{\TwoCond}$
sufficiently large, since $m\ge\log n$.
The theorem bounds every output point probability by
\[
 2^{-(m-o(m))}+2^{-m}+2^{-r/2}\le2^{-3m/4}
\]
for sufficiently large $n$. Thus the min-entropy guarantee holds for
the output distribution itself, with no statistical error.

\begin{theorem}[One independent source and one block-wise source
{\cite[Theorem~1.4]{Li15threesource}}]
\label{thm:li}
For every sufficiently large $n$ and every $r$ with
$\log^{12}n\le r\le n$, there is an explicit polynomial-time function
\[
 \IExt_0\colon\zo^n\times\zo^n\times\zo^n\longrightarrow\zo^{0.9r}
\]
such that, whenever $\U$ is independent of $(\A,\B)$,
$\minH(\U)\ge r$, and $(\A,\B)$ is an $(r,r)$-block-wise source,
$\IExt_0(\U,\A,\B)$ is $2^{-r^{\Omega(1)}}$-close to uniform.
The constants in the error exponent and the polynomial running-time
bound are universal. In particular, there is a universal
$c_{\IExt}>1$ such that the error is at most $2^{-r^{1/c_{\IExt}}}$
for all sufficiently large admissible parameters.
\end{theorem}

Li proves the stronger statement in which either independent input
may also be revealed; the weaker guarantee above is all we use.
This is the component used in the construction of a multiplicative extractor
of~\cite[Section~4.3]{shaltiel25extract}.

\subsubsection{Two-source non-malleable extraction}
\label{sec:two-source-nonmalleable}

The correlation breaker uses an extractor whose output remains
close to uniform given outputs on modified versions of its inputs.

\begin{definition}[Two-source non-malleable extractor]
\label{def:two-source-nonmalleable}
A function $\NMExt\colon\zo^L\times\zo^L\to\zo^q$ is
$t$-non-malleable for min-entropy $k$ and error $\eps$ if the
following holds. Let $\X,\Y$ be independent sources with
min-entropy at least $k$. Let $I$ be an index set with $|I|\le t$,
and let $f_i,g_i\colon\zo^L\to\zo^L$, $i\in I$, be arbitrary
functions such that, for each $i$, at least one of $f_i,g_i$ has
no fixed point. Writing $\X_i=f_i(\X)$ and $\Y_i=g_i(\Y)$,
we require
\[
 \begin{aligned}
 \TV\Bigl(&
 \bigl(\NMExt(\X,\Y),(\NMExt(\X_i,\Y_i))_{i\in I}\bigr),\\
 &\bigl(\U_q,(\NMExt(\X_i,\Y_i))_{i\in I}\bigr)
 \Bigr)\le\eps.
 \end{aligned}
\]
The extractor is strong in its first input if the same bound holds
with $\X$ included in both tuples, and strong in its second input
if it holds with $\Y$ included. A function $f$ has no fixed point
if $f(x)\ne x$ for every $x$.
\end{definition}

\subsubsection{Samplable distributions}
\label{sec:samplable}

We consider distributions generated by deterministic sampling
circuits, possibly conditioned on an event recognized by a nondeterministic circuit.

\begin{definition}[Samplable distributions]
\label{def:samplable}
A distribution $\X$ on $\zo^n$ is samplable by a circuit of size at
most $s$ if there is a deterministic Boolean circuit
$\Samp\colon\zo^r\to\zo^n$ of size at most $s$ such that $\Samp(\U_r)$
has exactly the distribution $\X$. In particular, $r,n\le s$.
\end{definition}

\begin{definition}[Nondeterministic postselection]
\label{def:postselection}
Let $\NPSources_{n,S}$ be the class of distributions on $\zo^n$
of the form
\[
 \X=(\Samp(\bR)\mid\Post(\bR)=1),
 \qquad \bR\leftarrow\U_r,
\]
where $\Samp\colon\zo^r\to\zo^n$ is deterministic,
$\Post\colon\zo^r\to\zo$ is nondeterministic, their total circuit
size is at most $S$, and $\Pr[\Post(\bR)=1]>0$.
\end{definition}

% \begin{definition}[Deterministic postselection]
% \label{def:deterministic-postselection}\label{def:sampler}
% Let $\DetSources_{n,S}$ be the class of distributions on $\zo^n$
% of the following form, with $\bR\leftarrow\U_r$:
% \[
%  \X=(\Samp(\bR)\mid\Post(\bR)=1),
% \]
% where $(\Samp,\Post)$ is a deterministic Boolean circuit of total
% size at most $S$ and $\Pr[\Post(\bR)=1]>0$.
% We call these distributions samplable with deterministic
% postselection by a size-$S$ sampling circuit.
% The same definition applies to joint distributions, with $\Samp$
% outputting the whole tuple.
% \end{definition}

% This is a subclass of \cref{def:postselection}; here the size bound
% applies to the combined circuit. No lower bound on the positive
% postselection probability is required. The constructions receive
% no description of the sampling circuit.

\subsubsection{Extraction from samplable sources}
\label{sec:multiplicative-extraction}\label{sec:preliminary}

We use the following two ingredients
from~\cite[Theorem~2.3 and Lemma~4.5]{Shaltiel25multiplicative}.
The first constructs a seed-extending generator and the second
shows that its new output bits extract from samplable sources.

\begin{lemma}[Short-stretch seed-extending PRG]
\label{pre:lem:short-prg}
Assume $\NDH$. For every sufficiently large $s$, integer
$1\le m\le s$, and $2^{-s}\le\eta\le1/s$, there is a
$\poly(s)$-time computable function $K:\zo^d\to\zo^m$, where
$m\le d=O(m+\log(1/\eta))$, such that every size-$s$
nondeterministic circuit $D$ satisfies
\[
 \Pr[D(\U_d\concat K(\U_d))=1]
 \le e^{1/s}\Pr[D(\U_{d+m})=1]+\eta.
\]
Here both occurrences of $\U_d$ denote the same sample.
Furthermore, if you assume $\SH$, then the same guarantee holds for nondeterministic circuits with oracle gates to SAT.
\end{lemma}

\begin{lemma}[From pseudorandomness to extraction]
\label{pre:lem:prg-to-extractor}
There is an absolute constant $C>1$ with the following property.
Let $s\ge d\ge m$, $0<\eps\le1$, and $\Delta>0$.
Suppose $K:\zo^d\to\zo^m$ satisfies
\[
 \Pr[D(\U_d\concat K(\U_d))=1]
 \le e^\eps\Pr[D(\U_{d+m})=1]+\eta
\]
for every size-$s$ nondeterministic circuit $D$, where
$\eta\le\eps 2^{-(m+\Delta)}$.
If $(S_0/\eps)^C\le s$, then every
$\X\in\NPSources_{d,S_0}$ with
$\minH(\X)\ge d-\Delta$ satisfies
\[
 \Pr[K(\X)=z]\le e^{12\eps}2^{-m}
 \qquad\text{for every }z\in\zo^m.
\]
Furthermore, if you assume $\SH$, then the same guarantee holds for nondeterministic circuits with oracle gates to SAT.
\end{lemma}
Combining these ingredients gives the following multiplicative
extractor, with the source length and sampling circuit size
as separate parameters.
\begin{lemma}\label{pre:lem:preliminary}
Assume $\NDH$. There is a constant $C_0>0$ such that, whenever
$2\le L\le S$, $m\ge1$, $\Delta\ge0$, and $0<\nu\le1$ satisfy
\begin{equation}\label{pre:eq:preliminary-length}
 L\ge C_0\bigl(\Delta+m+\log S+\log(1/\nu)\bigr),
\end{equation}
there is a function $P\colon\zo^L\to\zo^m$, computable in time
$\poly(S,1/\nu)$, such that every $\V\in\NPSources_{L,S}$ with
$\minH(\V)\ge L-\Delta$ satisfies
\begin{equation}\label{pre:eq:preliminary-cap}
 \Pr[P(\V)=z]\le e^\nu2^{-m}\qquad(z\in\zo^m).
\end{equation}
\end{lemma}

\begin{proof}
We construct a generator using \cref{pre:lem:short-prg}
and apply \cref{pre:lem:prg-to-extractor} to a projection
of the source.

Set $\eps_0=\nu/12$ and $\Delta'=\Delta+1$.
Choose $s=\poly(S,1/\nu)$ sufficiently large, and let
\(
 \eta=\min\{\eps_0 2^{-(m+\Delta')},1/s\}.
\)
The length hypothesis and $L\le S$ give
$m+\Delta+\log(1/\nu)=O(S)$.
Thus we can ensure $2^{-s}\le\eta$, $1/s\le\eps_0$,
and the size requirement in
\cref{pre:lem:prg-to-extractor} for sampling circuits
of size $O(S)$.

By \cref{pre:lem:short-prg}, there is a function
$K:\zo^d\to\zo^m$, computable in time $\poly(S,1/\nu)$,
with
\(
 d=O(m+\Delta+\log S+\log(1/\nu))\le L,
\)
where the final inequality follows from
\eqref{pre:eq:preliminary-length}, taking $C_0$
sufficiently large.
Define $P(v)=K(v_{[d]})$, where $v_{[d]}$ denotes
the first $d$ coordinates of $v$.

For any source $\V$ in the statement, its projection
$\V_{[d]}$ has min-entropy at least $d-\Delta$
and belongs to $\NPSources_{d,O(S)}$.
Since $1/s\le\eps_0$ and
$\eta\le\eps_0 2^{-(m+\Delta')}$,
\cref{pre:lem:prg-to-extractor} gives
\[
 \Pr[P(\V)=z]
 =\Pr[K(\V_{[d]})=z]
 \le e^{12\eps_0}2^{-m}
 =e^\nu2^{-m},
\]
as required.
\end{proof}

Thus $P$ is an
$(L-\Delta,\nu)$-multiplicative extractor for $\NPSources_{L,S}$.
The multiplicative bound also gives small statistical distance from
uniform, with output length linear in the source length.

\begin{corollary}\label{pre:cor:long-preliminary}
Assume $\NDH$ and fix $a>0$. There are constants $c_0,C_a>0$
such that, whenever $2\le L\le S$ and $L\ge C_a(\Delta+\log S)$,
there is a function $P:\zo^L\to\zo^{\lfloor c_0L\rfloor}$,
computable in time $\poly(S)$, satisfying
$\TV(P(\V),\U_{\lfloor c_0L\rfloor})\le S^{-a}$ for every
$\V\in\NPSources_{L,S}$ with $\minH(\V)\ge L-\Delta$.
\end{corollary}

\begin{proof}
In \cref{pre:lem:preliminary}, take $m=\lfloor c_0L\rfloor$,
the deficiency bound $L/C_a$, and $\nu=S^{-a}/4$.
For sufficiently small $c_0$ and sufficiently large $C_a$,
the length condition holds. Summing the positive differences
between the output probabilities and $2^{-m}$ gives
$\TV(P(\V),\U_m)\le e^\nu-1\le2\nu\le S^{-a}$.
\end{proof}

\subsubsection{Robust errorless block condensers}
\label{sec:robust-definition}

\begin{definition}[Robust errorless block condenser]
\label{def:block-robust}
Fix $0<\lambda\le1$. A function
\[
 \rbCond\colon\zo^{n_y+n_x}\longrightarrow\zo^m
\]
is a $(\lambda,k_y,k_y^{\ge},k_x,k_{\mathrm{out}})$-robust errorless
block condenser for a class $\mathcal{YX}$ of joint distributions
on $\zo^{n_y+n_x}$ if the following holds.
For every $(\Y,\X)\in\mathcal{YX}$ and every decomposition
\[
 (\Y,\X)=\lambda'(\Y_{\mathrm{good}},\X_{\mathrm{good}})
       +(1-\lambda')(\Y_{\mathrm{rest}},\X_{\mathrm{rest}})
\]
as a mixture of joint distributions with $\lambda'\ge\lambda$, satisfying
\begin{itemize}
 \item $\minH(\Y_{\mathrm{good}})\ge k_y$;
 \item $\Pr[\Y_{\mathrm{good}}=y]\ge2^{-k_y^{\ge}}$ for every
       $y\in\supp(\Y_{\mathrm{good}})$;
 \item $\minH(\X)\ge k_x$;
\end{itemize}
we have
\[
 \minH\bigl(\rbCond(\Y_{\mathrm{good}},\X_{\mathrm{good}})\bigr)
 \ge k_{\mathrm{out}}.
\]
\end{definition}

\subsubsection{Output amplification}
We will use the results from \cite{Shaltiel08mileage} to amplify the number of output bits of an extractor, in particular, we use the following formulation for sources sampled with postselection.
\begin{lemma}[Output amplification~{\cite[Section~4.2]{shaltiel25extract}}]
\label{pre:lem:output-amplification}
For all constants $c,d>0$ and $\eta\in(0,1)$, there are constants $C,c_0,d_0>0$ such that the following holds for all sufficiently large $n$.

Let $t=C\log n$ and $k \ge t$. Suppose that
$ \Ext:\zo^n\to\zo^t $
is a $(k,n^{-d_0})$-extractor for distributions samplable by size-$n^{c_0}$ deterministic circuits with postselection. Then, for every $k'$ satisfying
$ \frac{2k}{\eta}\le k'\le n$, 
there is a $(k',n^{-d})$-extractor
$\Ext':\zo^n\to\zo^{\lfloor(1-\eta)k'\rfloor}$
for distributions samplable by size-$n^c$ deterministic circuits with postselection. Moreover, if $\Ext$ is computable in time $T(n)$, then $\Ext'$ is computable in time $T(n)+\poly(n)$.
\end{lemma}
The lemma also holds for sources with nondeterministic postselection, since the conditioning used in the proof only conjoins an efficient deterministic predicate to the original postselection predicate, preserving nondeterministic postselection with polynomial size overhead.

\subsection{Algebra and Coding theory tools}

\subsubsection{Low-degree extension}
\label{sec:lde-prelim}

The following is the standard low-degree extension, as stated
in~\cite[Lemma~3.9]{DS26}, with its individual-degree bound made explicit.

\begin{theorem}[Low-degree extension]
\label{thm:lde}
Let $f\colon\zo^\ell\to\{0,1\}$, and let $d\le h\le q$ be positive
integers such that $h^d=2^\ell$ and $q$ is a power of two.
Let $H\subseteq\F_q$ be the set of the first $h$ field elements, and let
\[
 \phi\colon\zo^\ell\to H^d
\]
be an efficiently computable bijection. There is a polynomial
\[
 \widehat f\colon\F_q^d\to\F_q
\]
of degree at most $h-1$ in each variable, and hence total degree at most
\[
 \hat h\coloneqq d(h-1),
\]
such that $\widehat f(\phi(x))=f(x)$ for every $x\in\zo^\ell$.
Furthermore, $\widehat f$ can be computed in time
$\poly(2^\ell,\log q)$, given oracle access to $f$.
\end{theorem}

\subsubsection{Parametrized lines}
\label{subsec:lines}

\begin{definition}[Parametrized lines and random lines]
\label{def:random-line}\leavevmode
\begin{enumerate}[label=(\arabic*)]
    \item A \emph{parametrized affine line} is a map
    $L\colon\F_q\to\F_q^d$ of the form
    \[
        L(t)=u+t(v-u),\qquad u,v\in\F_q^d.
    \]
    \item We call $L$ \emph{degenerate} if $u=v$, in which case $L$ is constant.
    \item The experiment $\bL\leftarrow\RandLine$ samples $\bu,\bv$ independently
    and uniformly from $\F_q^d$, and returns $\bL(t)=\bu+t(\bv-\bu)$.
\end{enumerate}
\end{definition}

In particular,
\[
    \Pr_{\bL\leftarrow\RandLine}[\bL\text{ is degenerate}]=q^{-d}.
\]
We also write $L$ for its image, as in $w\in L$. For nondegenerate $L$,
let $\U_L$ denote the uniform distribution on its $q$ points, and let
$L^{-1}(w)$ be the unique $t\in\F_q$ with $L(t)=w$.

\begin{lemma}[Basic sampling facts]
\label{lem:line-sampling}
Let $\bL\leftarrow\RandLine$.
\begin{enumerate}[label=(\arabic*)]
    \item For each fixed $t\in\F_q$, $\bL(t)$ is uniform on $\F_q^d$.
    For distinct $t,t'$, the points $\bL(t)$ and $\bL(t')$ are independent.
    \item For every fixed $w\in\F_q^d$,
    \[
        \Pr[w\in\bL\mid\bL\textup{ is nondegenerate}]=q^{-(d-1)}.
    \]
\end{enumerate}
\end{lemma}

\begin{proof}
Write $\bL(t)=\bu+t\bb$, where $\bu,\bb$ are independent uniform points of $\F_q^d$.
For $t\ne t'$, the map $(u,b)\mapsto(u+tb,u+t'b)$ is invertible, proving
the first assertion. Conditioned on nondegeneracy, $\bb$ is uniform on
$\F_q^d\setminus\{0\}$ and independent of $\bu$. Each $\bL(t)$ is still
uniform, and the $q$ points are distinct, so the incidence probability is
$\sum_{t\in\F_q}\Pr[\bL(t)=w\mid\bb\ne0]=q\cdot q^{-d}$.
\end{proof}

\begin{lemma}[Averaging sampler property of a random line]
\label{lem:line-concentration}
For every $A\colon\F_q^d\to[0,1]$ and $\tau>0$,
\[
    \Pr_{\bL\leftarrow\RandLine}\!\left[
        \left|\E_{t\leftarrow\F_q}[A(\bL(t))]
        -\E_{w\leftarrow\F_q^d}[A(w)]\right|>\tau
    \right]
    \le\frac{1}{4q\tau^2}.
\]
\end{lemma}

\begin{proof}
By \cref{lem:line-sampling}, the variables $A(\bL(t))$ are pairwise
independent. Each has mean $\E_{w\leftarrow\F_q^d}[A(w)]$ and variance at
most $1/4$. Their average therefore has variance at most $1/(4q)$.
Apply Chebyshev's inequality.
\end{proof}

\subsubsection{List recovery}

\begin{lemma}[Johnson bound for list recovery]
\label{lem:johnson}
Suppose $S_t\subseteq\F_q$ and $|S_t|\le K$ for every $t$.
If $q\xi^2\ge2\hat hK$, the number of degree-at-most-$\hat h$
polynomials satisfying $p(t)\in S_t$ on at least a $\xi$ fraction
of field elements $t\in\F_q$ is at most $2K/\xi^2$.
\end{lemma}
This is a standard form of the Johnson bound for list recovery;
see~\cite[Lemma~5.2]{GKORS}. For completeness, we give a proof sketch.
\begin{proof}
Let the candidate family have size $N>0$, and let $N(t,a)$ count
its members taking value $a$ at $t$.
Distinct polynomials agree at most $\hat h$ times, so
$\sum_{t,a}N(t,a)^2\le Nq+N(N-1)\hat h$.
There are at most $qK$ pairs $(t,a)$ with $a\in S_t$.
Cauchy--Schwarz therefore gives
\[
 (Nq\xi)^2\le qK\bigl(Nq+N(N-1)\hat h\bigr).
\]
Dividing and using the hypothesis yields
$N\xi^2\le K+NK\hat h/q\le K+N\xi^2/2$,
which proves the claim.
\end{proof}

\subsection{Interactive Protocols}

\subsubsection{Approximate counting and pseudorandomness}
\label{sec:counting-pseudorandomness}

The hashing construction uses approximate counting to recognize
unsuitable hashes, then uses a pseudorandom generator to reduce
the number of hashes.

An $\mathsf{AM}$ protocol has two messages: Arthur sends a uniform
string $u$, and Merlin replies with $v$. A deterministic verifier
$V$ accepts according to $V(x,u,v)$, so the maximum acceptance
probability on input $x$ is $\Pr_u[\exists v:V(x,u,v)=1]$.

We use the following nondeterministic large-set formulation of the
Goldwasser--Sipser theorem~\cite{GoldwasserSipser}, as stated
in~\cite[Theorem~3.14 of the full version]{Shaltiel25multiplicative}.

\begin{theorem}[Goldwasser--Sipser for nondeterministic circuits]
\label{thm:gs-counting}
For every integer $s\ge1$ and $\lambda>0$, there is a
nondeterministic circuit $A(C,\gamma)$ of size
$\poly(s,1/\lambda)$ with the following guarantees for every
nondeterministic circuit $C$ of size at most $s$ and every rational
$\gamma\in(0,1]$ represented using at most $s$ bits:
\begin{align*}
 \Pr_{\rho}[C(\rho)=1]\ge\gamma
 &\quad\Longrightarrow\quad A(C,\gamma)=1,\\
 \Pr_{\rho}[C(\rho)=1]<\gamma e^{-\lambda}
 &\quad\Longrightarrow\quad A(C,\gamma)=0.
\end{align*}
Here $\rho$ is uniform over the ordinary input bits of $C$.
\end{theorem}

The Goldwasser--Sipser protocol extends to nondeterministic $C$ by
having Merlin supply a membership witness for each claimed accepted
input. Amplification and fixing the public coins give the stated
nondeterministic circuit, as in $\mathsf{AM}\subseteq\NP/\mathsf{poly}$.

We will apply this theorem to circuits $C_x$ obtained by hardwiring an
external input $x$ into a fixed circuit. Supplying the description of
$C_x$ and a computed threshold $\gamma(x)$ to $A$ gives one circuit
working for all $x$, with only polynomial overhead. The parameter $s$
is taken large enough to include the threshold's bit length.

\begin{lemma}[Shaltiel--Umans~\cite{SU05}]
\label{hash:lem:su}
Assume $\NDH$. For all constants $a,b>0$, there is an algorithm
that, for all sufficiently large $n$ and every integer
$1\le q\le n^a$, computes a generator
$G_{n,q}\colon\zo^d\to\zo^q$ on a given seed in time $\poly(n)$,
where $d=O(\log n)$, such that every nondeterministic circuit
$T\colon\zo^q\to\zo$ of size at most $n^a$ satisfies
\[
 \left|\Pr[T(G_{n,q}(\U_d))=1]-\Pr[T(\U_q)=1]\right|
 \le n^{-b}.
\]
\end{lemma}

\dobib

\part{Multiplicative Extractors}
\label{part: mult}

We here construct multiplicative extractors for $n^{0.01}$ entropy samplable sources, under nondeterministic hardness. Formally,

\begin{theorem}[Multiplicative extractors under nondeterministic hardness]
\label{thm:multiplicative}
Assume that $\sfE$ is hard for exponential-size nondeterministic
circuits. For every constant $0<\alpha\le1$ there exists a constant
$\alpha'>0$, depending only on $\alpha$, such that, for all constants
$c_{\mathrm{size}}>1$ and $c_\eps>0$, the following holds for all
sufficiently large $n$. There is an explicit polynomial-time computable function
\[
 \Ext\colon\zo^n\longrightarrow\zo^m,\qquad m=n^{\alpha'},
\]
that is an $(n^\alpha,n^{-c_\eps})$-multiplicative extractor for
distributions samplable by deterministic circuits with postselection
by nondeterministic circuits, both of size at most $n^{c_{\mathrm{size}}}$,
in the sense of \cref{def:multiplicative}.
One may take $\alpha'=\alpha\delta_{\mathrm{out}}^2/(8c_{\IExt})$,
where $\delta_{\mathrm{out}}$ and $c_{\IExt}$ are the universal constants
from \cref{thm:errorless-condenser,thm:li}.
\end{theorem}

Using this multiplicative extractor, we will also obtain an additive extractor, with similar requirements, under the same assumptions --- \cref{thm:additive from multiplicative}.

To construct such a multiplicative extractor, we will construct errorless condensers for $n^{0.01}$ entropy samplable sources under nondeterministic hardness. Formally,

\begin{theorem}[Errorless condensers under nondeterministic hardness]
\label{thm:errorless-condenser}
There exists a universal constant $0<\delta_{\mathrm{out}}<1/2$ such
that the following holds. Assume that $\sfE$ is hard for
exponential-size nondeterministic circuits. Then, for all constants
$0<\gamma_{\mathrm{in}}\le1$ and $c>1$, there is an explicit
polynomial-time function
\[
 \Cond_{\gamma_{\mathrm{in}},c}\colon\zo^n
 \longrightarrow\zo^{n^{\gamma_{\mathrm{in}}/2}}
\]
with the following guarantee for all sufficiently large $n$. For every
source $\X$ samplable by a deterministic circuit with postselection by
a nondeterministic circuit, both of size at most $n^c$,
\[
 \minH(\X)\ge n^{\gamma_{\mathrm{in}}}
 \quad\Longrightarrow\quad
 \minH\bigl(\Cond_{\gamma_{\mathrm{in}},c}(\X)\bigr)
 \ge n^{\gamma_{\mathrm{in}}\delta_{\mathrm{out}}}.
\]
Thus the function is an errorless
$(n^{\gamma_{\mathrm{in}}},n^{\gamma_{\mathrm{in}}\delta_{\mathrm{out}}})$-condenser
for this class of distributions, in the sense of
\cref{def:errorless-condenser}.
The polynomial running-time exponent may depend on $\gamma_{\mathrm{in}}$
and $c$, whereas $\delta_{\mathrm{out}}$ is independent of both.
\end{theorem}

The construction of this errorless condenser will rely on a related object which we call robust errorless block condenser, as defined in \cref{def:block-robust}. We will construct the following condenser.
\begin{theorem}[Robust errorless block condensers under nondeterministic hardness]
\label{thm:block-robust}
There exists a universal constant $\delta>0$ such that the following
holds. Assume that $\sfE$ is hard for exponential-size
nondeterministic circuits. Then, for every constant $0<\gamma\le1$
and any $c_{\mathrm{size}}>1$, the following holds for every sufficiently large $n$.
For every
\[
 n^\gamma\le k_y\le k_x\le n,
 \qquad (\log n)^2\le m\le k_y^\delta,
\]
there is an explicit polynomial-time computable function
\[
\rbCond\colon\zo^n\times\zo^n\longrightarrow\zo^m
\]
that is a
\[
 \bigl(\lambda=2^{-m/4},\ k_y,\ k_y^{\ge}=k_x/2,\ k_x,\
       k_{\mathrm{out}}=m/4\bigr)
\]
robust errorless block condenser for joint distributions samplable
by deterministic circuits with postselection by nondeterministic
circuits, both of size at most $n^{c_{\mathrm{size}}}$, in the sense of
\cref{def:postselection}.
\end{theorem}

The rest of the sections are organized as follows.
We will show how to obtain additive extractors from our multiplicative extractors in \cref{sec: additive extractors from multiplicative extractors}.
We show how to construct robust errorless block condensers (\cref{thm:block-robust}), as well as multiplicative extractors from errorless condensers (\cref{thm:multiplicative}) in \cref{sec:robust-proof} and \cref{sec:mult-proof} respectively.
Both these proofs will follow a common theme, and will be an elaborate extension of arguments in \cite{BDGM23} and the common components of these proofs are provided in \cref{sec:joint-preparations}.
We will also give an overview of the arguments involved in these constructions in \cref{sec:proof-overview-BGDM}.

We will show how to construct errorless condenser from robust errorless block condenser and obtain \cref{thm:errorless-condenser} in \cref{sec: construct errorless condenser}. We will do so by extending ideas appearing in \cite{shaltiel25extract}.

\section{Additive Extractors from Multiplicative Extractors}
\label{sec: additive extractors from multiplicative extractors}

In this section, we will show how using the multiplicative extractors from \cref{thm:multiplicative}, we can obtain additive extractors with similar guarantees.

\begin{theorem}[Additive extractors under nondeterministic hardness]
\label{thm:additive from multiplicative}
Assume that $\sfE$ is hard for exponential-size nondeterministic
circuits. For all constants $0<\alpha, \eta < 1$ and constants $c_{\mathrm{size}}>1$ and $c_\eps>0$, the following holds for all
sufficiently large $n$ and $k = n^{\alpha}$. There is an explicit polynomial-time computable function
\[
 \Ext\colon\zo^n\longrightarrow\zo^m,\qquad m= (1-\eta) k,
\]
that is a $(k,n^{-c_\eps})$-extractor for distributions samplable by deterministic circuits with postselection by nondeterministic circuits, both of size at most $n^{c_{\mathrm{size}}}$.
\end{theorem}

\begin{proof}
We set constants $C_0, C_1, C_2 > 0$ appropriately later and let $\Ext_0$ be an $(n^{0.99 \alpha}, n^{-C_0})$-multiplicative extractor for distributions samplable by size $n^{C_1}$ circuits, with postselection by size $n^{C_1}$ nondeterministic circuits, as guaranteed by \cref{thm:multiplicative}.

\paragraph{Construction}
On input $x\in \zo^n$ we let $y = \Ext_0(x)$, and let $y'$ be the length $C_2 \log(n)$ prefix of $y$ (we can take such a prefix since $y$ has length at least $n^{\delta}$ for some constant $\delta > 0$).
Using this, we instantiate \cref{pre:lem:output-amplification} to obtain $\Ext:\zo^n \to \zo^m$ which is an $(n^{\alpha}, n^{-c_{\eps}})$ extractor for sources samplable by size $n^{c_{\size}}$ circuits with postselection by size $n^{c_{\size}}$ nondeterministic circuits.

\paragraph{Analysis}
Let $\X$ be the input distribution and let $\Y, \Y'$ be the distributions of $y, y'$ on that source.
To obtain the desired guarantees on $\Ext$ using \cref{pre:lem:output-amplification}, it suffices to show that $\TV(\Y', \U_{C_2\log(n)}) \le n^{-C_3}$ and that circuit size parameter for $\Ext_0$ is at least $n^{C_4}$ and entropy parameter for $\Ext_0$ is at most $\eta k/2 = \eta\cdot n^{\alpha}/2$ where $C_2, C_3, C_4 > 0$ are constants supplied by \cref{pre:lem:output-amplification} (and depend on $\eta, c_{\size}, c_{\eps}$). Using this, we set $C_1 = C_4$ and $C_0 = C_2 + C_3 + 1$ to see that the parameter guarantees hold for $\Ext_0$, where we used the fact that $\eta\cdot n^{\alpha}/2 > n^{0.99 \alpha}$.
It remains to show that the desired statistical distance guarantees hold for $\Y'$.

Let $n_y = |\Y|$.
By guarantees of $\Ext_0$, we have that for all $y\in \zo^{n_y}$, setting $\eps = n^{-C_0}$,
\[
\Pr[\Y = y]\le e^{\eps}\cdot 2^{-n_y} \le (1 + 2\eps)\cdot 2^{-n_y}.
\]

As $\Y'$ is length $n_{y'} = C_2\log(n)$ prefix of $\Y$, we obtain that for all $y'\in \zo^{n_{y'}}$, 
\[
\Pr[\Y' = y']\le (1 + 2\eps)\cdot 2^{-n_{y'}}.
\]
From this we conclude that,
\[
\TV(\Y', \U_{n_{y'}}) \le 2^{n_{y'}}\cdot 2\eps \le n^{C_2}\cdot 2\cdot n^{-C_0}\le n^{-C_3}
\]
as desired, wherein the last inequality follows by our setting of $C_0$.
\end{proof}

\dobib

\section{Errorless condensers using robust errorless block condensers}
\label{sec: construct errorless condenser}

We here extend the errorless condensers of \cite{shaltiel25extract} under fewer computational assumptions and prove \cref{thm:errorless-condenser}.

% \begin{theorem}
% \label{thm: errorless condenser}
% Assume $\sfE$ is hard for nondeterministic exponential size circuits.
% There exists a constant $\delta_\out > 0$ such that the following holds.
% For all constants $\gamma_{\inn} > 0$, $c > 1$, and all $n\in \N$, there exists an explicit function $\Cond: \zo^n \to \zo^m$ such that for any source $\W$ samplable by size $n^c$ circuit with postselection by size $n^c$ nondeterministic circuit, with $\minH(\W) \ge n^{\gamma_\inn}$, we have that $\minH(\Cond(\W)) \ge n^{\gamma_\inn \cdot \delta_{\out}}$ where $m = n^{\gamma_\inn / 2}$.
% \end{theorem}

% We will construct such a condenser in \cref{sec: construct errorless condenser}.
Towards doing so, we will need two key ingredients.
First, we need the notion of a `robust' errorless block condenser which we defined in \cref{def:block-robust}.
We call such condensers robust since they actually condense from sub-sources of a samplable source.
Second, we need a classical structural result regarding chopping up any entropic source appropriately; we provide a self-contained statement and proof of this in our setting in \cref{sec: structure chopping result}. We will use these ingredients and carefully instantiate them to construct our errorless condensers.
To do that, we will use the robust errorless condenser from \cref{thm:block-robust} and construct a mild errorless condenser in \cref{subsec: basic errorless condenser}. Using this, we will construct our claimed errorless condenser in \cref{subsec: final errorless condenser}.

\subsection{Structural chopping result for any source}
\label{sec: structure chopping result}

We next prove the following structural chopping result for any source; such a result was introduced by \cite{SSZ98disperser, tashma1996extractor} and has been useful over the years in several works in extractor theory including \cite{SSZ98disperser, NT1999extracting, RSW06extractor, shaltiel25extract}.

\begin{lemma}
\label{lem: chop source convex entropy bounds}
For all $\eps > 0$ and $k, n, b, \ell\in \N$ such that $k\le n$ and $\ell\cdot b = n$, the following holds.
For any distribution $\X\sim \zo^n$ with $\minH(\X) \ge k$, consider the partition $\X = (\X_1,\dots, \X_{\ell})$ where $\X_i\sim \zo^b$. Then for all $0\le i\le \ell$, there exist sources $\X^{(i)} \sim \zo^n$ and $\lambda_i\ge 0$ such that $\sum_{i=0}^\ell \lambda_i = 1$, so that we can express $\X$ as a convex combination $\sum_{i=0}^{\ell} \lambda_i \X^{(i)}$ with the following properties:
\begin{enumerate}
\item 
For all $i\ge 0$ with $\lambda_i > 0$, each $\X^{(i)}$ has disjoint support, and hence partitions $\supp(\X)$.
Moreover, for $i\ge 1$, and $x\in \supp(\X)$, membership of $x\in \supp(\X^{(i)})$ is determined by the prefix $x_{[1, i]}$.

\item
$\lambda_0\le \eps$ and for all $i\ge 1$, either $\lambda_i \ge \frac{\eps}{2n}$ or $\lambda_i = 0$.

\item
For all $i\ge 1$ with $\lambda_i > 0$, we have that $\minH\left(\X^{(i)}_{[1, i]}\right) \ge \frac{k}{2} - \log(2n/\eps)$.

\item
For all $i\ge 1$ with $\lambda_i > 0$, and for all $x_i \in \supp\left(\X^{(i)}_{[1, i]}\right)$, we have that
$$\Pr\left[\X^{(i)}_{[1, i]} = x_i\right] \ge 2^{-k/2 - b - \log(2n/\eps)}.$$

\item
For all $i\ge 1$ with $\lambda_i > 0$, it holds that $\X^{(i)}_{[1, i]}\coloneqq \X_{[1, i]} \mid \left(\X \in \supp\left(\X^{(i)}\right)\right)$.
\end{enumerate}
\end{lemma}

\begin{proof}
We will first define several helper functions and sets so that we can define $\lambda_i$ and $\X^{(i)}$. Then, we will show how each of the properties holds for them.

\paragraph{Setup}
First, define an assignment function $\eta: \supp(\X) \to [\ell]$ as follows. Let $\eta(x)$ be the smallest index $i\in [\ell]$ such that $\Pr[\X_{[1, i]} = x_{[1, i]}] \le 2^{-k/2}$ (such an index always exists since $\Pr[\X = x] \le 2^{-k}$).

Next, for $i\in [\ell]$, let
\[
S_i = \{x\in \supp(\X): \eta(x) = i \land \Pr[\X_{[1, i]} = x_{[1, i]}] \ge 2^{-k/2-b-\log(2n/\eps)}\}.
\]

% \Cref{fig: chopping} illustrates these definitions and the resulting partition.

Then, for $i\in [\ell]$, let
\[
T_i =
\begin{cases}
S_i & \Pr[\X \in S_i] \ge \eps / 2n\\
\emptyset & \text{otherwise}
\end{cases}
\]
And then define
\[
T_0 = \supp(\X) \setminus \left(\bigcup_{i=1}^{\ell} T_i\right).
\]
Finally, for all $0\le i\le \ell$, we define $\X^{(i)} = \X \mid (\X\in T_i)$ and $\lambda_i = \Pr[\X\in T_i]$ (if $\lambda_i = 0$, then let $\X^{(i)}$ be arbitrary).

\paragraph{Proving Properties}
By definition, the sets $\{S_i\}_{i\ge 1}$ are disjoint and hence, the sets $\{T_i\}_{i\ge 0}$ partition the support of $\X$.
From this, we can indeed express $\X$ as the convex combination 
\[
\X = \sum_{i=0}^{\ell}\Pr[\X\in T_i]\cdot (\X \mid \X\in T_i) = \sum_{i=0}^{\ell} \lambda_i \X^{(i)}
\]
as desired.
Moreover, membership of $x\in T_i$ clearly depends only on $x_{[1, i]}$. This establishes the first property. We will prove each of the remaining properties one after the other.

\begin{description}
\item[Property 2]
By definition of $T_i$, for each $i\ge 1$ with $\lambda_i > 0$, we have that $\lambda_i \ge \frac{\eps}{2n}$ as desired.

For $\lambda_0$, we find it helpful to define 
\[
S_0 = \supp(\X) \setminus \left(\bigcup_{i=1}^{\ell} S_i\right).
\]
We then claim the following:
\begin{claim}
\label{claim: X in S0 small probability}
$\Pr[\X\in S_0] \le \frac{\eps}{2}$.
\end{claim}

We will prove this claim later. Let us see how our desired bound on $\lambda_0$ follows from this.
Let $L = \{i\in [\ell]: \Pr[\X\in S_i] < \frac{\eps}{2n}\}$.
We observe that $T_0 = S_0\cup \bigcup_{i\in L} S_i$.
From a union bound, we infer that
\[
\lambda_0
= \Pr[\X\in T_0]
\le \Pr[\X\in S_0] + \sum_{i\in L} \Pr[\X\in S_i]
\le \frac{\eps}{2} + \ell\cdot \frac{\eps}{2n}
\le \eps
\]
as desired.

We now prove our remaining claim regarding $S_0$.
\begin{proof}[Proof of \cref{claim: X in S0 small probability}]
For $i\in [\ell]$, define $R_i = \{x\in \supp(\X): \eta(x) = i\}$.
Then, $R_1, \dots, R_{\ell}$ partition $\supp(\X)$.
Also, observe that $S_i \subset R_i$ for all $i\ge 1$.
We claim that for all $i\ge 1$, $\Pr[\X\in (R_i\setminus S_i)] \le \frac{\eps}{2n}$.
This suffices to prove the claim since
\[
\Pr[\X\in S_0] = \sum_{i=1}^{\ell}\Pr[\X\in (R_i\setminus S_i)] \le \ell\cdot \frac{\eps}{2n}\le \frac{\eps}{2}
\]
as desired.

For the rest of this proof, we fix $j\in [\ell]$.
We will show that $\Pr[\X\in (R_j\setminus S_j)] \le \frac{\eps}{2n}$.
Fix any $x^*\in R_j\setminus S_j$.
As $x^*\in R_j$, we have that $\eta(x^*) = j$ and so $\Pr\left[\X_{[1, j-1]} = x^*_{[1, j-1]}\right] > 2^{-k/2}$.
Since $x^*\not\in S_j$ even though $\eta(x^*) = j$, we must have that $\Pr\left[\X_{[1, j]} = x^*_{[1, j]}\right] \le 2^{-k/2-b-\log(2n/\eps)}$.
Hence, we compute that
\[
\Pr\left[\X_{[1, j]} = x^*_{[1, j]} \mid \left(\X_{[1, j-1]} = x^*_{[1, j-1]}\right)\right]
= \frac{\Pr\left[\X_{[1, j]} = x^*_{[1, j]}\right]}{\Pr\left[\X_{[1, j-1]} = x^*_{[1, j-1]}\right]}
\le \frac{2^{-k/2-b-\log(2n/\eps)}}{2^{-k/2}} 
\le \frac{\eps\cdot 2^{-b}}{2n}.
\]
Since this is true for all $x^*\in R_j\setminus S_j$, we obtain the following: Let $V_j\subset \zo^{j\cdot b}$ be the projection of $R_j\setminus S_j$ onto the first $j$ blocks; then for all $v\in V_j$, we have that
\[
\Pr\left[\X_{[1, j]} = v \mid \left(\X_{[1, j-1]} = v_{[1, j-1]}\right)\right] \le \frac{\eps\cdot 2^{-b}}{2n} \numberthis\label{eq: lem partition prop 2 claim}.
\]
Using this, we compute that
\begin{align*}
\Pr[\X\in R_j\setminus S_j]
& \le \Pr[\X_{[1, j]}\in V_j]\\
& = \sum_{\alpha\in \zo^{(j-1)b}} \Pr[\X_{[1, j-1]} = \alpha]\cdot \sum_{\beta\in \zo^b: (\alpha, \beta)\in V_j} \Pr\left[\X_{[1, j]} = (\alpha, \beta) \mid \left(\X_{[1, j-1]} = \alpha\right)\right]\\
& \le \sum_{\alpha\in \zo^{(j-1)b}} \Pr[\X_{[1, j-1]} = \alpha]\cdot \sum_{\beta\in \zo^b: (\alpha, \beta)\in V_j} \frac{\eps\cdot 2^{-b}}{2n}\\
& \le \sum_{\alpha\in \zo^{(j-1)b}} \Pr[\X_{[1, j-1]} = \alpha]\cdot \frac{\eps}{2n}\\
& = \frac{\eps}{2n}
\end{align*}
as desired, where the first inequality follows since $V_j$ is a projection of $R_j\setminus S_j$, and the second inequality follows by \cref{eq: lem partition prop 2 claim}.
\end{proof}

\item[Properties 3 and 4]
Fix $j\ge 1$ with $\lambda_j>0$ (so that $T_j=S_j$) and let $v\in\supp(\X^{(j)}_{[1,j]})$ be arbitrary.
By Property 1, every $x\in\supp(\X)$ with $x_{[1,j]}=v$ lies in $T_j$ (since $v$ is the prefix of some element in $T_j$).
Therefore, $\Pr[\X^{(j)}_{[1,j]}=v]=\Pr[\X_{[1,j]}=v]/\lambda_j$.
By the definitions of $\eta$ and $S_j$, we have that $2^{-k/2-b-\log(2n/\eps)}\le\Pr[\X_{[1,j]}=v]\le 2^{-k/2}$.
Moreover, by Property 2, we have that $\eps/2n\le\lambda_j\le 1$.
Combining these three observations, and appropriately dividing, we infer that:
\[
  2^{-k/2-b-\log(2n/\eps)} \le \Pr\left[\X^{(j)}_{[1,j]}=v\right] \le 2^{-k/2}\cdot\frac{2n}{\eps}.
\]
This establishes Property 4, and also Property 3 after taking logarithms.

\item[Property 5]
This holds because for any $x\in T_i$, whether or not $x\in T_i$ can be decided solely by testing properties of $x_{[1, i]}$.
\end{description}
\end{proof}

\subsection{Constructing basic errorless Condenser}
\label{subsec: basic errorless condenser}

We here construct the following condenser that mildly shrinks and condenses from subsources of a samplable source, given access to another (potentially fully dependent) source.
\begin{lemma}
\label{lem: mild block errorless condenser by chopping}
There exists a universal constant $0 < \delta \le 1/4$ such that, assuming $\sfE$ is hard for nondeterministic exponential size circuits, the following holds.
For any constants $0 < \alpha, \gamma \le 1$, and $c > 1$, for all sufficiently large $n\in \N$, and all $n_a\in \N$ such that $n_a\le n^{\alpha}$, setting $k = n^{\gamma}$, there exists an explicit function $\Cond: \zo^{n_a} \times \zo^n \to \zo^m$ where $m = n^{\alpha - (1-\delta)\gamma}$ with the following properties.
Let $\A\sim \zo^{n_a}$ and $\W\sim \zo^n$ be such that the distribution $(\A, \W)$ is samplable by size $n^c$ circuits with postselection by size $n^c$ nondeterministic circuits, and $(\A, \W) = (1 - p)(\A^{(0)}, \W^{(0)}) + p(\A_\good, \W_{\good})$ for some $p \ge 2^{-n^{\delta\gamma } / 32}$ where $\minH(\A_\good) \ge k$ and $\minH(\W) \ge k$. Then, $\minH(\Cond(\A_\good, \W_\good)) \ge n^{\delta \gamma / 4}$.
\end{lemma}

\begin{proof}
Let $k_y = \frac{k}{16}$.
Let $0 < \delta_{\cref{thm:block-robust}}\le 1$ be the universal constant from \cref{thm:block-robust}. Then, we let $\delta = \delta_{\cref{thm:block-robust}}/4$.

We will construct such a function $\Cond$ and then analyze it.

\paragraph{Construction}
On input $a\sim \A, w\sim \W$, we proceed as follows.
\begin{itemize}
\item 
Let $b = 6k_y, \ell = \frac{n_a}{6k_y}$ and chop $a$ to obtain $a_1, \dots, a_{\ell}\in \zo^b$.

\item
Let $\rbCond: \zo^n \times \zo^n \to \zo^{n_1}$ be a $(2^{-k_1}, k_y/4, k/2, k, k_1)$ robust errorless block condenser for sources samplable by size $n^c$ circuits with postselection by size $n^c$ nondeterministic circuits from \cref{thm:block-robust}, with parameter $\gamma_{\cref{thm:block-robust}} = \gamma/2$, where $n_1 = k^{\delta}/4$ and $k_1 = n_1 / 4$ (we meet the output length requirement that $n_1\le (k_y/4)^{\delta_{\cref{thm:block-robust}}}$ by using the fact that $\delta \le \delta_{\cref{thm:block-robust}}/4$; the rest of the requirements are easily met).

\item
For $i\in [\ell]$, let $y_i = \rbCond(a_{[1, i]}, w)$ and output $y = (y_1, \dots, y_{\ell})$ and pad appropriately to obtain the desired output length.
\end{itemize}
% \Cref{fig: basic condenser} illustrates the construction.

\paragraph{Output Length}
First, we compute that the output length is
\[
m = \ell\cdot n_1 = \frac{n_a}{6k_y}\cdot \frac{k^{\delta}}{4}
\le \frac{1}{24}\cdot n^{\alpha + \gamma\delta}\cdot \frac{16}{n^{\gamma}}
= \frac{2}{3}\cdot n^{\alpha - (1-\delta)\gamma}
\le n^{\alpha - (1-\delta)\gamma}
\]
as desired.

\paragraph{Setting up random variables and applying the structure chopping result}
% Let $\W' = \W | (\A \equiv \X)$.
Let $\X = \A_\good$ so that
\[
(\A, \W) = (1 - p)(\A^{(0)}, \W^{(0)}) + p(\X, \W_{\good}).
\]
Let $\Y = \Cond(\A, \W) = (\Y_1, \dots, \Y_{\ell})$ with $\Y_i = \rbCond(\A_{[1, i]}, \W)$. Let $\Y' = \Cond(\X, \W_\good)$, and analogously define $\Y'_i$.
Our overall claim then becomes that $\minH(\Y') \ge n^{\delta \gamma / 4}$.

We apply \cref{lem: chop source convex entropy bounds} to $\X$ with values of $b, \ell, k_y$ as above with parameter $\eps = 2^{-k_1 / 4}$ and length parameter $n_a$ to write $\X$ as a convex combination of sources $\X = \sum_{i=0}^{\ell} \lambda_i \X^{(i)}$ with the properties guaranteed there.
When applying these properties below, we will repeatedly use that $n_a\le n$.

For all $0\le i\le \ell$, let $\Y^{(i)'} = \Y' \mid (\X\in \supp(\X^{(i)}))$ so that we can express $\Y'$ as the convex combination $\Y' = \sum_{i=0}^{\ell} \lambda_i \Y^{(i)'}$.
Let $G = \{i\in [\ell]: \lambda_i > 0\}$.
For $i \in G$, write $\Y^{(i)'}_i$ for the $i$-th block of $\Y^{(i)'}$, that is,
\[
\Y^{(i)'}_i = \Y'_i \mid \left(\X \in \supp(\X^{(i)})\right).
\]
We claim that for all $i\in G$, $\minH(\Y_i^{(i)'}) \ge k_1$.
This suffices to prove the desired claim as then for any $y\in \zo^m$, we will have that 
\[
    \Pr[\Y' = y]
    = \sum_{i=0}^{\ell} \lambda_i \Pr[\Y^{(i)'} = y]
    \le \lambda_0 + 2^{-k_1}
    \le \eps + 2^{-k_1}
    = 2^{-k_1 / 4} + 2^{-k_1}
    \le 2^{-k_1/8}.
\]
From this, we will be able to conclude that $\minH(\Y') \ge k_1 / 8 = n^{\delta \gamma} / 128 \ge n^{\delta \gamma / 4}$ as desired.

\paragraph{Applying robust errorless condenser}
It remains to show that $\minH(\Y_i^{(i)'}) \ge k_1$ for all $i\in G$.
Fix any such $i\in G$.
% Observe that $\Y_j^{(j)'} = \rbCond(\A_{[1, j]}, \W) | (\A\equiv \X, \X\in\supp(\X^{(j)}))$.
By Property 5 of \cref{lem: chop source convex entropy bounds}, we have that $\X_{[1, i]} \mid (\X\in \supp\X^{(i)})$ is exactly the same distribution as $\X_{[1, i]}^{(i)}$. Therefore, we can express $\X_{[1, i]}$ as the convex combination
\[
\X_{[1, i]} = \lambda_i \left(\X_{[1, i]} \mid \X\in \supp\left(\X^{(i)}\right)\right) + (1 - \lambda_i)\left(\X_{[1, i]} \mid \left(\X \not\in \supp\left(\X^{(i)}\right)\right)\right).
\]
Combining this with the fact that 
$\A= p\X + (1-p)\A^{(0)}$, we obtain
\[
\left(\A_{[1,i]}, \W\right)
=p\lambda_i\left(\X_{[1,i]}, \W_\good \mid \X\in \supp\left(\X^{(i)}\right)\right)
+(1-p\lambda_i)\left(\widetilde \A_i, \widetilde \W_i\right)
\]
for some distribution $(\widetilde \A_i, \widetilde \W_i)$.

We write, $\Y_i^{(i)'} = \rbCond\left(\X_{[1, i]}, \W_\good \mid \left(\X\in\supp\left(\X^{(i)}\right)\right)\right)$ and apply \cref{thm:block-robust} to the pair $(\A_{[1, i]}, \W)$ to obtain the desired min-entropy bounds on $\Y_i^{(i)'}$.
Note that $(\A_{[1, i]}, \W)$ is samplable by size $n^c$ circuits with postselection by size $n^c$ nondeterministic circuits, as we can truncate the output of the sampler for $(\A, \W)$ and keep the same postselection circuit.

We now show that we meet all the requirements required to apply \cref{thm:block-robust}.
\begin{enumerate}
\item 
First, we show that we meet the requirement on $p\lambda_i$, the weight with which $\X_{[1, i]}^{(i)}$ appears as a subsource of $\A_{[1, i]}$.
Since $i\in G$, we have that
\begin{align*}
p\cdot \lambda_i
& \ge p\cdot \frac{\eps}{2n}\\
& \ge 2^{-n^{\gamma \delta} / 32}\cdot \frac{2^{-k_1 / 4}}{2n}\\
& = 2^{-n^{\gamma \delta} / 32}\cdot 2^{-(n^{\gamma \delta} / 64) - 1 - \log(n)}\\
& \ge 2^{-n^{\gamma \delta} / 16}\\
& = 2^{-k_1}
\end{align*}
as desired.

\item
We next show that $\X_{[1, i]}^{(i)}$ has sufficient entropy.
Indeed, by Property 3 of \cref{lem: chop source convex entropy bounds}, we have that
\[
\minH\left(\X_{[1, i]}^{(i)}\right)
\ge \frac{k_y}{2} - \log(2n/\eps)
= \frac{k_y}{2} - \frac{k_1}{4} - \log(n) - 1
\ge \frac{k_y}{4}
\]
as desired.

\item
We now lower bound the probability of each outcome of $\X_{[1, i]}^{(i)}$.
Indeed, by Property 4 of \cref{lem: chop source convex entropy bounds}, for any $x_i\in \supp\left(\X^{(i)}_{[1, i]}\right)$ we have that
\[
\Pr\left[\X_{[1, i]}^{(i)} = x_i\right] 
\ge 2^{-k_y/2 - b - \log(2n/\eps)}
= 2^{-(k/32) - (3k/8) - (k^{\delta} / 64) - \log(n) - 1}
\ge 2^{-k/2}
\]
as desired.

\item
Lastly, we check that $\W$ has sufficient min-entropy since by assumption, $\minH(\W) \ge k$.

\end{enumerate}

\end{proof}

\subsection{Constructing the Errorless Condenser}
\label{subsec: final errorless condenser}

We finally construct our desired errorless condenser using the basic condenser from \cref{lem: mild block errorless condenser by chopping} and robust errorless block condenser from \cref{thm:block-robust}.

\begin{proof}
[Proof of \cref{thm:errorless-condenser}]
We first construct the condenser and then analyze it. The construction repeatedly shrinks $\W$ using the basic condenser of \cref{lem: mild block errorless condenser by chopping}, and at every stage applies the robust condenser of \cref{thm:block-robust} to the resultant string, together with $\W$.

\paragraph{Construction}
On input $w\in \zo^n$, we proceed as follows.
\begin{itemize}
\item
Let $\gamma = \gamma_\inn/4$ and let $0 < \delta \le \frac{1}{4}$ be the universal constant from \cref{lem: mild block errorless condenser by chopping}.

\item
Let $k = n^{\gamma}$ and for $i\ge 0$, let $n_i = \frac{n}{k^{i\cdot (1-\delta)}}$.

\item
Let $t$ be the earliest index for which $n_t < k$. We immediately see that $k^{\delta} \le n_t < k$.

\item
For $i\in [t]$, let $\Cond_i: \zo^{n_{i-1}} \times \zo^n \to \zo^{n_i}$ be the condenser from \cref{lem: mild block errorless condenser by chopping} for input entropy parameter $k$ and for sources samplable by size $n^{c_i}$ circuits with postselection by size $n^{c_i}$ nondeterministic circuits, where we show how we recursively set $c_i$ in the Explicitness part below. We check that $n_{i-1} \ge k$ since $t$ is the earliest index $j$ for which $n_j < k$; also by output length guarantees of $\Cond_i$, we have that $n_i = n_{i-1} / k^{1-\delta}$, which by induction equals $n / k^{i\cdot (1-\delta)}$, the value that we defined $n_i$ to be.

\item
Let $a_0 = w$ and for $i\in [t]$, iteratively define $a_i = \Cond_i(a_{i-1}, w)$.

\item
Let $\rbCond^*: \zo^n \times \zo^n \to \zo^{m^*}$ be the $(2^{-m^{*}/4}, k_y, k_x/2, k_x, m^*/4)$ robust errorless block condenser from \cref{thm:block-robust} for input parameter $k_y = n^{\gamma \delta / 8}, k_x = n^{4\gamma}$ and for sources samplable by size $n^{c^*}$ circuits with postselection by size $n^{c^*}$ nondeterministic circuits, where $m^* = n^{\gamma \delta\cdot \delta' / 8}$ ($\delta'$ is the universal constant from \cref{thm:block-robust}) and we set $c^*$ to be large enough, dependent on $c_t$.

\item
For $i\in [t]$, let $y_i = \rbCond^*(a_i, w)$.

\item
Output $y = (y_1, \dots, y_t)$, appropriately padded with zeroes to obtain the desired output length.
\end{itemize}
% \Cref{fig: final construction} illustrates the construction.

\paragraph{Explicitness}
We here argue that our construction is polynomial time computable.

We first show that $(\A_i, \W)$ are sources that can be generated by polynomial sized circuits. More concretely, we can show that the pair $(\A_i, \W)$ (defined below) is samplable by polynomial size circuits with postselection by the same size $n^c$ nondeterministic circuit used for $\W$: we sample $\W$, and then compute $\A_i$ from it using $\Cond_1, \dots, \Cond_{i}$.
To establish this, we need to define the size parameter $c_i$, that is used to define $\Cond_i$.
We associate a related quantity $d_i\in \N$ that we also recursively define.
We set $c_1 = c+1$ and let $d_1\in \N$ be such that $\Cond_1$ can be implemented by size $n^{d_1}$ circuits (since $\Cond_1$ runs in polynomial time, such $d_1$ exists).
With this, we let $c_i$ be defined as $c_i = c_{i-1} + d_{i-1}$ so that one can generate the source $\A_{i-1}$ by size $n^{c_i}$ circuits. With this, we let $d_i$ be such that $\Cond_i$ can be implemented by size $n^{d_i}$ circuits.
We will set $c^* = c_t + d_t$ so that $\A_t$ can be generated by size $n^{c^*}$ circuits.

We easily see that $t = O(1)$ is a constant, and so we iterate only a constant number of times. Using this, we have that each $c_i$ will be a constant, and each $\Cond_i$ will be polynomial time computable. 
From this, we have that $c^*$ will be a constant, and the resultant $\rbCond^*$ will be polynomial time computable, making our overall function explicit as desired.

\paragraph{Output length}
We now show that our output length is as claimed.
As the length of $y$ is $t\cdot n^{\gamma \delta\cdot \delta' / 8} \le n^{\gamma / 2} \le n^{\gamma_\inn/2}$, we can indeed pad the output to obtain the desired output length.

\paragraph{Setup for Entropy bound}
It remains to prove our output min-entropy bound and this is what we do for the rest of the proof.
Towards proving this, we define the following distributions.
The input distribution is $\W$. 
We let $\A_0 = \W$ and for $i\in [t]$, iteratively define $\A_i = \Cond_i(\A_{i-1}, \W)$.
Then, for $i\in [t]$, let $\Y_i = \rbCond^*(\A_i, \W)$ and let $\Y = (\Y_1,\dots, \Y_t)$.
To prove our output min-entropy bound, it suffices to show that $\minH(\Y) \ge n^{\gamma_{\inn}\cdot \delta_\out}$ where we will set $\delta_\out = \delta\cdot \delta'/256$.

While the above quantities correspond to the input, we also introduce the following quantities for the purpose of analysis. 
We will recursively partition the support of $\W$ into various sets. Towards that, we will introduce some sets $P_1, \dots, P_t, Q_1, \dots, Q_t \subset \zo^n$.
They will maintain the invariant that for all $i\in [t]$, $Q_1, \dots, Q_i, P_i$ partition the support of $\W$.
To construct $Q_{i+1}$, we will further split $P_i$, and then the set $P_{i+1}$ will equal $P_i\setminus Q_{i+1}$.

We now show how to form such a partition of $\supp(\W)$. 
First, for any input $w\in \zo^n$, we let $a_i(w)$ be the quantity $a_i$ when run on input $w$ in the above algorithm.
For the base case of our new definitions, let $\X_0 = \A_0' = \A_0 (= \W)$. Let $L_0 = P_0 = \supp(\W)$ and $M_0 = \emptyset$. Then, for $i\in [t]$, define the following quantities.
\begin{itemize}
\item
Let $\A_i' = \A_i \mid (\W\in P_{i-1})$. This is the distribution that we want to condense from and is the `remaining' distribution.

\item
Let $M_i = \{x\in \zo^{n_i}: \Pr[\A_i' = x] \ge 2^{-2k}\}$, the set of outcomes with medium weight under $\A_i'$.

\item
Let $L_i = \supp(\A_i')\setminus M_i$, the set of outcomes with light weight under $\A_i'$.

\item
Let $P_i = \{w\in P_{i-1}: a_i(w)\in L_i\}$ and let $p_i = \Pr[\W \in P_i]$. So, this is the part of the support of $\W$ that is given light weight under $\A'_i$.

\item
Let $Q_i = \{w\in P_{i-1}: a_i(w)\not\in L_i\}$ and let $q_i = \Pr[\W\in Q_i]$. So, this is the part of the support of $\W$ that is given medium weight under $\A'_i$.

\item
Let $\X_i = \A_i' \mid (\A_i' \in L_i)$. 
Equivalently, $\X_i = \A_i \mid \W\in P_i$.
This will be the good subsource of $\A_i$, on which $\Cond_{i+1}$ will act.

\item
Let $\Y'_i = \Y_i \mid (\W \in Q_i)$.
This will be the good subsource of $\Y_i$, and we will apply $\rbCond^*$ on it.
\end{itemize}

Let $\tau = n^{\gamma\cdot \delta\cdot \delta'/32}$.

\paragraph{Defining the cutoff index $j$}
Let $j\in [t]$ be the earliest index such that $p_j \le 2^{-\tau}$.
Since $p_{i-1}\ge p_i$ for all $i\in [t]$, to show that such a $j$ exists, it suffices to prove that $p_t \le 2^{-\tau}$.
By definition of $L_t$, and the fact that $n_t\le k$, we have that $\Pr[\A'_t\in L_t] \le 2^{n_t}\cdot 2^{-2k} \le 2^{-k}$. From this, we directly compute that
\[
p_t
= p_{t-1}\cdot \Pr[\A'_t\in L_t]
\le 1\cdot 2^{-k}
\le 2^{-\tau}
\]
as desired where for the last inequality we used the fact that $k = n^{\gamma} \ge n^{\gamma\cdot \delta\cdot \delta' / 32} = \tau$.

\paragraph{Overview of the remaining analysis}
We will prove lower bounds on entropy for $\A'_i, \Y'_i$ for $1\le i \le j$, and using them obtain our lower bound on output entropy.
We call $M_i$ and $L_i$ the medium and light parts of $\A'_i$;
% \cref{fig: analysis partition} shows how the inputs split into these parts, stage by stage.
we first claim that the light part at each stage has enough entropy to satisfy the requirements of the basic condenser $\Cond_{i+1}$, so that the conditioned sources stay entropic up to stage $j$. Formally,
\begin{claim}
\label{claim: thm proof A'_i has entropy}
For every $0 \le i < j$, we have $\minH(\A'_{i+1}) \ge n^{\delta\gamma/4}$.
\end{claim}

Next, we will show that $\rbCond^*$ condenses from every medium part of non-negligible weight. Formally,
\begin{claim}
\label{claim: thm proof Y'_i has entropy}
For every $1 \le i \le j$ with $q_i \ge 2^{-\tau}$, we have $\minH(\Y'_i) \ge m^*/4$.
\end{claim}

We will prove these claims one after the other towards the end. Let us first obtain the theorem assuming them.

\paragraph{Lower bounding output entropy}
We will lower bound the entropy of our output distribution $\Y$.
Let $\Low_q = \{i\in [j]: q_i < 2^{-\tau}\}$.
As $Q_1, \dots, Q_j, P_j$ partition the support of the input distribution $\W$, we have that $p_j + \sum_{i=1}^j q_i = 1$.
Using this, and that for $i\not\in\Low_q$, \cref{claim: thm proof Y'_i has entropy} shows $\minH(\Y'_i) \ge m^*/4$, we obtain that for any outcome $y\in \zo^m$: 
\begin{align*}
\Pr[\Y = y] 
& \le p_j + \sum_{i\in \Low_q} q_i + \sum_{i\not\in \Low_q} q_i\cdot 2^{-m^*/4}\\
& \le 2^{-\tau} + \abs{\Low_q}\cdot 2^{-\tau} + 2^{-m^* / 4}\\
& \le (j+2)\cdot 2^{-\tau}
\end{align*}
where for the last inequality, we used the fact that $m^*/4 \ge \tau$.

From this, we conclude that 
\[
\minH(\Y)
\ge \tau - \log(j+2)
\ge \tau / 2
\ge n^{\gamma_\inn \delta \delta' / 256}.
\]
As $\delta, \delta'$ are both universal constants, we set $\delta_{\out} = \delta\delta' / 256$, to obtain the desired result.

\paragraph{Lower bounding entropy of $\A'_i$ for $1\le i \le j$}

Next, we show that $\minH(\A'_{i+1}) \ge n^{\delta \gamma / 4}$ for $i < j$.
\begin{proof}[Proof of \cref{claim: thm proof A'_i has entropy}]
This holds for $i = 0$ since $\A'_0 = \W$, $p_0 = 1$, and by the guarantees of $\Cond_1$ from \cref{lem: mild block errorless condenser by chopping}.
Fix $i\ge 1$.
We will verify the hypotheses of the basic condenser $\Cond_{i+1}$ for $(\A_i,\W)$, with good component $(\A_i, \W) \mid (\W\in P_i)$ of weight $p_i$, and show that its output guarantees yield our entropy bound.
Also recall that $\X_i = \A_i \mid (\W\in P_i)$.

First, by assumption on $j$, we have that the weight of $\X_i$ in $\A_i$, $p_i$ is such that 
\[
p_i \ge 2^{-\tau} = 2^{-n^{\gamma\cdot \delta\cdot \delta'/32}} \ge 2^{-n^{\delta \gamma} / 32}.
\]
% Observe that $\A_{i+1}' = \Cond_{i+1}(\A_i, \W) \mid (\W \in P_i)$.
and so we satisfy the weight requirement of $\Cond_{i+1}$.
We also easily satisfy the requirement that $\minH(\W) \ge n^{\gamma_\inn} \ge n^{\gamma} = k$.

We now show that we satisfy the remaining requirement, that $\minH(\X_i) \ge n^{\gamma} = k$.
By definition of $L_i$, for all $x\in L_i$, we have that $\Pr[\A'_i = x] \le 2^{-2k}$.
Also, by definition of $\A'_i$, we see that 
\[
\Pr[\A'_i \in L_i] = \frac{p_i}{p_{i-1}} \ge p_i \ge 2^{-\tau}
\]
where we used the fact that $i < j$ for the last inequality.
As $\X_i = \A_i' \mid (\A_i' \in L_i)$, we conclude that 
\[
    \minH(\X_i) \ge 2k - \tau \ge k
\]
as desired.
Hence, we satisfy all the requirements to apply \cref{lem: mild block errorless condenser by chopping} and indeed conclude that $\minH(\A'_{i+1}) \ge n^{\delta \gamma / 4}$.
\end{proof}

\paragraph{Lower bounding entropy of $\Y'_i$ for $1 \le i \le j$}
We next show entropic bounds on $\Y'_i$.
\begin{proof}[Proof of \cref{claim: thm proof Y'_i has entropy}]
For that, we check that we satisfy the requirements to apply \cref{thm:block-robust}. This suffices to prove our claim since the output guarantees of $\rbCond^*$ will show that $\minH(\Y'_i) \ge m^*/4$.

Recall that we apply $\rbCond^*$ to the source $(\A_i,\W)$, which is samplable by size $n^{c^*}$ circuits with postselection by size $n^{c^*}$ nondeterministic circuits, and take
$(\A_i, \W) \mid (\W\in Q_i)$ as the good subsource, which will have weight $\Pr[\W\in Q_i] = q_i$.
Also recall that $\A_i\mid(\W\in Q_i) = \A'_i \mid (\A'_i\in M_i)$.
The corresponding conditioned output is exactly $\Y'_i$, i.e., $\Y'_i = \Y_i \mid (\W\in Q_i)$.

To apply $\rbCond^*$, we first require that $q_i \ge 2^{-m^*/4}$. 
By assumption on $i$, $q_i \ge 2^{-\tau}$. As $\tau = n^{\gamma\cdot \delta\cdot \delta'/32} \le \frac{1}{4}\cdot n^{\gamma\cdot \delta\cdot \delta'/8} = m^*/4$, the desired lower bound on $q_i$ follows.

We now show that we satisfy the min-entropy requirements for $\rbCond^*$.
We satisfy the requirement for $\W$, since by assumption, $\minH(\W) \ge n^{4\gamma} = n^{\gamma_\inn}$.
The remaining entropy requirement is that $\minH(\A'_i \mid (\A'_i\in M_i)) \ge n^{\delta \gamma / 8}$. 
From \cref{claim: thm proof A'_i has entropy}, we have that $\minH(\A'_i) \ge n^{\delta \gamma / 4}$.
Using the fact that $\Pr[\A'_i \in M_i] = \frac{q_i}{p_{i-1}} \ge q_i \ge 2^{-\tau} = 2^{- n^{\delta \delta' \gamma / 32}}$, we obtain that
\[
\minH(\A'_i \mid (\A'_i\in M_i))
\ge n^{\delta \gamma / 4} - n^{\delta \delta' \gamma / 32}
\ge n^{\delta \gamma / 8}
\]
as desired.

Recall that for all $x\in M_i$, we have that $\Pr[\A'_i = x] \ge 2^{-2\cdot n^{\gamma}}$.
Therefore, for any such $x$, we infer that
\[
\Pr[\A'_i = x \mid (\A'_i\in M_i) ]
= \frac{\Pr[\A'_i = x]}{\Pr[\A'_i\in M_i]}
\ge \Pr[\A'_i = x]
\ge 2^{-2\cdot n^{\gamma}}
\ge 2^{-n^{4\gamma}/2}
\]
satisfying the probability lower bound requirement of $\rbCond^*$.
Hence, we satisfy all requirements of $\rbCond^*$ and conclude that $\minH(\Y'_i) \ge m^*/4$.
\end{proof}

\end{proof}

\dobib

\section{Overview of the proofs of \texorpdfstring{\cref{thm:block-robust} and \cref{thm:multiplicative}}{Theorem~\ref{thm:block-robust} and Theorem~\ref{thm:multiplicative}}}
\label{sec:proof-overview-BGDM}

In this section, we give an informal overview of the technique used to
prove \cref{thm:block-robust,thm:multiplicative}.
This overview is not used in the formal proofs. Our approach builds on
past work on extractors for samplable distributions
\cite{TV00,AASY16,BDGM23,Shaltiel25multiplicative,BSS25,shaltiel25extract}, and it will be
helpful to begin by recalling some of this work.

\subsection{The extractor of \texorpdfstring{\cite{TV00}}{TV00} and the analysis of \texorpdfstring{\cite{BDGM23}}{BDGM23}}
\label{sec:overview-tv-bdgm}

\paragraph{The extractor construction of \texorpdfstring{\cite{TV00}}{TV00}.}
In their seminal paper, Trevisan and Vadhan~\cite{TV00} constructed an
extractor for samplable distributions for very high min-entropy, using the following
construction.\footnote{Some of the description here uses later observations
by~\cite{AASY16,BSS25}.}
For an input random variable $\X$ on $\zo^n$, the construction is as follows.
\begin{itemize}
\item The construction starts from a hard function
$f\colon\zo^{d\log n}\to\zo$, for a sufficiently large constant $d$.
It uses the low-degree extension (stated
precisely in \cref{thm:lde}) to convert $f$ into a $d$-variate
polynomial $\widehat f\colon\F_q^d\to\F_q$ of degree roughly $n$,
where $q=2^{\eta n/d}$ for some very
small constant $\eta>0$. This choice gives that the input length
of $\widehat f$ in bits is $d\log q=\eta n$.
\item Let $\Cond(\X)$ denote the first $\eta n$ bits of $\X$,
identified with an element of $\F_q^d$.
\item The construction also requires a
$(k_1,k_2,\eps_{\TExt})$-two-source extractor
$\TExt\colon\F_q\times\zo^n\to\zo^m$.
Although we aim for large $m$, for now think of $m=1$.
\end{itemize}
The final construction is
\[
 \Ext(\X)=\TExt(\widehat f(\W),\X),\qquad \W=\Cond(\X).
\]

Trevisan and Vadhan showed that if $f$ is hard for exponential size
$\Sigma_6$ circuits, then $\Ext$ is an extractor for samplable
distributions with very high min-entropy, namely $k=(1-\eta')n$ for some constant
$\eta'>0$ which is sufficiently smaller than $\eta/d$.

\paragraph{Improved analysis by \texorpdfstring{\cite{BDGM23}}{BDGM23}.}
Subsequent work by Ball, Dachman-Soled, Goldin and Mutreja~\cite{BDGM23}
improved the analysis, and showed that this construction works, even if
$f$ is only assumed to be hard for exponential size
\emph{nondeterministic} circuits.

In this paper, we simplify and generalize the analysis of~\cite{BDGM23}.
This allows us to tailor the construction and analysis to the case of
low min-entropy, and is a major component in the proofs of
\cref{thm:block-robust,thm:multiplicative}.
We start by reviewing the approach of~\cite{BDGM23}.

\paragraph{High level description of the analysis of \texorpdfstring{\cite{BDGM23}}{BDGM23}.}
Assume that $\Ext$ is not a good extractor. Then there exists a
samplable distribution $\X$ with $\minH(\X)\ge k$, and an output
$\sigma\in\zo^m$, such that
$\Pr[\Ext(\X)=\sigma]>(1+\eps)2^{-m}$.
We will derive a contradiction by showing that there exists a small
nondeterministic circuit $C$ such that for every $x\in\F_q^d$,
$C(x)=\widehat f(x)$, contradicting the hardness assumption.\footnote{As
$\widehat f$ is not a Boolean function, we consider \emph{single-valued}
nondeterministic circuits; see \cref{def:sv-circuits}.}
Here, small means size $\operatorname{poly}(n)$ for a polynomial that
does not depend on $d$. Note that $q$ is exponential in $n$, and so
the circuit $C$ needs to have size that is much smaller than $q$.

In fact, by previous work, it is sufficient to construct such a $C$ that only succeeds
well on average. That is,
$\Pr_{x\leftarrow\F_q^d}[C(x)=\widehat f(x)]\ge15/16$, as such a
circuit can be converted into a slightly larger circuit that computes
$f$ correctly on every input.\footnote{This is standard for deterministic
circuits~\cite{GLRSW91}, and there are subtleties in generalizing this argument for
nondeterministic circuits. Jumping ahead, we mention that in the formal
treatment in \cref{sec:reconstruction}, we need to consider
single-valued nondeterministic circuits. We ignore this technicality in
this high-level overview.}

Following~\cite{STV01PRG}, the circuit $C$ will work as follows: when
receiving a random point $x\leftarrow\F_q^d$ as input, $C$ will choose another
random point $z\leftarrow\F_q^d$, and consider the line
$L\colon\F_q\to\F_q^d$ that passes through these two points, say with
$L(0)=x$ and $L(1)=z$. Note that the univariate polynomial
$\widehat f\circ L\colon\F_q\to\F_q$ is of degree roughly $n$, and
$(\widehat f\circ L)(0)=\widehat f(x)$, which is our desired output.

The nondeterministic circuit $C$ will guess a degree-$n$
univariate polynomial $p\colon\F_q\to\F_q$. If we can devise a small nondeterministic
circuit $T_L(p)$ that accepts the correct polynomial $p=\widehat f\circ L$
and rejects all
other polynomials, then $C$ can test whether $p$ is correct and output
$p(0)$.

In fact, following~\cite{STV01PRG}, the task of $T_L$ can be relaxed.
It is sufficient to devise a test $T_L(p)$ such that the correct
polynomial is accepted, and the total number of accepted
degree-$n$ polynomials is significantly smaller than $q$.
We do not explain this step in this overview, and the precise details
appear in \cref{sec:reconstruction}.

Following the previous discussion, we have reduced the task of
constructing the circuit $C$ to the task of devising a test $T_L(p)$
such that for a random line $\bL\colon\F_q\to\F_q^d$, we have the following
properties with high probability:
\begin{description}
\item[Efficiency:] $T_{\bL}$ can be computed by a small nondeterministic circuit.
\item[Completeness:] $T_{\bL}$ accepts the correct polynomial $\widehat f\circ\bL$.
\item[Soundness:] The number of degree-$n$ polynomials that
$T_{\bL}$ accepts is significantly smaller than $q$.
\end{description}

\subsection{A simpler test that can be generalized to low min-entropy}
A significant technical contribution of~\cite{BDGM23} is devising such
a test. In this paper, we give a simplified version of this test.
(We will later need to modify the construction and generalize the test
in order to handle low min-entropy.)

\paragraph{A simplified version of the test of \texorpdfstring{\cite{BDGM23}}{BDGM23}.}
We present our simplified version, starting with the setup of~\cite{BDGM23}.
Recall that $\W=\Cond(\X)$. For $w\in\F_q^d$ and $a\in\F_q$, we define:
\[
 J(w,a)=\Pr[\W=w,\ \TExt(a,\X)=\sigma].
\]
This definition is made so that for every $w\in\F_q^d$,
\[
 J(w,\widehat f(w))
 =\Pr[\W=w,\ \TExt(\widehat f(\W),\X)=\sigma]
 =\Pr[\W=w,\ \Ext(\X)=\sigma].
\]
This gives that:
\[
 \E_{w\leftarrow\F_q^d}[J(w,\widehat f(w))]
 =\frac{1}{q^d}\sum_{w\in\F_q^d}\Pr[\W=w,\ \Ext(\X)=\sigma]
 =\frac{1}{q^d}\Pr[\Ext(\X)=\sigma].
\]
We stress that the expectation is with respect to a \emph{uniform}
$w\leftarrow\F_q^d$.
We then define
\[
 \Test_L(p)=\E_{t\leftarrow\F_q}[J(L(t),p(t))],
 \qquad
 T_L(p)=1\iff\Test_L(p)\ge\frac{1}{q^d}(1+\eps/2)2^{-m}.
\]

\paragraph{Efficiency.}
Efficiency follows because, writing $\X=\Samp(\bR)$, we can define a
deterministic circuit $A_{L,p}(t,R)$ that accepts if
$\Cond(\Samp(R))=L(t)$ and $\TExt(p(t),\Samp(R))=\sigma$.
Its acceptance probability over uniform $(t,R)$ is exactly $\Test_L(p)$.
(A careful implementation of this approach appears in
\cref{sec:sum-of-maximums}.) While $C$ does not have time to go
over all pairs $(t,R)$ and compute $A_{L,p}(t,R)$, it can use the
set-lower-bound protocol of Goldwasser and Sipser~\cite{GoldwasserSipser}
(see \cref{thm:gs-counting} for a precise statement) to
approximately count the fraction of pairs $(t,R)$ that $A_{L,p}$
accepts. (It should be noted that the fraction is typically very small,
and so it is critical to use a multiplicative approximation, as in the
Goldwasser--Sipser protocol, rather than an additive approximation as
given by standard random sampling.)
A technicality that we ignore in this high-level overview is that the
Goldwasser--Sipser protocol has a (multiplicative) approximation error.

\paragraph{Completeness.}
Completeness follows directly from the fact that a random line is a
good averaging sampler. More precisely, for the correct polynomial
$p(t)=(\widehat f\circ L)(t)$,
\[
 \E_{t\leftarrow\F_q}[J(L(t),p(t))]
 =\E_{t\leftarrow\F_q}[J(L(t),\widehat f(L(t)))],
\]
and for a random line $\bL$, with high probability this quantity is close to
\[
 \E_{w\leftarrow\F_q^d}[J(w,\widehat f(w))]
 =\frac{\Pr[\Ext(\X)=\sigma]}{q^d}
 >\frac{(1+\eps)2^{-m}}{q^d}.
\]

We remark that we need the two quantities to be very close because the
latter is tiny and can be smaller than $\frac{1}{q^d}$. To compensate for
that, one must require that $\minH(\W)$ is very large (say
$\minH(\W)>(d-1)\log q+m$, with sufficient slack). To achieve this for the function
$\W=\Cond(\X)$ that is considered by~\cite{TV00} (namely, $\Cond(\X)$ is a
prefix of $\X$), it suffices to require that $\minH(\X)>n-(\log q-m)$,
again with sufficient slack,
and so this approach can only produce extractors for samplable
distributions for very high min-entropy $k\gg n/2$.
The focus of this paper is constructing extractors for much lower
min-entropy $k$, and so we will need to modify the construction and analysis.

\paragraph{Soundness.}
For every $w\in\F_q^d$ with
$\Pr[\W=w]>0$, we define
\[
 R_w=\{a\in\F_q:\Pr[\TExt(a,\X)=\sigma\mid \W=w]
 >2^{-m}+\eps_{\TExt}\}.
\]
We say that $w\in\F_q^d$ is \emph{valid} if $\Pr[\W=w]>0$ and
$\minH(\X\mid \W=w)\ge k_2$. That is, if the source
$(\X\mid \W=w)$ meets the requirement that $\TExt$ makes of its
second source. For zero-probability points $w$, we set $R_w=\varnothing$
and regard $w$ as invalid. For every valid $w$, we have that
$|R_w|\le2^{k_1}$. This is because otherwise, the two independent
sources $\X_1$, uniform on $R_w$, and $\X_2$, distributed as
$(\X\mid \W=w)$, violate the guarantee of the two-source extractor, as
\[
 \Pr[\TExt(\X_1,\X_2)=\sigma]-\Pr[\U_m=\sigma]
 >\eps_{\TExt}.
\]

In order to bound the number of degree-$n$ polynomials $p$ such that
$\Test_L(p)\ge(1+\eps/2)2^{-m}/q^d$, we will show that every
such $p$ has a sufficiently large fraction of $t\in\F_q$ such that
$L(t)$ is valid and $p(t)\in R_{L(t)}$.
(This is sufficient, as by the list-recovery properties of Reed--Solomon
codes (see, e.g.,~\cite{Sudan97}), applied to the lists at valid points,
the number of such polynomials is small.)

We divide the points $t\in\F_q$ into three categories:
\begin{itemize}
\item $t$ such that $L(t)$ is not valid.
\item $t$ such that $L(t)$ is valid and $p(t)\notin R_{L(t)}$.
\item $t$ such that $L(t)$ is valid and $p(t)\in R_{L(t)}$.
\end{itemize}

The quantity $\Test_L(p)=\E_{t\leftarrow\F_q}[J(L(t),p(t))]$
averages the contributions of points in these three categories.
By an appropriate choice of $k_2$, the probability that $\W$ is invalid
can be made sufficiently small compared with $\eps2^{-m}$.
For the first category,
\[
 \E_{t\leftarrow\F_q}\!\left[
 J(L(t),p(t))\mathbf1_{\{L(t)\text{ is invalid}\}}\right]
 \le
 \E_{t\leftarrow\F_q}\!\left[
 \Pr[\W=L(t)]\mathbf1_{\{L(t)\text{ is invalid}\}}\right].
\]
The expectation of the right-hand side over a random line $\bL$ is
\[
 \frac{\Pr[\W\text{ is invalid}]}{q^d}
 \ll\frac{\eps2^{-m}}{q^d}.
\]
By Markov's inequality, with high probability over $\bL$, the contribution
from the first category is small compared with $\eps2^{-m}/q^d$.

For the second category, by definition,
\begin{align*}
 J(L(t),p(t))
 &=\Pr[\W=L(t),\ \TExt(p(t),\X)=\sigma]\\
 &=\Pr[\TExt(p(t),\X)=\sigma\mid \W=L(t)]\Pr[\W=L(t)]\\
 &\le(2^{-m}+\eps_{\TExt})\Pr[\W=L(t)].
\end{align*}
Consequently, the total contribution from the second category is bounded by
\[
 \E_{t\leftarrow\F_q}\!\left[
 J(L(t),p(t))\mathbf{1}_{\{t\text{ is in the second category}\}}
 \right]
 \le(2^{-m}+\eps_{\TExt})
 \E_{t\leftarrow\F_q}\Pr[\W=L(t)].
\]
The sampling property of a random line
gives that, with high probability,
\[
 \E_{t\leftarrow\F_q}\Pr[\W=\bL(t)]
 \quad\text{is close to}\quad
 \E_{w\leftarrow\F_q^d}\Pr[\W=w]=\frac{1}{q^d}.
\]
By picking $\eps_{\TExt}\ll\eps2^{-m}$ and controlling
the sampling error, the total contribution from the second category
is at most the baseline $2^{-m}/q^d$ plus an amount that is small
compared with $\eps2^{-m}/q^d$.

However, as $\Test_L(p)\ge(1+\eps/2)2^{-m}/q^d$, the third
category must contribute an amount of order
$\eps2^{-m}/q^d$. For every $t$, we have
\[
 J(L(t),p(t))\le\Pr[\W=L(t)]\le2^{-\minH(\W)}.
\]
We choose the min-entropy deficiency $d\log q-\minH(\W)$,
the output length $m$, and $k_1$ sufficiently small relative to $\log q$.
Consequently, there must be a sufficiently large fraction of $t$
such that $L(t)$ is valid and $p(t)\in R_{L(t)}$.
With an appropriate choice of parameters, list recovery then bounds
the number of accepted polynomials by a quantity significantly
smaller than $q$.

\begin{remark}[Comparison with the test of {\cite{BDGM23}}]
Ball et al.~\cite{BDGM23} use a test based on the following quantity:
\[
 \Test^{\mathrm{BDGM}}_L(p)
 =\Pr[\TExt(p(L^{-1}(\W)),\X)=\sigma\mid \W\in L].
\]
This test is more complicated than the one we consider here.
For a nondegenerate $L$ with $\Pr[\W\in L]>0$, we have
\begin{align*}
 \Test_L(p)
 &=\frac1q\sum_{t\in\F_q}J(L(t),p(t))\\
 &=\frac1q\sum_{t\in\F_q}
   \Pr[\W=L(t),\ \TExt(p(t),\X)=\sigma]\\
 &=\frac1q\Pr[\W\in L,\ \TExt(p(L^{-1}(\W)),\X)=\sigma]\\
 &=\frac{\Pr[\W\in L]}q
   \Pr[\TExt(p(L^{-1}(\W)),\X)=\sigma\mid \W\in L]\\
 &=\frac{\Pr[\W\in L]}q\cdot\Test^{\mathrm{BDGM}}_L(p).
\end{align*}
Our test avoids the normalization by $\Pr[\W\in L]$.
The analysis used by~\cite{BDGM23} is significantly more complicated
than the one used here, and is difficult to generalize to the
low min-entropy case.
\end{remark}

\subsection{Extending the construction and test to the case of low min-entropy}
\label{sec:overview-low-entropy}

Ball, Shaltiel and Silbak~\cite{BSS25} extended the approach
of~\cite{TV00} to lower min-entropy. Rather than taking $\Cond(\X)$
to be a prefix of $\X$, they use an errorless condenser for samplable
distributions (as in \cref{def:errorless-condenser}).
Unfortunately, the errorless condensers constructed in that paper
and the subsequent paper by Shaltiel~\cite{shaltiel25extract} do not guarantee
that $\Cond(\X)$ has small min-entropy deficiency, and are therefore
not compatible with the argument above.

In order to use the approach above, we modify the construction as
follows: given $\X$, we compute $\Y=\Cond(\X)$ of length $n_y$, and then
use a strong seeded extractor
$\SExt\colon\zo^{n_y}\times[D]\to\zo^{d\log q}$, with
$D=\operatorname{poly}(n)$, that is set up to extract randomness from
$\Y=\Cond(\X)$. Here, we choose $q\gg 2^{k_1}$ with $d\log q$
sufficiently smaller than the min-entropy of $\Y$, where $k_1$ is the
min-entropy required by $\TExt$ in its first source.
For every $i\in[D]$, we define $\W_i=\SExt(\Y,i)$ and then
\[
 \Ext(\X)=\TExt\bigl(\widehat f(\W_1)\concat\cdots\concat\widehat f(\W_D),\X\bigr).
\]
A complication is that we currently do not have explicit two-source
extractors $\TExt$ with the low-entropy and low-error parameters
needed here, and so, in the proofs of
\cref{thm:block-robust,thm:multiplicative}, we use
other objects. Specifically, in \cref{thm:block-robust}, we
use a low-error two-source condenser by Ben-Aroya, Cohen, Doron and
Ta-Shma~\cite{BCDT19condense}, and in \cref{thm:multiplicative}, we
use a low-error two-source extractor by Li~\cite{Li15threesource}, which makes
the additional requirement that its second source is a block-wise
source. We briefly discuss these modifications below
(\cref{sec:overview-robust-modifications,sec:overview-mult-modifications}).

The intuition behind this construction is that the guarantee of a
strong seeded extractor gives that there exists $i^*\in[D]$ such
that $\W_{i^*}=\SExt(\Y,i^*)$ is close to uniform. For simplicity of
notation, assume without loss of generality that $i^*=D$.
The hope is that we can try to repeat the argument above,
with $\W_D$ playing the role of the condenser output in the previous
argument. Indeed, we now have that $\W_D$ is close to uniform.

A major problem is that it seems that when the test checks whether
a polynomial $p$ is $\widehat f\circ L$ using the last coordinate
$D$, it will need to somehow compute $\widehat f$ on the remaining
$D-1$ coordinates (which it cannot do). To overcome this problem,
we define the following generalization of $J(w,a)$:
\[
 J(w,a)=\sum_{y:\SExt(y,D)=w}
 \max_{v\in\F_q^{D-1}}
 \Pr[\Y=y,\ \TExt((v,a),\X)=\sigma].
\]
The placement of the maximum in this expression allows $v$ to
depend on $y$, but not on the specific $x\leftarrow \X$.
Choosing $v=\widehat f(\W_1)\concat\cdots\concat\widehat f(\W_{D-1})$
in the maximum gives
\begin{equation}
\label{eq:overview-correct-J}
 J(w,\widehat f(w))\ge\Pr[\W_D=w,\ \Ext(\X)=\sigma],
\end{equation}
which was key for the previous completeness analysis.

\paragraph{Soundness.}
The main difficulty in extending the soundness argument is that the
guarantee that $\W_D$ is close to uniform is somewhat different from
the previous guarantee that $\W$ has high min-entropy. In particular,
$\W_D$ may have elements with large probability. To keep the number
of seeds $D$, and hence the running time, polynomial in $n$, we
choose $\eps_{\SExt}=1/n^{O(1)}$. We aim for $m\gg\log n$,
and therefore want to allow $\eps_{\SExt}$ to be much larger
than $2^{-m}$. In this regime, elements $t$ with large
$\Pr[\W_D=L(t)]$ can completely dominate the outcome of the averaging
test. Consequently, a large average need not imply agreement with the
candidate lists at sufficiently many points.

In this scenario, it is better to change the test from averaging
$J(L(t),p(t))$ over $t\leftarrow\F_q$ to a ``sum of thresholds'' as
follows:
\begin{align*}
 \Test_L(p)
 &=\E_{t\leftarrow\F_q}\!\left[
 \mathbf{1}_{\left\{J(L(t),p(t))\ge
 \frac{(1+\eps/2)2^{-m}}{q^d}\right\}}
 \right]\\
 &=\Pr_{t\leftarrow\F_q}\!\left[
 J(L(t),p(t))\ge\frac{(1+\eps/2)2^{-m}}{q^d}
 \right].
\end{align*}
The test accepts if a sufficiently large constant fraction of the
points on the line pass the threshold. The advantage is that this
test gives ``equal weight'' to all points on the line, so that
points $w$ with atypically large $\Pr[\W_D=w]$ do not get to
significantly change the value of the test.

We can now imitate the previous soundness analysis. Whereas
previously we used the fact that conditioning on the condenser
output $\W$ fixes the first input of $\TExt$, we will now use that
conditioning on $\{\Y=y\}$ fixes the first input of $\TExt$. Each
valid value $y$ gives a small list of candidate first inputs to
$\TExt$, and hence a small list of their last coordinates. Different
values $y$ that map to the same $w$ may give different lists.

Here we use the unconditional closeness of $\W_D$ to uniform: for all
but a small fraction of points $w$, measured under the uniform
distribution on $\F_q^d$, we have $\Pr[\W_D=w]\approx1/q^d$.
Moreover, $J(w,a)\le\Pr[\W_D=w]$. Outside a small exceptional set of
points $w$, a value $a$ can pass the threshold only if it belongs to
the lists of sufficiently much of the probability mass mapping to
$w$. Since each individual list is small, a counting argument shows
that only a small number of values $a$ can pass. We can then apply
the list-recovery argument as before.

\paragraph{Completeness.}
We can guarantee that for the good seed $i^*=D$, $\W_D$ is close to
uniform even conditioned on $\Ext(\X)=\sigma$. This helps the
completeness argument: recall from~\cref{eq:overview-correct-J} that
\begin{align*}
 J(w,\widehat f(w))
 &\ge\Pr[\W_D=w,\ \Ext(\X)=\sigma]\\
 &=\Pr[\Ext(\X)=\sigma]\cdot
   \Pr[\W_D=w\mid\Ext(\X)=\sigma].
\end{align*}
We now have a handle on both probabilities: the first is larger than
$(1+\eps)2^{-m}$ by assumption, and the second is close to
$1/q^d$ for most $w$ under the uniform distribution. This will be
used to ensure that with high probability over $\bL$, the correct
polynomial passes the threshold at most points on the line.

\paragraph{Efficiency.}
This ``sum of thresholds'' test can also be implemented by a
nondeterministic circuit using an additional layer of the
Goldwasser--Sipser protocol. More specifically, we first apply
Goldwasser--Sipser to obtain a nondeterministic circuit $\Thr(w,a)$
that checks whether $J(w,a)>(1+\eps/2)2^{-m}/q^d$, up to a
small gap around this threshold. We then apply Goldwasser--Sipser
again to check whether the fraction of $t\in\F_q$ for which
$\Thr(L(t),p(t))=1$ is sufficiently large.

The proofs of both \cref{thm:block-robust,thm:multiplicative} follow this outline. However, there are
some special modifications that we make for each of the two settings.

\subsection{Special modifications for \texorpdfstring{\Cref{thm:block-robust}}{Theorem~\ref{thm:block-robust}}}
\label{sec:overview-robust-modifications}

In \cref{thm:block-robust}, we aim to construct a robust block
condenser rather than an extractor. Such an object receives $\Y$ in
addition to $\X$, and the construction is
\[
 \rbCond(\Y,\X)=\TwoCond\bigl(\widehat f(\W_1)\concat\cdots\concat
 \widehat f(\W_D),\X\bigr),
\]
where $\W_i=\SExt(\Y,i)$ for every $i\in[D]$, as before, and $\TwoCond$
is the low-error two-source condenser of~\cite{BCDT19condense}.

An additional complication is that, in a robust block condenser,
the test has access to the distribution $(\Y,\X)$, whereas some of
the entropy guarantees apply to a subsource
$(\Y_{\mathrm{good}},\X_{\mathrm{good}})$. This dictates some changes
in the test and the argument that are explained in detail in
\cref{sec:robust-proof}.

\subsection{Special modifications for \texorpdfstring{\Cref{thm:multiplicative}}{Theorem~\ref{thm:multiplicative}}}
\label{sec:overview-mult-modifications}

In \cref{thm:multiplicative}, we aim to construct a
multiplicative extractor for low min-entropy with large output length.
For this purpose, the argument described above requires a two-source
extractor $\TExt$ with error $2^{-n^{\Omega(1)}}$. Such explicit
constructions are not known for low min-entropy, and we therefore
replace $\TExt$ with a two-source extractor $\IExt$ by Li~\cite{Li15threesource}
which achieves these parameters, but requires the second source to
be a block-wise source. We will imitate the approach of
Shaltiel~\cite{shaltiel25extract} that constructed such an extractor under
stronger hardness assumptions.

More specifically, we will apply the condenser $\Cond(\X)$ twice as
follows: We set $\Y=\Cond(\Cond(\X)), \W_i = \SExt(\Y, i)$ for $i\in [D]$, and
\[
 \Ext(\X)=\IExt\bigl(\widehat f(\W_1)\concat\cdots\concat
 \widehat f(\W_D),(\Cond(\X),\X)\bigr).
\]
The high-level intuition is that, for most values $y$ of $\Y$, the
conditional distribution $(\Cond(\X),\X)\mid \Y=y$ is close to a
block-wise source: $\Y$ is significantly shorter than the min-entropy
of $\Cond(\X)$, and $\Cond(\X)$ is significantly shorter than the
min-entropy of $\X$. This generalizes the construction described above,
and precise details are given in \cref{sec:mult-proof}.

\section{Joint preparations}
\label{sec:joint-preparations}

In this section, we set up some machinery that will be used in the
proofs of both \cref{thm:block-robust,thm:multiplicative}.

\subsection{A nondeterministic circuit for counting the sum of maximums}
\label{sec:sum-of-maximums}

In this subsection, we prove a general lemma that will be used to argue
that the tests defined in \cref{sec:robust-proof,sec:mult-proof} can be implemented by a small nondeterministic
circuit, establishing the efficiency property of these tests.

\begin{lemma}[Counting a sum of maximums]
\label{lem:sum-of-maximums}
Let $C(u,y,v,r)$ be a nondeterministic circuit, with inputs of length
$n_u,n_y,n_v,n_r$, respectively, and define $S(u)$ as follows:
\[
    S(u)\coloneqq\sum_{y\in\zo^{n_y}}\max_{v\in\zo^{n_v}}
    \Pr_{r\leftarrow\zo^{n_r}}[C(u,y,v,r)=1].
\]
For every rational $0<\Theta\le2^{n_y}$ and $0<\eta\le1/2$, there is
a nondeterministic circuit $V(u)$ such that for every $u\in\zo^{n_u}$,
\begin{align*}
    S(u)\ge\Theta&\quad\Longrightarrow\quad V(u)=1,\\
    S(u)\le(1-\eta)\Theta&\quad\Longrightarrow\quad V(u)=0.
\end{align*}
Its size is polynomial in the size of $C$, the input lengths, the bit
length of $\Theta$, and $1/\eta$.
\end{lemma}

\begin{proof}
Set $M_u(y)=\max_v\Pr_r[C(u,y,v,r)=1]$.
Choose a threshold grid
\[
    G=\{j2^{-n_r}:1\le j\le2^{n_r}\}.
\]
Note that $M_u(y)$ is an integer multiple of $2^{-n_r}$.

For every $u,y,v$, define the nondeterministic circuit
$C_{u,y,v}(r)\coloneqq C(u,y,v,r)$.
Apply \cref{thm:gs-counting} with $e^\lambda=1+\eta/8$
to obtain a nondeterministic circuit $A(u,y,v,\tau)$ such that
for every $u,y,v$ and every $\tau\in G$,
\begin{align*}
    \Pr_r[C_{u,y,v}(r)=1]\ge\tau
    &\quad\Longrightarrow\quad A(u,y,v,\tau)=1,\\
    \Pr_r[C_{u,y,v}(r)=1]<\frac{\tau}{1+\eta/8}
    &\quad\Longrightarrow\quad A(u,y,v,\tau)=0.
\end{align*}
Let $B(u,y,\tau)$ be the nondeterministic circuit that guesses $v$
and simulates $A(u,y,v,\tau)$.
For each fixed $u,y$, $B(u,y,\tau)$ accepts every $\tau\in G$
satisfying $\tau\le M_u(y)$, and rejects every $\tau\in G$
satisfying $\tau>(1+\eta/8)M_u(y)$. Consequently,
\[
    M_u(y)
    \le\Pr_{\tau\leftarrow G}[B(u,y,\tau)=1]
    \le(1+\eta/8)M_u(y).
\]

Averaging over $y\leftarrow\zo^{n_y}$ gives, for every $u$,
\[
    2^{-n_y}S(u)
    \le\Pr_{y\leftarrow\zo^{n_y},\,\tau\leftarrow G}[B(u,y,\tau)=1]
    \le(1+\eta/8)2^{-n_y}S(u).
\]
Thus the two cases in the statement imply, respectively,
\begin{align*}
    S(u)\ge\Theta
    &\quad\Longrightarrow\quad
    \Pr_{y\leftarrow\zo^{n_y},\,\tau\leftarrow G}[B(u,y,\tau)=1]\ge2^{-n_y}\Theta,\\
    S(u)\le(1-\eta)\Theta
    &\quad\Longrightarrow\quad
    \Pr_{y\leftarrow\zo^{n_y},\,\tau\leftarrow G}[B(u,y,\tau)=1]\le(1-7\eta/8)2^{-n_y}\Theta.
\end{align*}
For every $u$, define the nondeterministic circuit
$B_u(y,\tau)\coloneqq B(u,y,\tau)$, where $y\in\zo^{n_y}$ and $\tau\in G$ are its inputs.
Apply \cref{thm:gs-counting} to $B_u$ with threshold $2^{-n_y}\Theta$
and the same $\lambda$ to obtain a nondeterministic circuit $V(u)$.
It satisfies the required guarantees because
\[
    (1-7\eta/8)2^{-n_y}\Theta
    <\frac{2^{-n_y}\Theta}{1+\eta/8}.
\]
Both applications have polynomial size in the stated parameters.
\end{proof}

\subsection{Reconstruction from a polynomial test}
\label{sec:reconstruction}

As we explained in \cref{sec:proof-overview-BGDM}, the proofs of
\cref{thm:block-robust,thm:multiplicative} will use the
overall approach of~\cite{STV01PRG,BDGM23}. More specifically, in both proofs,
the construction converts the supplied hard function into its low-degree
extension $\widehat f\colon\F_q^d\to\F_q$, which is a polynomial of total degree
at most $\hat h$ for some parameter $\hat h$. The construction of the
required object will use $\widehat f$ as a component.
The correctness argument assumes that the constructed object
fails, and obtains a contradiction by constructing a small nondeterministic
circuit $C$ that computes $\widehat f$. In both proofs, this will be done
in three steps:
\begin{itemize}
 \item Design a small nondeterministic circuit $\Thr(w,a)$ and use it
 to define a test $\Test_L(p)$. (In the terminology of
 \cref{sec:proof-overview-BGDM}, $\Thr(w,a)$ checks whether $J(w,a)$
 is larger than some threshold.)
 \item Prove completeness and soundness with high probability over a
 random line $\bL\colon\F_q\to\F_q^d$.
 \item Use the test to construct the desired nondeterministic circuit
 $C$ that computes $\widehat f$.
\end{itemize}
The next lemma states the (by now standard) third step in this approach.
More specifically, given a small nondeterministic circuit $\Thr$, suppose
that with high probability over a random line $\bL$, the induced test
$\Test_{\bL}(p)$ satisfies completeness
($\Test_{\bL}(\widehat f\circ\bL)\ge7/8$) and soundness
(the number of $p$ of degree at most $\hat h$ with $\Test_{\bL}(p)>3/4$ is small).
The lemma gives a small single-valued nondeterministic circuit $C$
that computes $\widehat f$. This will be used to contradict the hardness
assumption.

\begin{lemma}[Reconstruction from a polynomial test]
\label{lem:reconstruction}
Let $q$ be a power of two, let $d\ge1$ and $\hat h\ge0$ be integers,
and let $\widehat f\colon\F_q^d\to\F_q$ be a polynomial of total degree at most
$\hat h$. Let $\Thr(w,a)$ be a nondeterministic circuit of size $S$,
with inputs $w\in\F_q^d$ and $a\in\F_q$.
For every nondegenerate line $L$ and every $p\in\F_q[t]$ of degree at most
$\hat h$, define
\[
 \Test_L(p)=\Pr_{t\leftarrow\F_q}[\Thr(L(t),p(t))=1].
\]
Set $\Test_L(p)=0$ for degenerate $L$, and define
\[
 \mathcal P_L=\{p\in\F_q[t]:\deg p\le\hat h,\ \Test_L(p)>3/4\}.
\]
Suppose that
\[
 \Pr_{\bL\leftarrow\RandLine}\!\left[
 \Test_{\bL}(\widehat f\circ\bL)\ge\frac78,\quad
 |\mathcal P_{\bL}|\le\frac{q}{32(\hat h+1)}
 \right]\ge\frac{31}{32}.
\]
There is a single-valued nondeterministic circuit $C$ of size
$\operatorname{poly}(S,\hat h,d,\log q)$ such that
\[
 \ACCEPT_C(x)=\{\widehat f(x)\}
 \qquad\textup{for every }x\in\F_q^d.
\]
\end{lemma}

\subsection*{Proof of \Cref{lem:reconstruction}}

We follow the reconstruction strategy of Sudan--Trevisan--Vadhan~\cite{STV01PRG}
and Ball et al.~\cite{BDGM23}: pin a polynomial using one correct evaluation,
and then correct the remaining errors.

\subsubsection{Pinning the correct polynomial}
\label{sec:pinning}

For every nondegenerate $L$ and every $p\in\F_q[t]$ of degree at most $\hat h$,
define the nondeterministic circuit
$D_{L,p}(t)\coloneqq\Thr(L(t),p(t))$.
Apply \cref{thm:gs-counting} with threshold $7/8$ and $e^\zeta=9/8$
to obtain a nondeterministic circuit $T(L,p)$ such that
\begin{align*}
 \Test_L(p)\ge7/8&\quad\Longrightarrow\quad T(L,p)=1,\\
 \Test_L(p)\le3/4&\quad\Longrightarrow\quad T(L,p)=0,
\end{align*}
where $3/4<7/9=(7/8)e^{-\zeta}$.
The inputs encode $L$ by its endpoints and $p$ by its coefficients.
Set $T(L,p)=0$ for degenerate $L$.
The size of $T$ is polynomial in $S,\hat h,d,\log q$.

\begin{definition}[The circuit $C_{z,b}$]
\label{def:pinned-circuit}
For every $z\in\F_q^d$ and $b\in\F_q$, define the nondeterministic circuit
$C_{z,b}$ as follows. On input $x\in\F_q^d$:
\begin{enumerate}[label=(\arabic*)]
 \item Set $L_{x,z}(t)=x+t(z-x)$.
 \item Guess a polynomial $p\in\F_q[t]$ of degree at most $\hat h$
 and a witness $\mathrm{wit}_T$ for $T$.
 \item If $T(L_{x,z},p;\mathrm{wit}_T)=1$ and $p(1)=b$,
 accept and output $p(0)$; otherwise, reject.
\end{enumerate}
Thus,
\[
 \ACCEPT_{C_{z,b}}(x)
 =\{p(0):\deg p\le\hat h,\ p(1)=b,\ T(L_{x,z},p)=1\}.
\]
\end{definition}

\begin{lemma}[Pinning]
\label{lem:pinning}
Under the hypotheses of \cref{lem:reconstruction},
\[
 \Pr_{x,z\leftarrow\F_q^d}\!\left[
 \ACCEPT_{C_{z,\widehat f(z)}}(x)=\{\widehat f(x)\}
 \right]\ge\frac{15}{16}.
\]
Consequently, there is $z^*\in\F_q^d$ such that
\[
 \Pr_{x\leftarrow\F_q^d}\!\left[
 \ACCEPT_{C_{z^*,\widehat f(z^*)}}(x)=\{\widehat f(x)\}
 \right]\ge\frac{15}{16}.
\]
The size of $C_{z,b}$ is polynomial in $S,\hat h,d,\log q$,
with an absolute polynomial bound.
\end{lemma}

\begin{proof}
For every nondegenerate $L$ and affine bijection $\psi\colon\F_q\to\F_q$,
\[
 \Test_{L\circ\psi}(p\circ\psi)
 =\Pr_{t\leftarrow\F_q}[\Thr(L(\psi(t)),p(\psi(t)))=1]
 =\Test_L(p).
\]
Hence
\[
 \mathcal P_{L\circ\psi}=\{p\circ\psi:p\in\mathcal P_L\}.
\]

Fix a nondegenerate $L$ with $|\mathcal P_L|\le q/(32(\hat h+1))$.
Each $p\ne\widehat f\circ L$ agrees with $\widehat f\circ L$ at at most
$\hat h$ points. Therefore,
\[
 \Pr_{t\leftarrow\F_q}\!\left[
 \exists p\in\mathcal P_L\setminus\{\widehat f\circ L\}:
 p(t)=\widehat f(L(t))
 \right]
 \le\frac{|\mathcal P_L|\hat h}{q}\le\frac1{32}.
\]
Conditioned on $x\ne z$, we can imagine first choosing a uniform
geometric line, fixing any parametrization $L$, and then choosing a
uniformly random affine bijection $\psi\colon\F_q\to\F_q$. Set
\[
 x=L(\psi(0)),\qquad z=L(\psi(1)).
\]
Then $L_{x,z}=L\circ\psi$. Since $\psi(1)$ is uniform in $\F_q$,
the invariance of $\mathcal P_L$ under reparametrization and the
preceding bound give
\[
 \Pr_{x,z}\!\left[
 \begin{gathered}
 x\ne z,\quad |\mathcal P_{L_{x,z}}|\le q/(32(\hat h+1)),\\
 \exists p\in\mathcal P_{L_{x,z}}\setminus\{\widehat f\circ L_{x,z}\}:
 p(1)=\widehat f(z)
 \end{gathered}
 \right]\le\frac1{32}.
\]
By the hypothesis of \cref{lem:reconstruction}, with probability at least
$31/32$ over $x,z$, the line $L_{x,z}$ is nondegenerate,
$T(L_{x,z},\widehat f\circ L_{x,z})=1$, and
$|\mathcal P_{L_{x,z}}|\le q/(32(\hat h+1))$.
Every polynomial accepted by $T$ belongs to $\mathcal P_{L_{x,z}}$.
Thus,
\[
 \Pr_{x,z}\!\left[
 \ACCEPT_{C_{z,\widehat f(z)}}(x)\ne\{\widehat f(x)\}
 \right]\le\frac1{32}+\frac1{32}=\frac1{16}.
\]
Averaging over $z$ gives $z^*$. The size bound follows from the size of
$T$ and the cost of evaluating $p$.
\end{proof}

\subsubsection{Correcting a nondeterministic approximation}
\label{sec:correction}

The next lemma converts a nondeterministic circuit that is correct and
single-valued on a $15/16$-fraction of inputs into one that is correct
and single-valued on every input.

\begin{lemma}[Nondeterministic local correction]
\label{lem:nondeterministic-correction}
Let $g\colon\F_q^d\to\F_q$ be a polynomial of total degree at most $\hat h$,
where $q-1\ge4(\hat h+1)$. Suppose a nondeterministic circuit $C$ of
size $S$ satisfies
\[
    \Pr_{x\leftarrow\F_q^d}[\ACCEPT_C(x)=\{g(x)\}]\ge15/16.
\]
Then there is a single-valued nondeterministic circuit $C^*$ satisfying
$\ACCEPT_{C^*}(x)=\{g(x)\}$ for every $x\in\F_q^d$, of size polynomial in
$S,\hat h,d,\log q$, with an absolute polynomial bound.
\end{lemma}

\begin{proof}
Let $G=\{x:\ACCEPT_C(x)=\{g(x)\}\}$, so $|G|\ge(15/16)q^d$.

\medskip
\noindent\textbf{Correction at one point.}
Set $s=4(\hat h+1)$ and fix distinct nonzero field elements
$t_1,\ldots,t_s\in\F_q$.
For a target $x$, choose $y\leftarrow\F_q^d$ uniformly and set
$L(t)=x+ty$. Guess a polynomial $p$ of degree at most $\hat h$,
a set of $3s/4$ distinct indices, and, for each selected index $i$, a
successful witness certifying $p(t_i)\in\ACCEPT_C(L(t_i))$.
Reject if any certificate fails; otherwise, accept and output $p(0)$.

Define
\[
 G_x=\{y\in\F_q^d:|\{i\in[s]:x+t_i y\in G\}|\ge3s/4\}.
\]
For each $i$, the point $x+t_i y$ is uniform on $\F_q^d$. Thus the
expected number of sampled points outside $G$ is at most $s/16$.
By Markov's inequality,
\[
 \Pr_{y\leftarrow\F_q^d}[y\in G_x]\ge3/4.
\]
For every $y\in G_x$, $p^*=g\circ L$ has successful witnesses on at least
$3s/4$ positions. Any other polynomial can have witnesses only at the
at most $s/4$ bad positions and at the at most $\hat h$ field elements where
it agrees with $p^*$. Since
\[
    s/4+\hat h<3s/4,
\]
no wrong polynomial is accepted. Consequently, for every $y\in G_x$, the
set of possible outputs is exactly $\{g(x)\}$.

\medskip
\noindent\textbf{Amplification and fixing $y_1,\ldots,y_R$.}
Set $R=16(d\log q+1)$ and choose $y_1,\ldots,y_R$ independently and
uniformly from $\F_q^d$. Guess a common output
$a\in\F_q$ and successful correction witnesses producing $a$ for
more than half of the repetitions. Reject if any selected certificate
fails; otherwise, accept and output $a$. If
$|\{j\in[R]:y_j\in G_x\}|>R/2$, this circuit has output set exactly
$\{g(x)\}$: the repetitions with $y_j\in G_x$ certify $g(x)$,
whereas any other value can be certified only when $y_j\notin G_x$.
We require no witnesses for the remaining repetitions.

For each fixed $x$, a Chernoff bound gives
\[
 \Pr\!\left[|\{j\in[R]:y_j\in G_x\}|\le R/2\right]\le e^{-R/8}.
\]
Hence
\[
    \Pr\!\left[\exists x\in\F_q^d:
    |\{j\in[R]:y_j\in G_x\}|\le R/2\right]
    \le q^d e^{-R/8}<1.
\]
Fix $y_1,\ldots,y_R$ such that
$|\{j\in[R]:y_j\in G_x\}|>R/2$ for every $x$, and hardwire them.
The resulting circuit is $C^*$.

Each repetition uses at most $s$ copies of the verifier for $C$, field
arithmetic, and polynomial evaluation. There are $O(d\log q+1)$
repetitions. All guesses and their witnesses form one existential
witness, and the total size is polynomial in the stated parameters.
\end{proof}

\begin{proof}[Proof of \Cref{lem:reconstruction}]
The hypothesis gives some $L$ with
$1\le|\mathcal P_L|\le q/(32(\hat h+1))$, so
$q\ge32(\hat h+1)$ and $q-1\ge4(\hat h+1)$.
Apply \cref{lem:nondeterministic-correction} to the circuit
$C_{z^*,\widehat f(z^*)}$ from \cref{lem:pinning} to obtain the
required circuit of size $\operatorname{poly}(S,\hat h,d,\log q)$.
\end{proof}

\section{Proof of the robust errorless block condenser theorem}
\label{sec:robust-proof}

In this section we prove \cref{thm:block-robust}.
The construction (that appears in \cref{sec:robust-construction})
and analysis (that appears in \cref{sec:robust-analysis}) closely
follow the high level description of \cref{sec:proof-overview-BGDM}.

\subsection{Construction}
\label{sec:robust-construction}

Fix constants $c_{\mathrm{size}}>1$ and $0<\gamma\le1$, and let
\[
 n^\gamma\le k_y\le k_x\le n,
 \qquad (\log n)^2\le m\le k_y^\delta.
\]
The universal constant $\delta>0$ will be chosen sufficiently small
in the two-source condenser item of \cref{sec:robust-components}.
Our goal is to construct an explicit function
\[
 \rbCond\colon\zo^n\times\zo^n\longrightarrow\zo^m
\]
that is computable in polynomial time and satisfies the robust errorless
block condenser guarantee of \cref{thm:block-robust} for
joint distributions samplable by deterministic circuits with
postselection by nondeterministic circuits, both of size at most $n^{c_{\mathrm{size}}}$.
We assume that $\sfE$ is hard for exponential-size nondeterministic
circuits.

\subsubsection{Components}
\label{sec:robust-components}

We rely on the following components:
\begin{itemize}[leftmargin=*]

\item \textbf{Hard function and low-degree extension.}
We assume that $\sfE$ is hard for exponential-size nondeterministic
circuits. By \cref{def:nd-hardness}, this gives a function
$f\in\sfE$ and a constant $\beta_{\mathrm{hard}}>0$ such that,
for every sufficiently large input length $\ell$, no nondeterministic
circuit of size at most $2^{\beta_{\mathrm{hard}}\ell}$ computes $f$
on all inputs in $\zo^\ell$.
Let $d\ge2$ be a sufficiently large integer that will be chosen later
in \cref{sec:robust-reconstruction-consequences}, and set
\[
 \ell=d\log n,\qquad h=2^{\ell/d}=n,\qquad q=2^{k_y^{1/2}}.
\]

Let $\F_q$ be the field with $q$ elements, which exists because
$q$ is a power of two. For sufficiently large $n$, $d\le h\le q$.
Let $\phi\colon\zo^\ell\to H^d$ be the natural bijection,
where $H$ is as in \cref{thm:lde}.
By \cref{thm:lde}, $f$ has a low-degree extension
\[
 \widehat f\colon\F_q^d\longrightarrow\F_q
\]
with $\widehat f(\phi(x))=f(x)$ for every $x\in\zo^\ell$, and total degree at most
\[
 \hat h\coloneqq d(h-1)<d\cdot n.
\]
Thus, for every affine line $L\colon\F_q\to\F_q^d$,
$\widehat f\circ L$ is a univariate polynomial of degree at
most $\hat h$. Since $f\in\sfE$, \cref{thm:lde} gives that
$\widehat f$ can be computed in time polynomial in
$2^\ell$ and $\log q$, and hence in time polynomial in $n$ for every
fixed $d$.

\item \textbf{Seeded extractor.}
We use
\[
 \SExt\colon\zo^{n_y}\times[D]\longrightarrow\F_q^d,
\]
with
\[
 n_y=n,\qquad k_{\SExt}=k_y-m/4,
 \qquad \eps_{\SExt}=1/8192.
\]

\begin{remark}[$D$ and $d$ are unrelated]
Note that the parameters $d$ and $D$ are unrelated, and in particular,
it is not the case that $D=2^d$. We use $d$ to measure the dimension
of the vector space over which $\widehat f$ is defined. We use $D$
as the number of seeds for the seeded extractor $\SExt$.
\end{remark}
The choice of $\delta$ in the next item will satisfy $\delta<1/2$.
Consequently, for every fixed $d$ and sufficiently large $n$,
\[
 \tfrac34k_{\SExt}-O(\log n)
 =\tfrac34(k_y-m/4)-O(\log n)
 \ge d\cdot k_y^{1/2}=d\log q.
\]
\Cref{thm:guv}, followed by truncation, supplies the stated
output length.
The seed length satisfies
\[
 \log D=O\!\left(\log n_y+\log(1/\eps_{\SExt})\right)
 =O(\log n).
\]
We write $D=n^{c_D}$, allowing unused seed bits if needed.
The constant $c_D$ is absolute and does not depend on $d$ or $m$.

\item \textbf{Two-source condenser.}
Set
\[
 k_1=k_2=k_y^{1/4}=(\log q)^{1/2},
 \qquad \delta=\min\{1/16,\,1/(8C_{\TwoCond})\},
\]
where $C_{\TwoCond}$ is the universal constant in
\cref{thm:two-source-condenser}.
By \cref{thm:two-source-condenser}, we obtain, after padding the inputs,
\[
 \TwoCond\colon\F_q^D\times\zo^n\longrightarrow\zo^m.
\]
The function $\TwoCond$ is an errorless $(k_1,k_2,3m/4)$-two-source
condenser (\cref{def:two-source-condenser}).

Indeed, the first input has bit length
\[
 D\log q=n^{c_D}\cdot k_y^{1/2}\le n^{c_D+1/2},
\]
so padding both inputs to $N=n^{c_D+1}$ suffices.
For sufficiently large $n$, we have $m\ge(\log n)^2\ge\log N$ and
\[
 (m\log N)^{C_{\TwoCond}}
 \le k_y^{1/8}\cdot(\log N)^{C_{\TwoCond}}
 =o(k_y^{1/4})=o(k_1),
\]
where we used $m\le k_y^\delta$ and $k_y\ge n^\gamma$.
Thus \cref{thm:two-source-condenser} applies with $r=k_1=k_2$.
The choice of $\delta$ is universal and does not depend on $\gamma$
or on the sampler-size exponent $c_{\mathrm{size}}$.
\end{itemize}

\subsubsection{Operation}
\label{sec:robust-operation}

For every $y\in\zo^{n_y}$ and $i\in[D]$, define
\[
 w_i(y)\coloneqq\SExt(y,i).
\]
Then define
\[
 g(y)\coloneqq\bigl(\widehat f(w_1(y)),\ldots,
                 \widehat f(w_D(y))\bigr)\in\F_q^D.
\]
The robust errorless block condenser construction is the function
\[
 \rbCond\colon\zo^n\times\zo^n\longrightarrow\zo^m
\]
defined by
\[
 \boxed{\rbCond(y,x)=\TwoCond(g(y),x).}
\]
Since $\ell=O(\log n)$ and $f\in\sfE$, the truth table of $f$
can be computed in polynomial time. Evaluating $g$ and the two-source
condenser therefore takes polynomial time once $d$ is fixed.

\subsection{Analysis of the construction}
\label{sec:robust-analysis}

\subsubsection{Preparations}
\label{sec:robust-preparations}

Assume for contradiction that the construction fails for a joint
distribution
\[
 (\Y,\X)=(\Samp(\bR)\mid\Post(\bR)=1),\qquad \bR\leftarrow \U_r,
\]
where $\Samp$ is deterministic, $\Post$ is nondeterministic, and
both have size at most $n^{c_{\mathrm{size}}}$. Set
\[
 p_{\mathrm{post}}\coloneqq\Pr_{\bR}[\Post(\bR)=1]>0.
\]
Set $\lambda=2^{-m/4}$ and fix a decomposition
\[
 (\Y,\X)=\lambda'(\Y_{\mathrm{good}},\X_{\mathrm{good}})
       +(1-\lambda')(\Y_{\mathrm{rest}},\X_{\mathrm{rest}}),
 \qquad \lambda'\ge\lambda,
\]
satisfying the hypotheses of \cref{def:block-robust} with
$k_y^{\ge}=k_x/2$. Thus
\[
 \minH(\Y_{\mathrm{good}})\ge k_y,\qquad
 \minH(\X)\ge k_x,\qquad
 \Pr[\Y_{\mathrm{good}}=y]\ge2^{-k_x/2}
 \quad\text{for every }y\in\supp(\Y_{\mathrm{good}}).
\]
Set $\mu=2^{-m/4}$. Failure of the output min-entropy
guarantee gives some $\sigma\in\zo^m$ with
\[
 \Pr[\rbCond(\Y_{\mathrm{good}},\X_{\mathrm{good}})=\sigma]>\mu.
\]
Fix this joint distribution, decomposition, and output. Write
\[
 \Y_\sigma=
 \bigl(\Y_{\mathrm{good}}\mid
       \rbCond(\Y_{\mathrm{good}},\X_{\mathrm{good}})=\sigma\bigr).
\]

\begin{lemma}[A seed good for the conditioned distribution]
\label{lem:robust-good-seed}
There is $i^*\in[D]$ such that
\[
 \TV\bigl(\SExt(\Y_\sigma,i^*),\U_{\F_q^d}\bigr)
 \le\eps_{\SExt}=1/8192.
\]
\end{lemma}
\begin{proof}
By \cref{lem:conditioning},
\begin{align*}
 \minH(\Y_\sigma)
 &\ge\minH(\Y_{\mathrm{good}})
 +\log\Pr[\rbCond(\Y_{\mathrm{good}},\X_{\mathrm{good}})=\sigma]\\
 &>k_y-m/4=k_{\SExt}.
\end{align*}
Strong extraction bounds the average over a uniform seed of the
statistical distance in the statement by $\eps_{\SExt}$.
Some seed has distance no greater than this average.
\end{proof}

In the analysis, in order to simplify the notation,
we will assume without loss of generality that $i^*=D$. This assumption is made only
for convenience in the proof. The choice of $i^*$ can be supplied to
the circuits that we construct as nonuniform advice, and we relabel
the seeds and the corresponding first-input coordinates of $\TwoCond$
accordingly.
Define
\[
 w(y)=\SExt(y,D),\qquad \W=w(\Y),\qquad \W_\sigma=w(\Y_\sigma),
\]
\[
 P(w)=\Pr[\W=w],\qquad Q(w)=\Pr[\W_\sigma=w].
\]
The distribution $Q$ is within distance $1/8192$ of uniform.
We do not assert that $P$ is close to uniform.

\paragraph{Valid values and the set $G$.}
We call $y\in\supp(\Y)$ valid if $\minH(\X\mid \Y=y)\ge k_2$.
We will construct a set $G$ whose membership can be verified by a small
nondeterministic circuit and such that
$\supp(\Y_{\mathrm{good}})\subseteq G\subseteq\mathcal Y_{\mathrm{valid}}$.

Set
\[
 \mathcal Y_{\mathrm{valid}}
 =\{y\in\supp(\Y):\minH(\X\mid \Y=y)\ge k_2\}.
\]

\begin{remark}[Changes in the definition of valid $y$]
\label{rem:robust-valid-definition}
In the definition above, we modify the notion of validity considered
in \cref{sec:overview-low-entropy}. The most natural
generalization would be to define validity in terms of the
min-entropy of $(\X_{\mathrm{good}}\mid \Y_{\mathrm{good}}=y)$.
However, we only have a sampler with postselection for the original
joint distribution $(\Y,\X)$, rather than for its good component.
We therefore define validity in terms of $(\X\mid \Y=y)$. In the
argument below, we will show that every
$y\in\supp(\Y_{\mathrm{good}})$ remains valid under this definition:
the mixture decomposition gives a lower bound on $\Pr[\Y=y]$, which,
together with the min-entropy guarantee for $\X$, ensures that
$(\X\mid \Y=y)$ has sufficiently high min-entropy.
\end{remark}

For every $y\in\supp(\Y_{\mathrm{good}})$, we have the lower bound
\[
 \Pr[\Y_{\mathrm{good}}=y]\ge2^{-k_y^{\ge}}=2^{-k_x/2},
\]
and hence
\[
 \Pr[\Y=y]\ge\lambda\cdot2^{-k_x/2}.
\]
We can use Goldwasser--Sipser (\cref{thm:gs-counting}) to
recognize, up to an approximation gap, values $y\in\zo^{n_y}$ for
which $\Pr[\Y=y]$ is large.
Write $\Samp(R)=(\Samp_Y(R),\Samp_X(R))$, where $\Samp_Y$ and
$\Samp_X$ denote the two output components of $\Samp$.
For every $y\in\zo^{n_y}$, define a nondeterministic circuit $A_y$
with input $R\in\zo^r$ as follows. Compute $\Samp_Y(R)$,
and guess a witness $\mathrm{wit}_{\mathrm{post}}$ for $\Post$.
Accept if and only if $\Samp_Y(R)=y$ and $\Post(R;\mathrm{wit}_{\mathrm{post}})=1$.
The circuit has size
\[
 \operatorname{size}(A_y)=O(n^{c_{\mathrm{size}}}),
\]
and its acceptance probability satisfies
\[
\begin{aligned}
 \Pr_{\bR\leftarrow \U_r}[A_y(\bR)=1]
 &=\Pr_{\bR\leftarrow \U_r}[\Samp_Y(\bR)=y\ \wedge\ \Post(\bR)=1]\\
 &=p_{\mathrm{post}}\cdot
   \Pr_{\bR\leftarrow \U_r}[\Samp_Y(\bR)=y\mid\Post(\bR)=1]\\
 &=p_{\mathrm{post}}\cdot\Pr[\Y=y].
\end{aligned}
\]

Let $A_{\mathrm{GS}}$ be the nondeterministic circuit supplied by
\cref{thm:gs-counting}, with the gap parameter $\lambda$
of that theorem set to $1/2$ and size parameter large enough to accommodate $A_y$ and the threshold
$p_{\mathrm{post}}\cdot \lambda\cdot2^{-k_x/2}$. Define the nondeterministic circuit
$A_{\mathrm{mass}}$ by
\[
 A_{\mathrm{mass}}(y)\coloneqq
 A_{\mathrm{GS}}(A_y,p_{\mathrm{post}}\cdot \lambda\cdot2^{-k_x/2}).
\]
Here $p_{\mathrm{post}}$ is hardwired into $A_{\mathrm{mass}}$ as
nonuniform advice. Note that $p_{\mathrm{post}}$ can be described using
only $O(r+1)$ bits, where $r\le n^{c_{\mathrm{size}}}$.
The threshold has bit length
$O(r+m+k_x+1)=O(n^{c_{\mathrm{size}}})$. Consequently,
\[
 \operatorname{size}(A_{\mathrm{GS}})
 =\operatorname{poly}(n^{c_{\mathrm{size}}}),
 \qquad
 \operatorname{size}(A_{\mathrm{mass}})
 =\operatorname{poly}(n^{c_{\mathrm{size}}}),
\]
with absolute polynomial exponents independent of $d$.
Since $1/2<e^{-1/2}$, the circuit satisfies
\begin{align*}
 \Pr[\Y=y]\ge\lambda\cdot2^{-k_x/2}
 &\quad\Longrightarrow\quad A_{\mathrm{mass}}(y)=1,\\
 \Pr[\Y=y]\le\lambda\cdot2^{-k_x/2-1}
 &\quad\Longrightarrow\quad A_{\mathrm{mass}}(y)=0.
\end{align*}
Define
\[
 G\coloneqq\{y\in\zo^{n_y}:A_{\mathrm{mass}}(y)=1\}.
\]
\begin{lemma}[The set $G$ contains the good component and only valid values]
\label{lem:robust-valid-set}
For sufficiently large $n$,
\[
 \supp(\Y_{\mathrm{good}})\subseteq G
 \subseteq\mathcal Y_{\mathrm{valid}}.
\]
\end{lemma}
\begin{proof}
For every $y\in\supp(\Y_{\mathrm{good}})$, the mixture decomposition,
$\lambda'\ge\lambda$, and the lower bound on its probability in
$\Y_{\mathrm{good}}$ give
\[
 \Pr[\Y=y]\ge\lambda\cdot\Pr[\Y_{\mathrm{good}}=y]
 \ge\lambda\cdot2^{-k_x/2}.
\]
Thus $y\in G$.

To prove the second inclusion, let $y\in G$. Then $\Pr[\Y=y]>\lambda\cdot2^{-k_x/2-1}>0$.
For every $x$,
\[
 \Pr[\X=x\mid \Y=y]
 =\frac{\Pr[\X=x\ \wedge\ \Y=y]}{\Pr[\Y=y]}
 \le\frac{\Pr[\X=x]}{\Pr[\Y=y]}
 <\frac{2^{-k_x}}{\lambda\cdot2^{-k_x/2-1}}
 =2^{-k_x/2+m/4+1}.
\]
Consequently,
\[
 \minH(\X\mid \Y=y)>k_x/2-m/4-1\ge k_2.
\]
The last inequality holds for sufficiently large $n$, since
$k_x\ge k_y$, $m\le k_y^\delta=o(k_y)$, and
$k_2=k_y^{1/4}=o(k_y)$. This proves $y\in\mathcal Y_{\mathrm{valid}}$.
\end{proof}

\paragraph{Candidate lists.}
Set $\Theta\coloneqq2^{-3m/4}$, the bound on an output point probability
supplied by the two-source condenser.

\begin{definition}
\label{def:robust-candidate-vectors}
For every $y\in\mathcal Y_{\mathrm{valid}}$, define
\begin{equation}
\label{eq:robust-candidate-vectors}
 \mathcal V_y=\left\{u\in\F_q^D:
 \Pr[\TwoCond(u,\X)=\sigma\mid \Y=y]>\Theta\right\}.
\end{equation}
\end{definition}

\begin{lemma}
\label{lem:robust-small-candidates}
For every $y\in\mathcal Y_{\mathrm{valid}}$, $|\mathcal V_y|<2^{k_1}$.
\end{lemma}
\begin{proof}
Otherwise take a uniform vector $\U_y$ from $\mathcal V_y$, independently
of a random variable $\X_y$ with distribution $(\X\mid \Y=y)$.
Then $\minH(\U_y)\ge k_1$ and $\minH(\X_y)\ge k_2$.
\Cref{thm:two-source-condenser} and its instantiation in
\cref{sec:robust-components} give
\[
 \Pr[\TwoCond(\U_y,\X_y)=\sigma]\le2^{-3m/4}=\Theta.
\]
On the other hand, averaging the strict inequality in the definition
of $\mathcal V_y$ over $\U_y$ gives
$\Pr[\TwoCond(\U_y,\X_y)=\sigma]>\Theta$, a contradiction.
\end{proof}

\begin{definition}
\label{def:robust-projected-list}
For every $y\in\mathcal Y_{\mathrm{valid}}$, define
\begin{equation}
\label{eq:robust-projected-list}
 R_y=\{u_D:u\in\mathcal V_y\},\qquad |R_y|\le2^{k_1}.
\end{equation}
\end{definition}

\subsubsection{The test}
\label{sec:robust-test}

Given a line $L$, we seek a test for whether a degree-at-most-$\hat h$
polynomial $p$ is the correct restriction $\widehat f\circ L$.
We first define a joint contribution at a point, then a pointwise
predicate, and finally the score of a candidate along a line.

\begin{definition}[Joint contribution at a point]
\label{def:robust-J}
For a point $w\in\F_q^d$ and a field element $a\in\F_q$, set
\begin{equation}
\label{eq:robust-J}
 J_G(w,a)=\sum_{\substack{y\in G\\w(y)=w}}
 \max_{v\in\F_q^{D-1}}
 \Pr_{(\Y,\X)}\bigl[\Y=y\ \land\ \TwoCond((v,a),\X)=\sigma\bigr].
\end{equation}
\end{definition}

\begin{remark}[Using the original joint source]
\label{rem:robust-joint-contribution-changes}
In \cref{def:robust-J}, we consider a slightly different
notion of joint contribution at a point from the one considered in
\cref{sec:overview-low-entropy}. More specifically, rather
than summing over all $y$ with $w(y)=w$, we restrict the sum to those
$y\in G$. Intuitively, we would like to restrict the sum to
$\supp(\Y_{\mathrm{good}})$, but we only have access to a sampler for
the original joint distribution $(\Y,\X)$. We instead use the set $G$,
which is recognized by the small nondeterministic circuit
$A_{\mathrm{mass}}$. \Cref{lem:robust-valid-set} guarantees that
$\supp(\Y_{\mathrm{good}})\subseteq G$ and every $y\in G$ is valid.
For completeness, the mixture decomposition allows us to lower-bound
the contribution from the good component, and the additional
nonnegative terms corresponding to
$y\in G\setminus\supp(\Y_{\mathrm{good}})$ cannot hurt. For soundness,
we use the fact that every $y\in G$ is valid, which is the property
we already used in the soundness analysis in
\cref{sec:overview-low-entropy}.
\end{remark}

Set
\begin{equation}
\label{eq:robust-point-thresholds}
 \theta_+=\frac{\lambda\mu}{2q^d},
 \qquad \theta_-=\frac{\lambda\mu}{4q^d}.
\end{equation}
Recall that $\lambda=\mu=2^{-m/4}$, so
$\lambda\mu=2^{-m/2}$.
These are thresholds for the joint contribution $J_G(w,a)$, whereas
$\Theta=2^{-3m/4}$ is the conditional-probability threshold used to
define the candidate lists in \cref{def:robust-candidate-vectors}.
Fix any Boolean predicate $\Thr(w,a)$ satisfying
\begin{align}
 J_G(w,a)\ge\theta_+&\ \Longrightarrow\ \Thr(w,a)=1,
 \label{eq:robust-point-high}\\
 J_G(w,a)\le\theta_-&\ \Longrightarrow\ \Thr(w,a)=0.
 \label{eq:robust-point-low}
\end{align}
Its behavior between the thresholds is unrestricted. \Cref{sec:robust-test-efficiency} constructs such a predicate as a small
nondeterministic circuit. Completeness and soundness will hold for every fixed predicate
with these two properties, and hence for that implementation.

\begin{definition}[The polynomial test]
\label{def:robust-polynomial-test}
For a nondegenerate line $L$ and a polynomial $p\in\F_q[t]$ of degree
at most $\hat h$, define
\begin{equation}
\label{eq:robust-score}
 \Test_L(p)=\Pr_{t\leftarrow\F_q}[\Thr(L(t),p(t))=1].
\end{equation}
Set $\Test_L(p)=0$ for degenerate lines.
\end{definition}

We prove completeness and soundness as required by
\cref{lem:reconstruction}, and then construct $\Thr$.

\subsubsection{Completeness}
\label{sec:robust-test-completeness}

\begin{lemma}[Completeness]
\label{lem:robust-test-completeness}
For all sufficiently
large $n$, with probability at least $1-\frac{1}{3n}$ over
$\bL\leftarrow\RandLine$, the line is nondegenerate and
\[
 \Test_{\bL}(\widehat f\circ\bL)\ge7/8.
\]
\end{lemma}
\begin{proof}
\textbf{Correct evaluations pass at almost all points.}
Recall that
\[
 Q(w)=\Pr\bigl[w(\Y_{\mathrm{good}})=w\mid
        \rbCond(\Y_{\mathrm{good}},\X_{\mathrm{good}})=\sigma\bigr].
\]
Let
\[
 \mathcal A\coloneqq\{w\in\F_q^d:Q(w)<1/(2q^d)\}.
\]
Since $\TV(Q,\U_{\F_q^d})\le1/8192$, \cref{lem:density} implies
\[
 \Pr_{w\leftarrow\F_q^d}[w\in\mathcal A]
 \le\frac{1/8192}{1/2}=\frac1{4096}\le\frac1{128}.
\]
We will now show that, for every $w\notin\mathcal A$,
$J_G(w,\widehat f(w))\ge\theta_+$, which implies that
$\Thr(w,\widehat f(w))=1$.

For every $w\notin\mathcal A$, in each summand of $J_G$,
choose $v$ to be the first $D-1$ coordinates of $g(y)$. When $w(y)=w$,
its last coordinate is $\widehat f(w)$. Consequently,
\begin{align*}
 J_G(w,\widehat f(w))
 &\ge\sum_{\substack{y\in G\\w(y)=w}}
   \Pr[\Y=y\ \land\ \rbCond(\Y,\X)=\sigma]\\
 &\ge\lambda\cdot\sum_{\substack{y\in G\\w(y)=w}}
   \Pr[\Y_{\mathrm{good}}=y\ \land\
        \rbCond(\Y_{\mathrm{good}},\X_{\mathrm{good}})=\sigma]\\
 &=\lambda\cdot
   \Pr[w(\Y_{\mathrm{good}})=w\ \land\
        \rbCond(\Y_{\mathrm{good}},\X_{\mathrm{good}})=\sigma]\\
 &=\lambda\cdot
   \Pr[\rbCond(\Y_{\mathrm{good}},\X_{\mathrm{good}})=\sigma]\cdot
   \Pr[w(\Y_{\mathrm{good}})=w\mid
        \rbCond(\Y_{\mathrm{good}},\X_{\mathrm{good}})=\sigma]\\
 &=\lambda\cdot
   \Pr[\rbCond(\Y_{\mathrm{good}},\X_{\mathrm{good}})=\sigma]\cdot
   Q(w)\\
 &>\frac{\lambda\mu}{2q^d}=\theta_+.
\end{align*}
The second inequality uses the mixture decomposition of $(\Y,\X)$
and $\lambda'\ge\lambda$.
The first equality uses
$\supp(\Y_{\mathrm{good}})\subseteq G$, from
\cref{lem:robust-valid-set}. The final inequality uses
$\Pr[\rbCond(\Y_{\mathrm{good}},\X_{\mathrm{good}})=\sigma]>\mu$ and $Q(w)\ge1/(2q^d)$.
This gives that, for every $w\notin\mathcal A$,
$\Thr(w,\widehat f(w))=1$. Thus the correct evaluation passes at all
but at most a $1/128$ fraction of the points.

\Needspace{16\baselineskip}
\medskip
\noindent\textbf{Restriction to a random line.}
Let $B(w)=\mathbf1\{\Thr(w,\widehat f(w))=1\}$. The bound above gives $\E_{w\leftarrow\F_q^d}[B(w)]\ge127/128$.
Applying \cref{lem:line-concentration} with
$127/128-7/8=15/128$ gives
\[
 \Pr_{\bL\leftarrow\RandLine}\!\left[
 \E_{t\leftarrow\F_q}[B(\bL(t))]<7/8\right]
 \le\frac{1}{4q(15/128)^2}<\frac{21}{q}.
\]
For a nondegenerate line, $\Test_L(\widehat f\circ L)
=\E_{t\leftarrow\F_q}[B(L(t))]$. Adding degeneracy, the correct
polynomial satisfies $\Test_{\bL}(\widehat f\circ\bL)\ge7/8$ except with probability at most
$q^{-d}+21/q=o(1/n)\le\frac{1}{3n}$.
\end{proof}

\subsubsection{Soundness}
\label{sec:robust-test-soundness}

\begin{lemma}[Soundness]
\label{lem:robust-test-soundness}
Set
\begin{equation}
\label{eq:robust-K}
 K=\frac{1024\cdot2^{k_1}}{\lambda\mu}
   =1024\cdot2^{k_1+m/2}.
\end{equation}
For all sufficiently
large $n$, with probability at least $1-\frac{1}{3n}$ over
$\bL\leftarrow\RandLine$, the line is nondegenerate and the list
\[
 \mathcal P_{\bL}=\{p\in\F_q[t]:\deg p\le\hat h,\ \Test_{\bL}(p)>3/4\}
\]
has size at most $8\cdot K=o(q/\hat h)$.
\end{lemma}

\begin{proof}
\textbf{Good points for soundness.}
Every $y\in G$ is valid, by \cref{lem:robust-valid-set}.
Define
\[
 \mathcal H\coloneqq\{w\in\F_q^d:P(w)>128/q^d\}.
\]
Since $\sum_wP(w)=1$, Markov's inequality gives
\[
 \Pr_{w\leftarrow\F_q^d}[w\in\mathcal H]
 \le\frac{\E_{w\leftarrow\F_q^d}[P(w)]}{128/q^d}
 =\frac{1/q^d}{128/q^d}=\frac1{128}.
\]
Thus $\mathcal H$ contains at most a $1/128$-fraction of the points.
This bound does not require $P$ to be close to uniform.

\medskip
\noindent\textbf{Few field elements pass at a good point.}
Set
\[
 \mathcal S_w=\{a\in\F_q:\Thr(w,a)=1\}.
\]
By~\cref{eq:robust-point-low}, every $a\in\mathcal S_w$ satisfies
$J_G(w,a)>\theta_-$. We show that $|\mathcal S_w|\le K$
for every $w\notin\mathcal H$.

Fix $w\notin\mathcal H$. For $a\in\F_q$, define
\[
 E_w(a)=\sum_{\substack{y\in G\\w(y)=w\\a\in R_y}}
                   \Pr[\Y=y].
\]
For every $y\in G\subseteq\mathcal Y_{\mathrm{valid}}$, every
$v\in\F_q^{D-1}$, and every $a\in\F_q\setminus R_y$,
\Cref{def:robust-candidate-vectors,def:robust-projected-list} give
\[
 \Pr[\TwoCond((v,a),\X)=\sigma\mid \Y=y]\le\Theta=2^{-3m/4}.
\]
For every $y\in G$, every $v\in\F_q^{D-1}$, and every $a\in R_y$,
we use the trivial bound
\[
 \Pr[\TwoCond((v,a),\X)=\sigma\mid \Y=y]\le1.
\]
These two groups give
\begin{equation}
\label{eq:robust-J-bound}
\begin{aligned}
 J_G(w,a)
 &=\sum_{\substack{y\in G\\w(y)=w}}\max_{v\in\F_q^{D-1}}
   \Pr_{(\Y,\X)}\bigl[\Y=y\ \land\ \TwoCond((v,a),\X)=\sigma\bigr]\\
 &\le\Theta\cdot
   \sum_{\substack{y\in G\\w(y)=w\\a\notin R_y}}\Pr[\Y=y]
   +\sum_{\substack{y\in G\\w(y)=w\\a\in R_y}}\Pr[\Y=y]\\
 &\le\Theta\cdot P(w)+E_w(a).
\end{aligned}
\end{equation}
The last inequality bounds the mass of the first group by all of $P(w)$.
Since $w\notin\mathcal H$, we have
\begin{align*}
 \Theta\cdot P(w)
 &\le\frac{128\cdot2^{-3m/4}}{q^d}\\
 &=\frac{128\cdot2^{-m/4}\cdot\lambda\mu}{q^d}\\
 &\le\frac{\lambda\mu}{8q^d}.
\end{align*}
The last inequality holds for sufficiently large $n$, since
$m\ge(\log n)^2$ and hence $128\cdot2^{-m/4}\le1/8$.
Recall that $\theta_-=\lambda\mu/(4q^d)$.
For every field element $a\in\mathcal S_w$, we therefore have
\begin{align*}
 E_w(a)
 &\ge J_G(w,a)-\Theta\cdot P(w)\\
 &>\theta_--\frac{\lambda\mu}{8q^d}\\
 &=\frac{\lambda\mu}{4q^d}-\frac{\lambda\mu}{8q^d}
 =\frac{\lambda\mu}{8q^d}.
\end{align*}
\Needspace{16\baselineskip}
On the other hand, expanding the definition of $E_w(a)$,
interchanging the sums, and using $|R_y|\le2^{k_1}$, we have
\begin{align*}
 \sum_{a\in\F_q}E_w(a)
 &=\sum_{\substack{y\in G\\w(y)=w}}
   \Pr[\Y=y]\cdot|R_y|\\
 &\le2^{k_1}\cdot
   \sum_{\substack{y\in G\\w(y)=w}}\Pr[\Y=y]\\
 &\le2^{k_1}\cdot\Pr[w(\Y)=w]\\
 &=2^{k_1}\cdot P(w)\\
 &\le\frac{128\cdot2^{k_1}}{q^d}.
\end{align*}
The last inequality uses $w\notin\mathcal H$.
Therefore,
\[
 |\mathcal S_w|\cdot\frac{\lambda\mu}{8q^d}
 \le\sum_{a\in\mathcal S_w}E_w(a)
 \le\sum_{a\in\F_q}E_w(a)
 \le\frac{128\cdot2^{k_1}}{q^d},
\]
and consequently
\[
 |\mathcal S_w|
 \le\frac{1024\cdot2^{k_1}}{\lambda\mu}=K.
\]

\Needspace{12\baselineskip}
\medskip
\noindent\textbf{Restriction to a random line and list recovery.}
Since $\mathcal H$ contains at most a $1/128$-fraction of the points,
applying \cref{lem:line-concentration} to its indicator gives
\[
 \Pr_{\bL\leftarrow\RandLine}\!\left[
 \Pr_{t\leftarrow\F_q}[\bL(t)\in\mathcal H]>\frac18
 \right]
 \le\frac{1}{4q(1/8-1/128)^2}<\frac{21}{q}.
\]
Fix a nondegenerate line $L$ satisfying
\[
 \Pr_{t\leftarrow\F_q}[L(t)\in\mathcal H]\le\frac18.
\]
Recall that $\mathcal S_{L(t)}=\{a\in\F_q:\Thr(L(t),a)=1\}$
and that $|\mathcal S_{L(t)}|\le K$ whenever $L(t)\notin\mathcal H$.
Define
\[
 \mathcal S_t=\begin{cases}
 \mathcal S_{L(t)},&L(t)\notin\mathcal H,\\
 \varnothing,&L(t)\in\mathcal H.
 \end{cases}
\]
Every list has size at most $K$. A polynomial $p$ such that
$\Test_L(p)>3/4$ agrees with these lists on more than a
$3/4-1/8=5/8$ fraction of the field elements $t\in\F_q$,
hence on at least half. \Cref{lem:johnson} with
$\xi=1/2$ bounds their number by $8\cdot K$, provided $q\ge8\cdot\hat h\cdot K$.
This condition holds for sufficiently large $n$, because
\[
 \log(8\cdot\hat h\cdot K)=k_y^{1/4}+m/2+O(\log n)
 =o(k_y^{1/2})=o(\log q).
\]
Here we used $m\le k_y^\delta$ with $\delta\le1/16$ and
$k_y\ge n^\gamma$. The same estimate gives $8\cdot K=o(q/\hat h)$.
By a union bound,
\[
 \Pr_{\bL\leftarrow\RandLine}\!\left[
 \bL\text{ is degenerate}\ \lor\
 \Pr_{t\leftarrow\F_q}[\bL(t)\in\mathcal H]>\frac18
 \right]
 \le q^{-d}+\frac{21}{q}=o(1/n)\le\frac{1}{3n}.
\]
\end{proof}

\subsubsection{Constructing \texorpdfstring{$\Thr$}{Thr}}
\label{sec:robust-test-efficiency}

\begin{lemma}[A small threshold circuit]
\label{lem:robust-test-efficiency}
There is a constant
$C_{\Thr}>0$, depending on $c_{\mathrm{size}}$ and the fixed component constructions but
independent of $d$, such that the following holds for all sufficiently
large $n$. There is a nondeterministic circuit $\Thr(w,a)$ of size at most
$n^{C_{\Thr}}$ satisfying \cref{eq:robust-point-high}--\cref{eq:robust-point-low}
for every $w,a$. This circuit may depend on the circuits $\Samp$ and $\Post$, the probability $p_{\mathrm{post}}$,
the circuit $A_{\mathrm{mass}}$, the output $\sigma$,
and the selected seed $i^*$ from \cref{sec:robust-preparations}.
\end{lemma}

\begin{proof}
Recall that $\Samp(R)=(\Samp_Y(R),\Samp_X(R))$.
With external input $(w,a)$ and uniform sampler coins $R\leftarrow \U_r$,
define a nondeterministic circuit $C$ by
\[
 C((w,a),y,v,R)=1\quad\Longleftrightarrow\quad
 \begin{cases}
 \Post(R)=1,\\
 A_{\mathrm{mass}}(y)=1,\\
 \SExt(y,D)=w,\\
 \Samp_Y(R)=y,\\
 \TwoCond((v,a),\Samp_X(R))=\sigma.
 \end{cases}
\]
The circuit $C$ guesses accepting witnesses for $\Post(R)$ and
$A_{\mathrm{mass}}(y)$.
We will apply \cref{lem:sum-of-maximums} to the circuit $C$,
with $u=(w,a)$.
By \cref{def:robust-J} and the definition of $G$,
\[
 S((w,a))=p_{\mathrm{post}}\cdot J_G(w,a),
\]
where $S$ is defined in \cref{lem:sum-of-maximums}.
Indeed, a value $y\notin G$ contributes zero, and for $y\in G$,
conditioning on $\Post(R)=1$ gives the corresponding summand of $J_G$.

Apply \cref{lem:sum-of-maximums} with threshold
$p_{\mathrm{post}}\theta_+$ and $\eta=1/4$.
Since $\theta_-=\theta_+/2<(1-\eta)\theta_+$, the resulting circuit
$\Thr$ satisfies \cref{eq:robust-point-high}--\cref{eq:robust-point-low}.
The threshold can be represented using $O(r+m+d\log q+1)$ bits.
Here $p_{\mathrm{post}}$ is the hardwired probability from
\cref{sec:robust-preparations}. Since $p_{\mathrm{post}}$ is an integer
multiple of $2^{-r}$, it can be described using $O(r+1)$ bits.

\medskip
\noindent\textbf{Circuit size.}
The circuit $C$ evaluates $\Samp$, the verifier of $A_{\mathrm{mass}}$,
$\SExt$, and $\TwoCond$, and verifies $\Post$; it does not evaluate
$g$ or $\widehat f$. \Cref{lem:sum-of-maximums} gives a fixed
polynomial bound in these circuit sizes, the input and witness lengths,
and the threshold bit length.
The circuit $A_{\mathrm{mass}}$ already has size polynomial in $n$
with an exponent independent of $d$. For every fixed $d$, the same
is true of all the other quantities: $r\le n^{c_{\mathrm{size}}}$,
$\log q=k_y^{1/2}\le n^{1/2}$, $D=n^{c_D}$, and $d\log q\le d n^{1/2}$.
Absorbing factors depending on $d$ into the sufficiently large-$n$
threshold gives the claimed exponent $C_{\Thr}$ independent of $d$.
\end{proof}

\subsubsection{Reconstruction and completion of the proof}
\label{sec:robust-reconstruction-consequences}

\paragraph{Applying the reconstruction lemma.}
By \cref{lem:robust-test-completeness,lem:robust-test-soundness},
with probability at least $1-\frac{2}{3n}$ over $\bL\leftarrow\RandLine$,
\[
 \Test_{\bL}(\widehat f\circ\bL)\ge7/8,
 \qquad |\mathcal P_{\bL}|\le8\cdot K,
\]
where $\mathcal P_L$ is defined in \cref{lem:reconstruction}.
Since $8\cdot K=o(q/\hat h)$, for sufficiently large $n$,
\[
 \frac{2}{3n}\le\frac1{32},
 \qquad 8\cdot K\le\frac{q}{32(\hat h+1)}.
\]
Applying \cref{lem:reconstruction} to the circuit $\Thr$ from
\cref{lem:robust-test-efficiency}, there is a single-valued
nondeterministic circuit $C$ of size
$\operatorname{poly}(\operatorname{size}(\Thr),\hat h,d,\log q)$ such that
\[
 \ACCEPT_C(x)=\{\widehat f(x)\}
 \qquad\text{for every }x\in\F_q^d.
\]

\paragraph{Restricting to the hard function.}
Define $C_f(x)$ to simulate $C(\phi(x))$ and accept if it outputs $1$.
Since $\ACCEPT_C(\phi(x))=\{\widehat f(\phi(x))\}=\{f(x)\}$,
$C_f$ computes $f$ on $\zo^\ell$.

The natural encoding $\phi$ (from \cref{thm:lde}) has size polynomial in $\ell$ and $\log q$,
with an absolute polynomial bound. Since $\ell=d\log n$ and
$\hat h=d(n-1)$, incorporating this encoding gives absolute constants
$C_0,a_0>0$ such that
\[
 \operatorname{size}(C_f)
 \le C_0(\operatorname{size}(\Thr)+\hat h+d+\log q+1)^{a_0}.
\]
By \cref{lem:robust-test-efficiency},
$\operatorname{size}(\Thr)\le n^{C_{\Thr}}$ with $C_{\Thr}$ independent of $d$.
For each fixed $d$, using $\hat h<d\cdot n$ and $\log q\le n$,
the quantity in parentheses is at most $n^{C_{\Thr}+3}$ for sufficiently
large $n$. Thus
\[
 \operatorname{size}(C_f)\le n^{C_{\mathrm{rec}}},
 \qquad C_{\mathrm{rec}}=a_0(C_{\Thr}+3)+1,
\]
where $C_{\mathrm{rec}}$ is independent of $d$.

\paragraph{Choosing the dimension and deriving the contradiction.}
The remaining component parameters were fixed independently of $d$.
We can therefore first obtain $C_{\Thr}$ and $C_{\mathrm{rec}}$, and then
choose a sufficiently large integer $d\ge2$ such that
\[
 \beta_{\mathrm{hard}}\cdot d>C_{\mathrm{rec}}+1.
\]
With this choice of $d$ fixed, all the preceding requirements hold for
sufficiently large $n$. Since $\ell=d\log n$,
\[
 \operatorname{size}(C_f)\le n^{C_{\mathrm{rec}}}
 <n^{\beta_{\mathrm{hard}}\cdot d}
 =2^{\beta_{\mathrm{hard}}\ell},
\]
contradicting the hardness of $f$. Thus, for every admissible joint
distribution and decomposition, no output of
$\rbCond(\Y_{\mathrm{good}},\X_{\mathrm{good}})$ has probability greater
than $2^{-m/4}$. Its min-entropy is at least $m/4$, proving
\cref{thm:block-robust}.

\section{Proof of the multiplicative extractor theorem}
\label{sec:mult-proof}

In this section we prove \cref{thm:multiplicative}.
The construction (that appears in \cref{sec:mult-construction})
and analysis (that appears in \cref{sec:mult-analysis}) closely
follow the high level description of \cref{sec:proof-overview-BGDM}.

\subsection{Construction}
\label{sec:mult-construction}

Fix constants $c_{\mathrm{size}}>1$, $0<\alpha\le1$, and
$c_{\eps}>0$, and set
\[
 k=n^\alpha,\qquad \eps=n^{-c_{\eps}}.
\]
Our goal is to construct an explicit function
\[
 \Ext\colon\zo^n\longrightarrow\zo^m,\qquad m=n^{\alpha'},
\]
computable in polynomial time, that is a $(k,\eps)$-multiplicative
extractor for distributions samplable by deterministic circuits with
postselection by nondeterministic circuits, both of size at most
$n^{c_{\mathrm{size}}}$, as required by \cref{thm:multiplicative}.
The constant $\alpha'>0$, depending only on $\alpha$, will be chosen
sufficiently small in the two-source extractor item of
\cref{sec:mult-components}.
We assume that $\sfE$ is hard for exponential-size nondeterministic
circuits.

\subsubsection{Components}
\label{sec:mult-components}

We rely on the following components:
\begin{itemize}[leftmargin=*]
\item \textbf{Errorless condenser.}
\Cref{thm:errorless-condenser} supplies a universal constant
$0<\delta_{\mathrm{out}}<1/2$ and errorless condensers
$\Cond_{\gamma_{\mathrm{in}},c}$ for the required sampling class.
We instantiate this theorem twice, using common input length $n$ and
padding shorter inputs with zeros. First set
\[
 \gamma_{\mathrm{in},1}=\alpha,\qquad c_1=c_{\mathrm{size}},
 \qquad \Cond_1=\Cond_{\gamma_{\mathrm{in},1},c_1}.
\]
Fix an exponent $e_1$ bounding both the running time and the Boolean
circuit size of $\Cond_1$ by $n^{e_1}$, enlarging it if necessary.
Choose
\[
 c_2>\max\{c_{\mathrm{size}},e_1,1\}+1,
 \qquad \gamma_{\mathrm{in},2}=\alpha\delta_{\mathrm{out}},
 \qquad \Cond_2=\Cond_{\gamma_{\mathrm{in},2},c_2}.
\]
The size bound $n^{c_2}$ accommodates the deterministic sampler for
$\X$ followed by $\Cond_1$ and zero-padding, and also the original
nondeterministic postselection circuit. Indeed, if
$\X=(\Samp(\bR)\mid\Post(\bR)=1)$, then
\[
 \Cond_1(\X)=(\Cond_1(\Samp(\bR))\mid\Post(\bR)=1).
\]
Let $n^{e_2}$ bound the time and circuit size of $\Cond_2$.
Both instances are fixed before the dimension $d$ is chosen; their
size bounds are independent of that dimension.

Write $m_{\Cond,i}$ and $k_{\Cond,i}$ for the output length and
guaranteed output min-entropy of the two instances:
\[
\begin{array}{c|c|c}
 \text{Instance}&\text{Output length}&\text{Output min-entropy at least}\\ \hline
 \Cond_1&m_{\Cond,1}=n^{\alpha/2}&k_{\Cond,1}=n^{\alpha\delta_{\mathrm{out}}}\\
 \Cond_2&m_{\Cond,2}=n^{\alpha\delta_{\mathrm{out}}/2}
        &k_{\Cond,2}=n^{\alpha\delta_{\mathrm{out}}^2}
\end{array}
\]
We will denote the distribution obtained after the second condenser
application by $\Y=\Cond_2(\Cond_1(\X))$, and set
\[
 n_y\coloneqq m_{\Cond,2}=n^{\alpha\delta_{\mathrm{out}}/2},\qquad
 k_y\coloneqq k_{\Cond,2}=n^{\alpha\delta_{\mathrm{out}}^2}.
\]
Thus $n_y$ is the length of $\Y$, and $k_y$ is its guaranteed min-entropy
for every source $\X$ in the specified class with $\minH(\X)\ge k$.

\item \textbf{Hard function and low-degree extension.}
We assume that $\sfE$ is hard for exponential-size nondeterministic
circuits. By \cref{def:nd-hardness}, this gives a function
$f\in\sfE$ and a constant $\beta_{\mathrm{hard}}>0$ such that,
for every sufficiently large input length $\ell$, no nondeterministic
circuit of size at most $2^{\beta_{\mathrm{hard}}\ell}$ computes $f$
on all inputs in $\zo^\ell$.
Let $d\ge2$ be a sufficiently large integer that will be chosen later
in \cref{sec:mult-reconstruction-consequences}, and set
\[
 \ell=d\log n,\qquad h=2^{\ell/d}=n,\qquad q=2^{k_y^{1/2}}.
\]

Let $\F_q$ be the field with $q$ elements, which exists because
$q$ is a power of two. For sufficiently large $n$, $d\le h\le q$.
Let $\phi\colon\zo^\ell\to H^d$ be the natural bijection,
where $H$ is as in \cref{thm:lde}.
By \cref{thm:lde}, $f$ has a low-degree extension
\[
 \widehat f\colon\F_q^d\longrightarrow\F_q
\]
with $\widehat f(\phi(x))=f(x)$ for every $x\in\zo^\ell$, and total degree at most
\[
 \hat h\coloneqq d(h-1)<d\cdot n.
\]
Thus, for every affine line $L\colon\F_q\to\F_q^d$,
$\widehat f\circ L$ is a univariate polynomial of degree at
most $\hat h$. Since $f\in\sfE$, \cref{thm:lde} gives that
$\widehat f$ can be computed in time polynomial in
$2^\ell$ and $\log q$, and hence in time polynomial in $n$ for every
fixed $d$.

\item \textbf{Seeded extractor.}
We use
\[
 \SExt\colon\zo^{n_y}\times[D]\longrightarrow\F_q^d,
\]
with
\[
 k_{\SExt}=k_y-m,
 \qquad \eps_{\SExt}=\eps/8192.
\]

\begin{remark}[$D$ and $d$ are unrelated]
Note that the parameters $d$ and $D$ are unrelated, and in particular,
it is not the case that $D=2^d$. We use $d$ to measure the dimension
of the vector space over which $\widehat f$ is defined. We use $D$
as the number of seeds for the seeded extractor $\SExt$.
\end{remark}
Here $m=n^{\alpha'}$, where $\alpha'>0$ will be chosen sufficiently
small in the next item; that choice will satisfy
$\alpha'<\alpha\delta_{\mathrm{out}}^2/2$.
Consequently, for every fixed $d$ and sufficiently large $n$,
\[
 \tfrac34k_{\SExt}-O(\log n)
 =\tfrac34(k_y-m)-O(\log n)
 \ge d\cdot k_y^{1/2}=d\log q.
\]
\Cref{thm:guv}, followed by truncation, supplies the stated
output length.
The seed length satisfies
\[
 \log D=O\!\left(\log n_y+\log(1/\eps_{\SExt})\right)
 =O(\log n).
\]
We write $D=n^{c_D}$, allowing unused seed bits if needed.
The constant $c_D$ depends on $c_\eps$, because the number of
seeds depends on the seeded extractor's error. It does not depend on
$d$ or $m$.
%\rnote{This may be a good point to elaborate on the choice of parameters.}
The seeded-extractor error depends on $\eps$, not on $m$.

\item \textbf{Two-source extractor.}
Following Shaltiel~\cite{shaltiel25extract}, we will generalize the
approach of~\cite{BSS25} to work with a two-source extractor with
a weaker guarantee. More specifically, the extractor works only
when its second source is a block-wise source
(\cref{def:block-source}). The advantage is that
Li~\cite{Li15threesource} constructed such an extractor with very low error,
whereas known constructions of ordinary two-source extractors in
the low-min-entropy regime do not achieve the error needed here.
Set
\[
 k_1=k_2=k_y^{1/4}=(\log q)^{1/2},\qquad
 \alpha'=\frac{\alpha\delta_{\mathrm{out}}^2}{8c_{\IExt}},
 \qquad m=n^{\alpha'}.
\]
\Cref{thm:li}, followed by padding and truncation, supplies
\[
 \IExt\colon\F_q^D\times\zo^{m_{\Cond,1}}\times\zo^n
           \longrightarrow\zo^m,
 \qquad \eps_{\IExt}\coloneqq\frac{\eps\cdot2^{-m}}{n^2}.
\]
The function $\IExt$ is a $(k_1,k_2,k_2,\eps_{\IExt})$-extractor
for one source and one block-wise source
(\cref{def:independent-block-extractor}).
Indeed, the first input has bit length
\[
 D\log q=n^{c_D}\cdot k_y^{1/2}
 =n^{c_D+\alpha\delta_{\mathrm{out}}^2/2}\le n^{c_D+1/2},
\]
so padding all three inputs to $N=n^{c_D+1}$ suffices.
We have $k_1\ge\log^{12}N$ and $m<0.9k_1$ for large $n$.
Moreover, $k_1^{1/c_{\IExt}}=m^2$, and hence
\[
 2^{-k_1^{1/c_{\IExt}}}=2^{-m^2}
 \le\eps\cdot2^{-m}/n^2.
\]
This holds for every fixed $c_\eps$. The output exponent
$\alpha'$ is independent of $c_\eps$ and $c_{\mathrm{size}}$;
with the component constants fixed, it depends only on $\alpha$.
\end{itemize}

\subsubsection{Operation}
\label{sec:mult-operation}

For every $y\in\zo^{n_y}$ and $i\in[D]$, define
\[
 w_i(y)\coloneqq\SExt(y,i).
\]
Then define
\[
 g(y)\coloneqq\bigl(\widehat f(w_1(y)),\ldots,
                 \widehat f(w_D(y))\bigr)\in\F_q^D.
\]
For $x\in\zo^n$, define
\[
 F_1(x)\coloneqq\Cond_1(x),\qquad F_2(x)\coloneqq\Cond_2(F_1(x)),
\]
where the input to $\Cond_2$ is padded with zeros to length $n$.
Write $\Src(x)=(F_1(x),x)$, and use the shorthand
$\IExt(u,\Src(x))\coloneqq\IExt(u,F_1(x),x)$.
The extractor construction is the function
\[
 \Ext\colon\zo^n\longrightarrow\zo^m
\]
defined by
\[
 \boxed{\Ext(x)=\IExt(g(F_2(x)),F_1(x),x)
                  =\IExt(g(F_2(x)),\Src(x)).}
\]
Thus the distribution $\Y$ introduced above is $\Y=F_2(\X)$.
Since $\ell=O(\log n)$ and $f\in\sfE$, the truth
table of $f$ can be computed in polynomial time. Evaluating $g$ and all
other components therefore takes polynomial time once $d$ is fixed.

\subsection{Analysis of the construction}
\label{sec:mult-analysis}

\subsubsection{Preparations}
\label{sec:mult-preparations}

Assume for contradiction that the construction fails for a source $\X$
on $\zo^n$ with $\minH(\X)\ge k$, where
\[
 \X=(\Samp(\bR)\mid\Post(\bR)=1),\qquad \bR\leftarrow \U_r,
\]
$\Samp$ is deterministic, $\Post$ is nondeterministic, and both
have size at most $n^{c_{\mathrm{size}}}$. Set
\[
 p_{\mathrm{post}}\coloneqq\Pr_{\bR}[\Post(\bR)=1]>0,\qquad
 \mu=2^{-m}.
\]
Failure of multiplicative extraction gives some $\sigma\in\zo^m$ with
\[
 \Pr[\Ext(\X)=\sigma]>e^\eps\mu
                                      \ge(1+\eps)\mu.
\]
Fix this source and output. Write
\[
 \Y=F_2(\X),\qquad \Y_\sigma=(\Y\mid\Ext(\X)=\sigma).
\]

\begin{lemma}[A seed good for both distributions]
\label{lem:mult-good-seed}
For all sufficiently large $n$, there is $i^*\in[D]$ such that
\[
 \TV(\SExt(\Y,i^*),\U_{\F_q^d})\le\frac{\eps}{4096}
 \qquad\textup{and}\qquad
 \TV(\SExt(\Y_\sigma,i^*),\U_{\F_q^d})\le\frac{\eps}{4096}.
\]
\end{lemma}
\begin{proof}
The second errorless condenser gives $\minH(\Y)\ge k_y$.
By \cref{lem:conditioning},
\[
 \minH(\Y_\sigma)\ge k_y+\log\Pr[\Ext(\X)=\sigma]
                      \ge k_y-m=k_{\SExt}.
\]
Both distributions meet the seeded-extractor min-entropy threshold.
Strong extraction bounds the average over a uniform seed of the sum
of the two distances by $2\eps_{\SExt}=\eps/4096$.
Some seed has sum no greater than this average, so each of the two
distances is at most $\eps/4096$.
\end{proof}

In the analysis, in order to simplify the notation,
we will assume without loss of generality that $i^*=D$. This assumption is made only
for convenience in the proof. The choice of $i^*$ can be supplied to
the circuits that we construct as nonuniform advice, and we relabel
the seeds and the corresponding first-input coordinates of $\IExt$
accordingly.
Define
\[
 w(y)=\SExt(y,D),\qquad \W=w(\Y),\qquad \W_\sigma=w(\Y_\sigma),
\]
\[
 P(w)=\Pr[\W=w],\qquad Q(w)=\Pr[\W_\sigma=w].
\]
Thus each of $P,Q$ is within distance $\eps/4096$ of uniform.

\paragraph{Valid condenser outputs and their candidate lists.}
We will now define a set of ``valid $y$'s''. These are the values
of $y$ for which $(\Src(\X)\mid \Y=y)$ is close to a source meeting
the requirements that $\IExt$ imposes on its second source. Set
\[
 \eta_{\mathrm{blk}}\coloneqq\frac{\eps\mu}{n^2},
\]
and define
\[
 \mathcal Y_{\mathrm{valid}}=
 \left\{y\in\supp(\Y):\begin{array}{l}
 (\Src(\X)\mid \Y=y)\text{ is }\eta_{\mathrm{blk}}\text{-close to a}\\
 (k_2,k_2)\text{-block-wise source}
 \end{array}\right\}.
\]
\begin{remark}[Valid and useful condenser outputs]
\label{rem:mult-valid-useful}
In the notation used here, the approach of~\cite{TV00,BSS25}
applies a two-source extractor $\TExt(g(\Cond(\X)),\X)$, where $\Cond$
is an errorless condenser. We replace $\TExt$ with $\IExt$, which
expects its second input to be a block-wise source. This is the task
of the function $\Src(x)=(F_1(x),x)$. In other words, we can think of
\cite{TV00,BSS25} as using the function $\Src(x)=x$.

In this proof, we say that $y$ is valid if
$(\Src(\X)\mid \Y=y)$ meets the requirements that the two-source
extractor imposes on its second source. Here, unlike
in~\cite{TV00,BSS25}, we allow this conditional distribution to be
statistically close to the required source type.
\end{remark}

\begin{lemma}[The conditional second input is close to a block-wise source]
\label{lem:mult-conditional-block}
With
\[
 \beta\coloneqq2^{m_{\Cond,2}+k_2-k_{\Cond,1}}
       +\frac{2^{m_{\Cond,1}+k_2-k}}{\eta_{\mathrm{blk}}},
\]
we have, for sufficiently large $n$,
\begin{equation}
 \label{eq:mult-bad-y}
 \Pr[\Y\notin\mathcal Y_{\mathrm{valid}}]\le\beta
 \le\frac{\eps\mu}{256n}
 \le\frac{\eps\mu}{4096}.
\end{equation}
\end{lemma}
\begin{proof}
Write $\A=F_1(\X)$; then $\Y=\Cond_2(\A)$ is a deterministic function of
$\A$. We have $\minH(\A)\ge k_{\Cond,1}$ and $\minH(\X)\ge k$.
First, any $y$ with $\Pr[\Y=y]\ge2^{k_2-k_{\Cond,1}}$ satisfies
\[
 \Pr[\A=a\mid \Y=y]\le\frac{2^{-k_{\Cond,1}}}{\Pr[\Y=y]}
 \le2^{-k_2},
\]
so $\minH(\A\mid \Y=y)\ge k_2$.
The total probability of the remaining values of $y$ is at most
$2^{m_{\Cond,2}+k_2-k_{\Cond,1}}$.

Next let
\[
 \mathcal B=\{a\in\supp(\A):\Pr[\A=a]<2^{k_2-k}\}.
\]
Its probability is at most $2^{m_{\Cond,1}+k_2-k}$.
For $a\notin\mathcal B$, conditioning gives
$\minH(\X\mid \A=a)\ge k_2$.
By Markov's inequality, the set of $y$ with
$\Pr[\A\in\mathcal B\mid \Y=y]>\eta_{\mathrm{blk}}$ has probability
at most $2^{m_{\Cond,1}+k_2-k}/\eta_{\mathrm{blk}}$.

Fix a $y$ outside the two exceptional sets.
Keep the marginal distribution $(\A\mid \Y=y)$ exactly as it is.
For each $a\notin\mathcal B$, keep $(\X\mid \A=a,\Y=y)$ unchanged.
Because $\Y$ is a function of $\A$, this conditional distribution equals
$(\X\mid \A=a)$ and has min-entropy at least $k_2$.
For $a\in\mathcal B$, replace the conditional second block by a
uniform $n$-bit string. The resulting pair is a $(k_2,k_2)$-block-wise
source. Its statistical distance from $(\A,\X)\mid \Y=y$ is at most
$\Pr[\A\in\mathcal B\mid \Y=y]\le\eta_{\mathrm{blk}}$.
This proves the first inequality in \cref{eq:mult-bad-y}.

For the parameter bound, the choices in \cref{sec:mult-construction} give
$m_{\Cond,2}=o(k_{\Cond,1})$, $k_2=o(k_{\Cond,1})$, and
$m=o(k_{\Cond,1})$. Thus the first term of $\beta$ is at most
$2^{-k_{\Cond,1}/2}=o(\eps\mu/n)$.
Similarly, $m_{\Cond,1},k_2,m,\log(1/\eps),\log n=o(k)$,
so the second term is at most
\[
 2^{m_{\Cond,1}+k_2-k+m+(c_\eps+2)\log n}
 \le2^{-k/2}=o(\eps\mu/n).
\]
Their sum is at most $\eps\mu/(256n)$ for sufficiently large
$n$, proving the remaining inequalities.
\end{proof}

\paragraph{Candidate lists.}
Set $\Theta\coloneqq(1+\eps/8)\mu$.

\begin{definition}
\label{def:mult-candidate-vectors}
For every $y\in\mathcal Y_{\mathrm{valid}}$, define
\begin{equation}
\label{eq:mult-candidate-vectors}
 \mathcal V_y=\left\{u\in\F_q^D:
 \Pr[\IExt(u,\Src(\X))=\sigma\mid \Y=y]>\Theta\right\}.
\end{equation}
\end{definition}

\begin{lemma}
\label{lem:mult-small-candidates}
For every $y\in\mathcal Y_{\mathrm{valid}}$, $|\mathcal V_y|<2^{k_1}$.
\end{lemma}
\begin{proof}
Otherwise take a uniform vector $\U_y$ from $\mathcal V_y$, independently
of $(\Src(\X)\mid \Y=y)$. Its min-entropy is at least $k_1$.
The conditional second input is $\eta_{\mathrm{blk}}$-close to a
$(k_2,k_2)$-block-wise source, and tensoring with the independent $\U_y$
preserves this distance. \Cref{thm:li} and contraction of
statistical distance therefore give
\[
 \Pr[\IExt(\U_y,\Src(\X))=\sigma\mid \Y=y]
 \le\mu+\eps_{\IExt}+\eta_{\mathrm{blk}}
 =\mu+2\eps\mu/n^2<(1+\eps/8)\mu=\Theta
\]
for sufficiently large $n$, contradicting the definition of every
vector in $\mathcal V_y$. This is the only use of the extraction
property of $\IExt$ in the polynomial-test analysis.
\end{proof}

\begin{definition}
\label{def:mult-projected-list}
For every $y\in\mathcal Y_{\mathrm{valid}}$, define
\begin{equation}
\label{eq:mult-projected-list}
 R_y=\{u_D:u\in\mathcal V_y\},\qquad |R_y|\le2^{k_1}.
\end{equation}
\end{definition}

\subsubsection{The test}
\label{sec:mult-test}

Given a line $L$, we seek a test for whether a degree-at-most-$\hat h$
polynomial $p$ is the correct restriction $\widehat f\circ L$.
We first define a joint contribution at a point, then a pointwise
predicate, and finally the score of a candidate along a line.

\begin{definition}[Joint contribution at a point]
\label{def:mult-J}
For a point $w\in\F_q^d$ and a field element $a\in\F_q$, set
\begin{equation}
\label{eq:mult-J}
 J(w,a)=\sum_{y:w(y)=w}\max_{v\in\F_q^{D-1}}
 \Pr_{(\Y,\X)}[\Y=y\ \land\
                 \IExt((v,a),\Src(\X))=\sigma].
\end{equation}
\end{definition}

\begin{remark}[Why we sum over all condenser outputs]
In \cref{sec:robust-test}, we restrict the contribution $J_G$
to values in $G$, because invalid values may carry substantial mass.
Here, \cref{lem:mult-conditional-block} shows that the total
probability of invalid values is very small. We can therefore use the
simpler contribution $J$, summing over all values of $y$, and account
for the contribution of invalid values in the soundness analysis.
\end{remark}

Set
\begin{equation}
\label{eq:mult-point-thresholds}
 \theta_+=(1+3\eps/4)\mu/q^d,
 \qquad\theta_-=(1+\eps/2)\mu/q^d.
\end{equation}
Fix any Boolean predicate $\Thr(w,a)$ satisfying
\begin{align}
 J(w,a)\ge\theta_+&\ \Longrightarrow\ \Thr(w,a)=1,\label{eq:mult-point-high}\\
 J(w,a)\le\theta_-&\ \Longrightarrow\ \Thr(w,a)=0.\label{eq:mult-point-low}
\end{align}
Its behavior between the thresholds is unrestricted. \Cref{sec:mult-test-efficiency} constructs such a predicate as a small nondeterministic circuit.
Completeness and soundness below hold for every fixed predicate with
these two properties, and hence for that implementation.

\begin{definition}[The polynomial test]
\label{def:mult-polynomial-test}
For a nondegenerate line $L$ and a polynomial $p\in\F_q[t]$ of degree
at most $\hat h$, define
\begin{equation}
\label{eq:mult-score}
 \Test_L(p)=\Pr_{t\leftarrow\F_q}[\Thr(L(t),p(t))=1].
\end{equation}
Set $\Test_L(p)=0$ for degenerate lines.
\end{definition}

We prove completeness and soundness as required by
\cref{lem:reconstruction}, and then construct $\Thr$.

\subsubsection{Completeness}
\label{sec:mult-test-completeness}

\begin{lemma}[Completeness]
\label{lem:mult-test-completeness}
For all sufficiently
large $n$, with probability at least $1-\frac{1}{3n}$ over
$\bL\leftarrow\RandLine$, the line is nondegenerate and
\[
 \Test_{\bL}(\widehat f\circ\bL)\ge7/8.
\]
\end{lemma}
\begin{proof}
\textbf{Correct evaluations pass at almost all points.}
Recall that $Q(w)=\Pr[\W=w\mid\Ext(\X)=\sigma]$. Let
\[
 \mathcal A\coloneqq\left\{w\in\F_q^d:
 Q(w)<\frac{1-\eps/16}{q^d}\right\}.
\]
Since $\TV(Q,\U)\le\eps/2048$, \cref{lem:density} implies
\[
 \Pr_{w\leftarrow\F_q^d}[w\in\mathcal A]
 \le\frac{\eps/2048}{\eps/16}=\frac1{128}.
\]
We will now show that, for every $w\notin\mathcal A$,
$J(w,\widehat f(w))\ge\theta_+$, which implies that
$\Thr(w,\widehat f(w))=1$.

For every $w\notin\mathcal A$, in each summand of $J$,
choose $v$ to be the first $D-1$ coordinates of $g(y)$. When $w(y)=w$,
its last coordinate is $\widehat f(w)$. Consequently,
\begin{align*}
 J(w,\widehat f(w))
 &\ge\sum_{y:w(y)=w}\Pr[\Y=y\ \land\ \Ext(\X)=\sigma]\\
 &=\Pr[w(\Y)=w\ \land\ \Ext(\X)=\sigma]\\
 &=\Pr[\Ext(\X)=\sigma]\cdot
   \Pr[w(\Y)=w\mid\Ext(\X)=\sigma]\\
 &=\Pr[\Ext(\X)=\sigma]\cdot Q(w)\\
 &>(1+\eps)\cdot(1-\eps/16)\mu/q^d\\
 &\ge(1+7\eps/8)\mu/q^d\ge\theta_+.
\end{align*}
The penultimate inequality uses $0<\eps\le1$.
This gives that, for every $w\notin\mathcal A$,
$\Thr(w,\widehat f(w))=1$. Thus the correct evaluation passes at all
but at most a $1/128$ fraction of the points. This is where extraction from the conditioned source
$\Y_\sigma$ spreads the heavy-output mass.

\medskip
\noindent\textbf{Restriction to a random line.}
Let $B(w)=\mathbf1\{\Thr(w,\widehat f(w))=1\}$. The bound above gives $\E_{w\leftarrow\F_q^d}[B(w)]\ge127/128$.
Applying \cref{lem:line-concentration} with
$127/128-7/8=15/128$ gives
\[
 \Pr_{\bL\leftarrow\RandLine}\!\left[
 \E_{t\leftarrow\F_q}[B(\bL(t))]<7/8\right]
 \le\frac{1}{4q(15/128)^2}<\frac{21}{q}.
\]
For a nondegenerate line, $\Test_L(\widehat f\circ L)
=\E_{t\leftarrow\F_q}[B(L(t))]$. Adding degeneracy, the correct
polynomial satisfies $\Test_{\bL}(\widehat f\circ\bL)\ge7/8$ except with probability at most
$q^{-d}+21/q=o(1/n)\le\frac{1}{3n}$.
\end{proof}

\subsubsection{Soundness}
\label{sec:mult-test-soundness}

\begin{lemma}[Soundness]
\label{lem:mult-test-soundness}
Set
\begin{equation}
\label{eq:mult-K}
 K=\frac{8\cdot2^{k_1}}{\eps\mu}
   =\frac{8\cdot2^{k_1+m}}{\eps}.
\end{equation}
For all sufficiently
large $n$, with probability at least $1-\frac{1}{3n}$ over
$\bL\leftarrow\RandLine$, the line is nondegenerate and the list
\[
 \mathcal P_{\bL}=\{p\in\F_q[t]:\deg p\le\hat h,\ \Test_{\bL}(p)>3/4\}
\]
has size at most $8\cdot K=o(q/\hat h)$.
\end{lemma}
\begin{proof}
\textbf{Good points for soundness.}
We will try to identify a large subset of points $w\in\F_q^d$
that are ``good for soundness.'' For this purpose, we consider how much
of the weight
\[
 P(w)=\Pr[\W=w]=\Pr[w(\Y)=w]
\]
comes from $y\notin\mathcal Y_{\mathrm{valid}}$. Set
\[
 b(w)=\Pr[\Y\notin\mathcal Y_{\mathrm{valid}}\ \land\ \W=w].
\]
A point $w\in\F_q^d$ is good for soundness if $P(w)$ is not too
large and the weight contributed by non-valid $y$'s is small.
More specifically, we require that
\begin{equation}
\label{eq:mult-good-point}
 P(w)\le(1+\eps/16)/q^d,
 \qquad b(w)\le\eps\mu/(32q^d).
\end{equation}
Let $\mathcal H$ be the set of points failing either condition.
The first fails on at most a $1/128$ fraction of the points, by
\cref{lem:density}. Since $\sum_w b(w)\le\beta$, by Markov's
inequality the second fails on a fraction at most
$32\beta/(\eps\mu)\le1/128$.
Thus $\mathcal H$ contains at most a $1/64$-fraction of the points.

\medskip
\noindent\textbf{Few field elements pass at a good point.}
Set
\[
 \mathcal S_w=\{a\in\F_q:\Thr(w,a)=1\}.
\]
By~\cref{eq:mult-point-low}, every $a\in\mathcal S_w$ satisfies
$J(w,a)>\theta_-$. We show that $|\mathcal S_w|\le K$
for every $w\notin\mathcal H$.

Fix $w\notin\mathcal H$.
For $a\in\F_q$, define
\[
 E_w(a)=\sum_{\substack{y\in\mathcal Y_{\mathrm{valid}}\\w(y)=w\\a\in R_y}}
                   \Pr[\Y=y].
\]
For every $y\in\mathcal Y_{\mathrm{valid}}$, every
$v\in\F_q^{D-1}$, and every $a\in\F_q\setminus R_y$,
\Cref{def:mult-candidate-vectors,def:mult-projected-list} give
\[
 \Pr[\IExt((v,a),\Src(\X))=\sigma\mid \Y=y]
 \le\Theta=(1+\eps/8)\mu.
\]
For every $y\in\mathcal Y_{\mathrm{valid}}$, every
$v\in\F_q^{D-1}$, and every $a\in R_y$, we use the trivial bound
\[
 \Pr[\IExt((v,a),\Src(\X))=\sigma\mid \Y=y]\le1.
\]
For every $y\in\supp(\Y)\setminus\mathcal Y_{\mathrm{valid}}$,
every $v\in\F_q^{D-1}$, and every $a\in\F_q$, we use the trivial bound
\[
 \Pr[\IExt((v,a),\Src(\X))=\sigma\mid \Y=y]\le1.
\]
These three groups give
\begin{equation}
\label{eq:mult-J-bound}
\begin{aligned}
 J(w,a)
 &=\sum_{y:w(y)=w}\max_{v\in\F_q^{D-1}}
   \Pr_{(\Y,\X)}\!\left[\Y=y\ \land\
   \IExt((v,a),\Src(\X))=\sigma\right]\\
 &\le\Theta\cdot
   \sum_{\substack{y\in\mathcal Y_{\mathrm{valid}}\\w(y)=w\\a\notin R_y}}\Pr[\Y=y]
   +\sum_{\substack{y\in\mathcal Y_{\mathrm{valid}}\\w(y)=w\\a\in R_y}}\Pr[\Y=y]\\
 &\qquad+\sum_{\substack{y\in\supp(\Y)\setminus\mathcal Y_{\mathrm{valid}}\\w(y)=w}}\Pr[\Y=y]\\
 &\le\Theta\cdot P(w)+E_w(a)+b(w).
\end{aligned}
\end{equation}
The last inequality bounds the mass of the first group by all of $P(w)$.
Since $w\notin\mathcal H$, we have
\begin{align*}
 \Theta\cdot P(w)+b(w)
 &\le\left((1+\eps/8)\cdot(1+\eps/16)
                          +\eps/32\right)\mu/q^d\\
 &=\left(1+7\eps/32+\eps^2/128\right)\mu/q^d
 \le(1+\eps/4)\mu/q^d.
\end{align*}
The last inequality uses $\eps\le1$.
Recall that $\theta_-=(1+\eps/2)\mu/q^d$.
For every field element $a\in\mathcal S_w$, we therefore have
\begin{align*}
 E_w(a)
 &\ge J(w,a)-\bigl(\Theta\cdot P(w)+b(w)\bigr)\\
 &>\theta_--(1+\eps/4)\frac{\mu}{q^d}\\
 &=\left((1+\eps/2)-(1+\eps/4)\right)\frac{\mu}{q^d}\\
 &=\frac{\eps\mu}{4q^d}.
\end{align*}
On the other hand, expanding the definition of $E_w(a)$,
interchanging the sums, and using $|R_y|\le2^{k_1}$, we have
\begin{align*}
 \sum_{a\in\F_q}E_w(a)
 &=\sum_{\substack{y\in\mathcal Y_{\mathrm{valid}}\\w(y)=w}}
   \Pr[\Y=y]\cdot|R_y|\\
 &\le2^{k_1}\cdot
   \sum_{\substack{y\in\mathcal Y_{\mathrm{valid}}\\w(y)=w}}\Pr[\Y=y]\\
 &\le2^{k_1}\cdot\Pr[w(\Y)=w]\\
 &=2^{k_1}\cdot P(w)\\
 &\le2^{k_1}\cdot\frac{1+\eps/16}{q^d}\\
 &\le\frac{17\cdot2^{k_1}}{16q^d}.
\end{align*}
The last two inequalities use $w\notin\mathcal H$ and $\eps\le1$,
respectively.

Therefore,
\[
 |\mathcal S_w|\cdot\frac{\eps\mu}{4q^d}
 \le\sum_{a\in\mathcal S_w}E_w(a)
 \le\sum_{a\in\F_q}E_w(a)
 \le\frac{17\cdot2^{k_1}}{16q^d},
\]
and consequently
\[
 |\mathcal S_w|\le\frac{17\cdot2^{k_1}}{4\eps\mu}\le K.
\]

\medskip
\noindent\textbf{Restriction to a random line and list recovery.}
Since $\mathcal H$ contains at most a $1/64$-fraction of the points,
applying \cref{lem:line-concentration} to its indicator gives
\[
 \Pr_{\bL\leftarrow\RandLine}\!\left[
 \Pr_{t\leftarrow\F_q}[\bL(t)\in\mathcal H]>\frac18
 \right]
 \le\frac{1}{4q(1/8-1/64)^2}<\frac{21}{q}.
\]
Fix a nondegenerate line $L$ satisfying
\[
 \Pr_{t\leftarrow\F_q}[L(t)\in\mathcal H]\le\frac18.
\]
Recall that $\mathcal S_{L(t)}=\{a\in\F_q:\Thr(L(t),a)=1\}$
and that $|\mathcal S_{L(t)}|\le K$ whenever $L(t)\notin\mathcal H$.
Define
\[
 \mathcal S_t=\begin{cases}
 \mathcal S_{L(t)},&L(t)\notin\mathcal H,\\
 \varnothing,&L(t)\in\mathcal H.
 \end{cases}
\]
Every list has size at most $K$. A polynomial $p$ such that
$\Test_L(p)>3/4$ agrees with these lists on more than a
$3/4-1/8=5/8$ fraction of the field elements $t\in\F_q$,
hence on at least half. \Cref{lem:johnson} with
$\xi=1/2$ bounds their number by $8\cdot K$, provided $q\ge8\cdot\hat h\cdot K$.
This condition holds for sufficiently large $n$, because
\[
 \log(8\cdot\hat h\cdot K)=k_y^{1/4}+m+O(\log n)
                 =o(\log q).
\]
The same estimate gives $8\cdot K=o(q/\hat h)$. By a union bound,
\[
 \Pr_{\bL\leftarrow\RandLine}\!\left[
 \bL\text{ is degenerate}\ \lor\
 \Pr_{t\leftarrow\F_q}[\bL(t)\in\mathcal H]>\frac18
 \right]
 \le q^{-d}+\frac{21}{q}=o(1/n)\le\frac{1}{3n}.
\]
\end{proof}

\subsubsection{Constructing \texorpdfstring{$\Thr$}{Thr}}
\label{sec:mult-test-efficiency}

\begin{lemma}[A small threshold circuit]
\label{lem:mult-test-efficiency}
There is a constant
$C_{\Thr}>0$, depending on $\alpha,c_{\eps},c_{\mathrm{size}},e_1,e_2$
but independent of $d$, such that the
following holds for all sufficiently large $n$.
There is a nondeterministic circuit $\Thr(w,a)$ of size at most
$n^{C_{\Thr}}$ satisfying \cref{eq:mult-point-high}--\cref{eq:mult-point-low}
for every $w,a$. This circuit may depend on the circuits $\Samp$
and $\Post$, the probability $p_{\mathrm{post}}$, the output $\sigma$,
and the selected seed $i^*$ from \cref{sec:mult-preparations}.
\end{lemma}

\begin{proof}
With external input $(w,a)$ and uniform sampler coins $R\leftarrow \U_r$,
define a nondeterministic circuit $C$ by
\[
 C((w,a),y,v,R)=1\quad\Longleftrightarrow\quad
 \begin{cases}
 \Post(R)=1,\\
 \SExt(y,D)=w,\\
 F_2(\Samp(R))=y,\\
 \IExt((v,a),\Src(\Samp(R)))=\sigma.
 \end{cases}
\]
The circuit $C$ guesses an accepting witness for $\Post(R)$.
We will apply \cref{lem:sum-of-maximums} to the circuit $C$,
with $u=(w,a)$.
By \cref{def:mult-J},
\[
 S((w,a))=p_{\mathrm{post}}\cdot J(w,a),
\]
where $S$ is defined in \cref{lem:sum-of-maximums}.
Indeed, conditioning on $\Post(R)=1$ gives the corresponding summand of $J$.

Apply \cref{lem:sum-of-maximums} with threshold
$p_{\mathrm{post}}\theta_+$ and $\eta=\eps/16$.
For $0<\eps\le1$,
\[
 (1-\eta)\theta_+
 =(1-\eps/16)(1+3\eps/4)\mu/q^d
 >(1+\eps/2)\mu/q^d=\theta_-.
\]
The resulting circuit $\Thr$ therefore satisfies
\cref{eq:mult-point-high}--\cref{eq:mult-point-low}.
The threshold can be represented using
$O(r+m+d\log q+\log(1/\eps)+1)$ bits.
Here $p_{\mathrm{post}}$ is the hardwired probability from
\cref{sec:mult-preparations}. Since $p_{\mathrm{post}}$ is an integer
multiple of $2^{-r}$, it can be described using $O(r+1)$ bits.

\medskip
\noindent\textbf{Circuit size.}
The circuit $C$ evaluates $\Samp,F_1,F_2,\SExt,\IExt$ and verifies
$\Post$; it does not evaluate $g$ or $\widehat f$.
\Cref{lem:sum-of-maximums} gives a fixed polynomial bound in these
circuit sizes, the input and witness lengths, the threshold bit
length, and $1/\eps$.
For every fixed $d$, these quantities are polynomial in $n$ with
exponents independent of $d$: $r\le n^{c_{\mathrm{size}}}$,
$\log q=k_y^{1/2}\le n^{1/2}$, $D=n^{c_D}$, $d\log q\le d n^{1/2}$,
and $1/\eps=n^{c_\eps}$. The condenser circuit sizes
were fixed independently of $d$ in \cref{sec:mult-components}.
Absorbing factors depending on $d$ into the sufficiently large-$n$
threshold gives the claimed exponent $C_{\Thr}$ independent of $d$.
\end{proof}

\subsubsection{Reconstruction and completion of the proof}
\label{sec:mult-reconstruction-consequences}

The argument used here is identical to that of
\cref{sec:robust-reconstruction-consequences}, and we repeat it for completeness.

\paragraph{Applying the reconstruction lemma.}
By \cref{lem:mult-test-completeness,lem:mult-test-soundness},
with probability at least $1-\frac{2}{3n}$ over $\bL\leftarrow\RandLine$,
\[
 \Test_{\bL}(\widehat f\circ\bL)\ge7/8,
 \qquad |\mathcal P_{\bL}|\le8\cdot K,
\]
where $\mathcal P_L$ is defined in \cref{lem:reconstruction}.
Since $8\cdot K=o(q/\hat h)$, for sufficiently large $n$,
\[
 \frac{2}{3n}\le\frac1{32},
 \qquad 8\cdot K\le\frac{q}{32(\hat h+1)}.
\]
Applying \cref{lem:reconstruction} to the circuit $\Thr$ from
\cref{lem:mult-test-efficiency}, there is a single-valued
nondeterministic circuit $C$ of size
$\operatorname{poly}(\operatorname{size}(\Thr),\hat h,d,\log q)$ such that
\[
 \ACCEPT_C(x)=\{\widehat f(x)\}
 \qquad\text{for every }x\in\F_q^d.
\]

\paragraph{Restricting to the hard function.}
Define $C_f(x)$ to simulate $C(\phi(x))$ and accept if it outputs $1$.
Since $\ACCEPT_C(\phi(x))=\{\widehat f(\phi(x))\}=\{f(x)\}$,
$C_f$ computes $f$ on $\zo^\ell$.

The natural encoding $\phi$ (from \cref{thm:lde}) has size polynomial in $\ell$ and $\log q$,
with an absolute polynomial bound. Since $\ell=d\log n$ and
$\hat h=d(n-1)$, incorporating this encoding gives absolute constants
$C_0,a_0>0$ such that
\[
 \operatorname{size}(C_f)
 \le C_0(\operatorname{size}(\Thr)+\hat h+d+\log q+1)^{a_0}.
\]
By \cref{lem:mult-test-efficiency},
$\operatorname{size}(\Thr)\le n^{C_{\Thr}}$ with $C_{\Thr}$ independent of $d$.
For each fixed $d$, using $\hat h<d\cdot n$ and $\log q\le n$,
the quantity in parentheses is at most $n^{C_{\Thr}+3}$ for sufficiently
large $n$. Thus
\[
 \operatorname{size}(C_f)\le n^{C_{\mathrm{rec}}},
 \qquad C_{\mathrm{rec}}=a_0(C_{\Thr}+3)+1,
\]
where $C_{\mathrm{rec}}$ is independent of $d$.

\paragraph{Choosing the dimension and deriving the contradiction.}
The remaining component parameters were fixed independently of $d$.
We can therefore first obtain $C_{\Thr}$ and $C_{\mathrm{rec}}$, and then
choose a sufficiently large integer $d\ge2$ such that
\[
 \beta_{\mathrm{hard}}\cdot d>C_{\mathrm{rec}}+1.
\]
With this choice of $d$ fixed, all the preceding requirements hold for
sufficiently large $n$. Since $\ell=d\log n$,
\[
 \operatorname{size}(C_f)\le n^{C_{\mathrm{rec}}}
 <n^{\beta_{\mathrm{hard}}\cdot d}
 =2^{\beta_{\mathrm{hard}}\ell},
\]
contradicting the hardness of $f$. Thus, for every source in the
specified class with min-entropy at least $k$ and every output $\sigma$,
$\Pr[\Ext(\X)=\sigma]\le e^\eps\cdot2^{-m}$.
This proves \cref{thm:multiplicative}, using the errorless
condenser supplied by \cref{thm:errorless-condenser}.

\dobib

\part{Additive Extractors}
\label{part: additive}
In this part, we construct the following additive extractors for $\NPSources_{n,n^c}$:
\begin{itemize}
    \item \textbf{One-bit extraction from logarithmic entropy} (\cref{thm:log-extractor}): Under $\NDH$, a polynomial-time extractor producing one bit from entropy $C\log n$, with arbitrarily small constant error.

    \item \textbf{Almost-all extraction from polylogarithmic entropy} (\cref{many:thm:almost-all}): Under $\NDH$, an extractor producing $\lfloor(1-\eta)k\rfloor$ bits from entropy $k>(\log n)^A$, with error $n^{-e}$ and running time $2^{O(\log^2 n/\log k)}$.

    \item \textbf{Polynomial-time almost-all extraction} (\cref{poly:thm:main}): Under $\SH$, a polynomial-time extractor producing $\lfloor(1-\eta)k\rfloor$ bits from entropy $k>(\log n)^A$, with error $n^{-e}$.
\end{itemize}

% \begin{theorem}\label{thm:log-extractor}
% Assume $\NDH$. For all constants $c,\eps>0$, there is a constant $C$ and a polynomial-time computable $(C\log n,\eps)$-extractor $\Ext:\zo^n\to\zo$ for $\NPSources_{n,n^c}$.
% \end{theorem}

% \begin{theorem}\label{many:thm:almost-all}
% Assume $\NDH$. There is a constant $A>0$ such that, for all constants $c,e>0$ and $\eta\in(0,1)$ and all sufficiently large $n$ (depending on $c,e,\eta$), the following holds. For every $(\log n)^A<k\le n$, there is a $(k,n^{-e})$-extractor $\Ext:\zo^n\to\zo^{\lfloor(1-\eta)k\rfloor}$ for $\NPSources_{n,n^c}$, computable in time $2^{O(\log^2 n/\log k)}$. The implicit constant may depend on $c,e,\eta$.
% \end{theorem}

% \begin{theorem}\label{poly:thm:main}
% Assume $\SH$. There is a constant $A>0$ such that, for all constants $c,e>0$ and $\eta\in(0,1)$ and all sufficiently large $n$ (depending on $c,e,\eta$), the following holds. For every $(\log n)^A<k\le n$, there is a polynomial-time computable $(k,n^{-e})$-extractor $\Ext:\zo^n\to\zo^{\lfloor(1-\eta)k\rfloor}$ for $\NPSources_{n,n^c}$. The polynomial running-time bound may depend on $c,e,\eta$.
% \end{theorem}

\Cref{part: additive} is organized as follows:
In \cref{sec:hash}, we construct an errorless strong seeded condenser. \Cref{sec:decorrelation} develops additive and multiplicative correlation breakers for jointly samplable sources. We use these tools in \cref{sec:log-construction} to construct a polynomial-time one-bit extractor for sources with logarithmic min-entropy under $\NDH$. In \cref{sec:many-construction}, we use lightest-bin selection and correlation breaking to extract from sources with polylogarithmic min-entropy under the same assumption. Finally, \cref{sec:poly-construction} improves the running time of this construction to polynomial under the stronger assumption $\SH$.

\section{Errorless seeded condensation by hashing}\label{sec:hash}

In this section, we construct an errorless strong seeded condenser (\cref{def:errorless-strong-seeded-condenser}). Enumerating its seeds
gives polynomially many jointly samplable rows, most of which have
almost full min-entropy. This is going to be the first step of all our extractor constructions.

\begin{lemma}[Errorless strong seeded condensation]\label{hash:lem:rows}
Assume $\NDH$ and fix a constant $c>0$. There are constants $C>1$,
$\eta>0$, and an integer $\Delta\ge1$ such that, for all sufficiently
large $n$ and every integer $k$ satisfying $C\log n\le k\le n$,
there is a uniformly polynomial-time computable errorless strong
seeded $(k,k-\log k,R^{-\eta})$-condenser
$\Cond:\zo^n\times[R]\to\zo^m$ for $\NPSources_{n,n^c}$,
where $n\le R=\poly(n)$ and
$m=\lfloor k-\log k\rfloor+\Delta$.
\end{lemma}

Write $h_i(x)=\Cond(x,i)$. For every such source $\X$, the lemma
gives a set $B_{\X}\subseteq[R]$ with $|B_{\X}|\le R^{1-\eta}$
such that $\minH(h_i(\X))\ge k-\log k\ge m-\Delta$ for every
$i\in[R]\setminus B_{\X}$.

\begin{proof}
We first prove the entropy guarantee for a random hash, then use
the counting theorem and pseudorandom generator from Preliminaries
to derandomize the choice.

\paragraph{Step 1: Random hashing.}
Fix a sufficiently large integer $\Delta$, and set $r=k$ and
$m=\lfloor k-\log k\rfloor+\Delta$. Let $\mathbf H$ be a uniformly
random hash from an $r$-wise independent family $\mathcal H$ of
maps $\zo^n\to\zo^m$.
Identify $\zo^n$ with a subset of $\F_{2^v}$, where
$v=\max\{n,m\}$, and evaluate a polynomial of degree at most
$r-1$ with independent uniform coefficients. Retaining $m$ fixed
binary coordinates of the output gives such a family. Its hashes
are computable in polynomial time and indexed by $q=rv=O(n^2)$
random bits.

Fix $\X\in\NPSources_{n,n^c}$ with $\minH(\X)\ge k$. For
independent copies $\X_1,\ldots,\X_r$, define
\[
 W(h)=\Pr\bigl[\X_1,\ldots,\X_r\text{ are distinct and }
                  h(\X_1)=\cdots=h(\X_r)\bigr].
\]
By $r$-wise independence,
$\E_{\mathbf H}[W(\mathbf H)]\le2^{-m(r-1)}$.
Set $\tau_H=r!\,2^{-rk}$ and $\tau_L=\tau_H/2$. Then
\begin{align*}
 \Pr[W(\mathbf H)>\tau_L]
 &\le \frac{\E_{\mathbf H}[W(\mathbf H)]}{\tau_L}\\
 &\le \frac{2^{1+rk-m(r-1)}}{r!}\\
 &\le \frac{2^{1+k^2-(k-\log k+\Delta-1)(k-1)}}{k!}\\
 &\le \frac{2^{k+1-(\Delta-1)(k-1)}k^{k-1}}{(k/e)^k}\\
 &= \frac{2^\Delta}{k}\bigl(e\,2^{2-\Delta}\bigr)^k
 \le 2^{-\Omega(k)}
 \le n^{-2},
\end{align*}
where the last two inequalities hold for sufficiently large
$\Delta$ and $C$.

Now fix $h\in\mathcal H$ and suppose
$p_z=\Pr[h(\X)=z]>r2^{-k}$ for some $z\in\zo^m$.
Given distinct values $x_1,\ldots,x_j$ with
$h(x_1)=\cdots=h(x_j)=z$, the next independent sample has hash
value $z$ and avoids $x_1,\ldots,x_j$ with probability at least
$p_z-j2^{-k}$. Hence
\[
 W(h)\ge p_z(p_z-2^{-k})\cdots(p_z-(r-1)2^{-k})
 >r!\,2^{-rk}=\tau_H.
\]
Thus $W(h)\le\tau_H$ implies
$\minH(h(\X))\ge k-\log r=k-\log k$.

\paragraph{Step 2: Derandomizing the construction.}
We use \cref{thm:gs-counting} to test whether the collision
probability exceeds $\tau_H$, with rejection guaranteed below
$\tau_L$, and then apply \cref{hash:lem:su}.

Let $(\Samp,\Post)$ be a deterministic sampling circuit with
nondeterministic postselection of total size at most $n^c$ for $\X$. Write $\ell$ for
its input length and $p_{\mathrm{post}}=\Pr[\Post(\U_\ell)=1]>0$.
For each hash description $u\in\zo^q$, let $C_u$ accept an ordered
$r$-tuple $(\rho_1,\ldots,\rho_r)\in(\zo^\ell)^r$ exactly when
\[
 \begin{gathered}
 \Post(\rho_j)=1\quad\text{for every }j\in[r],\\
 \Samp(\rho_1),\ldots,\Samp(\rho_r)\text{ are distinct},\\
 h_u(\Samp(\rho_1))=\cdots=h_u(\Samp(\rho_r)).
 \end{gathered}
\]
The circuit $C_u$ is nondeterministic, which is allowed by
\cref{thm:gs-counting}. Its witnesses certify the postselection tests
and are not counted as ordinary inputs. With $(\Samp,\Post)$ fixed, the description of $C_u$ is computable
from $u$ in polynomial time. Conditioning on all $r$ inputs
passing $\Post$ gives $r$ independent copies of $\X$, so
\[
 \Pr[C_u(\U_{r\ell})=1]=p_{\mathrm{post}}^r W(h_u).
\]

Apply \cref{thm:gs-counting} with
$\gamma=p_{\mathrm{post}}^r\tau_H$ and $\lambda=1/2$.
The threshold $\gamma$ is positive, at most one, and has
polynomially many bits: $p_{\mathrm{post}}$ has denominator
$2^\ell$, and $\tau_H=r!\,2^{-rk}$.
Since $\tau_L=\tau_H/2<e^{-1/2}\tau_H$, the resulting
nondeterministic test $T_{\Samp,\Post}$ satisfies
\[
 \begin{aligned}
 W(h_u)>\tau_H&\ \Longrightarrow\ T_{\Samp,\Post}(u)=1,\\
 W(h_u)\le\tau_L&\ \Longrightarrow\ T_{\Samp,\Post}(u)=0.
 \end{aligned}
\]
The sampling circuit and $\gamma$ are fixed in the test's
description; the test need not compute the postselection
probability. Its size is bounded by a common polynomial in $n$,
over all sampling circuits of size at most $n^c$ and all $k\le n$.

Use \cref{hash:lem:su} for this common size bound and error
$n^{-2}$. This gives a generator $G:\zo^d\to\zo^q$ with
$d=O(\log n)$. Pad the seed with unused bits if necessary so
that $d\ge\lceil\log n\rceil$. Put $R=2^d$, enumerate the seeds
as $\sigma_1,\ldots,\sigma_R$, and define
$\Cond(x,i)=h_{G(\sigma_i)}(x)$.

Since $T_{\Samp,\Post}$ rejects whenever $W(h_u)\le\tau_L$,
Step 1 gives $\Pr[T_{\Samp,\Post}(\U_q)=1]\le n^{-2}$.
For $B_{\X}=\{i:T_{\Samp,\Post}(G(\sigma_i))=1\}$, we have
\begin{align*}
 \frac{|B_{\X}|}{R}
 &=\Pr[T_{\Samp,\Post}(G(\U_d))=1]\\
 &\le\Pr[T_{\Samp,\Post}(\U_q)=1]+n^{-2}
 \le 2n^{-2}.
\end{align*}
Choose a constant $\eta>0$ such that $R^\eta\le n$.
Then $|B_{\X}|\le R^{1-\eta}$ for sufficiently large $n$.
For every $i\notin B_{\X}$, the test's acceptance guarantee implies
$W(h_{G(\sigma_i)})\le\tau_H$, and Step 1 gives
$\minH(\Cond(\X,i))\ge k-\log k$.
\end{proof}

\dobib

\section{Decorrelation of correlated samplable sources}\label{sec:decorrelation}

\subsection{Section Overview}
We construct correlation breakers for indexed sources/rows that are jointly
samplable with postselection. The correlation breaker will ultimately allow us to convert indexed (high-entropy) correlated rows into something resembling a non-oblivious bit-fixing source, on which we can then invoke techniques from the two-source extractor literature. To do this, the correlation breaker can be locally applied to each individual source. The resulting outputs have the property that outputs on good sources are effectively independent of the rest.

The construction follows the precise template of Ball, Dachman-Soled, and Loss~\cite{BDL22,BD25}'s approach to constructing non-malleable extractors. Essentially, Ball and Dachman-Soled~\cite{BD25} used their non-malleable extractor to instantiate this precise object---without naming it---for their application in black-box barriers.
The present construction instantiates their recipe with stronger ingredients and provides a better analysis that yields the result from better assumptions. In particular, we construct decorrelators with
additive error under $\NDH$, and pointwise multiplicative error with
an additive slack under $\SH$.

The recipe of Ball and Dachman-Soled is as follows:
\begin{enumerate}
    \item Extract a short seed, $\sigma$ from a row, $v$, using an extractor for post-selecting sources, $\Ext_{\mathrm{pre}}$: $\sigma = \Ext_{\mathrm{pre}}(v)$.
    
        For our application, we will promise ``good'' rows $v$ have extremely high min-entropy (just small poly deficiency), so it suffices to use Shaltiel's extractor~\cite{Shaltiel25multiplicative}, or the construction from \cref{sec:multiplicative-extraction}. Both of these constructions follow from the $\NDH$ assumption. For the additive-error correlation breaker, an additive-error extractor suffices.
        
    \item Expand the seed, $\sigma$, with a \emph{seed-extending multiplicative pseudorandom generator}, $G$: $G(\sigma) = \sigma \concat y$. For the additive version, we require $G$ to be secure against nondeterministic circuits.

    A seed-extending multiplicative pseudorandom generator for nondeterministic circuits is a PRG with special properties: (a) the seed is output as a prefix, (b) fooling is with respect to one-sided multiplicative error: for any distinguisher* $D$,
        \[ \Pr[D(G(\U_k))=1]\le e^\eps\Pr[D(\U_N)=1]+\delta.\]
    In this inequality, $\eps$ is inverse-polynomial and $\delta$ can be substantially smaller.\footnote{The seed length grows as $O(\log(1/\delta))$; consequently, polynomial stretch is compatible with additive slack exponentially small in the seed length.}
    To get the desired properties and parameters, we compose a seed-extending short stretch PRG~\cite{Shaltiel25multiplicative} with a long stretch PRG~\cite{DS26}.
        
    *For the additive error variant, we require security against bounded-polynomial-size nondeterministic circuits, which follows from $\NDH$. For the multiplicative variant, we require security against nondeterministic NP-circuits, which follows from $\SH$.
        
    \item Apply a two-source $t$-non-malleable extractor, $\NMExt$, to the expanded seed $(\sigma,y)$, and the original indexed row $v$: $z=\NMExt((\sigma,y),v)$.
    
    A two-source $t$-non-malleable extractor extracts randomness from two \emph{independent} high min-entropy sources, even given its output on $t$ tampered variants of this pair. Specifically, for independent high min-entropy $\X,\Y$, and arbitrary fixed-point-free\footnote{For each $i$, either $f_i$ or $g_i$ has no fixed points.} pairs of functions $(f_i,g_i)$:
    \[
        \TV\Bigl(\bigl(\NMExt(\X,\Y),(\NMExt(f_i(\X),g_i(\Y)))_{i=1}^t\bigr)
        ,
        \bigl(\U_q,(\NMExt(f_i(\X),g_i(\Y)))_{i=1}^t\bigr) \Bigr) < \eps.
    \]

We use two constructions for $\NMExt$. For constant $t$ and one
output bit, Li's construction~\cite{li2023two}, combined with
the reduction of Chattopadhyay and Liao~\cite{CL22}, gives
error $2^{-\Omega(L)}$. This is the ingredient used in our
one-bit extractor at logarithmic entropy.
When $t$ or the output length $q$ grows with $L$, we use
Chattopadhyay, Goyal, and Li~\cite{CGL20}, obtaining error
$2^{-cL^\gamma}$ whenever $(t+1)q\le cL^\gamma$, for absolute
constants $c,\gamma>0$.
In both cases, we apply the standard conversion to make
the extractor strong in either input.
\end{enumerate}

An obvious obstacle is that the two inputs to the non-malleable extractor
are highly dependent: one is in fact completely determined by the other! However, following Ball and Dachman-Soled's analysis~\cite{BDL22,BD25}, the computational guarantee will grant us enough ``pseudo-independence'' to exploit the $t$-non-malleability of $\NMExt$.
More precisely, we turn a failure of
correlation breaking into a distinguisher for the pseudorandom generator.

The challenge in this setting is that any such reduction seemingly must run the tampering experiment, but that involves evaluating the PRG on adversarially correlated outputs. For the PRGs used, if an efficient reduction (within the circuit-size bound) could evaluate them then they would be insecure. Instead, to escape this contradiction, \cite{BDL22,BD25} realized it suffices to allow an untrusted Merlin to provide the tampered PRG evaluations. In particular, it suffices to use the failure of correlation breaking to construct a constant-round, private-coin interactive proof that distinguishes the PRG from random. This is sufficient, because constant-round interactive proofs can be simulated by nondeterministic circuits: $\IP[O(1)]\subseteq\NP/\poly$~\cite{GoldwasserSipser,BM88,Adleman1978}.

We sketch the argument at a high level before proceeding.
To distinguish a generated pair $(\sigma,y)= G(\sigma)$ from a uniform pair $(\sigma,y)\gets \U$, we assume for contradiction that the correlation breaker doesn't work. Namely, if $g$ is the index of a ``good'' (i.e.~high entropy) row
\[ (\CB(\V_g),(\CB(\V_i))_{i\neq g})\not\approx (\U,(\CB(\V_i))_{i\neq g}).\]
To exploit this, our interactive proof reduction will use $(\sigma,y)$ to simulate either the LHS (if $(\sigma,y)$ is pseudorandom) or the RHS (if $(\sigma,y)$ is truly random).

Concretely, the verifier samples the
rows privately, giving up if the ``good'' pre-extracted sample misses $\sigma$.\footnote{The event of hitting $\sigma$ happens with very low probability, but to collapse to $\NP/\poly(n)$ we just need a multiplicative gap around a fixed threshold (due to the power of the Goldwasser--Sipser transformation~\cite{GoldwasserSipser,GL20}).}
Then the prover supplies the other rows' purported
generator outputs. From here, we can use an additive or multiplicative test on the distance between LHS and RHS above. 
Critically, Arthur does not actually evaluate the PRG himself, and so can be much more efficient (with Merlin's assistance of course).

Why does this work? Suppose that $(\sigma,y)$ is pseudorandom. In this case, there is a gap between LHS and RHS, and Merlin can realize it by simply responding with the actual PRG outputs on the correlated seeds sent by Arthur. On the other hand, if $(\sigma,y)$ is truly random, then the total amount of information sent by Arthur is small and thus most of the entropy of $\V_g$ is retained even if we condition on a given $\sigma$ and Arthur message. We will thus think about $\V_g$ and $y$ being the inputs to the two-source non-malleable extractor, both of which have high entropy (recall that in this case $y$ is uniform and independent of $\sigma$). We thus can show that any Merlin strategy here that convinces Arthur to accept (conditioned on hitting $\sigma$ with $\Ext_{\mathrm{pre}}$ and the initial message from Arthur of correlated seeds) would yield a split-state tampering strategy on the two source extractor. The left tamperings on the $y$ portion are simply the Merlin replies. The tamperings on the $\V_g$ portion are induced by conditional sampling $\V_i$ consistent with the transcript sent (and conditioned on). Thus, because Arthur and Merlin represent a valid split-state $t$-tampering strategy, and the non-malleable extractor resists this, the LHS experiment (with random $y$, instead of pseudorandom $y$) and the RHS experiment (again with random $y$, instead of pseudorandom $y$) are close. Thus, the final test distinguishing LHS from RHS will only work (on average) when the input is pseudorandom and hence allows Arthur to efficiently distinguish PRG outputs from random (with Merlin's help).

\subsection{Computational preliminaries}
\label{sec:computational}\label{app:cb-auxiliary}\label{sec:mult-tools}

We first recall some useful background. 
\paragraph{Nondeterminism and NP gates/oracles.}

A nondeterministic circuit $C$ has an associated deterministic circuit $D$ (with additional inputs for a witness: $C(x)=1$ iff there exists a witness $a$ for which $D(x,a)=1$). An NP-circuit is allowed $\SAT$-gates. A nondeterministic NP-circuit is a nondeterministic circuit where the underlying deterministic circuit is allowed NP-gates. Nondeterministic NP-circuits are equivalent, up to polynomial  size overhead, to $\Sigma_2$ circuits.

Write $\mathcal O\in\{\varnothing,\SAT\}$ for these two verifier models. The corresponding hardness assumption is $\NDH$ or $\SH$ from \cref{sec:hardness}: in both cases the hard function belongs to ordinary $\sfE$, and the assumed lower bound is against nondeterministic circuits with access to $\mathcal O$.
Fixing inputs, deterministic composition, and conjunction preserve the relevant circuit class. In the oracle model, existential guesses and conjunctions can be combined into one existential block and one universal block.

\paragraph{Seed-extending pseudorandomness.}
A generator is seed-extending if $G(\sigma)=\sigma\concat G'(\sigma)$.
The visible seed lets a verifier match its privately sampled row
to the input without evaluating $G$.
The constructions of~\cite{Shaltiel25multiplicative,DS26} and their
reconstruction arguments give the following two versions of the
same ingredient. We verify the oracle extension and the composition
in \cref{sec:decorrelation-prg-details}.

\begin{lemma}[Seed-extending multiplicative PRG]
\label{mtools:lem:generator}
Fix $\mathcal O\in\{\varnothing,\SAT\}$ and assume $\NDH$ when
$\mathcal O=\varnothing$, or $\SH$ when $\mathcal O=\SAT$.
There is a constant $C$ such that, whenever $s\ge2$,
$0<\delta\le1/s$ and $C\log(1/\delta)\le L\le s$, there is a
uniformly $\poly(s)$-time computable function
$G:\zo^\ell\to\zo^L$, with $\ell=O(\log(1/\delta))$ and
$G(\sigma)=\sigma\concat G'(\sigma)$, such that every size-$s$
nondeterministic $\mathcal O$-oracle circuit $D$ satisfies
\[
 \Pr[D(G(\U_\ell))=1]
 \le e^{1/s}\Pr[D(\U_L)=1]+\delta.
\]
The algorithm evaluating $G$ uses no oracle; its constants may depend
on the fixed hard function in the relevant assumption. An arbitrary
error bound $\delta$ is implemented using the integer precision
$\lceil\log(1/\delta)\rceil$ and the smaller power-of-two error
$2^{-\lceil\log(1/\delta)\rceil}$.
\end{lemma}

\begin{corollary}[Additive guarantee]
\label[corollary]{comp:lem:seed-extending-prg}\label[corollary]{comp:lem:prg-wrapper}
Assume $\NDH$. For every fixed $a,b>0$, there is a constant $C$
such that, whenever $C\log S\le L\le S$, there is a uniformly
$\poly(S)$-time computable seed-extending generator
$G:\zo^\ell\to\zo^L$ with $\ell=O(\log S)$ such that every
nondeterministic circuit $D$ of size at most $S^a$ satisfies
\[
 \Pr[D(G(\U_\ell))=1]\le\Pr[D(\U_L)=1]+S^{-b}.
\]
\end{corollary}
\begin{proof}
Choose a fixed integer $K\ge\max\{1,a,b+3\}$, set
$s=\lceil S^K\rceil$, and use \cref{mtools:lem:generator} with
$\mathcal O=\varnothing$ and $\delta=1/s$. Since
$e^{1/s}-1+1/s\le3/s\le S^{-b}$ for $S\ge2$, the claim follows.
\end{proof}

\paragraph{Interactive proofs and round collapse.}
A private-coin protocol $\mathcal{P}$ has an unbounded prover and a randomized verifier. For an input $w$, let
$\operatorname{val}_{\mathcal P}(w)$ be its maximum acceptance
probability on $w$, over all prover strategies. When the protocol is clear from context, we abbreviate $\operatorname{val}_{\mathcal P}(w)$ to $\operatorname{val}(w)$. A public-coin
verifier sends fresh uniform random bits in each message.
In an $\AM$ protocol Arthur sends one random string and
Merlin replies once.

 The following lemma demonstrates that a constant-round private coin interactive proof with a sufficiently large constant \emph{multiplicative} gap between acceptance probabilities in the case of yes vs no instances suffices for compilation down to a nondeterministic circuit. The resulting nondeterministic circuit is only polynomially larger than the complexity of the private coin Arthur. This process follows from composing classical reductions: Goldwasser--Sipser public-coin simulation~\cite{GoldwasserSipser},
in the small-acceptance-probability formulation of
Goldreich--Leshkowitz~\cite{GL20}, Babai--Moran round
reduction~\cite{BM88}, and, finally, Adleman's trick~\cite{Adleman1978}. We state the lemma without formally proving it.

\begin{lemma}[Collapse of a constant-round protocol]
\label{comp:lem:protocol}\label{mtools:lem:protocol}
For every fixed integer $r\ge1$, there are a constant $B>1$
and a polynomial $p_r$ with the following property, for either
$\mathcal O\in\{\varnothing,\SAT\}$.
Let $\mathcal P$ be a private-coin protocol with at most $r$ rounds,
whose verifier is a deterministic $\mathcal O$-oracle circuit once
its random tape is fixed. Suppose its verifier size, input length,
total randomness, and total communication are at most $T$.
Let $c_N,c_Y$ be rationals with descriptions of at most $T$ bits,
where $0\le c_N<c_Y\le1$ and $c_Y\ge B^r c_N$.
There is a nondeterministic $\mathcal O$-oracle circuit of size
$p_r(T)$ accepting every $w$ with
$\operatorname{val}_{\mathcal P}(w)\ge c_Y$ and rejecting every
$w$ with $\operatorname{val}_{\mathcal P}(w)\le c_N$.
In particular, the size has no dependence on $1/c_Y$ or $1/c_N$.
\end{lemma}

In the multiplicative reduction of \cref{mcb:lem:ip}, the verifier uses a $\SAT$ oracle for approximate counting. Applying \cref{comp:lem:protocol}
with $\mathcal O=\SAT$ therefore produces a nondeterministic
$\SAT$-oracle distinguisher, equivalently a $\Sigma_2$ circuit up
to polynomial size overhead. This is why the multiplicative
construction uses the $\mathcal O=\SAT$ version of
\cref{mtools:lem:generator}, obtained under $\SH$.

\subsection{Correlation Breaker Ingredients}
\label{sec:nm}\label{sec:cb}

We use the following setup for convenience.

An indexed row is a pair $\V_i=(\alpha_i,\W_i)$, identified with
the concatenation $\alpha_i\concat\W_i\in\zo^L$, where
$\alpha_i\in\zo^h$ is fixed and the labels are distinct, and
$\W_i\in\zo^{L-h}$ is the random part. Thus $L$ denotes the total
indexed-row length, including the label.
Throughout, the entire tuple is sampled by a deterministic
circuit $\Samp$ with a nondeterministic postselection circuit $\Post$,
of total size at most $S$, conditioned on $\Post(\U_{r_0})=1$,
where $r_0\le S$ and the conditioning event has positive probability.
Write $\Post(R)=1$ iff $\exists a_0\;Q(R,a_0)=1$, where $Q$ is
deterministic and its size and witness length are at most $S$. There is no lower bound on that probability. 
A good row $\V_g$ has $\minH(\V_g)\ge L-\Delta$; this entropy
already includes the loss from its fixed index.

\paragraph{Preliminary extraction.}
We use \cref{pre:lem:preliminary} to construct a multiplicative extractor for postselecting sources:
\[\Ext_{\mathrm{pre}}:\zo^L\to\zo^\ell.\]
This map has the property that under $\NDH$, a good row can
be mapped to $\ell$ bits with point probabilities at most
$e^\nu2^{-\ell}$, provided
$L\ge C_0(\Delta+\ell+\log S+\log(1/\nu))$.
For the additive case, \cref{pre:cor:long-preliminary} gives
$\ell=\lfloor c_0L\rfloor$ bits within inverse-polynomial statistical
distance of uniform. For the multiplicative case we use the
pointwise bound with $\nu=1/100$. (Both guarantees are preserved
by taking an output prefix.)

\paragraph{Non-malleability for correlated rows.} We call all the inputs besides the good input the ``companion rows.''
The distinct indices ensure that each companion input differs
from the good input. The following lemma shows how to apply
ordinary two-source non-malleability even when the companion
rows are not deterministic functions of the good row.

\begin{lemma}\label{cb:lem:correlated}
Let $\NMExt:\zo^L\times\zo^L\to\zo^q$ be $t$-non-malleable for
min-entropy $k_{\mathsf{NM}}$ and error $\eps_{\mathsf{NM}}$, strong in its first
input. Let $(\V_i)_{i\in I}$ be at most $t+1$ rows with fixed,
distinct indices, fix $g\in I$, and let $\ZZ\in\zo^L$ be
independent of this tuple. If $\minH(\ZZ),\minH(\V_g)\ge k_{\mathsf{NM}}$, then
for arbitrary functions $f_i:\zo^L\to\zo^L$,
\[
 \begin{aligned}
 \TV\Bigl(&
   \bigl(\NMExt(\ZZ,\V_g),\ZZ,(\NMExt(f_i(\ZZ),\V_i))_{i\ne g}\bigr),\\
 &\bigl(\U_q,\ZZ,(\NMExt(f_i(\ZZ),\V_i))_{i\ne g}\bigr)
 \Bigr)\le\eps_{\mathsf{NM}}.
 \end{aligned}
\]
\end{lemma}
To see this, for each possible value $v$ of $\V_g$, choose a tuple
with the conditional distribution of $(\V_i)_{i\ne g}$ given
$\V_g=v$. Make all these choices independently of $(\ZZ,\V_g)$.
Fixing the choices expresses each $\V_i$ as a deterministic function
of $\V_g$. On the support of $\V_g$, this function changes its fixed
index to the distinct index of $\V_i$, so it has no fixed point.
Extend it outside the support without fixed points. Non-malleability
now applies to the independent sources $\ZZ,\V_g$ with the fixed
functions, and averaging over the choices gives the lemma.

For a constant number of rows, we use the following quantitative
consequence of~\cite{CL22,li2023two}.
The extension to $t$ tamperings follows from the standard
advice-generator construction using the correlation breaker
of~\cite[Theorem~7.12]{li2023two}, based on the reduction
of~\cite{CL22}, with advice length $O(\lambda)$ and error
parameter $2^{-\lambda}$, where
$\lambda=\alpha L/(t+1)^3$ for a sufficiently small absolute
constant $\alpha>0$.
The standard conversion of~\cite[Lemma~54]{ADNOPRS19} makes
the extractor strong in either input, after adjusting the
absolute constants.

\begin{theorem}
\label{nm:thm:constant-nm}
\label{nm:sec:constant-nm}
There are absolute constants $c,C>0$ such that, for all integers
$t,L\ge1$ satisfying $L\ge C(t+1)^3(t+\log_2 L)$, there is an
explicit function $\NMExt:\zo^L\times\zo^L\to\zo$ that is
$t$-non-malleable for min-entropy $L-cL/(t+1)^3$ and error
$2^{-cL/(t+1)^3}$, strong in either input and computable in time
polynomial in $L$ and $t$.
\end{theorem}

For more rows and output bits, we apply the same conversion
to the extractor of~\cite{CGL20}.
The restriction on $Dq$ accounts for the factor $2^{Dq}$
in the error bound of the conversion.

\begin{lemma}[Strong non-malleability for many tamperings]
\label{nm:lem:many-nm}
There are absolute constants $\gamma,c>0$ such that, for all
sufficiently large $L$ and integers $D,q\ge1$ satisfying
$Dq\le cL^\gamma$, there is an explicit polynomial-time function
$\NMExt:\zo^L\times\zo^L\to\zo^q$ that is $(D-1)$-non-malleable for
min-entropy $L-cL^\gamma$ and error $2^{-cL^\gamma}$, strong
in either input.
\end{lemma}

In particular, fix positive constants $\rho_t,\rho_o,\rho_k,\rho_e$
satisfying $\rho_t+\rho_o<\gamma$, $\rho_k<\gamma$, and
$\rho_e<\gamma$. For all sufficiently large $L$,
\cref{nm:lem:many-nm} applies whenever $D-1\le L^{\rho_t}$
and $q\le L^{\rho_o}$, with min-entropy requirement
$L-L^{\rho_k}$ and error $2^{-L^{\rho_e}}$.

\subsection{Correlation Breaker Construction Template}
Given $\Ext_{\mathrm{pre}}:\zo^L\to\zo^\ell$, a seed-extending generator
$G:\zo^\ell\to\zo^L$, and a two-source non-malleable extractor
$\NMExt:\zo^L\times\zo^L\to\zo^q$, define
$\CB:\zo^h\times\zo^{L-h}\to\zo^q$ by
\begin{equation}\label{cb:eq:construction}
 \CB(\alpha,v)
   =\NMExt\bigl(G(\Ext_{\mathrm{pre}}(\alpha\concat v)),\alpha\concat v\bigr).
\end{equation}
Here $\alpha\in\zo^h$ is the fixed row label and
$v\in\zo^{L-h}$ is the random part. The label remains explicit in
$\NMExt$'s second input. Moving to a companion row changes this label,
which supplies the no-fixed-point condition in
\cref{cb:lem:correlated}. No such condition is needed on the left
tampering, so $\Ext_{\mathrm{pre}}$ may process the label without preserving it
in its output. Since the label is fixed, it contributes no entropy.
For a complete indexed row $u=\alpha\concat v$, we also write
$\CB(u)=\CB(\alpha,v)=\NMExt(G(\Ext_{\mathrm{pre}}(u)),u)$; in particular,
$\CB(\V_i)$ below includes the label $\alpha_i$ on the right.
The notation $\CBmult$ denotes this same construction with
ingredients and parameter choices chosen for the multiplicative guarantee.

\paragraph{Parameter key.}
There are numerous parameters involved in this construction. We introduce them in the order in which we set them. We write $D$
for the maximum number of rows, denoted $D_*$ in the multiplicative
case.
\begin{enumerate}[label=(\roman*),leftmargin=*]
\item \emph{Source and output parameters: $L,S,h,\Delta,D,q$.}
Each row has length $L$, its fixed index has length $h$, and the
joint sampling-and-postselection circuit has size at most $S$.
A good row has min-entropy at least $L-\Delta$, already accounting
for its fixed index. The correlation breaker outputs $q$ bits per row.

\item \emph{Error parameters: $\theta,\eta$ or $\vartheta,\xi,\nu$.}
For additive error, fix $b>0$, set $\theta=S^{-b}$, and take
$\eta=\theta^3/2^{20}$ as the error bound for both extraction
ingredients; see \cref{cb:lem:additive-ip}.
For multiplicative error, the target factor $e^\vartheta$ and
pointwise slack $\xi$ satisfy \eqref{mcb:eq:errors}.
Take $\nu=1/100$ for the preliminary extractor's pointwise cap
and require non-malleable-extractor error at most $\xi/100$;
see \cref{mcb:lem:ip}.

\item \emph{Extractor parameters: $m_P$ and $t$.}
First choose a preliminary extractor $P_0$ with
$m_P=\lfloor c_0L\rfloor$ output bits, for a sufficiently small
fixed $c_0>0$. Its length and accuracy constraints are given in \cref{pre:lem:preliminary} and \cref{pre:cor:long-preliminary}.
Choose $\NMExt$ with $t=D-1$ tamperings and $q$ output bits.
The available entropy, output-length, and error guarantees are
\cref{nm:thm:constant-nm,nm:lem:many-nm}.

\item \emph{PRG parameters: $s,\delta_G,\ell$.}
With $P_0$ and $\NMExt$ fixed, bound the interactive verifier's size
and its size after collapse. Choose $G$ to fool circuits of that
size. In the additive case, require additive error at most
$\theta/4$, as in \cref{cb:lem:transfer}; then
$\ell=O(\log S)$. In the multiplicative case, choose $s$ to
dominate the collapsed circuit size and take
$\delta_G=\min\{\xi/256,1/s\}$ in
\cref{mtools:lem:generator}; then
$\ell=O(\log S+\log(1/\xi))$.
Define $\Ext_{\mathrm{pre}}$ to be the first $\ell$ output bits of $P_0$,
requiring $\ell\le m_P$.

\item \emph{Remaining entropy: $L-B_*$.}
After $\ell$ is fixed, compute the conditioning loss $B_*$ from
\eqref{cb:eq:deficiencies} in the additive case or from
\cref{mcb:lem:ip} in the multiplicative case. Check that the
chosen $\NMExt$ works at min-entropy $L-B_*$ with the error required
above. The sufficient length bound for constant $D$ and $q=1$
is in \cref{cb:lem:cb}; for growing $D,q$ it is
\eqref{cb:eq:cb-budget}; for the multiplicative guarantee it is
\eqref{mcb:eq:budget}. Their proofs verify these constraints
and $\ell\le m_P$ for the choices just described.
\end{enumerate}
% The verifier uses the sampling circuit, $P_0$, and $\NMExt$, together
% with the appropriate additive or counting test. The prover supplies
% the generator suffixes, so the circuit-size bound can be fixed
% before choosing $G$.

\subsection{Reduction Template}
\label{sec:cb-distinguishing}

% Fix a tuple $(\V_i)_{i\in I}$ of indexed $L$-bit rows with fixed,
% distinct $h$-bit indices, jointly samplable with nondeterministic
% postselection by a circuit $(\Samp,\Post)$ of total size at most $S$.
% Thus, for $R\leftarrow\U_{r_0}$ with $r_0\le S$,
% \[
%  (\V_i)_{i\in I}\equiv
%  (\Samp(R)\mid\Post(R)=1).
% \]
% Fix a ``good'' index $g\in I$ with $\minH(\V_g)\ge L-\Delta$.
% Write $J=I\setminus\{g\}$, $\bP_i=\Ext_{\mathrm{pre}}(\V_i)$,
% $\bP=(\bP_i)_{i\in I}$, and
% $p_0=\Pr[\Post(\U_{r_0})=1]>0$.
% Probabilities involving $\V_i$ or $\bP$ refer to the postselected
% distribution.
Fix a tuple $(\V_i)_{i\in I}$ of indexed rows and a good index
$g\in I$ as in \cref{sec:nm}.
Write $J=I\setminus\{g\}$, $\bP_i=\Ext_{\mathrm{pre}}(\V_i)$,
$\bP=(\bP_i)_{i\in I}$, and
$p_0=\Pr[\Post(\U_{r_0})=1]$. We write $\CB(\V_J)=(\CB(\V_i))_{i\in J}$. 

Suppose that this tuple and good index violate the desired
decorrelation guarantee. In the additive case, this means
\[
 \TV\bigl((\CB(\V_g),\CB(\V_J)),
          (\U_q,\CB(\V_J))\bigr)>\theta.
\]
In the multiplicative case, additionally fix an output tuple $z=(z_i)_{i\in I}$
at which one of the pointwise comparisons in
\eqref{mcb:eq:conclusion} fails.

\begin{figure}[htbp]
\centering
\begin{tikzpicture}[x=1cm,y=1cm,>=Stealth,font=\small]
 \node[font=\bfseries] at (2.9,0) {Verifier};
 \node[font=\bfseries] at (10.5,0) {Prover};
 \node at (6.5,-0.55) {Common input $w=\sigma\concat y$};

 \node[draw,rounded corners,anchor=north,align=center,
       text width=6cm,inner sep=5pt] (setup)
   at (2.9,-0.95) {Privately sample rows and seeds.\\
   Reject unless $p_g=\sigma$.\\
   Multiplicative case: also check $\Post(R)=1$ using SAT.};

 \path (setup.south) ++(0,-0.65) coordinate (seeds);
 \draw[->] (seeds)--node[above]
   {$p_J$: companion seeds} ++(7.6,0);

 \path (seeds) ++(0,-0.75) coordinate (suffixes);
 \draw[<-] (suffixes)--node[above]
   {$a_J$: proposed generator suffixes} ++(7.6,0);

 \path (suffixes) ++(3.6,-0.65) coordinate (timing);
 \node[align=center] at (timing)
   {Suffixes are fixed before the test's fresh random bits.};

 \path (timing) ++(0,-0.4) coordinate (tests);
 \node[draw,rounded corners,anchor=north,align=left,
       text width=12.7cm,inner sep=7pt]
   at (tests) {\textbf{Additive test.} Choose a hidden bit $c$.
   Send the companion outputs and either $\NMExt(w,v_g)$ or an independent
   uniform string. After the prover guesses $c$, reveal $R$
   and request $a_0$. Accept iff the guess is correct and
   $Q(R,a_0)=1$.\\[5pt]
   \textbf{Multiplicative test.} With fresh private coins and a SAT
   oracle, compare the counts defining the real and ideal output
   probabilities. Accept if the chosen comparison test accepts.};
\end{tikzpicture}
\caption{The common interaction and its two tests. The verifier never
computes $G$. On a generator input, the honest prover supplies
$a_i=G'(p_i)$; soundness allows arbitrary suffix replies.}
\label{fig:decorrelation-interaction}
\end{figure}
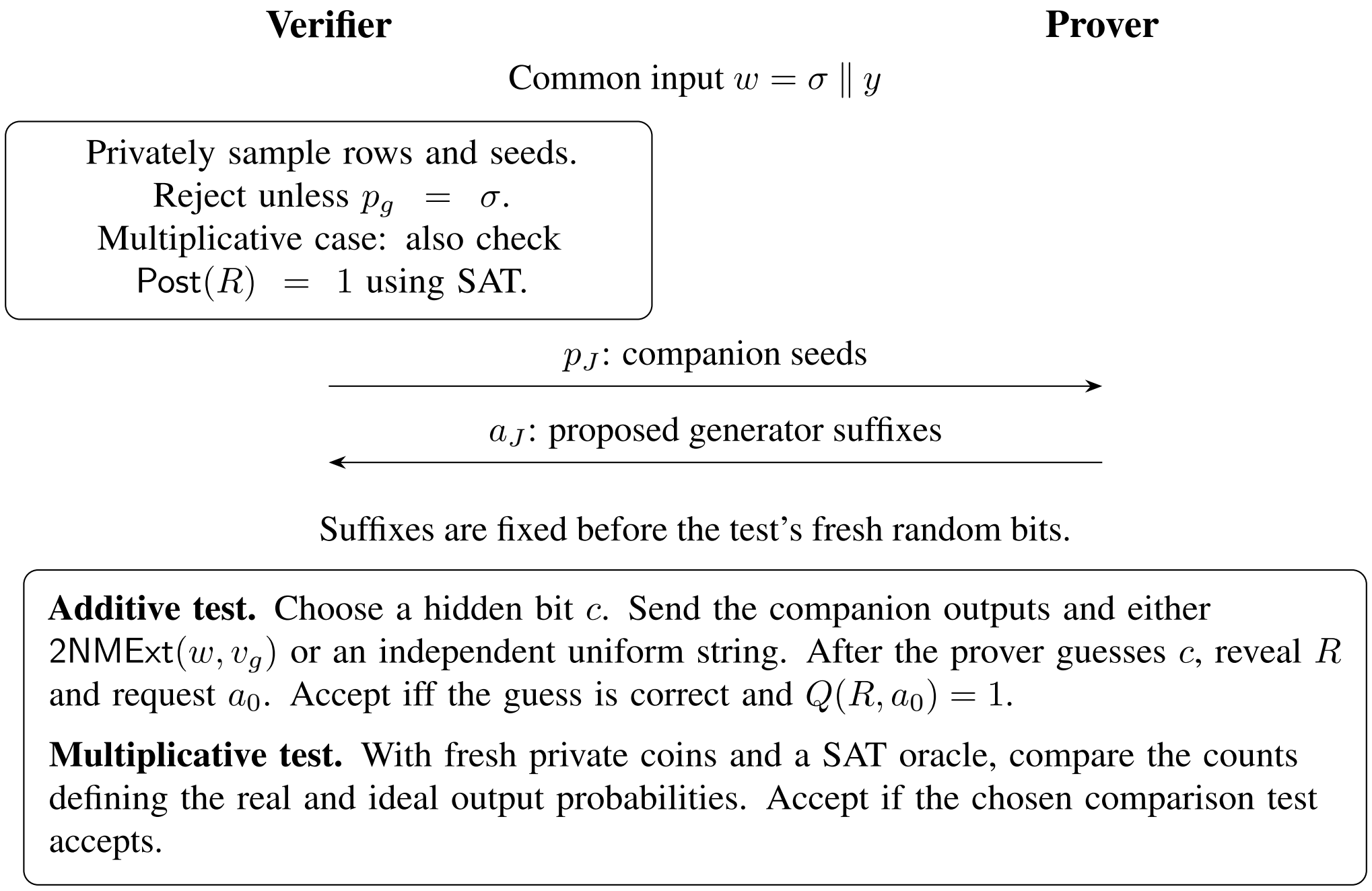

On common input $w=\sigma\concat y\in\zo^L$, the verifier privately samples $R\leftarrow\U_{r_0}$, computes $(v_i)_{i\in I}=\Samp(R)$, and sets $p_i=\Ext_{\mathrm{pre}}(v_i)$ for each $i\in I$. It rejects unless $p_g=\sigma$.
In the multiplicative case, it also checks $\Post(R)=1$ using
its SAT oracle before continuing.
It sends the companion seeds $p_J$ to the prover, which returns
suffixes $a=(a_i)_{i\in J}$. These define proposed first inputs
$p_i\concat a_i$ to $\NMExt$. The verifier then runs the test specified
below, using fresh randomness. In the additive case,
the rows and sampling tape remain private until the prover has
sent its hidden-bit guess; only then does the verifier reveal $R$
and request a witness $a_0$, accepting only if $Q(R,a_0)=1$.
In the multiplicative case, the rows and sampling tape remain
private throughout. \cref{fig:decorrelation-interaction}
shows the two possible continuations.

% \begin{figure}[htbp]
% \centering
% \begin{tikzpicture}[x=1cm,y=1cm,>=Stealth,font=\small]
%  \node[font=\bfseries] at (2.5,0) {Verifier};
%  \node[font=\bfseries] at (10.5,0) {Prover};
%  \node at (6.5,-0.55) {Common input $w=\sigma\concat y$};
%  \node[draw,rounded corners,align=center,text width=6cm,inner sep=5pt]
%    at (2.9,-1.55) {Privately sample rows and seeds.\\
%    Reject unless $\Post=1$ and $p_g=\sigma$.};
%  \draw[->] (3,-2.75)--node[above] {$p_J$: companion seeds} (10.5,-2.75);
%  \draw[<-] (3,-3.5)--node[above] {$a_J$: proposed generator suffixes} (10.5,-3.5);
%  \node[align=center] at (6.5,-4.15)
%    {Suffixes are fixed before the test's fresh random bits.};
%  \node[draw,rounded corners,anchor=north,align=left,text width=12.7cm,inner sep=7pt]
%    at (6.5,-4.55) {\textbf{Additive test.} Choose a hidden bit $c$.
%    Send the companion outputs and either $\NMExt(w,v_g)$ or an independent
%    uniform string. The prover guesses $c$; accept a correct guess.\\[5pt]
%    \textbf{Multiplicative test.} With fresh private coins and a SAT oracle,
%    compare the counts defining the real and ideal output probabilities.
%    Accept if the chosen comparison test accepts.};
% \end{tikzpicture}
% \caption{The common interaction and its two tests. The verifier never
% computes $G$. On a generator input, the honest prover supplies
% $a_i=G'(p_i)$; soundness allows arbitrary suffix replies.}
% \label{fig:decorrelation-interaction}
% \end{figure}

On generator inputs, a successful prover can supply the actual
suffixes $G'(p_i)$, so failure of decorrelation becomes detectable.
For uniform inputs, fixing the seed tuple makes the uniform suffix
independent of the conditioned rows. Non-malleability then limits
the success of every prover. The next observation isolates the
common final step, after these two acceptance estimates have been proved.

\begin{lemma}[From an interactive reduction to a contradiction]
\label{cb:lem:ip-contradiction}
Let $G$ satisfy
$\Pr[D(G(\U_\ell))=1]\le e^{\varepsilon_G}\Pr[D(\U_L)=1]+\delta_G$
for nondeterministic $\mathcal O$-oracle circuits of size $p_r(T)$,
where $p_r$ is from \cref{comp:lem:protocol}.
There cannot be an $r$-round private-coin protocol obeying that
lemma's resource and threshold bounds such that, writing
$v(w)=\operatorname{val}_{\mathcal P}(w)$,
\[
 \Pr_{\sigma\leftarrow\U_\ell}[v(G(\sigma))\ge c_Y]\ge\gamma_Y,
 \qquad
 \Pr_{w\leftarrow\U_L}[v(w)>c_N]\le\gamma_N,
\]
with $c_Y\ge B^r c_N$ and
$\gamma_Y>e^{\varepsilon_G}\gamma_N+\delta_G$.
\end{lemma}
\begin{proof}
Apply \cref{comp:lem:protocol} to the promise sets
$\{w:v(w)\ge c_Y\}$ and $\{w:v(w)\le c_N\}$.
Its circuit accepts at least a $\gamma_Y$ fraction of generator
outputs and at most a $\gamma_N$ fraction of uniform inputs,
contradicting the stated pseudorandomness inequality.
\end{proof}

To apply \Cref{cb:lem:ip-contradiction}, we establish the required
acceptance bounds in \Cref{cb:lem:additive-ip,mcb:lem:ip}.
Both reductions apply to any seed-extending map $G$, without assuming
pseudorandomness or requiring the verifier to evaluate $G$.

\subsection{Additive decorrelation}
\label{sec:local-cb-proof}\label{app:cb-proof}

The additive test asks the prover to distinguish the good row's
output from an independent uniform string, given the companion
outputs. Its advantage is exactly half the statistical distance.
We first establish an interactive reduction for arbitrary $G$;
only the ensuing corollary assumes pseudorandomness.

\begin{lemma}[Interactive reduction for additive error]
\label{cb:lem:additive-ip}
For fixed integers $a,b\ge1$, let $2\le L,D\le S$, $1\le q\le S$,
$1\le\ell\le L$, and $\Delta\ge0$. Put $\theta=S^{-b}$,
$\eta=\theta^3/2^{20}$, and
\begin{equation}\label[equation]{cb:eq:deficiencies}
 B_* = \max\{\ell,\Delta+D\ell+\log(1/\eta)\}.
\end{equation}
Suppose that $\Ext_{\mathrm{pre}}:\zo^L\to\zo^\ell$ and
$\NMExt:\zo^L\times\zo^L\to\zo^q$ have circuit size at most $S^a$,
that $\TV(\Ext_{\mathrm{pre}}(\X),\U_\ell)\le\eta$ for every
$\X\in\NPSources_{L,S}$ with $\minH(\X)\ge L-\Delta$, and that
$\NMExt$ is $(D-1)$-non-malleable for min-entropy $L-B_*$ and error
$\eta$, strong in its first input.
Let $G:\zo^\ell\to\zo^L$ be any seed-extending map.

If some tuple of at most $D$ indexed rows and good index $g$ as
above satisfies
\[
 \TV\bigl((\CB(\V_g),\CB(\V_J)),
          (\U_q,\CB(\V_J))\bigr)>\theta,
\]
then there is a constant-round private-coin protocol with an ordinary
verifier, resource bound $T=\poly_{a,b}(S)$, and thresholds
$t_Y,t_N$ of polynomial description length satisfying
$t_Y\ge B^r t_N$, where $r$ is its number of rounds, such that
\[
 \Pr_{\sigma\leftarrow\U_\ell}
   [\operatorname{val}(G(\sigma))\ge t_Y]\ge3\theta/4,
 \qquad
 \Pr_{w\leftarrow\U_L}
   [\operatorname{val}(w)>t_N]\le\theta/8.
\]
\end{lemma}

\begin{proof}
Fix a violating tuple $(\V_i)_{i\in I}$ and its good index $g$.
Use the notation of \cref{sec:cb-distinguishing}, writing $p=p_0$.

For the analysis, consider the ideal protocol that also
checks $\Post(R)=1$ before sending the companion seeds in
\cref{sec:cb-distinguishing}; this check need not be efficient.
After its initial sampling check and the prover's suffix replies
$a_J$, run the following additive test on input $w=\sigma\concat y$:
\begin{enumerate}[label=(\roman*)]
\item Choose an independent uniform bit $c$. If $c=0$, choose
$u\leftarrow\U_q$ independently; if $c=1$, set $u=\NMExt(w,v_g)$.
Send $\bigl(u,(\NMExt(p_i\concat a_i,v_i))_{i\in J}\bigr)$ to the prover.
\item Accept if the prover correctly guesses $c$.
\end{enumerate}

The ordinary verifier omits the initial $\Post$ check.
After receiving the guess $\widehat c$, it reveals $R$, requests
$a_0$, and accepts iff $\widehat c=c$ and $Q(R,a_0)=1$.
For every fixed strategy up to the guess, optimizing the final
witness reply gives acceptance probability
\[
 \Pr[\Post(R)=1,\ p_g=\sigma,\ \widehat c=c].
\]
The two protocols allow the same suffix replies and guesses on
valid tapes, and invalid tapes cannot accept. Hence their optimal
values agree on every input. We analyze the ideal protocol below.

\paragraph{Completeness: generator inputs.}
For $\sigma$ with $\Pr[\bP_g=\sigma]>0$, let $d_\sigma$ be the
statistical distance, conditioned on $\bP_g=\sigma$, between
\[
 \bigl(\bP_J,\NMExt(G(\sigma),\V_g),
             (\NMExt(G(\bP_i),\V_i))_{i\in J}\bigr)
\]
and the same tuple with its second entry replaced by an independent
$\U_q$. Set $d_\sigma=1$ when $\Pr[\bP_g=\sigma]=0$.
Revealing $\bP$ can only increase the distance assumed to
exceed $\theta$, so
$\E_{\sigma\leftarrow \bP_g}d_\sigma>\theta$.
Since $\TV(\bP_g,\U_\ell)\le\eta$, this gives
$\E_{\sigma\leftarrow \U_\ell}d_\sigma>\theta-\eta$.

Put $\beta=\theta/8$ and $\alpha=\theta/128$.
Call $\sigma$ bad if
$|\Pr[\bP_g=\sigma]-2^{-\ell}|>\alpha2^{-\ell}$, and good otherwise.
By \cref{lem:density}, the uniform fraction of bad seeds is at most
\begin{equation}\label{cb:eq:bad-seeds}
 b_0=2\eta/\alpha=\theta^2/4096.
\end{equation}
Using $0\le d_\sigma\le1$, at least a $3\theta/4$ fraction of
uniform seeds are good and satisfy $d_\sigma>\beta$.

For each such seed, the input $G(\sigma)$ passes the first check
with probability $p\Pr[\bP_g=\sigma]\ge p2^{-\ell}(1-\alpha)$.
The prover can reply with $a_i=G'(p_i)$ and then distinguish the
two challenge distributions with success probability
$(1+d_\sigma)/2$. Thus the protocol's maximum acceptance
probability on $G(\sigma)$ is at least
\begin{equation}\label{cb:eq:yes}
 c_Y=p\,2^{-\ell-1}(1-\alpha)(1+\beta).
\end{equation}

\paragraph{Soundness: uniform inputs.}
We condition on the extracted seeds while controlling the
min-entropy of the good row. Discard values $\mathbf p$ of
$\bP$ with
$\Pr[\bP=\mathbf p]<\eta2^{-D\ell}$.
There are at most $2^{D\ell}$ such values, so their total
probability is at most $\eta$. For every remaining value, \cref{lem:conditioning} gives
\begin{equation}\label{cb:eq:conditional-entropy}
 \minH(\V_g\mid\bP=\mathbf p)
 \ge L-\Delta-D\ell-\log(1/\eta).
\end{equation}

Fix a deterministic prover and a remaining value
$\mathbf p=(p_i)_{i\in I}$. Its first replies are functions
$a_i(w,p_J)$, so put $f_i(w)=p_i\concat a_i(w,p_J)$.
The source $\ZZ=p_g\concat \U_{L-\ell}$ is independent of the rows
conditioned on $\bP=\mathbf p$, and has min-entropy $L-\ell$.
By \cref{cb:eq:deficiencies,cb:eq:conditional-entropy}, both
$\ZZ$ and the conditioned good row have min-entropy at least $L-B_*$.
Applying \cref{cb:lem:correlated}, even given $\ZZ$, the two challenge
distributions have statistical distance at most $\eta$.
Consequently, for this fixed $\mathbf p$, the prover's average
probability of guessing $c$, over the uniform suffix of $\ZZ$, is
at most $(1+\eta)/2$.

For each input $w$, let $d^*(w)$ be twice the maximum probability
of correctly guessing $c$, conditioned on passing the first check,
minus one. Set $d^*(w)=1$ when that check has probability zero.
A single deterministic prover can attain these maxima for every
input and first message, since the prover has no computational
restriction. Apply the preceding non-malleability bound to this prover.

When averaging over $w\leftarrow \U_L$, bad prefixes contribute
at most $b_0$ to $\E_w d^*(w)$. For a good prefix $\sigma$,
conditioning on passing the first check gives the row distribution
$(\V_i)_{i\in I}\mid \bP_g=\sigma$.
Averaging over uniform $\sigma$ multiplies the original probability
of each such row tuple by
$1/(2^\ell\Pr[\bP_g=\sigma])\le(1-\alpha)^{-1}$.
Thus discarded values of $\bP$ contribute at most
$\eta/(1-\alpha)$. For every remaining $\mathbf p$, the average
advantage over the uniform suffix is at most $\eta$. Hence
\[
 \E_{w\leftarrow \U_L}d^*(w)
 \le\eta+b_0+\frac{\eta}{1-\alpha}
 \le\frac{\theta^2}{2048}.
\]
Markov's inequality gives $d^*(w)\le\alpha$ except on uniform
probability at most $\theta/16$. If the prefix $\sigma$ is also
good, the first check passes with probability at most
$p2^{-\ell}(1+\alpha)$. Combining this with
\cref{cb:eq:bad-seeds}, on at least a $1-\theta/8$ fraction
of uniform inputs every prover is accepted with probability at most
\begin{equation}\label{cb:eq:no}
 c_N=p\,2^{-\ell-1}(1+\alpha)^2.
\end{equation}

\paragraph{Amplifying the acceptance ratio.}
The acceptance bounds satisfy
\[
 \frac{c_Y}{c_N}
 =\frac{(1-\alpha)(1+\beta)}{(1+\alpha)^2}
 \ge1+\beta/2.
\]
Repeat the ordinary protocol independently $N=O(1/\theta)$
times in parallel, receiving all hidden-bit guesses before revealing
any sampling tapes and requesting the postselection witnesses.
Accept if every copy accepts. By the same argument, its value equals
that of $N$ parallel copies of the ideal protocol, which is exactly
the $N$th power of the one-copy value. For fixed first verifier messages and prover
replies, the hidden rows and challenge experiments remain
independent across copies. For each vector of second messages,
the most likely challenge-bit vector is obtained
by choosing the most likely bit in each copy. Thus the maximum
probability of guessing all challenge bits is the product of the
individual maxima. Optimizing the first replies separately and
averaging over the independent first messages preserves this product.

Choose $N$ so that $(c_Y/c_N)^N\ge B^{r_{\rm IP}}$, where $r_{\rm IP}$ is the
constant number of rounds including the final witness round and $B$ is the constant in
\cref{comp:lem:protocol}. The repeated protocol still has $r_{\rm IP}$ rounds,
and its verifier, messages, and random inputs have size $\poly(S)$.
Since the sampling circuit has at most $S$ input bits and $p>0$,
we have $\log(1/p)\le S$. Thus $c_Y^N$ and $c_N^N$ have
polynomial-length descriptions.

Set $t_Y=c_Y^N$ and $t_N=c_N^N$. The repeated protocol has
value equal to the $N$th power of the original value, so the two
input-density bounds are unchanged. Its verifier uses only
 $\Samp$ and $Q$, $\Ext_{\mathrm{pre}}$, and $\NMExt$, and its threshold descriptions
have polynomial length.    
\end{proof}

\begin{corollary}[Additive transfer]\label[corollary]{cb:lem:transfer}
For every fixed $a,b\ge1$ there is a constant $A$ such that, under
the extraction and parameter hypotheses of \cref{cb:lem:additive-ip},
if $G$ satisfies
\[
 \Pr[D(G(\U_\ell))=1]\le\Pr[D(\U_L)=1]+\theta/4
\]
for every nondeterministic circuit of size at most $S^A$, then every
admissible tuple and good index satisfy
\[
 \TV\bigl((\CB(\V_g),\CB(\V_J)),
          (\U_q,\CB(\V_J))\bigr)\le\theta.
\]
\end{corollary}
\begin{proof}
Choose $A$ to bound the collapsed circuit size for the protocol
in \cref{cb:lem:additive-ip}. If the conclusion failed, that lemma
and \cref{cb:lem:ip-contradiction} would give
$3\theta/4\le\theta/8+\theta/4$, a contradiction.
\end{proof}

\subsubsection{A constant number of rows and one output bit}

We now choose the ingredients to obtain the logarithmic-length
correlation breaker used in the one-bit extractor.

\begin{lemma}[Local correlation breaker]\label{cb:lem:cb}
Assume $\NDH$. For every fixed integer $D\ge1$ and constant $b>0$,
there is a constant $C$ such that, for all parameters satisfying
$\Delta\ge0$, $ h\le L$, and $C(\Delta+\log S)\le L\le S$,
there is a $\poly(S)$-time computable function
$\CB:\zo^L\to\zo$ with the following property.

Let $(\ZZ_1,\ldots,\ZZ_d)$ be a tuple of at most $D$ indexed rows
with $h$-bit indices, jointly samplable with nondeterministic
postselection by a size-$S$ circuit. For every $g\in[d]$ with
$\minH(\ZZ_g)\ge L-\Delta$, we have
\[
 \TV\Bigl(
   \bigl(\CB(\ZZ_g),(\CB(\ZZ_i))_{i\ne g}\bigr),
   \bigl(\U_1,(\CB(\ZZ_i))_{i\ne g}\bigr)
 \Bigr)\le S^{-b}.
\]
\end{lemma}

\begin{proof}[Proof of \cref{cb:lem:cb}]
Increase $b$ to an integer. For $D=1$, take one bit from \cref{pre:cor:long-preliminary}, increasing the length constant
if necessary. Suppose $D\ge2$, and put $\theta=S^{-b}$ and
$\eta=\theta^3/2^{20}$.

Choose $P$ from \cref{pre:cor:long-preliminary} with error at most
$\eta$ and $\lfloor c_0L\rfloor$ output bits, and choose the
$(D-1)$-non-malleable extractor $\NMExt$ from
\cref{nm:thm:constant-nm}. Their circuit sizes are bounded by $S^a$
for a fixed integer $a$. With the exponent $A$ from
\cref{cb:lem:transfer}, choose the generator $G$ from
\cref{comp:lem:seed-extending-prg} with error at most $\theta/4$.
Its seed length is $\ell=O(\log S)$. Define $\Ext_{\mathrm{pre}}$ to be the
first $\ell$ bits of $P$.

Since $D$ is fixed, $B_*$ in \cref{cb:eq:deficiencies} is
$O(\Delta+\log S)$. Taking $C$ sufficiently large in
$L\ge C(\Delta+\log S)$ ensures
$\ell\le\lfloor c_0L\rfloor$, $B_*\le \frac{cL}{D^3}$, and
$2^{-cL/D^3}\le\eta$, where $c$ is the constant in \cref{nm:thm:constant-nm}.
Thus \cref{cb:lem:transfer} applies with $q=1$ and proves the
claimed local guarantee. 
\end{proof}

\subsubsection{More rows and output bits}
\label{sec:many-cb-proof}

The same interactive reduction applies when the number of rows and
the output length grow. Only the parameters of $\NMExt$ and the entropy
budget change.

\begin{lemma}[Multi-bit local correlation breaker]
\label{cb:lem:cb-growing}
Assume $\NDH$. There is a constant $0<\rho<1$ such that, for every
constant $b>0$, there is a constant $C$ with the following property.
Let $D,q\ge1$, $h\le L\le S$, and $\Delta\ge0$ satisfy
\begin{equation}\label[equation]{cb:eq:cb-budget}
 \Delta+Dq+C(D+1)\log S\le L^\rho.
\end{equation}
There is a function $\CB:\zo^L\to\zo^q$, computable in time
$\poly(S)$, with the following guarantee.

Let $(\ZZ_1,\ldots,\ZZ_d)$ be a tuple of at most $D$ indexed rows
with $h$-bit indices, jointly samplable with nondeterministic
postselection by a size-$S$ circuit. For every $g\in[d]$ with
$\minH(\ZZ_g)\ge L-\Delta$, we have
\[
 \TV\Bigl(
   \bigl(\CB(\ZZ_g),(\CB(\ZZ_i))_{i\ne g}\bigr),
   \bigl(\U_q,(\CB(\ZZ_i))_{i\ne g}\bigr)
 \Bigr)\le S^{-b}.
\]
\end{lemma}

\begin{proof}[Proof of \cref{cb:lem:cb-growing}]
Fix $0<\rho<\min\{1,\rho_t,\rho_o,\rho_k,\rho_e\}$, where the
four exponents are from \cref{nm:lem:many-nm}.
Increase $b$ to an integer if necessary, and put $\theta=S^{-b}$
and $\eta=\theta^3/2^{20}$.
Use \cref{pre:cor:long-preliminary} with error exponent $3b+20$
to obtain $P:\zo^L\to\zo^{\lfloor c_0L\rfloor}$ with error at
most $S^{-(3b+20)}\le\eta$.
Its length requirement follows from \cref{cb:eq:cb-budget} after
increasing $C$: the budget gives $\Delta+\log S\le L^\rho$,
and $L^\rho=o(L)$. Increasing $C$ also makes $L$ sufficiently
large for all the bounds below. If $D=1$, the first $q$ output
bits of $P$ prove the claim, since $q\le L^\rho\le\lfloor c_0L\rfloor$.

Suppose $D\ge2$. By \cref{nm:lem:many-nm}, there is an extractor
$\NMExt:\zo^L\times\zo^L\to\zo^q$ that is $(D-1)$-non-malleable
for min-entropy $L-L^{\rho_k}$ and error $2^{-L^{\rho_e}}$,
strong in either input.
Fix an integer $a_0$ bounding the circuit sizes of $P$ and $\NMExt$
by $S^{a_0}$. The constant in \cref{cb:lem:transfer} then
determines the circuit size to be fooled by the generator.
Choose $G$ from \cref{comp:lem:seed-extending-prg} for that size
and error at most $\theta/4$. Its seed length is
$\ell=O(\log S)$, and it has the form
$G(\sigma)=\sigma\concat G'(\sigma)$.
Increasing $C$ ensures $\ell\le\lfloor c_0L\rfloor$; let
$\Ext_{\mathrm{pre}}$ be this prefix of $P$.

Choose $C$ large enough that $\ell\le C\log S$ and
$\log(1/\eta)\le C\log S$. The budget then gives
\[
 B_*:=\max\{\ell,\Delta+D\ell+\log(1/\eta)\}
 \le L^\rho\le L^{\rho_k},
 \qquad
 \log(1/\eta)\le L^\rho\le L^{\rho_e}.
\]
Thus $\NMExt$ works at min-entropy $L-B_*$ with error at most $\eta$.
All hypotheses of \cref{cb:lem:transfer} hold, so
$\CB(v)=\NMExt(G(\Ext_{\mathrm{pre}}(v)),v)$ has the required guarantee.
\end{proof}

\subsection{Multiplicative decorrelation}
\label{sec:mult-cb}

For distributions $A,B$ on the same finite space, we write  $A\approx^{\rm pt}_{\vartheta,\xi}B$ if, for every outcome $z$,
\[
 \Pr[A=z]\le e^\vartheta\Pr[B=z]+\xi,
 \qquad
 \Pr[B=z]\le e^\vartheta\Pr[A=z]+\xi.
\]
This expresses relative error with an additive slack for each
individual outcome. It resembles the comparison used in differential
privacy, but here summing over an event of size $m$ gives slack
$m\xi$. We use only the pointwise guarantee.

We use the same construction \eqref{cb:eq:construction}, now with a
preliminary extractor satisfying a pointwise seed bound and a PRG
against nondeterministic NP-oracle circuits. To test a failure at a
fixed output tuple, the verifier compares its real and ideal
probabilities using approximate counting. The following theorem
states the guarantee locally; its proof follows the interactive
lemma below.

\begin{theorem}\label{mcb:thm:cb}
Assume $\SH$. There is a constant $\rho_m>0$ such that, for every
constant $b>0$, there is a constant $C$ with the following property.
Let $D_*,q\ge1$, $ h\le L\le S$, and $\Delta\ge0$. Suppose
\begin{equation}\label{mcb:eq:errors}
 0<\xi\le S^{-b}\le\vartheta\le1/10
\end{equation}
and
\begin{equation}\label{mcb:eq:budget}
 \Delta+D_*q+
 C\bigl(D_*\log S+(D_*+1)\log(1/\xi)\bigr)
 \le L^{\rho_m}.
\end{equation}
There is a $\poly(S)$-time computable function
$\CBmult:\zo^L\to\zo^q$ with the following guarantee.

Let $(\V_i)_{i\in I}$ be a tuple of at most $D_*$ indexed rows
with $h$-bit indices, jointly samplable with nondeterministic
postselection by a size-$S$ sampling circuit.
For every $g\in I$ with $\minH(\V_g)\ge L-\Delta$ and every
output tuple $z=(z_i)_{i\in I}$, the probabilities
\[
 \begin{aligned}
 P_z&=\Pr[\CBmult(\V_i)=z_i
          \text{ for every }i\in I],\\
 Q_z&=2^{-q}\Pr[\CBmult(\V_i)=z_i
          \text{ for every }i\ne g]
 \end{aligned}
\]
satisfy
\begin{equation}\label{mcb:eq:conclusion}
 P_z\le e^\vartheta Q_z+\xi,
 \qquad Q_z\le e^\vartheta P_z+\xi.
\end{equation}
\end{theorem}

Equivalently, the two output distributions in the theorem are
 $\approx^{\rm pt}_{\vartheta,\xi}$-close. %To give thresholds short finite descriptions, replace accuracy
% parameters by powers of two:
% $\widehat\xi=2^{-\lceil\log(1/\xi)\rceil}$ and
% $\widehat\vartheta=2^{-\lceil\log(1/\vartheta)\rceil}$.
% Each lies between half the original parameter and the parameter
% itself. This preserves their order and changes only constants in
% the final entropy budget. The interactive lemma uses these short
% parameters, with the hats suppressed.

\subsubsection{The interactive lemma}

\begin{lemma}[Interactive reduction for multiplicative error]
\label{mcb:lem:ip}
Let $2\le L\le S$, $1\le D_*,q\le S$, $1\le\ell<L$,
$\Delta\ge0$, and $0<\xi\le\vartheta\le1/10$, where
$\xi$ and $\vartheta$ are powers of two.
Set $\zeta=\eps_c=\xi^8$,
$\Lambda=\lceil8\log(1/\xi)\rceil+2$, and
\[
 B_* = \max\{\ell+\Lambda,
                 \Delta+D_*\ell+\log(1/\zeta)\}.
\]
Assume $\ell+\Lambda\le L$.
Suppose $\Ext_{\mathrm{pre}}:\zo^L\to\zo^\ell$ and
$\NMExt:\zo^L\times\zo^L\to\zo^q$ have size at most $S^a$ for a
fixed $a$, and $\NMExt$ is $(D_*-1)$-non-malleable for min-entropy
$L-B_*$ and error at most $\xi/100$, strong in its first input.
Let $G$ be any seed-extending map and define $\CB$ by
\eqref{cb:eq:construction}.

Fix an admissible tuple of at most $D_*$ rows and a good index $g$
such that
\begin{equation}\label{mcb:eq:seed-cap}
 \Pr[\Ext_{\mathrm{pre}}(\V_g)=\sigma]\le e^{1/100}2^{-\ell}
 \quad\text{for every }\sigma.
\end{equation}
If either comparison in \eqref{mcb:eq:conclusion} fails at some
output tuple $z$, there is a constant-round private-coin protocol
with a SAT-oracle verifier and resource bound
\[
 T=\poly_a(S,D_*,q,1/\vartheta,\log(1/\xi))
\]
whose optimal acceptance probability $v$ satisfies
\[
 \Pr_{\sigma\leftarrow\U_\ell}[v(G(\sigma))\ge c_Y]\ge\xi/32,
 \qquad
 \Pr_{w\leftarrow\U_L}[v(w)>c_N]\le768K\xi^6.
\]
Here $K$ is a fixed integer at least $B^r$ for the constant number
$r$ of rounds, $c_Y=p_0 2^{-\ell}\xi^2/256$, and $c_N=c_Y/K$.
 \end{lemma}

\begin{proof}
Fix the violating sampler, good index, output tuple, and direction.
Use the common outer interaction of
\cref{sec:cb-distinguishing}. After receiving the suffix tuple $a$,
the verifier runs the count-comparison test described below with
fresh private random bits. We first define the probabilities it
compares, then analyze its acceptance on generator and uniform inputs.

\paragraph{The real and ideal events.}
Write $d=|I|\le D_*$ and $\bP=(\Ext_{\mathrm{pre}}(\V_i))_{i\in I}$.
For a seed tuple $p$ of positive probability, a suffix $y$, and
companion suffixes $a$, put
\begin{align*}
 A_p(y,a)&=\Pr[\NMExt(p_g\concat y,\V_g)=z_g,\NMExt(p_i\concat a_i,\V_i)=z_i\ (i\ne g)\mid\bP=p],\\
 B_p(a)&=2^{-q}\Pr[\NMExt(p_i\concat a_i,\V_i)=z_i\ (i\ne g)
                    \mid\bP=p].
\end{align*}
Let $N_p$ count the original sampler inputs with $\Post=1$ and
$\bP=p$. Let $N_A$ count those additionally satisfying the full
output event in $A_p(y,a)$, and let $N_B$ count those satisfying
the companion output event. Then
$A_p(y,a)=N_A/N_p$ and $B_p(a)=2^{-q}N_B/N_p$.
Each count has a nondeterministic membership circuit
that guesses $a_0$ and evaluates only $Q$, $\Samp$, $\Ext_{\mathrm{pre}}$, and $\NMExt$.

\paragraph{The comparison test.}
Stockmeyer's approximate counting algorithm~\cite{Sto85} uses a
SAT oracle to estimate the number of accepting inputs of a Boolean
circuit within $e^{\pm\tau}$, with failure probability $\beta$,
in time polynomial in the circuit size, $1/\tau$, and
$\log(1/\beta)$. The same algorithm counts the distinct
sampling tapes accepted by our nondeterministic membership circuits:
its hash constraints apply only to $R$, and each existence query
guesses both $R$ and a witness $a_0$. Thus these queries remain in NP;
postselection witnesses are not counted. Zero counts are recognized exactly by SAT queries.
Reject when $N_p=0$. Otherwise estimate each of $N_A,N_B,N_p$
with $\tau=\vartheta/100$ and failure probability at most
$\eps_c/3$. Choose  $t$ with
$|\ln t-5\vartheta/8|\le\vartheta/100$, and accept the forward
test exactly when
\[
 \widehat N_A>t\,2^{-q}\widehat N_B+(5\xi/8)\widehat N_p.
\]
 
On the event that all three estimates are accurate, this test accepts if
\begin{equation}\label{mcb:eq:test-yes}
 A_p(y,a)>e^{3\vartheta/4}B_p(a)+3\xi/4,
\end{equation}
and rejects if
\begin{equation}\label{mcb:eq:test-no}
 A_p(y,a)\le e^{\vartheta/2}B_p(a)+\xi/2.
\end{equation}
Indeed, after allowing two estimation errors and threshold rounding,
the multiplicative margins are at least
$(1/8-3/100)\vartheta>0$. The additive margins follow from
$e^{2\vartheta/100}<6/5$ and $e^{2\vartheta/100}<5/4$.
Thus each promised behavior holds with probability at least
$1-\eps_c$. Interchange $N_A$ and $2^{-q}N_B$ for the reverse
test.

\paragraph{Completeness: generator inputs.}
Suppose first that $P_z>e^\vartheta Q_z+\xi$.
On input $G(\sigma)$, the prover replies $a_i=G'(p_i)$.
With these honest suffixes define
\[
 F(p)=A_p(G'(p_g),G'(p_J))
       -e^{3\vartheta/4}B_p(G'(p_J))-3\xi/4.
\]
Averaging over $\bP$ gives $\E F(\bP)>\xi/4$.
Since $F\le1$, the event $\mathcal E=\{p:F(p)>0\}$ has
probability greater than $\xi/4$.
Let $h_\sigma=\Pr[\mathcal E\mid\bP_g=\sigma]$ on seeds of
positive mass. The seeds with $h_\sigma\ge\xi/8$ have total
$\bP_g$-mass greater than $\xi/8$: outside this set the
contribution to $\E h_{\bP_g}$ is at most $\xi/8$.
Remove seeds with
$\Pr[\bP_g=\sigma]<(\xi/16)2^{-\ell}$; their total mass is
at most $\xi/16$. The remaining set has mass greater than
$\xi/16$, hence uniform seed density at least $\xi/32$ by
\eqref{mcb:eq:seed-cap}.

For every remaining seed the initial sampling check succeeds with
probability at least $p_0(\xi/16)2^{-\ell}$.
Conditioned on that check, \eqref{mcb:eq:test-yes} holds with
probability at least $\xi/8$, and the test then accepts with
probability at least $1-\eps_c\ge1/2$. The optimal acceptance
probability is therefore at least
$c_Y=p_0 2^{-\ell}\xi^2/256$.
The reverse violation $Q_z>e^\vartheta P_z+\xi$ is identical
with $A_p$ and $B_p$ interchanged.  

\paragraph{Soundness: uniform inputs.}
Discard seed tuples with probability below $\zeta2^{-d\ell}$;
their total probability is at most $\zeta$.
For every remaining $p$, \cref{lem:conditioning} gives
\[
 \minH(\V_g\mid\bP=p)
 \ge L-\Delta-d\ell-\log(1/\zeta).
\]
The following consequence of non-malleability bounds the suffixes
on which some prover response can cause the test to accept outside
its error event.

\begin{claim}\label[claim]{mcb:lem:small-set}
For every remaining $p$, each of the sets
\begin{align*}
 \mathcal L_p^+&=\{y:\exists a,\
       A_p(y,a)>e^{\vartheta/2}B_p(a)+\xi/2\},\\
 \mathcal L_p^-&=\{y:\exists a,\
       B_p(a)>e^{\vartheta/2}A_p(y,a)+\xi/2\}
\end{align*}
has size less than $2^{L-\ell-\Lambda}$.
\end{claim}
\begin{proof}
If either set were that large, choose $\Y$ uniformly on a subset
of exactly that cardinality and a witnessing tuple $a(y)$ for
each $y$. This cardinality is an integer because
$\Lambda\le L-\ell$.
The independent sources $p_g\concat\Y$ and
$\V_g\mid\bP=p$ have deficiency at most $B_*$.
Apply \cref{cb:lem:correlated} with tampered first inputs
$p_i\concat a_i(\Y)$, extended arbitrarily outside the support.
Testing the output tuple $z$ yields
\[
 \left|\E_{\Y}[A_p(\Y,a(\Y))-B_p(a(\Y))]\right|\le\xi/100.
\]
For every $y$ in the support of $\Y$, the difference
$A_p(y,a(y))-B_p(a(y))$ is greater than $\xi/2$ in the
$\mathcal L_p^+$ case and less than $-\xi/2$ in the
$\mathcal L_p^-$ case. Thus the absolute value of its expectation is greater than $\xi/2$, contradicting the above bound.
\end{proof}

Fix an optimal deterministic prover response for every input and
visible seed message. Average over $w\leftarrow\U_L$ and the
sampler tape. Passing postselection contributes $p_0$, and matching
the independent uniform prefix contributes $2^{-\ell}$.
For each retained seed tuple the fraction of suffixes in the relevant
violating set is at most $2^{-\Lambda}$. Outside it,
\eqref{mcb:eq:test-no} holds for every suffix response, so the test
accepts with probability at most $\eps_c$. Its coins are drawn
after the response is fixed. Including discarded tuples gives
\[
 \E_{w\leftarrow\U_L}v(w)
 \le p_0 2^{-\ell}(2^{-\Lambda}+\eps_c+\zeta)
 \le 3p_0 2^{-\ell}\xi^8.
\]
With $c_N=c_Y/K$, Markov's inequality gives
$\Pr[v(\U_L)>c_N]\le768K\xi^6$.
The factor $p_0$ cancels in the ratio $c_Y/c_N=K$.
Finally, $p_0$ is an integer multiple of $2^{-r_0}$ with
$r_0\le S$, so the thresholds have polynomial-length descriptions.
This proves all the interactive lemma's claims.
\end{proof}

\subsubsection{Construction and analysis}

\begin{proof}[Proof of \Cref{mcb:thm:cb}]
Choose $\rho_m$ strictly smaller than $1$ and the four exponents
in \Cref{nm:lem:many-nm}. We first assume that $\xi$ and
$\vartheta$ are powers of two. For arbitrary values, round both
down to the nearest powers of two. This strengthens the required
guarantee, increases $\log(1/\xi)$ by at most one, and changes
the lower bound on $\vartheta$ by at most a factor of two.

We first choose the ingredients.
\begin{itemize}
\item
Use \Cref{pre:lem:preliminary} with $\nu=1/100$ to obtain
$P_0$ with $m_P=\lfloor c_0L\rfloor$ output bits, for a
sufficiently small fixed $c_0>0$. Use \Cref{nm:lem:many-nm}
to choose $\NMExt$ for $D_*-1$ tamperings and $q$ output bits.
Fix $a_0$ so that both circuit sizes are at most $S^{a_0}$.

\item
The verifier's resource bound in \Cref{mcb:lem:ip}, and the
circuit size after applying \Cref{comp:lem:protocol}, are bounded
by a fixed polynomial in $S$, $1/\vartheta$, and $\log(1/\xi)$,
uniformly over all prefix lengths $\ell\le m_P$.
Choose an integer
$s=\poly(S,1/\vartheta,\log(1/\xi))$ dominating this size and $S$.
Set $\delta_G=\min\{\xi/256,1/s\}$, choosing $s$ large enough
that $2^{-s}\le\delta_G$, and apply
\Cref{mtools:lem:generator} with $\mathcal O=\SAT$ and output
length $L$. The resulting seed length satisfies
\[
 \ell=O(\log(1/\delta_G))
      =O(\log S+\log(1/\xi)),
\]
since $1/\vartheta=O(S^b)$.
Let $\Ext_{\mathrm{pre}}$ be the first $\ell$ bits of $P_0$, and define
$\CBmult$ by \eqref{cb:eq:construction}.
\end{itemize}

We now verify the parameter conditions and prove the claimed
guarantee. Set $\zeta=\eps_c=\xi^8$ and
$\Lambda=\lceil8\log(1/\xi)\rceil+2$ as in \Cref{mcb:lem:ip}.
By \eqref{mcb:eq:budget}, we have
\[
 B_*:=\max\{\ell+\Lambda,
           \Delta+D_*\ell+\log(1/\zeta)\}
      =O(L^{\rho_m}),
 \qquad
 \log(1/\xi)=O(L^{\rho_m}).
\]
Since $\rho_m<\rho_k,\rho_e$, for sufficiently large $L$ we have
\[
 B_*\le L^{\rho_k},
 \qquad
 2^{-L^{\rho_e}}\le\xi/100.
\]
Thus $\NMExt$ meets the entropy and error requirements of
\Cref{mcb:lem:ip}. The same condition ensures
$D_*-1\le L^{\rho_t}$, $q\le L^{\rho_o}$,
$\ell\le m_P$, $\ell+\Lambda\le L$, and the length conditions
for $P_0$ and $G$. Increasing $C$ excludes the remaining
bounded values of $L$. Taking a prefix of $P_0$ preserves
the pointwise seed cap \eqref{mcb:eq:seed-cap}.

Suppose that \eqref{mcb:eq:conclusion} fails.
By \Cref{mcb:lem:ip}, there is a protocol satisfying the resource
and acceptance-ratio conditions of
\Cref{cb:lem:ip-contradiction}, with
\[
 \gamma_Y=\xi/32,
 \qquad
 \gamma_N=768K\xi^6.
\]
We show that these bounds contradict the guarantee of $G$.

Since $\xi\le1/10$, \eqref{mcb:eq:budget} implies
$C\le L^{\rho_m}\le S^{\rho_m}$, and therefore
\[
 \xi\le S^{-b}\le C^{-b/\rho_m}.
\]
As $K$ is fixed and $e^{1/s}<2$, taking $C$ sufficiently large
ensures $e^{1/s}768K\xi^6\le\xi/256$. Together with
$\delta_G\le\xi/256$, this gives
\[
 e^{1/s}\gamma_N+\delta_G
 \le \xi/128
 < \xi/32
 = \gamma_Y.
\]
Thus \Cref{cb:lem:ip-contradiction} rules out this protocol,
proving \eqref{mcb:eq:conclusion}.

Since $1/\vartheta=O(S^b)$ and \eqref{mcb:eq:budget} gives
$\log(1/\xi)=O(S^{\rho_m})$, all ingredients, and hence
$\CBmult$, are computable in $\poly(S)$ time.
\end{proof}

% \begin{remark}[Relative error and computational tampering]
% The same statements apply when the companion rows are efficiently
% computed tamperings of one samplable row, provided the resulting
% indexed tuple fits the joint sampling and entropy hypotheses.
% The multiplicative theorem then gives both directions of a
% pointwise relative-error non-malleability guarantee, with additive
% slack $\xi$. This is the precise sense in which the analysis
% upgrades to relative error. Applying it to a key-derivation or
% leakage-resilience construction additionally requires matching its
% source, tampering, leakage, and security-event models; no such
% application is needed for the extraction results here.
% \end{remark}

\subsection{Relativization of the generator constructions}
\label{sec:decorrelation-prg-details}

We justify the oracle extension in \cref{mtools:lem:generator}.
The ordinary nondeterministic versions are the published results
of~\cite{Shaltiel25multiplicative,DS26}; the nondeterministic
SAT-oracle version follows by preserving the oracle in their
reconstructions. This is an adaptation of those proofs, not a
change to the algorithms evaluating the generators.

\paragraph{Lower-bound tests for acceptance probabilities.}
\label{sec:mult-closure}
The following oracle-preserving form of Goldwasser--Sipser is the
only extra counting ingredient in the reconstructions.

\begin{lemma}\label{mtools:lem:lower-set}
For $\mathcal O\in\{\varnothing,\SAT\}$, let $C$ be a size-$s$
nondeterministic $\mathcal O$-oracle circuit on $r\le s$ ordinary
input bits. Given $C$ and thresholds, the promise problem
\[
 \Pr[C(\U_r)=1]\ge\alpha
 \quad\text{versus}\quad
 \Pr[C(\U_r)=1]\le e^{-\lambda}\alpha
\]
has a nonuniform nondeterministic $\mathcal O$-oracle circuit of
size $\poly(s,B,1/\lambda)$, for $0<\alpha,\lambda\le1$.
Here $B$ bounds the binary description lengths of the thresholds;
real thresholds may be rounded within a constant fraction of
the gap. The dependence on $\alpha$ is only through $B$.
\end{lemma}

\begin{proof}
For $\mathcal O=\varnothing$, apply \cref{thm:gs-counting} with size bound $O(s+B)$ and gap
parameter $\lambda/2$. The slack accommodates the non-strict
inequality in the rejection promise and threshold rounding.

For $\mathcal O=\SAT$, the same Goldwasser--Sipser argument
applies. Merlin supplies a membership witness for each claimed
accepted input of $C$, and Arthur checks it using the
deterministic $\SAT$-oracle verifier of $C$.
The hashing analysis is unchanged. Amplification and fixing
public coins proceed as in \cref{thm:gs-counting},
preserving the oracle and giving the claimed size bound.
\end{proof}

\begin{proof}[Proof of \cref{mtools:lem:generator}]
Write $M=L$ for the output length. For $\mathcal O=\varnothing$
use the cited constructions directly. We explain the adaptation
for $\mathcal O=\SAT$ by checking that each reconstruction still
produces a nondeterministic SAT-oracle circuit, or equivalently an
$\Sigma_2$ circuit. Their evaluation algorithms are unchanged.

\paragraph{Shaltiel's reconstruction.}
For~\cite[Theorem~2.3]{Shaltiel25multiplicative}, the proof from Lemma~2.5 in
Section~2.3.1 guesses the coefficients of a polynomial $p$, checks
its values at fixed reference points, and tests a lower bound on
the acceptance probability of
\[
 (t,v)\longmapsto D(C_x(t),v,\operatorname{SExt}(p(t),v)).
\]
Here $C_x$ is the curve through the input $x$ used in that proof,
and $\operatorname{SExt}$ is its strong seeded extractor. These
maps have deterministic polynomial-size evaluations, so their
composition with $D$ is a $\Sigma_2$ circuit.
Apply \cref{mtools:lem:lower-set} to this circuit and conjoin the
result with the reference-point checks and $p(0)=1$.
The guess of $p$ joins the existential block, giving a $\Sigma_2$
circuit for the hard language.

\paragraph{Dermer--Shaltiel reconstruction.}
For~\cite[Theorem~4.3]{DS26}, the XOR and hybrid steps fix inputs
of the distinguisher or compose it with a fixed XOR. The test in
Figure~2 is
\[
 (t,v)\longmapsto
 D_j\bigl(\operatorname{SExt}(o_1(t),v),\ldots,
          \operatorname{SExt}(o_{m-1}(t),v),
          \operatorname{SExt}(p(t),v)\bigr),
\]
where the $o_j$ are restriction polynomials fixed by the
reconstruction. The test checks reference values and invokes a
lower-bound test of the form in \cref{mtools:lem:lower-set}.
Figure~3 requires two such tests to accept. Section~4.2.9 guesses
the polynomials at successive offsets and requires all verification
steps to accept. If the probability tests have the form
$\exists a_j\forall b_j R_j$, the final predicate is
\[
 \exists\bigl((p_j)_j,(a_j)_j\bigr)\,
 \forall(b_j)_j\quad
 [\text{all consistency checks hold}]
 \land\bigwedge_j R_j\land[p_{\rm final}(0)=1].
\]
The polynomials are guessed together; their successive verification
introduces no further quantifier block. Neither reconstruction
complements the distinguisher.

Where a test samples $t$ uniformly from the nonzero elements of
$\F_Q$, conjoin $t\ne0$ with the test and multiply both
probability thresholds by $(Q-1)/Q$. This expresses the test using
uniform bits and preserves its relative gap.
The curve-sampling and list-size arguments determine which reference
values and thresholds may be fixed nonuniformly; the reconstructed
circuit only uses those fixed values.
The thresholds can be rounded within the promised gaps using
$\poly(s)$ bits, since $\log(1/\delta)\le s$ and the relative
gaps are inverse-polynomial in $s$.

The size calculation in~\cite[Section~4.2.9]{DS26} bounds the
reconstruction by $(sd_0)^c$ for a fixed constant $c$, after absorbing
the polynomial cost of our oracle simulation. Here $d_0$ is a freely
chosen constant dimension. Choose $d_0>(c+1)/\beta_{\mathrm{hard}}$ and set
$u=d_0\log s$. For sufficiently large $s$,
$(sd_0)^c<2^{\beta_{\mathrm{hard}}u}$, contradicting $\SH$.
The algorithms use the hard language only on $u=O(\log s)$ bits,
so they run in $\poly(s)$ time even when $\log(1/\delta)=\Theta(s)$.

We therefore obtain the generator of~\cite[Theorem~4.3]{DS26}
against $\Sigma_2$ circuits, with seed length $O(\log(1/\delta))$
and the stated errors. The generator
of~\cite[Theorem~2.3]{Shaltiel25multiplicative}, against the same class, produces
$r$ additional bits from $O(r+\log(1/\delta))$ seed bits while
including its seed in the output.

\paragraph{Including the seed.}
Compose the two generators as follows. At size bound
$4s$, obtain $F:\zo^r\to\zo^M$ with multiplicative exponent
$1/(4s)$ and additive error $\delta/4$.
For $T=s^J$ with a sufficiently large fixed $J\ge2$, obtain
$A:\zo^\ell\to\zo^r$ such that
\[
 \Pr[D''(\U_\ell,A(\U_\ell))=1]
 \le e^{1/T}\Pr[D''(\U_\ell,\U_r)=1]+\frac\delta{4T}
\]
for tests of size $T$. Its seed length is
$\ell=O(r+\log(4T/\delta))=O(\log(1/\delta))$.
The hypothesis $C\log(1/\delta)\le M\le s$, with $C\ge1$,
implies $\delta\ge2^{-s}$. Consequently $\delta/4$ is in the
permitted range at circuit size $4s$, and $\delta/(4T)$ is in the
permitted range at size $T=s^J$ for sufficiently large $s$.
Increasing $C$ excludes the remaining bounded parameter values
and ensures $\ell\le M$. Now
put $H(z)=z\concat F(A(z))_{[M-\ell]}$.

For each fixed $z$, the test of $F$ remains a nondeterministic
$\mathcal O$-oracle circuit of size at most $4s$.
The circuit $(z,v)\mapsto D(z,F(v)_{[M-\ell]})$ is also a
nondeterministic $\mathcal O$-oracle circuit and has size at most
$T$ by the choice of $J$.
Applying the two guarantees gives
\begin{align*}
 \Pr[D(H(\U_\ell))=1]
 &\le e^{1/T+1/(4s)}\Pr[D(\U_M)=1]
       +e^{1/T}\frac\delta4+\frac\delta{4T}\\
 &\le e^{1/s}\Pr[D(\U_M)=1]+\delta.
\end{align*}
 \end{proof}

\section{One-bit extraction from logarithmic entropy}\label{sec:log-construction}

The condenser from \cref{sec:hash} gives polynomially many rows
that are jointly samplable with nondeterministic postselection, most
of which have almost full min-entropy. We use these rows to construct
a one-bit extractor for sources with logarithmic min-entropy. Following~\cite{CZ19,BDT17},
we transform the rows into bits that are almost $t$-wise independent
outside a small set of bad coordinates, and then apply a resilient
function.

Our main result is the following.
\begin{theorem}\label{thm:log-extractor}
Assume $\NDH$. For all constants $c,\eps>0$, there is a
constant $C$ and a polynomial-time computable
$(C\log n,\eps)$-extractor $\Ext:\zo^n\to\zo$
for $\NPSources_{n,n^c}$.
\end{theorem}

We collect the tools for the construction in
\cref{sec:log-tools}. We then present the extractor construction in
\cref{sec:log-extractor-construction} and prove
\cref{thm:log-extractor} in \cref{sec:log-analysis}.

\subsection{Tools}\label{sec:log-tools}

We use local correlation breaking and a grouping lemma to obtain
a non-oblivious bit-fixing source. We then extract one bit using majority.

\paragraph{Local correlation breaking.}
We use the correlation breaker that we developed in \cref{cb:lem:cb} to make the output on a good row close to a uniform
bit independent of the outputs on a constant number of other rows.

\paragraph{Non-oblivious bit-fixing sources.}
We apply the correlation breaker to indexed rows in fixed groups
of constant size and then XOR its outputs within each group to obtain
one bit per group. We will show that the resulting bits
from any $t$ groups outside a small fixed set are jointly close
to $\U_t$. The following definition captures this guarantee.

\begin{definition}[Non-oblivious bit-fixing source~{\cite{CZ19,BDT17}}]
\label{def:nobf}
Let $1\le t\le M-q$. A distribution $\Y=(\Y_1,\ldots,\Y_M)\in\zo^M$
is a $(q,t,\gamma)$-non-oblivious bit-fixing (NOBF) source if there
is a fixed set $\mathcal Q\subseteq[M]$ of at most $q$ bad
coordinates such that, for every $T\subseteq[M]\setminus\mathcal Q$
of size $t$, the tuple $\Y_T=(\Y_u)_{u\in T}$ satisfies
$\TV(\Y_T,\U_t)\le\gamma$. The bad bits may depend arbitrarily on
the good bits. When $\gamma=0$, we call $\Y$ a $(q,t)$-NOBF source.
\end{definition}

\paragraph{Grouping rows.}
Following~\cite{BDT17}, we use Zuckerman's disperser~\cite{Zuckerman07}
to form groups, all but a small number of which contain a good row.
The following lemma is a direct consequence of Zuckerman's
disperser~\cite{Zuckerman07}, in the form stated in
\cite[Lemma~4.24]{CGL21}. 

\begin{lemma}\label{group:lem:constant-grouping}
For every constant $0<\eta\le1/2$, there is a constant $d_0$ such
that, for every sufficiently large integer $R$, one can uniformly
construct in polynomial time a map
$\Gamma:[M]\times[d_0]\to[R]$, where $M=\Theta(R^8)$, with the
following property. For every $B\subseteq[R]$ with
$|B|\le R^{1-\eta}$, fewer than $M^{1/4}$ groups $u\in[M]$
satisfy $\Gamma(u,a)\in B$ for every $a\in[d_0]$.
\end{lemma}

\paragraph{Extraction by majority.}
To extract from the resulting NOBF source, we use the following
guarantee for majority under bounded independence~\cite{viola2014extractors},
in the form stated in~\cite[Lemma~4.28]{CGL21}.
Define $\Maj(y)=1$ if
$\sum_{u=1}^M y_u\ge\lceil M/2\rceil$, and $\Maj(y)=0$ otherwise.

\begin{lemma}
\label{group:lem:constant-majority}
There is a constant $C_{\rm Maj}$ such that every
$(q,t,\gamma)$-NOBF source $\Y\in\zo^M$ with $2\le t\le M-q$
satisfies
\[
 \TV(\Maj(\Y),\U_1)
 \le C_{\rm Maj}\left(\frac{\log t}{\sqrt t}+\frac q{\sqrt M}\right)
       +\gamma M^t.
\]
\end{lemma}

\subsection{The extractor construction}\label{sec:log-extractor-construction}

We now present the extractor construction, first specifying its
ingredients and their parameters.

Fix a constant $c>0$ and an integer $t\ge2$. Let $C_{\rm hash}$,
$\eta$, and $\Delta$ be the constants in \cref{hash:lem:rows},
decreasing $\eta$ to at most $1/2$ if necessary. We leave the
entropy constant $C$ to be fixed below.

\begin{itemize}
\item \textbf{Condenser rows.}
Apply \cref{hash:lem:rows} at threshold
$k_*=\lfloor C\log n\rfloor$. Write
$h_i(x)=\Cond(x,i)$ for the resulting row maps, indexed by
$i\in[R]$ and of length $m$. Every good row has min-entropy
at least $m-\Delta$.

\item \textbf{Groups.}
Choose $d_0$ and
$\Gamma:[M]\times[d_0]\to[R]$ from
\cref{group:lem:constant-grouping}, where $M=\Theta(R^8)$.
Group $u$ consists of the rows
$h_{\Gamma(u,1)},\ldots,h_{\Gamma(u,d_0)}$.
Put $D=d_0t$, the number of row positions in any $t$ groups.

\item \textbf{Indexed rows.}
For each appearance of a row, add an index $(u,a)$, where $u$
is the group and $a$ is its position within that group.
Set $h=\lceil\log(Md_0)\rceil$ and $L=m+h$, and define
\[
 Z_{u,a}(x)=(u,a)\concat h_{\Gamma(u,a)}(x),
\]
encoding $(u,a)$ in $h$ bits. Distinct positions have distinct
prefixes, even when they contain the same hash row.
If $h_{\Gamma(u,a)}(\X)$ is good, then
$\minH(Z_{u,a}(\X))\ge m-\Delta=L-(\Delta+h)$.

\item \textbf{Sampling circuit size.}
Choose a constant $s\ge2$ and put $S=n^s$ so that any at most
$D$ indexed rows are jointly samplable with nondeterministic
postselection by a size-$S$ circuit. For a source sampled by
$(\Samp,\Post)$, this circuit chooses a uniform input $\rho$,
computes the specified
hashes of $\Samp(\rho)$, appends their indices, and conditions on
$\Post(\rho)=1$.
The polynomial bounds in \cref{hash:lem:rows} let us
choose $s$ independently of $C$.

\item \textbf{Local correlation breaker.}
Since $M$ is polynomially bounded in $n$, choose a constant
$b>0$, also independently of $C$, such that
$tS^{-b}\le M^{-(t+1)}$, and put $\theta=S^{-b}$.
We use \cref{cb:lem:cb} with at most $D$ rows, input length $L$,
index length $h$, entropy deficiency $\Delta+h$, sampling circuit
size at most $S$, and error $\theta$.

Now fix $C$ sufficiently large that
$k_*\ge C_{\rm hash}\log n$ and the correlation breaker's length
requirement holds. We verify these bounds below.
Thus \cref{cb:lem:cb} supplies the required function $\CB:\zo^L\to\zo$.
\end{itemize}

\begin{construction}\label{log:construction}
Using these ingredients, define $\Ext:\zo^n\to\zo$ as follows
on input $x$.
\begin{enumerate}[label=(\roman*)]
\item Compute $h_i(x)$ for every $i\in[R]$ and form the indexed
rows $Z_{u,a}(x)$ for every $(u,a)\in[M]\times[d_0]$.
\item Apply the correlation breaker and XOR within each group:
\begin{equation}\label{eq:log-construction}
 Y_u(x)=\bigoplus_{a=1}^{d_0}\CB(Z_{u,a}(x)),\qquad
 F(x)=(Y_1(x),\ldots,Y_M(x)).
\end{equation}
\item Output $\Ext(x)=\Maj(F(x))$.
\end{enumerate}
\end{construction}

\subsection{Analysis of the construction}
\label{sec:log-analysis}\label{sec:log-nobf}

We show that the XORs from any $t$ groups that each contain a
good row are jointly close to $\U_t$. The grouping lemma then
makes $F(\X)$ a NOBF source, from which majority extracts one bit.

\begin{lemma}\label{log:lem:nobf-reduction}
Assume $\NDH$. For every constant $c>0$ and integer $t\ge2$,
there is a constant $C$ such that 
the map $F$ defined in \eqref{eq:log-construction} is
polynomial-time computable, has output length $n\le M=\poly(n)$,
and for every $\X\in\NPSources_{n,n^c}$ with $\minH(\X)\ge C\log n$, $F(\X)$ is a
$(M^{1/4},t,M^{-(t+1)})$-NOBF source.
\end{lemma}

\begin{proof}
We first check the length requirement in \cref{cb:lem:cb}.
Since $h=O(\log n)$, $\Delta$ is constant, and $S=n^s$,
we have $\Delta+h+\log S=O(\log n)$, with a constant independent
of $C$. Also $L\ge k_*-\log k_*\ge(C/2)\log n$ for sufficiently
large $n$. We can therefore choose $C$ so that
$k_*\ge C_{\rm hash}\log n$ and
$L\ge C_{\rm cb}(\Delta+h+\log S)$, where $C_{\rm cb}$ is the
constant in \cref{cb:lem:cb}. We also have $L=O(\log n)\le S$
and $k_*\le n$ for sufficiently large $n$.

Fix a source $\X\in\NPSources_{n,n^c}$ with $\minH(\X)\ge C\log n$. Apply the condenser guarantee from \cref{hash:lem:rows} at threshold $k_*$, and let $\mathcal G=[R]\setminus B_{\X}$ be its set of good indices. For every $i\in \mathcal G$, we have $\minH(h_i(\X))\ge m-\Delta$.
Write $\ZZ_{u,a}=Z_{u,a}(\X)$ and $\Y_u=Y_u(\X)$ for the indexed
rows and group outputs. Let $\mathcal Q\subseteq[M]$ be the set
of groups containing no row indexed by $\mathcal G$. By \cref{group:lem:constant-grouping},
$|\mathcal Q|<M^{1/4}$. For every $\Gamma(u,a)\in \mathcal G$, we have
$\minH(\ZZ_{u,a})\ge m-\Delta=L-(\Delta+h)$.

Fix any $t$ groups outside $\mathcal Q$ and choose one good row
in each. These groups give $D=d_0t$ indexed rows with distinct
prefixes $(u,a)$. By \cref{cb:lem:cb}, the output on each chosen
row is $\theta$-close to an independent uniform bit, jointly
with the outputs on all the other indexed rows in these groups.
Taking the XOR within each group preserves this guarantee:
each of the $t$ group outputs is $\theta$-close to an independent
uniform bit jointly with the other $t-1$ group outputs.
By \cref{pre:lem:joint-uniformity}, their joint distribution
is $t\theta$-close to $\U_t$. This completes the proof, recalling that by choice $t\theta\le M^{-(t+1)}$.
\end{proof}

\begin{proof}[Proof of \cref{thm:log-extractor}]
Choose an integer $t\ge2$ such that
$C_{\rm Maj}(\log t)/\sqrt t\le\eps/3$, and use \cref{log:construction} with the resulting constant $C$.
The result follows from \cref{log:lem:nobf-reduction,group:lem:constant-majority}.
\end{proof}

\section{Extraction from polylogarithmic entropy}\label{sec:many-construction}
In this section, we extend the construction of \cref{sec:log-construction} to extract $k^{\Omega(1)}$ bits with inverse-polynomial error when $k \ge \poly(\log n)$. The higher entropy allows the local correlation breaker to handle a growing number of rows. We replace the resilient function with a construction based on Feige's lightest bin protocol~\cite{Fei99}, using the correlation breaker twice: first to assign almost $t$-wise independent bin labels, and then to extract from the rows in a least populated bin. Finally, we apply Shaltiel's result~\cite{Shaltiel08mileage} to obtain the desired output length.

\begin{theorem}\label{many:thm:base}
Assume $\NDH$. There are constants $A,a>0$ such that, for every pair of constants $c,e>0$, there is $n_0=n_0(c,e)$ such that the following holds. For every $n\ge n_0$ and every $(\log n)^A\le k$, there is a $(k,n^{-e})$-extractor $ \Ext:\zo^n\to\zo^{\lfloor k^a\rfloor}$
for $\NPSources_{n,n^c}$, computable in time $2^{O(\log^2 n/\log k)}$. 
\end{theorem}

\begin{theorem}[Extracting almost all entropy] \label{many:thm:almost-all}
Assume $\NDH$. There is a constant $A>0$ such that, for every $\eta\in (0,1)$, and for every pair of constants $c,e>0$, there is $n_0=n_0(c,e,\eta)$ such that the following holds. For every $n\ge n_0$ and every $(\log n)^A\le k$, there is a $(k,n^{-e})$-extractor
$
\Ext:\zo^n\to\zo^{\lfloor(1-\eta)k\rfloor}
$
for $\NPSources_{n,n^c}$, computable in time $2^{O(\log^2 n/\log k)}$.
\end{theorem}

We first state the correlation-breaking and selection tools, then
set up the ingredients and prove \cref{many:thm:base}. We prove
\cref{many:thm:almost-all} in \cref{sec:amplification}.
In \cref{sec:poly-construction}, we improve the running time of our extractor to polynomial under $\SH$.

\subsection{Tools}\label{sec:many-local-cb}

The first tool is local correlation breaking for more rows
and output bits. The other two bound the probability that some
bin contains no good row.

\paragraph{Local correlation breaking.} 

We use the multi-bit local correlation breaker from \cref{cb:lem:cb-growing}, with the indexed-row convention of \cref{sec:cb}. The row length and entropy deficiency include the fixed index. We instantiate this result for bin labels and final extraction in \cref{cb:lem:lab-instantiation,cb:lem:ext-instantiation}.

\paragraph{Committee selection.}
We use Feige's lightest bin protocol for committee selection on the rows: we assign each row a bin label, then select a lightest bin.
The selected bin has at most $\lfloor R/T\rfloor$ rows. We prove a couple of lemmas upper bounding the probability that there's a bin with no good row.  Both bounds follow from standard
moment arguments. For completeness, we provide the proofs.

\begin{lemma}[A good row in every bin: statistical distance]
\label{select:lem:occupancy-statistical}
 Let $\Y_1,\ldots,\Y_R$ be jointly distributed labels in $[T]$,
where $1\le T\le R$. Fix $G\subseteq[R]$, let $g=|G|$ and $\mu=g/T$, and let
$\ZZ_z=\sum_{i\in G}\ind[\Y_i=z]$ for $z\in[T]$.
 Suppose $t$ is even with $2\le t\le\mu$. If every set of at most $t$ labels indexed by $G$ has statistical distance at most $\omega$ from independent
uniform labels in $[T]$, then
\[
 \Pr[\exists z\in[T]:\ZZ_z=0]
 \le T\left(\frac{64t}{\mu}\right)^{t/2}+2\omega T^{t+1}.
\]
\end{lemma}

\begin{lemma}[A good row in every bin: relative error]
\label{select:lem:occupancy-relative}
Let $\Y_1,\ldots,\Y_R$, $G$, $\mu$, and $\ZZ_z$ be as in
\cref{select:lem:occupancy-statistical}.
Suppose $t$ is even with $2\le t\le\mu$. If, for every
$1\le d\le t$, distinct $i_1,\ldots,i_d\in G$, and $z\in[T]$,
\[
 \left|\Pr[\Y_{i_1}=\cdots=\Y_{i_d}=z]-T^{-d}\right|
 \le\delta T^{-d},
\]
then
\[
 \Pr[\exists z\in[T]:\ZZ_z=0]
 \le T\left(\frac{64t}{\mu}\right)^{t/2}+3^tT\delta.
\]
\end{lemma}

\textbf{Setup for both proofs:} Fix a bin $z\in[T]$, put $p=1/T$, and let
$\xi_i=\ind[\Y_i=z]$ for $i\in G$.
Let $\xi'_i$, $i\in G$, be independent Bernoulli variables of
mean $p$. Write
$\ZZ_z=\sum_{i\in G}\xi_i$ and
$\ZZ'_z=\sum_{i\in G}\xi'_i$, so $\E\ZZ'_z=\mu=gp$.
The standard moment bound~\cite{BR94} gives, for even $t$ with
$2\le t\le\mu$,
\begin{equation}\label{select:eq:binomial-moment}
 \E[(\ZZ'_z-\mu)^t]\le(64\mu t)^{t/2}.
\end{equation}

\begin{proof}[Proof of \cref{select:lem:occupancy-statistical}]
Expand
\[
 (\ZZ_z-\mu)^t
 =\sum_{(i_1,\ldots,i_t)\in G^t}
   \prod_{j=1}^t(\xi_{i_j}-p),
\]
and similarly for $\ZZ'_z$. Each product depends on at most $t$
labels and takes values in $[-1,1]$. The statistical distance
hypothesis bounds the difference of corresponding expectations
by $2\omega$. There are $g^t$ products, so
\[
 \E[(\ZZ_z-\mu)^t]
 \le(64\mu t)^{t/2}+2\omega g^t.
\]
Since $\ZZ_z=0$ implies $(\ZZ_z-\mu)^t=\mu^t$, Markov's
inequality gives
\[
 \Pr[\ZZ_z=0]
 \le\left(\frac{64t}{\mu}\right)^{t/2}+2\omega T^t.
\]
A union bound over the $T$ bins proves the claim.
\end{proof}

\begin{proof}[Proof of \cref{select:lem:occupancy-relative}]
The case $T=1$ is immediate, so suppose $T\ge2$.
For every nonempty $I\subseteq G$ with $|I|\le t$, the hypothesis
gives
\[
 \left|\E\left[\prod_{i\in I}\xi_i\right]
       -\E\left[\prod_{i\in I}\xi'_i\right]\right|
 \le\delta p^{|I|}.
\]
We use these bounds to compare the expansions of
$(\ZZ_z-\mu)^t$ and $(\ZZ'_z-\mu)^t$.

Fix an ordered tuple $(i_1,\ldots,i_t)\in G^t$, let $J$ be its
set of distinct indices, and let $m_i$ be the multiplicity of
$i\in J$. Since $\xi_i\in\zo$,
\[
 (\xi_i-p)^{m_i}=\alpha_i+\beta_i\xi_i,
 \qquad \alpha_i=(-p)^{m_i},\qquad
 \beta_i=(1-p)^{m_i}-(-p)^{m_i}.
\]
Here $|\alpha_i|\le p$ and $|\beta_i|\le1$.
Writing $d=|J|$, the coefficient of
$\prod_{i\in I}\xi_i$ in the product over $J$ has absolute
value at most $p^{d-|I|}$. The constant terms cancel, so the
corresponding product expectations differ by at most
\[
 \delta\sum_{\varnothing\ne I\subseteq J}
 p^{d-|I|}p^{|I|}\le\delta(2p)^d.
\]

There are at most $\binom td g^d t^{t-d}$ ordered tuples with
$d$ distinct indices: choose the positions where new indices
first appear, their values, and an earlier index for each remaining
position. Summing the expectation differences gives
\begin{align*}
 \left|\E[(\ZZ_z-\mu)^t]-\E[(\ZZ'_z-\mu)^t]\right|
 &\le\delta\sum_{d=1}^t\binom td g^dt^{t-d}(2p)^d\\
 &\le\delta(2\mu+t)^t\le\delta(3\mu)^t.
\end{align*}
Together with \cref{select:eq:binomial-moment}, this yields
\[
 \E[(\ZZ_z-\mu)^t]\le(64\mu t)^{t/2}+\delta(3\mu)^t.
\]
Markov's inequality and a union bound over the $T$ bins now give
\[
 \Pr[\exists z\in[T]:\ZZ_z=0]
 \le T\left(\frac{64t}{\mu}\right)^{t/2}+3^tT\delta.
\]
\end{proof}

\subsection{The extractor construction}\label{sec:many-parameters}

We now present the extractor construction, first setting up its
ingredients. We split each hash row into two parts. The first part
determines its bin label; the second part supplies the entropy for
extraction after selection.

Fix $c,e>0$. Choose a constant $0 < \rho\le1/2$ no larger
than the exponent in \cref{cb:lem:cb-growing}, and put $a=\rho/8$.
We take the entropy exponent $A$ sufficiently large that $Aa>4$.
All bounds below hold for $(\log n)^A\le k\le n$. Parameters below with subscript $f$ denote parameters of the final extractor. 

\begin{itemize}
\item \textbf{Hash rows.}
Apply \cref{hash:lem:rows} at threshold $k$ to obtain
$h_1,\ldots,h_R:\zo^n\to\zo^m$, with $n\le R\le n^{C_H}$.
For every source under consideration, there is a fixed set
$G\subseteq[R]$ of size at least $R-R^{1-\eta_H}$ such that
$\minH(h_i(\X))\ge m-\Delta_h$ for $i\in G$, where
$\eta_H,\Delta_h>0$ are constants.
Write $h_i=h_i^{(0)}\concat h_i^{(1)}$, with part lengths
$m_0=\lfloor m/2\rfloor$ and $m_1=m-m_0$.
Both parts of every good row have deficiency at most $\Delta_h$.

\item \textbf{Bins and indices.}
Set $s=\lceil k^a\rceil$ and
$T=2^{\lfloor\log(R/s)\rfloor}$, so $s\le R/T<2s$.
Let $D_f=\lfloor R/T\rfloor$, $h_0=\lceil\log R\rceil$, and
$h_f=\lceil\log D_f\rceil$. Fix distinct $h_0$-bit encodings of
$[R]$ and distinct $h_f$-bit encodings of $[D_f]$.
The indexed row lengths are $L_0=m_0+h_0$ and $L_f=m_1+h_f$.

\item \textbf{Labeling parameters.}
Choose a sufficiently large constant $K$ and let $t$ be the least
even integer at least $K\log n/\log k$. Choose a constant
$C_s>e+6$ such that any $t$ indexed first parts are jointly
samplable with nondeterministic postselection by a circuit of size
at most $n^{C_s}$. Let $S_0$ be the least power of two at least
$n^{C_s}T^{2t}$, and set $q_0=\log T$.

\item \textbf{Extraction parameters.}
Set $q_1=\lfloor k^a\rfloor$ and $\zeta=n^{-(e+3)}$.
Let $S_f$ be the least power of two at least $(nS_0)^{C_f}$, for
a sufficiently large constant $C_f$, and put
\[
 \Delta_f=\Delta_h+h_f+
 \left\lceil\log\frac{D_fR}{\zeta}\right\rceil.
\]
The choice of $C_f$ covers the conditional sampling circuits in
\cref{select:lem:selected-extraction} and ensures $S_f^{-1}\le\zeta$.
\end{itemize}

The next two lemmas instantiate the correlation breaker with these
parameters.  

\begin{lemma}[Correlation breaking for bin labels]
\label{cb:lem:lab-instantiation}
Assume $\NDH$. With the preceding parameters, there is a
$\poly(S_0)$-time computable map
$\CBlab:\zo^{L_0}\to\zo^{q_0}$ such that, for every
$\X\in\NPSources_{n,n^c}$ with $\minH(\X)\ge k$, the labels
\[
 \Y_i=\CBlab(i\concat h_i^{(0)}(\X))\in[T]
\]
satisfy
\[
 \TV\bigl((\Y_{i_1},\ldots,\Y_{i_d}),\U_{[T]^d}\bigr)
 \le d/S_0
\]
for every $1\le d\le t$ and distinct $i_1,\ldots,i_d\in G$.
Here $\U_{[T]^d}$ denotes $d$ independent uniform labels in $[T]$.
\end{lemma}

\begin{proof}
Apply \cref{cb:lem:cb-growing} with $D=t$, output length $q_0$,
row length $L_0$, deficiency $\Delta_h+h_0$, and error $S_0^{-1}$.
The joint sampling circuit computes the specified hashes of
$\Samp(\bR)$, keeps their first parts, appends their indices, and
conditions on $\Post(\bR)=1$. Its size is at most $n^{C_s}\le S_0$.

We have $L_0\ge k/3$, $\log S_0=O(t\log n)$, and
$q_0=O(\log n)$. The left side of \cref{cb:eq:cb-budget} is
$O(t^2\log n)=O(\log^3 n)=o(k^\rho)$, and
$t,q_0\le L_0^\rho$. Thus the correlation breaker can be used.
By \cref{pre:lem:joint-uniformity}, any $d$ good labels are
jointly $dS_0^{-1}$-close to independent uniform labels.
\end{proof}

\begin{lemma}[Correlation breaking for the selected rows]
\label{cb:lem:ext-instantiation}
Assume $\NDH$. With the preceding parameters, there is a
$\poly(S_f)$-time computable map
$\CBext:\zo^{L_f}\to\zo^{q_1}$ with the following
property. For every tuple $(\ZZ_1,\ldots,\ZZ_d)$ of at most
$D_f$ indexed rows with distinct $h_f$-bit prefixes, jointly
samplable with nondeterministic postselection by a size-$S_f$ circuit,
and every $j\in[d]$ with $\minH(\ZZ_j)\ge L_f-\Delta_f$,
\[
 \begin{aligned}
 \TV\Bigl(&\bigl(\CBext(\ZZ_j),
               (\CBext(\ZZ_\ell))_{\ell\ne j}\bigr),\\
 &\bigl(\U_{q_1},
               (\CBext(\ZZ_\ell))_{\ell\ne j}\bigr)
 \Bigr)\le\zeta.
 \end{aligned}
\]
\end{lemma}

\begin{proof}
Apply \cref{cb:lem:cb-growing} with $D=D_f$, output length $q_1$,
row length $L_f$, deficiency $\Delta_f$, and error $S_f^{-1}$.
Here $L_f=\Theta(k)$, $\Delta_f=O(\log n)$,
$D_f=O(k^a)$, and $\log S_f=O(t\log n)$. Hence the left side of
\cref{cb:eq:cb-budget} is
$O(k^a\log^2 n+k^{2a})=o(k^\rho)$, and
$D_f,q_1\le L_f^\rho$. The resulting error is at most
$S_f^{-1}\le\zeta$.
\end{proof}

\begin{construction}\label{con:ext-construction-many}
Define
$\Ext:\zo^n\to\zo^{q_1}$ as follows: on input $x$,
\begin{enumerate}[label=(\roman*)]
\item Compute the hash rows and their bin labels
\[
 Y_i(x)=\CBlab(i\concat h_i^{(0)}(x))\in[T],
 \qquad i\in[R].
\]
\item Choose a least populated bin, breaking ties by its index,
and list its row indices as $i_1<\cdots<i_d$. Then $d\le D_f$.
\item Index the selected second parts by their positions and pad
with zeros:
\[
 W_j(x)=
 \begin{cases}
 j\concat h_{i_j}^{(1)}(x),&1\le j\le d,\\
 j\concat0^{m_1},&d<j\le D_f.
 \end{cases}
\]
\item Output
\begin{equation}\label{select:eq:construction}
 \Ext(x)=\bigoplus_{j=1}^{D_f}\CBext(W_j(x)).
\end{equation}
\end{enumerate}
\end{construction}

\subsection{Analysis}\label{sec:selection}

The proof has two steps: we first analyze extraction from the selected rows
assuming that every bin contains a good row with high probability,
and then we establish this condition using the almost-independence
of the good labels. We formulate the first step for arbitrary
row maps $h_i$ and an arbitrary labeling map $\CBlab$,
so that it also applies in \cref{sec:poly-construction}.

\begin{lemma}[Extraction from the selected rows]
\label{select:lem:selected-extraction}
There is a fixed polynomial $p$ with the following property.
Consider the procedure in \cref{con:ext-construction-many}
with arbitrary row maps $h_i=h_i^{(0)}\concat h_i^{(1)}$ of
lengths $m_0,m_1$ and an arbitrary labeling map
$\CBlab$. Suppose $\X$ is samplable with nondeterministic
postselection by a size-$s_X$ circuit, each row map has circuit
size at most $s_h$, and the labeling map has circuit size at most
$s_0$. Let $G\subseteq[R]$ satisfy
$\minH(h_i^{(1)}(\X))\ge m_1-\Delta_h$ for every $i\in G$,
and suppose every bin contains a row indexed by $G$ with probability
at least $1-\eps_0$.

For $0<\zeta\le1/2$, let $D_f=\lfloor R/T\rfloor$,
$h_f=\lceil\log D_f\rceil$, $L_f=m_1+h_f$, and
$\Delta_f=\Delta_h+h_f+\lceil\log(D_fR/\zeta)\rceil$.
Suppose $\CBext$ satisfies the local guarantee in
\cref{cb:lem:cb-growing} with error at most $\zeta$ for at most
$D_f$ indexed rows of length $L_f$, good-row deficiency at most
$\Delta_f$, and joint sampling circuit size at most $S_f$.
If $S_f\ge p(n,s_X,s_h,s_0,R,m_0+m_1)$, then
\[
 \TV(\Ext(\X),\U_{q_1})\le\eps_0+2\zeta.
\]
\end{lemma}

\begin{proof}
Let $\mathcal C$ be the event that every bin contains a good row.
For $j\in[D_f]$ and $i\in G$, let $E_{j,i}$ be the event that
$\mathcal C$ occurs and the first good selected row has original
index $i$ and selected position $j$. These events partition
$\mathcal C$. There are at most $D_fR$ events, so those with
probability below $\zeta/(D_fR)$ have total probability at most
$\zeta$.

Fix a remaining event $E_{j,i}$. By \cref{lem:conditioning},
\[
 \minH(h_i^{(1)}(\X)\mid E_{j,i})
 \ge m_1-\Delta_h-\log(D_fR/\zeta).
\]
On this event, $W_j(\X)=j\concat h_i^{(1)}(\X)$, so
$\minH(W_j(\X)\mid E_{j,i})\ge L_f-\Delta_f$.

Write the original sampling circuit as $(\Samp,\Post)$.
To sample the selected tuple conditioned on $E_{j,i}$, choose a
uniform input $\bR$, compute $x=\Samp(\bR)$ and
$(W_1(x),\ldots,W_{D_f}(x))$, and condition on
$\Post(\bR)=1$ and $E_{j,i}(x)$. The circuit tests $E_{j,i}$
using the fixed membership bits of $G$: it checks that every bin
contains a good row and identifies the original index and position
of the first good selected row.  Since $E_{j,i}$ can be checked deterministically, conditioning
further on $E_{j,i}$ preserves nondeterministic postselection.

The conditional sampling circuit has size at most
$p(n,s_X,s_h,s_0,R,m_0+m_1)\le S_f$. The rows have distinct
position prefixes, and the $j$th row has deficiency at most
$\Delta_f$. The local guarantee therefore allows us to replace
$\CBext(W_j(\X))$ by a uniform string independent
of the other outputs, changing their joint distribution by
statistical distance at most $\zeta$.
Note that taking XOR cannot increase this distance. After the replacement,
the XOR is uniform, so
\[
 \TV\bigl((\Ext(\X)\mid E_{j,i}),\U_{q_1}\bigr)\le\zeta.
\]

Let $\beta$ be the total probability of $\neg\mathcal C$ and
the discarded events. We have $\beta\le\eps_0+\zeta$.
On each remaining event, the conditional output is
$\zeta$-close to uniform; on the other events, we use the
trivial bound of $1$ on statistical distance. Averaging these
bounds gives
\[
 \TV(\Ext(\X),\U_{q_1})
 \le (1-\beta)\zeta+\beta
 \le \zeta+\eps_0+\zeta
 = \eps_0+2\zeta.
\]
\end{proof}

\begin{proof}[Proof of \cref{many:thm:base}]
Use the parameters from \cref{sec:many-parameters}, and let
$\X\in\NPSources_{n,n^c}$ have min-entropy at least $k$.
By \cref{cb:lem:lab-instantiation}, every set of at most $t$ good
labels is $t/S_0$-close to independent uniform labels.
Since $g=|G|\ge R/2$, we have $\mu=g/T\ge s/2$.
For sufficiently large $n$, the entropy threshold gives
$2\le t\le\mu$ and $\log(s/(128t))\ge(a/2)\log k$.
Thus \cref{select:lem:occupancy-statistical} gives
\begin{align*}
 \Pr[\neg\mathcal C]
 &\le T(128t/s)^{t/2}+2tT^{t+1}/S_0\\
 &\le n^{C_H-aK/4}+2t n^{-C_s}\le\zeta,
\end{align*}
for sufficiently large $K$ and $n$.

The choice of $C_f$ ensures that $S_f$ covers the conditional
sampling circuits in \cref{select:lem:selected-extraction}:
the original circuit, row maps, and number of rows have polynomial
size in $n$, and the labeling map has size $\poly(S_0)$.
By \cref{cb:lem:ext-instantiation,select:lem:selected-extraction}, with $\eps_0= \zeta$, we have
\[
 \TV(\Ext(\X),\U_{q_1})\le3\zeta\le n^{-e}.
\]

There are $R$ computations of $\CBlab$ and $D_f$
computations of $\CBext$, with polynomial overhead
for hashing and selection. The total running time is
\(
 \poly(n,S_0,S_f)=2^{O(t\log n)}
 =2^{O(\log^2 n/\log k)}.
\)
 \end{proof}

\subsection{Extracting almost all the entropy}\label{sec:amplification}
We use Shaltiel's result \cite{Shaltiel08mileage}, in the form stated in \cref{pre:lem:output-amplification} to extend the output to almost all the entropy.  
\begin{proof}[Proof of \cref{many:thm:almost-all}]
Take $C,c_0,d_0$ from \cref{pre:lem:output-amplification} with $d=e$, and set $k_0=\lfloor\eta k/2\rfloor$. For sufficiently large $A$ independent of $c,e$ and sufficiently large $n$, \cref{many:thm:base} with parameters $c_0,d_0$ gives an extractor at threshold $k_0$ whose output can be truncated to $t=\lceil C\log n\rceil\le k_0$ bits. The extension of \cref{pre:lem:output-amplification} to $\NPSources$ applies since $2k_0/\eta\le k$, giving the desired output length and error in time $2^{O(\log^2 n/\log k_0)}+\poly(n) =2^{O(\log^2 n/\log k)}$.
\end{proof}

\section{Polynomial-time extraction from polylogarithmic entropy}\label{sec:poly-construction}

We obtain polynomial running time under $\SH$ by using the multiplicative correlation breaker from \cref{mcb:thm:cb} for the bin labels in \cref{sec:many-construction}. Its pointwise guarantee controls the rare label events needed for selection with only polynomial cost. We then select $O(\log n)$ rows, extract using \cref{cb:lem:cb-growing,select:lem:selected-extraction}, and amplify the output using \cref{pre:lem:output-amplification}.

\begin{theorem}\label{poly:thm:main}
Assume $\SH$. There is a constant $A>0$, such that for every $\eta\in(0,1)$, and for all constants $c,e>0$, for every $k>(\log n)^A$,
there is a polynomial-time computable $(k,n^{-e})$-extractor
$\Ext:\zo^n\to\zo^{\lfloor(1-\eta)k\rfloor}$ for
$\NPSources_{n,n^c}$.
\end{theorem}
\subsection{Relative probabilities of bin labels}\label{poly:sec:labels}

Applied to any $d\le t$ good indexed rows and the output tuple $(z,\ldots,z)$, \cref{mcb:thm:cb} gives the comparisons in the following lemma.

\begin{lemma}[Relative probabilities of labels]\label{select:lem:relative-probabilities}
Let $\Y_1,\ldots,\Y_R$ be labels in $[T]$, with good index set $G$. Suppose that, for every ordered tuple of $1\le d\le t$ distinct indices in $G$ and every $z\in[T]$, the probabilities
\[
 P=\Pr[\Y_{i_1}=\cdots=\Y_{i_d}=z],\qquad
 Q=T^{-1}\Pr[\Y_{i_2}=\cdots=\Y_{i_d}=z]
\]
satisfy $P\le e^\vartheta Q+\xi$ and $Q\le e^\vartheta P+\xi$, with the empty conjunction having probability one. If $0<\delta\le1/2$, $\vartheta=\delta/(8t)$, and $\xi=\vartheta T^{-t}$, then
\[
 \left|\Pr[\Y_{i_1}=\cdots=\Y_{i_d}=z]-T^{-d}\right|\le\delta T^{-d}.
\]
\end{lemma}
\begin{proof}
Fix distinct indices $i_1,\ldots,i_d\in G$ and $z\in[T]$,
where $d\le t$. We replace the labels one at a time by independent
uniform labels. After replacing the first $j$ labels, the probability
that all labels equal $z$ is
$p_j=T^{-j}\Pr[\Y_{i_{j+1}}=\cdots=\Y_{i_d}=z]$.
Thus $p_0=P$ and $p_d=T^{-d}$.

For $0\le j<d$, the hypothesis gives
$p_j\le e^\vartheta p_{j+1}+T^{-j}\xi$ and
$p_{j+1}\le e^\vartheta p_j+T^{-j}\xi$.
Iterating these inequalities and using $T^{-j}\le1$ gives
\[
 e^{-d\vartheta}T^{-d}-d\xi
 \le P
 \le e^{d\vartheta}T^{-d}+de^{d\vartheta}\xi.
\]
Our choices ensure $d\vartheta\le\delta/8$ and
$d\xi T^d=d\vartheta T^{d-t}\le\delta/8$.
The upper bound is therefore at most
$e^{\delta/8}(1+\delta/8)T^{-d}\le(1+\delta)T^{-d}$,
and the lower bound is at least
$(e^{-\delta/8}-\delta/8)T^{-d}\ge(1-\delta)T^{-d}$,
since $0<\delta\le1/2$.
Thus $|P-T^{-d}|\le\delta T^{-d}$.
\end{proof}

\subsection{The polynomial-time construction}

\begin{theorem}\label{poly:thm:base}
Assume $\SH$. There exist constants $A,\gamma>0$, such that for all constants $c,e>0$, for every $k>(\log n)^A$, there is a
polynomial-time computable $(k,n^{-e})$-extractor
$B_0:\zo^n\to\zo^{\lfloor k^\gamma\rfloor}$ for
$\NPSources_{n,n^c}$.
\end{theorem}

\begin{proof}
Choose $0<\rho\le1/2$ no larger than the exponent in
\cref{cb:lem:cb-growing}, and let $\rho_m$ be the exponent in
\cref{mcb:thm:cb}. Set $\gamma=\rho/2$ and choose $A$ so that
$A\rho,A\rho_m>4$.

\paragraph{Ingredients.} 
\begin{itemize}
\item
Use the hash rows $h_i=h_i^{(0)}\concat h_i^{(1)}$ from
\cref{sec:many-parameters}, with $n\le R\le n^{C_H}$ and
$m_0,m_1=\Theta(k)$. Let $\Delta_H=O(1)$ denote the upper bound on the entropy deficiency of a good row.

\item
Choose an integer $e_0>e+3$ and set $\zeta=n^{-e_0}$.
Let $t$ be the least even integer at least $K\log n$, and let
$T$ be the largest power of two at most $R/(C_0t)$, for
sufficiently large constants $C_0,K$. Then
$D_f=\lfloor R/T\rfloor=O(\log n)$. Set
\begin{equation*} 
 \delta=\frac{\zeta}{3^{t+1}R},\qquad
 \vartheta=\frac{\delta}{8t},\qquad
 \xi=\vartheta T^{-t}.
\end{equation*}

\item
Let $h_0=\lceil\log R\rceil$, $L_0=m_0+h_0$,
$\Delta_0=\Delta_H+h_0$, and $q_0=\log T$.
Choose $S_0=n^{C_s}$, with $C_s\ge2$ large enough to cover
the joint sampler for any $t$ indexed first parts.
Take $\CBlab$ from \cref{mcb:thm:cb} with these
parameters, at most $t$ rows, and errors $\vartheta,\xi$.
Define $Y_i(x)=\CBlab(i\concat h_i^{(0)}(x))$.

\item
Use these labels in \cref{con:ext-construction-many},
and let $W_1,\ldots,W_{D_f}$ be the selected, reindexed, and
padded rows. Set $q_1=\lfloor k^\gamma\rfloor$,
$h_f=\lceil\log D_f\rceil$, and $L_f=m_1+h_f$. Let
\begin{equation*} 
 \Delta_f=\Delta_H+h_f+
 \left\lceil\log(D_fR/\zeta)\right\rceil.
\end{equation*}
Choose a polynomial bound $S_f\ge n^2$ covering the conditional
sampling circuits in \cref{select:lem:selected-extraction}
and satisfying $S_f^{-1}\le\zeta$.
Since $\SH$ implies $\NDH$, take $\CBext$ from
\cref{cb:lem:cb-growing} with row length $L_f$, at most $D_f$
rows, deficiency $\Delta_f$, output length $q_1$, and error
$S_f^{-1}$.
\end{itemize}
Define
\begin{equation*} 
 B_0(x)=\bigoplus_{j=1}^{D_f}\CBext(W_j(x)).
\end{equation*}

\paragraph{Analysis.}
We first check the parameter conditions for the correlation
breakers. For the labeling map, $1/\vartheta=n^{O(1)}$ and
$\xi=2^{-\Theta(\log^2 n)}$, so a fixed constant $b>0$
satisfies $\xi\le S_0^{-b}\le\vartheta\le1/10$.
The left-hand side of \eqref{mcb:eq:budget}, with
$D_*=t$, $q=q_0$, and $S=S_0$, is
$O(\log^3 n)=o(L_0^{\rho_m})$ because $A\rho_m>4$.
Thus \cref{mcb:thm:cb} supplies the required labeling map
in polynomial time.

Consequently, the conditional sampling circuits in
\cref{select:lem:selected-extraction} have polynomial size,
so we can choose $S_f=\poly(n)$.
We have $\Delta_f=O(\log n)$, $D_f=O(\log n)$, and
$\log S_f=O(\log n)$. The left-hand side of
\eqref{cb:eq:cb-budget} is therefore
$O(\log^2 n+k^\gamma\log n)=o(L_f^\rho)$, since
$\gamma=\rho/2$ and $A\rho>4$.
Thus \cref{cb:lem:cb-growing} gives a polynomial-time computable
$\CBext$ with error $S_f^{-1}\le\zeta$.

Now fix $\X\in\NPSources_{n,n^c}$ with $\minH(\X)\ge k$.
The hash guarantee gives a fixed set $G$ of at least $R/2$
good rows. Applying the pointwise guarantee of
$\CBlab$ and
\cref{select:lem:relative-probabilities} shows that the good
labels satisfy the hypothesis of
\cref{select:lem:occupancy-relative}.
Let $\mathcal C$ be the event that every bin contains a good row.
Since $\mu=|G|/T\ge C_0t/2$, that lemma gives
\[
 \Pr[\neg\mathcal C]
 \le T(64t/\mu)^{t/2}+3^tT\delta
 \le R2^{-t/2}+\zeta/3
 \le\zeta,
\]
where the last two inequalities follow by choosing $C_0$
and then $K$ sufficiently large.

By \cref{select:lem:selected-extraction}, with $\eps_0=\zeta$,
we obtain $\TV(B_0(\X),\U_{q_1})\le3\zeta\le n^{-e}$.
The construction makes polynomially many evaluations of
polynomial-time maps, so its running time is polynomial.
\end{proof}

\begin{proof}[Proof of \cref{poly:thm:main}]
Follows by combining \cref{pre:lem:output-amplification}
 and \cref{poly:thm:base}. The proof is the same as in \cref{many:thm:almost-all}.
\end{proof}

\paragraph{Acknowledgement}
Part of the work was done while the authors were visiting the Simons Institute for Theory of Computing, UC Berkeley. We would like to thank the Simons Institute for the stimulating research environment and the hospitality.

\printbibliography

\end{document}